\documentclass[submission, Phys]{SciPost}

\usepackage{graphicx}
\usepackage{dcolumn}
\usepackage{bm}
\usepackage{bbm}
\usepackage{multirow}

\usepackage{cancel}
\usepackage[normalem]{ulem}

\usepackage{enumerate}

\usepackage{amsmath}
\usepackage{amsthm}
\usepackage{amstext}
\usepackage{amssymb}
\usepackage{mathrsfs}
\usepackage{amsfonts}
\usepackage{amsbsy} 

\usepackage{braket}

\newcommand{\Ketbra}[2]{\Ket{#1}\!\!\Bra{#2}}

\usepackage[all,matrix,cmtip]{xy}
\usepackage{mathtools}

\usepackage{color}
\definecolor{red}{rgb}{1,0,0}
\definecolor{blue}{rgb}{0,0,1}
\definecolor{dblue}{rgb}{0,0,0.4}
\definecolor{green}{rgb}{0,1,0}
\definecolor{black}{rgb}{0,0,0}
\definecolor{white}{rgb}{1,1,1}
\definecolor{pastelblue}{RGB}{20,93,160}

\definecolor{brn}{rgb}{.8,.4,.0}
\definecolor{redo}{rgb}{1,.5,.0}
\definecolor{ddgrn}{rgb}{0,0.4,0}
\definecolor{dgrn}{rgb}{0,0.55,0}
\definecolor{dbl}{rgb}{0,0,0.5}

\hypersetup{
	colorlinks,
	citecolor=pastelblue,
	linkcolor=pastelblue,
	urlcolor=pastelblue
}

\newcommand{\one}{\mathbf{1}}
\newcommand{\two}{\mathbf{2}}

\newcommand{\Z}{\mathbb{Z}}

\newcommand{\G}{\mathbb{G}}

\renewcommand{\t}[1]{\widetilde{#1}} 
\newcommand{\ii}{\hspace{1pt}\mathrm{i}\hspace{1pt}}
\newcommand{\ee}{\hspace{1pt}\mathrm{e}}

\newcommand{\<}{\langle} 
\renewcommand{\>}{\rangle} 
\renewcommand{\=}{\coloneqq}

\newcommand{\Rf}[1]{Ref.~\cite{#1}}
\newcommand{\Rfs}[1]{Refs.~\cite{#1}}
\newcommand{\Eq}[1]{(\ref{#1})} 
 
\newcommand{\eqn}[1]{eqn.~(\ref{#1})} 
 
\newcommand{\scn}[1]{Sec.~\ref{#1}} 
\newcommand{\app}[1]{Appendix~\ref{#1}} 
\newcommand{\fig}[1]{Fig.~\ref{#1}}

\newcommand{\Tr}{{\rm Tr}}

\newcommand{\ie}{{\it i.e.},~} 
\newcommand{\eg}{{\it e.g.},~} 
\newcommand{\etc}{{\it etc.~}}

\newcommand{\bpm}{\begin{pmatrix}}
	\newcommand{\epm}{\end{pmatrix}}
\newcommand{\bmm}{\begin{matrix}}
	\newcommand{\emm}{\end{matrix}}
\newcommand{\bvm}{\begin{vmatrix}}
	\newcommand{\evm}{\end{vmatrix}}

\newcommand{\cA}{ {\cal A} } 
	
	\newcommand{\cC}{ {\cal C} } 
	\newcommand{\cD}{ {\cal D} }

	\newcommand{\cH}{ {\cal H} }

	\newcommand{\cR}{ {\cal R} } 
	 
	\newcommand{\cT}{ {\cal T} } 
	\newcommand{\cO}{ {\cal O} }

\usepackage{euscript}

	\newcommand\eD           {\EuScript{D}}

	\newcommand\eM          {\EuScript{M}}

	\newcommand\eZ         {\EuScript{Z}}

\newcommand{\al}{\alpha} 
	\newcommand{\bt}{\beta} 
	\newcommand{\del}{\delta} 
	\newcommand{\Del}{\Delta} 
	\newcommand{\eps}{\epsilon} 
	 
	\newcommand{\ga}{\gamma}

	\newcommand{\la}{\lambda} 
	 
	\newcommand{\om}{\omega} 
	 
	\renewcommand{\th}{\theta} 
	 
	\newcommand{\si}{\sigma}

\makeatletter
\newsavebox{\@brx}
\newcommand{\llangle}[1][]{\savebox{\@brx}{\(\m@th{#1\langle}\)}%
	\mathopen{\copy\@brx\kern-0.5\wd\@brx\usebox{\@brx}}}
\newcommand{\rrangle}[1][]{\savebox{\@brx}{\(\m@th{#1\rangle}\)}%
	\mathclose{\copy\@brx\kern-0.5\wd\@brx\usebox{\@brx}}}
\makeatother

\newcommand{\cRep}{\mathsf{Rep}}
\newcommand{\cVec}{\mathsf{Vec}}

\theoremstyle{definition}

\theoremstyle{remark}

\numberwithin{equation}{section}

\usepackage{subcaption}
\usepackage{caption}

\allowdisplaybreaks

\usepackage{orcidlink}

\usepackage{comment}

\begin{document}

\begin{center}{\Large \textbf{
Quantum Phases and Transitions in Spin Chains with Non-Invertible Symmetries
}}\end{center}

\begin{center}
Arkya Chatterjee\,\orcidlink{0000-0001-6735-0513}$^{\, *\, \mathfrak{s}}$,
\"Omer M. Aksoy\,\orcidlink{0000-0003-0234-0650}$^{\, * \, \mathfrak{r}}$,
Xiao-Gang Wen\,\orcidlink{0000-0002-5874-581X}
\end{center}

\begin{center}
Department of Physics, Massachusetts Institute of Technology, Cambridge, Massachusetts 02139, USA
\\
$ ^* $ These authors contributed equally to this work.\\
$ ^\mathfrak{s} $ achatt@mit.edu \quad
$ ^\mathfrak{r} $ omaksoy@mit.edu
\end{center}

% \begin{center}
% \today
% \end{center}

% For convenience during refereeing: line numbers
%\linenumbers

\section*{Abstract}
{
	\bf

Generalized symmetries often appear in the form of emergent symmetries in low energy effective descriptions of quantum many-body systems. Non-invertible symmetries are a particularly exotic class of generalized symmetries, in that they are implemented by transformations that do not form a group. 
% Such symmetries appear generically in gapless states of quantum matter constraining the low-energy dynamics. 
Such symmetries appear in large families of gapless states of
quantum matter and constrain their low-energy dynamics.
To provide a UV-complete description of such symmetries, it is useful to construct lattice models that respect these symmetries exactly. In this paper, we discuss two families of one-dimensional lattice Hamiltonians with finite on-site Hilbert spaces: one with (invertible) $S^{\,}_3$ symmetry and the other with non-invertible $\mathsf{Rep}(S^{\,}_3)$ symmetry. Our models are largely analytically tractable and demonstrate all possible spontaneous symmetry breaking patterns of these symmetries. Moreover, we use numerical techniques to study the nature of continuous phase transitions between the different symmetry-breaking gapped phases associated with both symmetries. Both models have self-dual lines, where the models are enriched by so-called intrinsically non-invertible symmetries generated by Kramers-Wannier-like duality transformations. We provide explicit lattice operators that generate these non-invertible self-duality symmetries. We show that the enhanced symmetry at the self-dual lines is described by a 2+1d symmetry-topological-order (SymTO) of type $\mathrm{JK}^{\,}_4\boxtimes \overline{\mathrm{JK}}^{\,}_4$. The condensable algebras of the SymTO determine the allowed gapped and gapless states of the self-dual $S^{\,}_3$-symmetric and $\mathsf{Rep}(S^{\,}_3)$-symmetric models.
}

\vspace{10pt}
\noindent\rule{\textwidth}{1pt}
\tableofcontents
\noindent\rule{\textwidth}{1pt}
\vspace{10pt}

\section{Introduction}
\label{sec:intro}

In quantum many-body physics, global internal symmetries are conventionally
represented by unitary (or anti-unitary) operators acting on all the degrees of
freedom constituting the system. The associated symmetry transformations can 
be composed following group-like multiplication rules. Symmetries allow the
decomposition of quantum states into sectors that are dynamically decoupled, \ie
under time-evolution generated by the Hamiltonian, states in a particular
sector do not develop overlaps with those in other sectors. These sectors are
labeled by symmetry charges, which correspond to irreducible representations of
the symmetry group.

This conventional picture of symmetries has been generalized considerably in recent 
years. Such ``generalized symmetries" have been studied from various perspectives in
the mathematical physics, high energy physics, and condensed matter physics 
communities for several decades. However, attempts at a unified understanding
are more recent. One class of generalized symmetries that is particularly exotic
and will form the main focus of this work are the so-called non-invertible 
symmetries, which were first studied in rational conformal field theories 
in the form of topological defect lines~\cite{FSh0607247,FS09095013}.\footnote{In this context, the word ``defect" simply refers to the fact that these are extended operators in the theory.}
The composition of these symmetry transformations are, in general, 
not described by a group, but by a fusion category (in the case that there are a finite number of them)~\cite{CLY180204445}.

Generalized symmetries may also be implemented by operators with support on  
higher co-dimension manifolds in spacetime~\cite{NOc0605316,NOc0702377},
instead of co-dimension 1 manifolds. 
These lead to a different generalization, known as higher-form 
symmetries~\cite{GKW14125148,KT13094721}.
For finite symmetries in $d+1$-dimensional spacetime, 
invertible and non-invertible $0$- and higher-form symmetries 
(also known as algebraic higher symmetries~\cite{KZ200514178}), are understood 
to be unified in the structure of a fusion $d$-category 
$\mathcal{C}$~\cite{FMT220907471}. 
Anomaly-free algebraic higher symmetries are classified by 
\emph{local} fusion $d$-categories~\cite{KZ200514178}.
In this work, we will focus only on
non-invertible $0$-form symmetries in $1+1$d so that the relevant mathematical 
structure is that of ordinary fusion categories~\cite{TW191202817}.
Symmetry operators in a fusion category $\mathcal{C}$ are labeled by the 
objects $\mathsf{a}$ of $\mathcal{C}$. All objects can be decomposed into a direct sum
of finitely many simple objects,\footnote{This is because fusion categories are semi-simple~\cite{ENOm0203060}.} so our labels can be allowed to take values in the set of
simple objects without loss of generality.
The composition of two symmetry operators $\widehat{W}^{\,}_{\mathsf{a}}$ and 
$\widehat{W}^{\,}_{\mathsf{b}}$, labeled by simple objects $\mathsf{a}$ and 
$\mathsf{b}$, can be decomposed in terms of simple objects as
\begin{align}
\widehat{W}^{\,}_{\mathsf{a}}\,
\widehat{W}^{\,}_{\mathsf{b}}
=
\sum_{\alpha}
\mathrm{N}^{\mathsf{c}}_{\mathsf{a}\,\mathsf{b}}\,
\widehat{W}^{\,}_{\mathsf{c}},
\end{align}
where non-negative integers $\mathrm{N}^{\mathsf{c}}_{\mathsf{a}\,\mathsf{b}}$ 
are known as \emph{fusion coefficients}.

Non-invertible symmetries have found various applications in the context of 
continuum QFTs -- comprehensive lists of references can be found, for example, in 
\Rfs{S230518296,S230800747}. Realizations of
non-invertible symmetries in lattice models and associated phases of matter 
are far less understood; however, see the discussion on related prior work in the next subsection. 
While the infrared (IR) limit of many-body quantum systems 
are often described by effective QFTs, which can accommodate non-invertible 
symmetries, their na\"ive lattice regularization may
break emergent symmetries of the continuum theory (see, \eg \Rf{GLS210301257}).
In this spirit, it is desirable to construct spin chains that respect
generalized symmetries, putting them on the same footing as ordinary
symmetries. To be precise, by ``spin chains'' here we mean local Hamiltonians
acting on a Hilbert space that is a tensor product of finite-dimensional local
Hilbert spaces.  

In this paper, we explore the possible gapped or gapless phases and continuous phase 
transitions realized in spin chains with non-invertible symmetries. In 
particular, we study the example of the smallest anomaly-free\footnote{ We call a symmetry anomaly-free if it allows a symmetric trivially gapped (\ie without degenerate ground states) phase. This generalizes the definition of anomaly used for invertible symmetry in a way that is also applicable to non-invertible symmetries.} non-invertible symmetry category: 
$\cRep(S_3)$. Building up to that,
in Sec.\ \ref{sec:S3model} we introduce 
a spin chain with $S^{\,}_{3}$ symmetry constructed out of qubit and qutrit degrees of freedom. 
In Sec.\ \ref{sec:RepS3model}, we show how gauging either the entire $S^{\,}_{3}$ symmetry or its non-normal
$\mathbb{Z}^{\,}_{2}$ subgroup delivers a spin chain with $\cRep(S^{\,}_{3})$ symmetry.
By studying appropriate limits, we identify fixed-point ground states corresponding to the four distinct $\cRep(S^{\,}_{3})$ spontaneous symmetry breaking (SSB) patterns.
We explore the phase diagrams of both spin chains using tensor network algorithms to verify the analytical predictions.
Section \ref{sec:symTO} is a synthesis
of the salient aspects of our results from the point of view of the  
symmetry-topological-order (SymTO) framework.  
In Sec.\ \ref{sec:disc}, we discuss connections 
of our results with more abstract approaches, propose order 
parameters that detect SSB patterns in our models,
and comment on an incommensurate gapless phase that our numerical calculations reveal.
We close with some comments on directions for future exploration in Sec.\ \ref{sec:conclusion}.

\subsection*{Relation to prior work}

The literature on non-invertible symmetries has a long history.
Topological defect lines in  $1+1$d rational conformal 
field theories (CFTs) have been studied since the 1980s~\cite{V8860,MS8977,FFIS0404051,FSh0607247,FGRS07053129,
FS09095013}. A general study of topological defects in 
topological quantum field theories (TQFTs) was 
carried out in~\cite{KK11045047,DKR11070495,CRS170506085}; see \Rf{C230716674} for a recent review. 
A study of invertible defects of various dimensions in the context of general quantum
field theories was carried out in great detail in
\Rfs{GKW14125148,KT170108264} under the name of higher-form symmetries.
It is interesting to note that, \Rfs{NOc0605316,NOc0702377} earlier discussed lattice analogues of higher-form symmetry 
transformations in the context of topologically 
ordered phases of quantum matter.
Finite non-invertible symmetries in $1+1$d and their anomalies were 
systematically studied in
\Rfs{BT170402330,T171209542,TW191202817}, and constraints on RG flows 
obtained in \Rf{CLY180204445}.

Parallel to these developments, 
(non-invertible) gravitational anomalies were classified
by topological orders in one higher dimension in \Rf{KW1458}.\footnote{In this context, a gravitational anomaly is an obstruction to realizing a $d$-dimensional theory in a $d$ dimensional Hamiltonian lattice model on a tensor product Hilbert space. If such a theory is realizable on the boundary of a non-invertible, or invertible, topological order defined on a tensor product Hilbert space in $d+1$ dimensions, the gravitational anomaly is referred to as non-invertible, or invertible, respectively.} 
An isomorphic holographic decomposition of a quantum field theory 
was introduced in \Rf{KZ150201690} to expose its hidden 
gravitational anomaly. 
It was later realized that a subclass of non-invertible gravitational
anomalies are nothing but generalized 
symmetries~\cite{JW190513279,JW191213492,KZ200308898,KZ200514178}. 
The aforementioned isomorphic holographic decomposition can be re-interpreted as a holographic
theory of generalized symmetry (see Fig. \ref{fig:QFTS3}), which was
described via the ``sandwich'' construction in \Rf{FMT220907471}. 
This holographic description of symmetries was also discovered in the 
context of superstring theory~\cite{ABS211202092}.
In the holographic approach, symmetry data is stored 
in a non-invertible field theory (or a topological order) in 
one higher dimension such that the physical
theory with generalized symmetries is realized as a boundary theory of the
former. This idea has various names in different parts of the theoretical
physics community: 
symmetry-topological-order (SymTO) 
correspondence~\cite{KZ150201690,KZ200308898,KZ200514178},\footnote{
SymTO was referred to as ``categorical symmetry'' in some early
papers~\cite{JW191213492,KZ200308898,KZ200514178}. In current 
literature, categorical symmetry usually refers to 
non-invertible symmetry (referred to as algebraic higher symmetry 
in \Rf{KZ200514178}). See \app{app:SymTO} for a brief
review of SymTO.} symmetry-topological-field-theory
(SymTFT)~\cite{ABS211202092}, topological symmetry \cite{FT220907471}, or
topological holography~\cite{MMT220710712,HC231016878,WYP240419004}. 

The holographic approach has many applications. 
It leads to a classification of anomaly-free generalized 
symmetries using local fusion higher categories $\cC^\vee$ that describe the 
fusion of corresponding
charged operators, which may be of arbitrary 
dimensionality~\cite{KZ200308898,KZ200514178}. 
There is also an equivalent classification by local fusion higher categories 
$\cC$ describing 
the fusion of the symmetry defects, instead of the charged operators.  
This approach also provides a
classification of invertible anomalies\footnote{Invertible ('t Hooft) anomalies are those for which the anomaly theory in one higher dimension is an invertible topological field theory.} for generalized symmetries in any dimensions~\cite{KZ200514178}.
A related discussion on the classification of
gravitational anomalies and anomalies of group-like symmetries can be found in \Rfs{W1313,KW1458}. 
Some anomalous non-invertible symmetries were also studied in 
\Rfs{ACL221214605,CRBSS230509713}.
Generalized symmetries, anomalous or not, are classified (up to holo-equivalence) by their SymTO in one higher dimension~\cite{KZ200514178}.

More importantly, the holographic approach allows the use of
emergent generalized symmetries to
constrain compatible gapless liquid states. 
In \Rfs{KZ170501087,KZ190504924,KZ191201760,KZ201102859}, 
the concept of topological Wick rotation was introduced to describe the 
canonical gapless liquids determined by a generalized symmetry,
via the gapless boundary of the corresponding SymTO that has no anyon 
condensation. 
Such canonical  gapless liquids for a SymTO were studied in~\Rfs{JW191213492,LS210108343,CW220506244,CJW221214432} 
using the holographic modular bootstrap.
The condensable algebras~\cite{K13078244,KZ240307813}
of the SymTO classify the different allowed phases~\cite{CW220506244,BS231217322}.
As long as the condensable algebra is non-Lagrangian, 
the corresponding state must be gapless ~\cite{KZ200514178,CW220506244}. 
The Lagrangian condensable algebras, on the other hand, classify the gapped states 
allowed by the SymTO.
Such a classification includes SSB, symmetry-protected topological (SPT), 
and symmetry-enriched topological (SET) phases~\cite{KZ200308898,KZ200514178}.
Related discussions about phases of $1+1$d systems can also be found in \Rfs{KWZ210808835,XZ220509656}.

% More importantly, the holographic approach allows the use of
% emergent generalized symmetries to
% constrain compatible gapless liquid states. 
% In \Rfs{KZ170501087,KZ190504924,KZ191201760,KZ201102859}, 
% the concept of topological Wick rotation was introduced to describe the 
% canonical gapless liquids determined by a generalized symmetry,
% via the gapless boundary of the corresponding SymTO that has no anyon 
% condensation (\ie does not ``break" the SymTO). 
% Such canonical  gapless liquids for a SymTO were studied in~\Rfs{JW191213492,LS210108343,CW220506244,CJW221214432} 
% using the holographic modular bootstrap.
% The condensable algebras~\cite{K13078244,KZ240307813}
% of the SymTO classify ``spontaneous SymTO breaking" patterns~\cite{CW220506244,BS231217322}.\footnote{We use this terminology to draw analogies with the spontaneous breaking of symmetries -- ``SymTO broken" phases are constrained by a reduced ``unbroken" SymTO, which is non-trivial only for gapless phases.}
% It was shown that, as long as the SymTO is not completely broken (\ie the condensable algebra is non-Lagrangian), 
% the corresponding state must be gapless ~\cite{KZ200514178,CW220506244}. The Lagrangian 
% condensable algebras are then said to completely break the SymTO and hence they classify gapped states allowed by the SymTO.
% Such a classification includes SSB, symmetry-protected topological (SPT), 
% and symmetry-enriched topological (SET) phases~\cite{KZ200308898,KZ200514178}.
% Related discussions about phases of $1+1$d systems can also be found in \Rfs{KWZ210808835,XZ220509656}.

It is worthwhile to note here that generalized symmetries can also be viewed 
from the perspective of the
algebra of a subset of all local operators. Given a set of symmetry 
transformations, the subset of local operators invariant under these 
transformations forms the algebra of local symmetric operators (also 
called a bond algebra~\cite{CON09070733,CON11032776}). 
One can turn this idea on its head and consider subsets of local operators 
as (indirectly) defining a generalized symmetry, provided the subset forms 
an algebra. \Rf{CW220303596} took this point view and showed that
isomorphic algebras of local symmetric operators correspond one-to-one to
topological orders in one higher dimension, by considering simple examples.
The commutant algebra of the subset of local operators contains operators
that implement (generalized) symmetry transformations~\cite{MM210810324,MM220903370}.
The structure of commutant algebras is rich enough to include the
above-mentioned non-invertible symmetries.
Notably, this structure is less rigid than that of fusion
(higher) categories since the fusion coefficients need not be non-negative
integers. Making contact between the commutant algebraic approach and
the topological defect approach of generalized symmetries is an 
interesting open question.

As an instance of generalized symmetries, higher-form symmetries have found various applications in condensed matter physics; 
see 
\Rf{M220403045} for a recent review.
For instance, it was found that even if microscopic lattice models do not have exact
higher-form symmetries, they can appear as emergent~\cite{W181202517}, or even 
\emph{exact emergent}~\cite{PW230105261,PW231008554}, symmetries at low energies.
In many ways, higher-form symmetries behave just like ordinary symmetries: they
can be spontaneously broken leading to degenerate ground states or Goldstone
bosons \cite{NOc0702377,L180207747}, depending on whether the symmetry is
discrete or continuous; they can have 't Hooft anomalies themselves, or have
mixed 't Hooft anomalies with crystalline symmetries leading to
Lieb-Schultz-Mattis (LSM)-type theorems \cite{KSR180505367}; they can lead to
new symmetry protected topological (SPT) phases \cite{KT13094721,Y150803468}.
A generic way to construct models with higher-form symmetries in $2+1$d and 
higher, is via gauging (some subgroup of) an ordinary symmetry
\cite{JW191213492,MAT230701266}.

Non-invertible symmetries also have a natural place in the condensed matter setting.
For instance, \Rf{P230805730} showed that these symmetries appear generically as
emergent symmetries in SSB phases of ordinary symmetries.
Another generic way to realize non-invertible symmetries is to
start from a model with 0-form non-Abelian (finite) $G$ symmetry and
gauge this symmetry.  In the resulting gauge theory, in $d+1$ spacetime dimensions, 
the Wilson
loops obey the fusion rules dictated by the representations of $G$, forming the
layer of $(d-1)$-morphisms of a fusion $d$-category, $d$-$\cRep(G)$. The fusion 
rules of the Wilson loops are not group-like whenever $G$ is
non-Abelian.  This strategy was used to construct various lattice models with
non-invertible symmetries~\cite{JW191213492,DT180210104,KZ200514178,LV211209091}.  
Another class of examples can be obtained through the so-called half-gauging
scheme. Namely, if gauging an invertible symmetry of a theory $\cT$ produce an
isomorphic dual symmetry, a defect constructed by gauging this symmetry 
on one half of spacetime, the interface can be thought of 
as a non-invertible self-duality defect~\cite{KOZ211101141,CCS211101139,CCS220409025}.
Building on this, it is also possible to construct new duality
by half-gauging non-invertible symmetries~\cite{DLW231117044,CLS231019867}.
On a related note, statistical mechanical models with general fusion 
category symmetries were proposed and studied in 
\Rfs{AFM200808598,AMF160107185}.
Recently, non-invertible self-duality symmetries in Hamiltonian lattice models have also 
been obtained by gauging internal symmetries that
participate in a mixed anomaly with translation symmetry such as in the case of
LSM anomalies~\cite{SS230702534,S230805151,SSS240112281}.

There has been an exciting flurry of recent work
\cite{OF191209464,KZ200308898,KZ200514178,I211012882,LV211209091,LDV221103777,DT230101259,EF230214081,
BBS231003784,FTA231209272,SSS240112281}
exploring phases of matter with fusion category symmetries.
A generalized Landau paradigm \cite{CW220506244,CJW221214432,BBS231217322},
classifying both gapped and gapless phases in systems with general fusion
category symmetries in 1+1d has been formulated based on condensable algebras in
the symmetry topological order (SymTO) in one higher dimension. 
Our work contributes to this rapidly developing literature by exploring simple examples in the spin chain context. The key results of this paper are summarized below.

\subsection*{Summary of key results}

\begin{enumerate}[(i)]
\item 
We show, through a microscopic calculation, that our spin chain \eqref{eq:def Ham gen S3} with 
$S^{\,}_3$ symmetry is dual to the $\cRep(S^{\,}_{3})$-symmetric spin chain \eqref{eq:def Ham gen RepS3}  
by gauging a 
$\Z^{\,}_{2}$ subgroup of $S^{\,}_{3}$.
As a consequence, phase diagrams of these spin chains can be mapped to each other in a one-to-one manner.

\item 
We find gapped phases realizing all four SSB patterns of both
$S^{\,}_{3}$ and $\cRep(S^{\,}_{3})$ symmetries which correspond to the four inequivalent
module categories over the corresponding fusion categories~\cite{TW191202817}.
We define order and disorder operators whose non-vanishing expectation values 
can be used to distinguish different SSB patterns. 
For the non-invertible $\cRep(S^{\,}_{3})$ symmetry, the SSB is detected 
by string order parameters as opposed to the invertible $S^{\,}_{3}$ symmetry.

\item 
We show that for special subspaces in the parameter space, our spin chains 
are both invariant under an exact, intrinsic~\cite{KOZ220911062,KZZ220501104}, non-invertible 
self-duality symmetry. 
We provide the lattice operators that implement the respective 
self-duality symmetry in the form of a sequential circuit.\footnote{Notably, 
\Rf{CDZ230701267} considered this class of sequential circuits as maps between
distinct gapped phases. See also \Rfs{SS230702534,MLSW240209520} for closely related constructions.}
In particular, for the
$\cRep(S^{\,}_{3})$-symmetric model \eqref{eq:def Ham gen RepS3},
this circuit implements a self-duality symmetry associated with gauging $\cRep(S^{\,}_{3})$
by the algebra object $\one\oplus\two$.

\item 
For both spin chains, the four gapped phases meet at a multi-critical
point that is symmetric under the respective non-invertible self-duality 
symmetry. For each multi-critical point, we identify three relevant perturbations,
two of which break the non-invertible self-duality symmetry explicitly. 
In particular, for the $S^{\,}_3$ spin chain, one of these relevant perturbations 
allows the realization of a (Landau-forbidden) direct continuous transition 
between SSB phases preserving $\Z^{\,}_3$ and $\Z^{\,}_2$ subgroups.

\item 
We find an extended gapless region in the 
parameter space which is consistent with an incommensurate gapless phase. 
% This 
% phase has low-energy excitations carrying incommensurate quasi-momenta. 
To better understand
this gapless phase, we draw an analogy 
with an exactly solvable spin-1/2 chain with 
exact $\Z^{\,}_{2}$ KW self-duality symmetry. 
This spin chain, also supports a 
gapless incommensurate phase~\footnote{ In a gapless incommensurate phase, the ground state carries non-vanishing momentum.
As a result, the correlation functions generically 
have an oscillatory component with a period 
that is an irrational multiple of the lattice constant~\cite{Ostlund1981}. See \Rf{ZCTH150205049} for a discussion in a model similar to our \eqref{eq:def Ham gen S3}.} with central charge $c=1$ 
and an anomalous chiral $\mathrm{U}(1)$ symmetry~\footnote{In the low-energy CFT description, only the left-movers 
carry the $\mathrm{U}(1)$ charge while the right-movers are described by two branches of Majorana fermion
fields with different velocities.} that emanates~\cite{CS221112543} from
lattice translation symmetry.\footnote{
Along the self-dual line, the Majorana representation of translation symmetry carries an
LSM anomaly since there are odd number of Majorana degrees of freedom per unit cell,
see Refs.\ \cite{HHG160408591,ATM210208389,SS230702534}.
}
This gapless phase is separated from the neighboring gapped phases by critical
lines with dynamical critical exponent $z>1$. 
We provide arguments on why this feature may be valid for incommensurate phases more generally.
% An incommensurate phase contains $U(1)$ current and has a central charge $(c,\bar c) \geq (1,1)$.
% The continuous transition between
% incommensurate phase and gapped commensurate phase is described by a critical
% point with dynamical critical exponent $z>1$.
We also discover a new type of continuous  transition in the incommensurate phase.
The new continuous transition has the same number of gapless modes at and away from the transition point.\footnote{The number of gapless mode can be measured experimentally, via the the thermal conductance of short clean sample.}
In our example, there is only one $U(1)$ current $J_L$ for left-movers at the transition point, while there are two $U(1)$ currents $J_L,J_R$ for for both left- and right-movers away from the transition point.

\item 
We identify the SymTO of our self-dual $S^{\,}_3$ and $\cRep(S^{\,}_3)$-symmetric spin chains to be the
2+1d $\mathrm{JK}^{\,}_{4}\boxtimes \overline{\mathrm{JK}^{\,}}_4$ topological order. 
We also obtain possible phases and phase transitions of the self-dual spin chains from the allowed boundary conditions of the SymTO. In particular, we show that
the self-dual spin chains do not allow gapped non-degenerate ground states, consistent with the fact that the non-invertible self-duality  symmetries are anomalous.

\end{enumerate}

\textbf{Note added:} While this manuscript was being completed, 
we became aware of potentially overlapping work in Refs.\ \cite{BBST240505302,BBST240505964}; 
we thank the authors for coordinating their arXiv submission with us.

\section{\texorpdfstring{$S^{\,}_{3}$}{S3}-symmetric spin chain}
\label{sec:S3model}

\subsection{Definitions}

We consider lattice $\Lambda$ in one spatial dimension with
$|\Lambda|=L$ sites. 
We associate a tensor product Hilbert space $\mathcal{H}$ 
with lattice $\Lambda$, 
where the  each site $i\in \Lambda$ supports an on-site Hilbert space
$\mathcal{H}^{\,}_{i}$ that is $12$-dimensional. 
We label the orthonormal basis vectors spanning $\mathcal{H}^{\,}_{i}$ by a 
$\mathbb{Z}^{\,}_{2}\times\mathbb{Z}^{\,}_{2}\times\mathbb{Z}^{\,}_{3}$-valued
triplet $(a^{\,}_{i},\,b^{\,}_{i},\,c^{\,}_{i})$, \ie
\begin{equation}
\cH = \otimes_{i=1}^{L} \cH^{\,}_i, \quad \cH^{\,}_i = 
\text{span}\{\Ket{a^{\,}_{i},\,b^{\,}_{i},\,c^{\,}_{i}}|\,
a^{\,}_{i}\in \mathbb{Z}^{\,}_{2},\,
b^{\,}_{i}\in \mathbb{Z}^{\,}_{2},\,
c^{\,}_{i}\in \mathbb{Z}^{\,}_{3}\}.
\label{eq:def Hilbert space}
\end{equation}
On each local Hilbert space $\cH^{\,}_{i}$, 
we define $\mathbb{Z}^{\,}_{2}$ (qubit) and $\mathbb{Z}^{\,}_{3}$ (qutrit)
clock operators that satisfy the algebras
\begin{subequations}
\begin{align}
&
\hat{\sigma}^{z}_{i}\,\hat{\sigma}^{x}_{j}
=
(-1)^{\delta^{\,}_{ij}}
\hat{\sigma}^{x}_{j}\,\hat{\sigma}^{z}_{i},
&&
\left(\hat{\sigma}^{z}_{i}\right)^{2}
=
\left(\hat{\sigma}^{x}_{i}\right)^{2}
=
\hat{\mathbbm{1}},
\\
&
\hat{\tau}^{z}_{i}\,\hat{\tau}^{x}_{j}
=
(-1)^{\delta^{\,}_{ij}}\,
\hat{\tau}^{x}_{j}\,\hat{\tau}^{z}_{i},
&&
\left(\hat{\tau}^{z}_{i}\right)^{2}
=
\left(\hat{\tau}^{x}_{i}\right)^{2}
=
\hat{\mathbbm{1}},
\\
&
\widehat{Z}^{\,}_{i}\,\widehat{X}^{\,}_{j}
=
\omega^{\delta^{\,}_{ij}}
\widehat{X}^{\,}_{j}\,\widehat{Z}^{\,}_{i},
&&
\left(\widehat{Z}^{\,}_{i}\right)^{3}
=
\left(\widehat{X}^{\,}_{i}\right)^{3}
=
\hat{\mathbbm{1}},
\end{align}
with $\omega=\mathrm{exp}\{\mathrm{i}2\pi/3\}$
such that the operators $\hat{\sigma}^{z}_{i}$, 
$\hat{\tau}^{z}_{i}$, and $\widehat{Z}^{\,}_{i}$ are 
diagonal in the basis
\eqref{eq:def Hilbert space}, \ie
\begin{equation}
\begin{split}
\hat{\sigma}^{z}_{i}
\ket{\bm{a},\bm{b},\bm{c}}
=
(-1)^{a^{\,}_{i}}
\ket{\bm{a},\bm{b},\bm{c}},
\quad
\hat{\tau}^{z}_{i}
\ket{\bm{a},\bm{b},\bm{c}}
=
(-1)^{b^{\,}_{i}}
\ket{\bm{a},\bm{b},\bm{c}},
\quad
\widehat{Z}^{\,}_{i}
\ket{\bm{a},\bm{b},\bm{c}}
=
\omega^{c^{\,}_{i}}
\ket{\bm{a},\bm{b},\bm{c}}.
\end{split}
\end{equation}
We impose periodic boundary conditions on the Hilbert space $\mathcal{H}$
by identifying operators at site $i$ with those at site $i+L$
\begin{equation}
\widehat{X}^{\,}_{i+L} 
\equiv
\widehat{X}^{\,}_{i},
\quad
\widehat{Z}^{\,}_{i+L}
\equiv
\widehat{Z}^{\,}_{i},
\quad
\hat{\sigma}^{x}_{i+L}
\equiv 
\hat{\sigma}^{x}_{i},
\quad
\hat{\sigma}^{z}_{i+L} 
\equiv
\hat{\sigma}^{z}_{i},
\quad
\hat{\tau}^{x}_{i+L} 
\equiv
\hat{\tau}^{x}_{i},
\qquad 
\hat{\tau}^{z}_{i+L}
\equiv
\hat{\tau}^{z}_{i}.
\end{equation}
\end{subequations}

\begin{figure}[t!]
\centering
\includegraphics[width=0.6\linewidth]{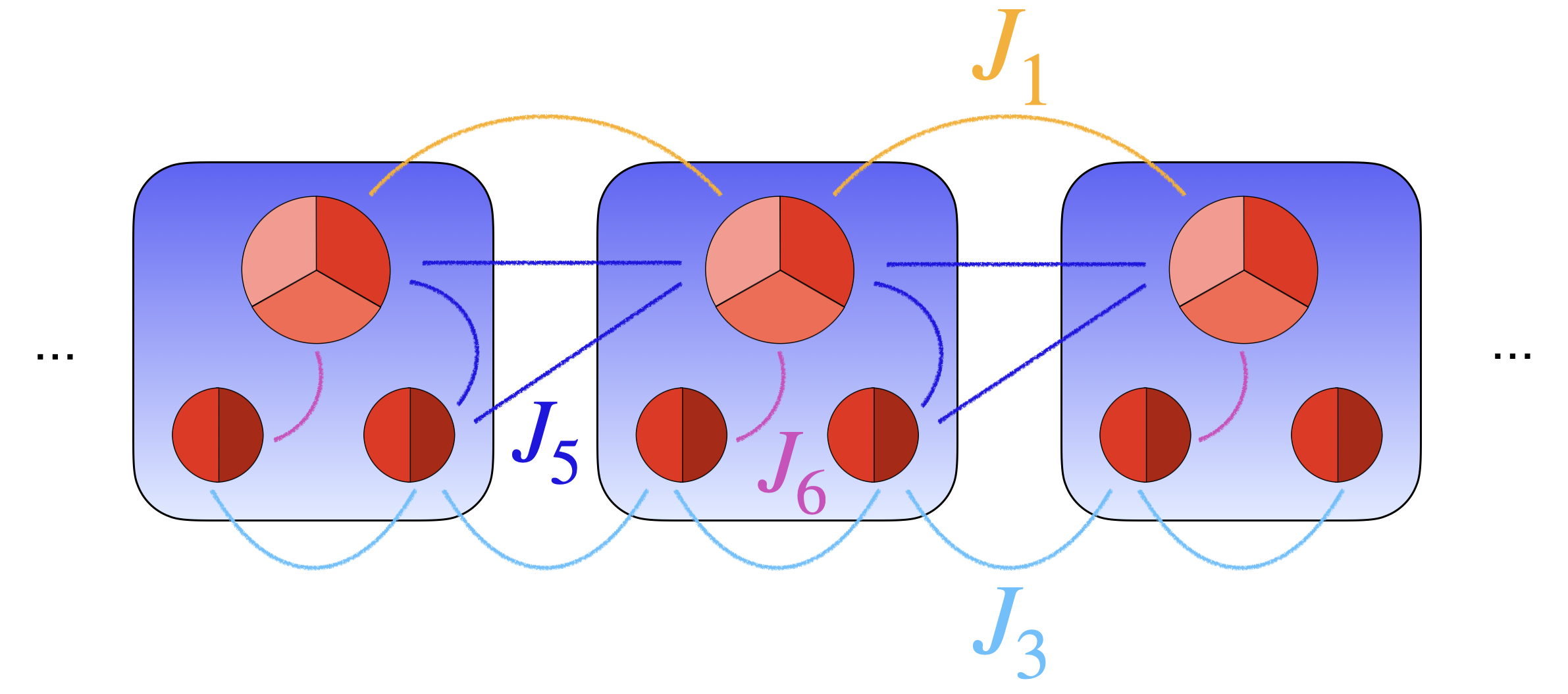}
\caption{Schematic of the Hamiltonian \eqref{eq:def Ham gen S3} 
showing the couplings between qutrit (depicted by a tripartitioned disk) 
and qubit (depicted by a bipartitioned disk) degrees of freedom. 
Single-body terms $ J^{\,}_2,J^{\,}_4 $ are suppressed.}
\label{fig:s3-chain-ham}
\end{figure}

Our starting point is the Hamiltonian
\begin{subequations}
\label{eq:def Ham gen S3}
\begin{align}
&
\widehat{H}^{\,}_{S^{\,}_{3}}
\=
\widehat{H}^{\,}_{\mathrm{P}}
+
\widehat{H}^{\,}_{\mathrm{I}}
+
\widehat{H}^{\,}_{\mathrm{PI}},
\label{eq:H s3}
\\
&
\widehat{H}^{\,}_{\mathrm{P}}
\=
-
J^{\,}_{1}\,
\sum_{i=1}^{L}
\left(
\widehat{Z}^{\,}_{i}\,
\widehat{Z}^{\dagger}_{i+1}
+
\widehat{Z}^{\dagger}_{i}\,
\widehat{Z}^{\,}_{i+1}
\right)
-
J^{\,}_{2}\,
\sum_{i=1}^{L}
\left(
\widehat{X}^{\,}_{i}
+
\widehat{X}^{\dagger}_{i}
\right),
\\
&
\widehat{H}^{\,}_{\mathrm{I}}
\=
-
J^{\,}_{3}\,
\sum_{i=1}^{L}
\left(
\hat{\sigma}^{z}_{i}\,
\hat{\tau}^{z}_{i}
+
\hat{\tau}^{z}_{i}\,
\hat{\sigma}^{z}_{i+1}
\right)
-
J^{\,}_{4}\,
\sum_{i=1}^{L}
\left(
\hat{\sigma}^{x}_{i}
+
\hat{\tau}^{x}_{i}
\right),
\\
&
\widehat{H}^{\,}_{\mathrm{PI}}
\=
-
J^{\,}_{5}\,
\sum_{i=1}^{L}
\mathrm{i}\,
\hat{\tau}^{z}_{i}
\left(
\widehat{Z}^{\,}_{i}\,
\widehat{Z}^{\dagger}_{i+1}
-
\widehat{Z}^{\dagger}_{i}\,
\widehat{Z}^{\,}_{i+1}
\right)
-
J^{\,}_{6}\,
\sum_{i=1}^{L}
\mathrm{i}\,
\hat{\sigma}^{z}_{i}
\left(
\widehat{X}^{\,}_{i}
-
\widehat{X}^{\dagger}_{i}
\right),
\end{align}
\end{subequations}
with six positive coupling constants $J^{\,}_{i}>0$ for $i=1,\cdots,6$.
Hamiltonians $\widehat{H}^{\,}_{\mathrm{P}}$ and $\widehat{H}^{\,}_{\mathrm{I}}$ 
describe the quantum three-state Potts model on a chain of $L$ sites 
and transverse-field Ising model defined on $2L$ sites, 
respectively~\footnote{The reason for choosing number of qubits to be twice that of qutrits
will be clear in Sec.\ \ref{sec:phase transitions S3} when we discuss 
the self-dual points in the phase diagram of Hamiltonian
\eqref{eq:def Ham gen S3}.}. 
The last Hamiltonian $\widehat{H}^{\,}_{\mathrm{PI}}$ 
then describes the coupling between qubits and qutrits.
A schematic description of the couplings in Hamiltonian \eqref{eq:def Ham gen S3}
is shown in Fig.\ \ref{fig:s3-chain-ham}.

Hamiltonian \eqref{eq:def Ham gen S3} is invariant under an
$S^{\,}_{3}$ symmetry generated by the unitary operators
\begin{align}
\widehat{U}^{\,}_{r}
\=
\prod_{i=1}^{L}
\widehat{X}^{\,}_{i},
\qquad
\widehat{U}^{\,}_{s}
\=
\prod_{i=1}^{L}
\hat{\sigma}^{x}_{i}\,
\hat{\tau}^{x}_{i}\,
\widehat{C}^{\,}_{i}\,,
\end{align}
where 
$ \widehat{C}^{\,}_{i}
\=
\sum_{\alpha=0}^{2}
\widehat{X}^{\alpha}_{i}\,
\widehat{P}^{Z^{\,}_{i}=\omega^{\alpha}} $ and
$\widehat{P}^{Z^{\,}_{i}=\omega^{\alpha}}$ is the projector onto the 
subspace of $\widehat{Z}^{\,}_{i}=\omega^{\alpha}$. 
The operator $\widehat{C}^{\,}_{i}$ implements the charge conjugation on the qutrits, \ie it maps $\widehat{X}^{\,}_{i}\mapsto \widehat{X}^{\dag}_{i}$
and $\widehat{Z}^{\,}_{i}\mapsto \widehat{Z}^{\dag}_{i}$.
On the local operators, the $ S^{\,}_3 $ symmetry generators, $\widehat{U}^{\,}_{r}$ 
and $\widehat{U}^{\,}_{s}$, implement the transformations
\begin{equation}
\begin{split}
&
\widehat{U}^{\,}_{r}\,
\begin{pmatrix}
\widehat{X}^{\,}_{i} &
\widehat{Z}^{\,}_{i} &
\hat{\sigma}^{x}_{i} &
\hat{\sigma}^{z}_{i} &
\hat{\tau}^{x}_{i} &
\hat{\tau}^{z}_{i}
\end{pmatrix}\,
\widehat{U}^{\dagger}_{r}
=
\begin{pmatrix}
+\widehat{X}^{\,}_{i} &
\omega^{2}\,\widehat{Z}^{\,}_{i} &
+\hat{\sigma}^{x}_{i} &
+\hat{\sigma}^{z}_{i} &
+\hat{\tau}^{x}_{i} &
+\hat{\tau}^{z}_{i}
\end{pmatrix},
\\
&
\widehat{U}^{\,}_{s}\,
\begin{pmatrix}
\widehat{X}^{\,}_{i} &
\widehat{Z}^{\,}_{i} &
\hat{\sigma}^{x}_{i} &
\hat{\sigma}^{z}_{i} &
\hat{\tau}^{x}_{i}   &
\hat{\tau}^{z}_{i}
\end{pmatrix}\,
\widehat{U}^{\dagger}_{s}
=
\begin{pmatrix}
+\widehat{X}^{\dagger}_{i} &
+\widehat{Z}^{\dagger}_{i} &
+\hat{\sigma}^{x}_{i} &
-\hat{\sigma}^{z}_{i} &
+\hat{\tau}^{x}_{i} &
-\hat{\tau}^{z}_{i}
\end{pmatrix},
\end{split}
\label{eq:S3 representation}
\end{equation}
respectively. Any operator that commutes with $\widehat{U}^{\,}_{r}$ and 
$\widehat{U}^{\,}_{s}$  can be written as 
linear combinations of products of eight local operators. 
These are precisely
those that appeared in the Hamiltonian \eqref{eq:def Ham gen S3}. 
Accordingly we define the bond algebra~\cite{CON09070733,CON11032776} of $S^{\,}_{3}$-symmetric operators
\begin{align}
\mathfrak{B}^{\,}_{S^{\,}_{3}}
\=
\Big\langle
&
\hat{\sigma}^{z}_{i}\,
\hat{\tau}^{z}_{i},\,
\hat{\tau}^{z}_{i}\,
\hat{\sigma}^{z}_{i+1},\,
\hat{\sigma}^{x}_{i},\,
\hat{\tau}^{x}_{i},\,
\left(\widehat{X}^{\,}_{i}+\widehat{X}^{\dagger}_{i}\right),\,
\left(\widehat{Z}^{\,}_{i}\,\widehat{Z}^{\dagger}_{i+1}
+\widehat{Z}^{\dagger}_{i}\,\widehat{Z}^{\,}_{i+1}\right),
\nonumber\\
&\,\,
\hat{\sigma}^{z}_{i}\,\left(\widehat{X}^{\,}_{i}-\widehat{X}^{\dagger}_{i}\right),\,
\hat{\tau}^{z}_{i}\left(\widehat{Z}^{\,}_{i}\,\widehat{Z}^{\dagger}_{i+1}
-\widehat{Z}^{\dagger}_{i}\,\widehat{Z}^{\,}_{i+1}\right)
\Big\vert 
\,\,
i\in \Lambda
\Big\rangle.
\label{eq:def S3 algebra}
\end{align}
We identify the $S^{\,}_{3}$ symmetry as the commutant algebra of 
$\mathfrak{B}^{\,}_{S^{\,}_{3}}$, \ie algebra of all operators that commute with 
all elements of $\mathfrak{B}^{\,}_{S^{\,}_{3}}$. 

In Sections \ref{sec:gauging Z3 in S3}, \ref{sec:gauging Z2 in S3}, and \ref{sec:gauging S3 in S3}, we are going to construct the dual bond algebras $\mathfrak{B}^{\,}_{S^{\,}_{3}/\mathbb{Z}^{\,}_{3}}$, 
$\mathfrak{B}^{\,}_{S^{\,}_{3}/\mathbb{Z}^{\,}_{2}}$, and 
$\mathfrak{B}^{\,}_{S^{\,}_{3}/S^{\,}_{3}}$, that are delivered by gauging the subgroups
$\mathbb{Z}^{\,}_{3}$, $\mathbb{Z}^{\,}_{2}$, and $S^{\,}_{3}$, respectively. 
As we shall see, the precise statement of the duality will then be expressed as isomorphisms between appropriately defined
``symmetric" subalgebras of these bond algebras. 
Therein, for each dual bond algebra, we will identify the corresponding commutant algebras, \ie
the corresponding dual symmetry structure.

\subsection{Gauging \texorpdfstring{$\Z^{\,}_3$}{Z3} subgroup: non-invertible self-duality symmetry}
\label{sec:gauging Z3 in S3}

Gauging the $\mathbb{Z}^{\,}_{3}$ subgroup is achieved in two steps. 
First, on the each link between sites $i$ and $i+1$, 
we introduce $\mathbb{Z}^{\,}_{3}$ clock operators $\{\hat{x}^{\,}_{i+1/2},\,\hat{z}^{\,}_{i+1/2}\}$ 
that satisfy the algebra
\begin{subequations}
\begin{equation}
\begin{split}
&
\hat{z}^{\,}_{i+1/2}\,\hat{x}^{\,}_{j+1/2}
=
\omega^{\delta^{\,}_{ij}}
\hat{x}^{\,}_{j+1/2}\,\hat{z}^{\,}_{i+1/2},
\quad
\left(\hat{z}^{\,}_{i+1/2}\right)^{3}
=
\left(\hat{x}^{\,}_{i+1/2}\right)^{3}
=
\hat{\mathbbm{1}},
\\
&
\hat{z}^{\,}_{i+1/2+L}
=
\hat{z}^{\,}_{i+1/2},
\quad
\hat{x}^{\,}_{i+1/2+L}
=
\hat{x}^{\,}_{i+1/2},
\end{split}
\end{equation}
where we imposed periodic boundary conditions.
This enlarges the dimension of the Hilbert space \eqref{eq:def Hilbert space}
by a factor of $3^{L}$. 
Second, we define the Gauss operators on every site
\begin{align}
\widehat{G}^{\mathbb{Z}^{\,}_{3}}_{i}
\=
\hat{z}^{\dagger}_{i-1/2}\,
\widehat{X}^{\,}_{i}\,
\hat{z}^{\,}_{i+1/2},
\qquad
\left[\widehat{G}^{\mathbb{Z}^{\,}_{3}}_{i}\right]^{3}
=
\hat{\mathbbm{1}}.
\label{eq:Gauss operator Z3 gauging}
\end{align}
\end{subequations}
Hereby, the link operators $\hat{z}^{\,}_{i+1/2}$ and $\hat{x}^{\,}_{i+1/2}$ 
take the roles of 
$\mathbb{Z}^{\,}_{3}$-valued electric field  
and $\mathbb{Z}^{\,}_{3}$-valued gauge field, respectively.
The physical Hilbert space consists of those states for which the Gauss constraint
$\widehat{G}^{\mathbb{Z}^{\,}_{3}}_{i}=1$ is satisfied. Imposing each of one of the $L$ Gauss constraints
reduces the dimension of the extended Hilbert space by a factor of 
$1/3$. The $S^{\,}_{3}$-symmetric algebra
\eqref{eq:def S3 algebra} is not invariant under local gauge transformations.
By minimally coupling it to the gauge field $\hat{x}^{\,}_{i+1/2}$, 
we define the gauge invariant extended algebra  
\begin{align}
\mathfrak{B}^{\mathrm{mc}}_{S^{\,}_{3}/\mathrm{Z}^{\,}_{3}}
\=
\Big\langle
&
\hat{\sigma}^{z}_{i}\,
\hat{\tau}^{z}_{i},\,
\hat{\tau}^{z}_{i}\,
\hat{\sigma}^{z}_{i+1},\,
\hat{\sigma}^{x}_{i},\,
\hat{\tau}^{x}_{i},
\left(\widehat{X}^{\,}_{i}+\widehat{X}^{\dagger}_{i}\right),\,
\left(\widehat{Z}^{\,}_{i}\,
\hat{x}^{\,}_{i+1/2}\,
\widehat{Z}^{\dagger}_{i+1}
+\widehat{Z}^{\dagger}_{i}\,
\hat{x}^{\dagger}_{i+1/2}\,
\widehat{Z}^{\,}_{i+1}\right),
\nonumber\\
&\,\,
\hat{\sigma}^{z}_{i}\,\left(\widehat{X}^{\,}_{i}-
\widehat{X}^{\dagger}_{i}\right),\,
\hat{\tau}^{z}_{i}\left(\widehat{Z}^{\,}_{i}\,
\hat{x}^{\,}_{i+1/2}\,
\widehat{Z}^{\dagger}_{i+1}
-\widehat{Z}^{\dagger}_{i}\,
\hat{x}^{\dagger}_{i+1/2}\,
\widehat{Z}^{\,}_{i+1}\right)
\Big\vert 
\,\,
\widehat{G}^{\mathbb{Z}^{\,}_{3}}_{i}=1,
\quad
i\in \Lambda
\Big\rangle.
\label{eq:def S3 Z3 minimally coupled algebra}
\end{align}
It is convenient to do a basis transformation to impose the Gauss constraint explicitly. 
To this end, we apply a unitary operator $\widehat{U}$ which implements the transformation
\begin{align}
&\nonumber
\widehat{U}\,
\hat{\sigma}^{x}_{i}\,
\widehat{U}^{\dagger}
=
\hat{\sigma}^{x}_{i},
\qquad
&&
\widehat{U}\,
\hat{\sigma}^{z}_{i}\,
\widehat{U}^{\dagger}
=
\hat{\sigma}^{z}_{i},
\\
&\nonumber
\widehat{U}\,
\hat{\tau}^{x}_{i}\,
\widehat{U}^{\dagger}
=
\hat{\tau}^{x}_{i},
\qquad
&&
\widehat{U}\,
\hat{\tau}^{z}_{i}\,
\widehat{U}^{\dagger}
=
\hat{\tau}^{z}_{i},
\\
&
\widehat{U}\,
\widehat{X}^{\,}_{i}\,
\widehat{U}^{\dagger}
=
\hat{z}^{\,}_{i-1/2}\,
\widehat{X}^{\,}_{i}\,
\hat{z}^{\dagger}_{i+1/2},
\qquad
&&
\widehat{U}\,
\widehat{Z}^{\,}_{i}\,
\widehat{U}^{\dagger}
=
\widehat{Z}^{\,}_{i},
\label{eq:Z3 gauging unitary}
\\
&\nonumber
\widehat{U}\,
\hat{x}^{\,}_{i+1/2}\,
\widehat{U}^{\dagger}
=
\widehat{Z}^{\dagger}_{i}\,
\hat{x}^{\,}_{i+1/2}\,
\widehat{Z}^{\,}_{i+1},
\qquad
&&
\widehat{U}\,
\hat{z}^{\,}_{i+1/2}\,
\widehat{U}^{\dagger}
=
\hat{z}^{\,}_{i+1/2}.
\end{align}
In particular, this unitary simplifies the Gauss operator to
$\widehat{U}\,\widehat{G}^{\mathbb{Z}^{\,}_{3}}_{i}\,\widehat{U}^{\dagger}=\widehat{X}^{\,}_{i}$.
After the unitary transformation, we project down to the $\widehat{X}^{\,}_{i}=1$ subspace
and relabel the link degrees of freedom by $i+1/2 \mapsto i+1$ for notational simplicity.
This delivers the dual bond algebra
\begin{align}
\mathfrak{B}^{\,}_{S^{\,}_{3}/\mathrm{Z}^{\,}_{3}}
&\=
\widehat{U}\,
\mathfrak{B}^{\mathrm{mc}}_{S^{\,}_{3}/\mathbb{Z}^{\,}_{3}}\,
\widehat{U}^{\dagger}\,
\Big\vert^{\,}_{\widehat{X}^{\,}_{i}=1}
\nonumber\\
&=
\Big\langle
\hat{\sigma}^{z}_{i}\,
\hat{\tau}^{z}_{i},\,
\hat{\tau}^{z}_{i}\,
\hat{\sigma}^{z}_{i+1},\,
\hat{\sigma}^{x}_{i},\,
\hat{\tau}^{x}_{i},
\left(\hat{z}^{\,}_{i}\,\hat{z}^{\dagger}_{i+1}
+
\hat{z}^{\dagger}_{i}\,\hat{z}^{\,}_{i+1}\right),\,
\left(\hat{x}^{\,}_{i}+\hat{x}^{\dagger}_{i}\right),
\nonumber\\
&\qquad
\hat{\sigma}^{z}_{i}\,\left(\hat{z}^{\,}_{i}\,\hat{z}^{\dagger}_{i+1}-
\hat{z}^{\dagger}_{i}\,\hat{z}^{\,}_{i+1}\right),\,
\hat{\tau}^{z}_{i}\left(\hat{x}^{\,}_{i+1}-\hat{x}^{\dagger}_{i+1}\right)
\Big\vert 
\,\,
i\in\Lambda
\Big\rangle.
\label{eq:def S3/Z3 algebra}
\end{align}
We note that the dual bond algebra contains the same type of terms 
as algebra \eqref{eq:def S3 algebra} and, hence, is the algebra of $S^{\vee}_{3}$-symmetric operators.\footnote{We use the superscript $\vee$ to differentiate the dual $S^{\vee}_{3}$ symmetry of 
the dual algebra \eqref{eq:def S3/Z3 algebra} from the $S^{\,}_{3}$ symmetry of 
the algebra \eqref{eq:def S3 algebra}.} The generators of dual $S^{\vee}_{3}$ symmetry 
are represented by the unitary operators~\footnote{
The dual symmetry $\widehat{U}^{\vee}_{s}$ is obtained from 
the operator $\widehat{U}^{\,}_{s}$ by demanding the covariance of the Gauss operator 
$\widehat{G}^{\mathbb{Z}^{\,}_{3}}_{i}$, \ie demanding 
\begin{align*}
\widehat{U}^{\mathrm{mc}}_{s}\,
\widehat{G}^{\mathbb{Z}^{\,}_{3}}_{i}\,
\left(\widehat{U}^{\mathrm{mc}}_{s}\right)^{\dagger}
=
\left(
\widehat{G}^{\mathbb{Z}^{\,}_{3}}_{i}\,
\right)^{\dagger}
,
\end{align*}
where $\widehat{U}^{\mathrm{mc}}_{s}$ is an operator acting on the extended Hilbert space 
and contains both site and link degrees of freedom. The dual symmetry $\widehat{U}^{\vee}_{s}$
is then obtained by applying the unitary transformation \eqref{eq:Z3 gauging unitary} and 
projecting to the $\widehat{X}^{\,}_{i}=1$ subspace. 
}
\begin{align}
\widehat{U}^{\vee}_{r}
\=
\prod_{i=1}^{L}
\hat{x}^{\,}_{i},
\qquad
\widehat{U}^{\vee}_{s}
\=
\prod_{i=1}^{L}
\hat{\sigma}^{x}_{i}\,
\hat{\tau}^{x}_{i}\,
\hat{c}^{\,}_{i},
\qquad
\hat{c}^{\,}_{i}
\=
\sum_{\alpha=0}^{2}
\hat{x}^{\alpha}_{i}\,
\widehat{P}^{z^{\,}_{i}=\omega^{\alpha}}.
\label{eq:Svee3 representations}
\end{align}

We note that the duality as we described does not hold between entirety of 
algebras $\mathfrak{B}^{\,}_{S^{\,}_{3}}$ and $\mathfrak{B}^{\,}_{S^{\,}_{3}/\mathbb{Z}^{\,}_{3}}$.
On the one hand, because we imposed periodic boundary conditions on the 
operators $\left\{\hat{x}^{\,}_{i},\,\hat{z}^{\,}_{i}\right\}$, 
the product of all Gauss operators is equal to the 
generator of global $\mathbb{Z}^{\,}_{3}$ transformations, \ie
\begin{subequations}
\label{eq:Z3 gauging dual subspaces}
\begin{align}
\prod_{i=1}^{L}
\widehat{G}^{\mathbb{Z}^{\,}_{3}}_{i}
=
\widehat{U}^{\,}_{r}
=
1.
\label{eq:Z3 gauging dual subspaces a}
\end{align}
On the other hand, since we imposed periodic boundary conditions 
on the operators $\left\{\widehat{X}^{\,}_{i},\,\widehat{Z}^{\,}_{i}\right\}$, 
the image of the product $\prod_{i=1}^{L}\widehat{Z}^{\,}_{i}\,\widehat{Z}^{\,}_{i+1}$,
which is the dual $\mathbb{Z}^{\vee}_{3}$ symmetry generator, must be equal to identity, \ie
\begin{align}
\prod_{i=1}^{L}
\widehat{Z}^{\,}_{i}\,\widehat{Z}^{\,}_{i+1}
\equiv
\widehat{U}^{\vee}_{r}
=
1.
\label{eq:Z3 gauging dual subspaces b}
\end{align}
Therefore, the duality holds when both conditions \eqref{eq:Z3 gauging dual subspaces a}
and \eqref{eq:Z3 gauging dual subspaces b} hold. In other words, the isomorphism 
\begin{align}
\mathfrak{B}^{\,}_{S^{\,}_{3}}\Big\vert^{\,}_{\widehat{U}^{\,}_{r}=1}
\cong 
\mathfrak{B}^{\,}_{S^{\,}_{3}/\mathbb{Z}^{\,}_{3}}\Big\vert^{\,}_{\widehat{U}^{\vee}_{r}=1},
\label{eq:Z3 gauging dual subspaces c}
\end{align}
\end{subequations}
holds.\footnote{We could have also gauged the $\mathbb{Z}^{\,}_{3}$ symmetry of the 
bond algebra \eqref{eq:def S3 algebra} in the presence of a $\mathbb{Z}^{\,}_{3}$ twist. 
However, such twisted boundary conditions lead to a reduced $ \Z^{\,}_3 $ symmetry due to the fact that $ \Z^{\,}_2 $ elements of $ S^{\,}_3 $ act nontrivially the $ \Z^{\,}_3 $ twist. 
Here, we keep the periodic boundary conditions on both sides of the gauging duality to ensure both bond algebras
$\mathfrak{B}^{\,}_{S^{\,}_{3}}$ and $\mathfrak{B}^{\,}_{S^{\,}_{3}/\mathbb{Z}^{\,}_{3}}$
have full $S^{\,}_{3}$ and $S^{\vee}_{3}$ symmetries, respectively.}

%\subsection{$S^{\,}_{3}$-symmetric Hamiltonians}
%\label{sec:S3 phase diagram}

Using the mapping between the two operator algebras $\mathfrak{B}^{\,}_{S^{\,}_{3}}$
and $\mathfrak{B}^{\,}_{S^{\,}_{3}/\mathbb{Z}^{\,}_{3}}$, we obtain the Hamiltonian
%\begin{subequations}
\begin{align}
\widehat{H}^{\,}_{S^{\vee}_{3}}
\=&
-
J^{\,}_{1}\,
\sum_{i=1}^{L}
\left(
\widehat{x}^{\,}_{i}
+
\widehat{x}^{\dagger}_{i}
\right)
-
J^{\,}_{2}\,
\sum_{i=1}^{L}
\left(
\widehat{z}^{\,}_{i}\,
\widehat{z}^{\dagger}_{i+1}
+
\widehat{z}^{\dagger}_{i}\,
\widehat{z}^{\,}_{i+1}
\right)
\nonumber\\
&
-
J^{\,}_{3}\,
\sum_{i=1}^{L}
\left(
\hat{\sigma}^{z}_{i}\,
\hat{\tau}^{z}_{i}
+
\hat{\tau}^{z}_{i}\,
\hat{\sigma}^{z}_{i+1}
\right)
-
J^{\,}_{4}\,
\sum_{i=1}^{L}
\left(
\hat{\sigma}^{x}_{i}
+
\hat{\tau}^{x}_{i}
\right)
\nonumber\\
&
-
J^{\,}_{5}\,
\sum_{i=1}^{L}
\mathrm{i}\,
\hat{\tau}^{z}_{i}
\left(
\widehat{x}^{\,}_{i+1}
-
\widehat{x}^{\dagger}_{i+1}
\right)
-
J^{\,}_{6}\,
\sum_{i=1}^{L}
\mathrm{i}\,
\hat{\sigma}^{z}_{i}
\left(
\widehat{z}^{\,}_{i}\,
\widehat{z}^{\dagger}_{i+1}
-
\widehat{z}^{\dagger}_{i}\,
\widehat{z}^{\,}_{i+1}
\right).
\label{eq:def Ham gen Svee3}
\end{align}
This Hamiltonian is unitarily equivalent to the Hamiltonian \eqref{eq:def Ham gen S3} 
under exchanging the couplings $J^{\,}_{1}$ and $J^{\,}_{2}$,
and the couplings $J^{\,}_{5}$ and $J^{\,}_{6}$.
The unitary transformation connecting the two Hamiltonians is 
a \emph{half-translation} of the qubits implemented by the unitary operator 
\begin{equation}
\hat{\mathfrak{t}}^{\,}_{\mathbb{Z}^{\,}_{2}}
\begin{pmatrix}
\hat{\tau}^{x}_{i} &
\hat{\tau}^{z}_{i} &
\hat{\sigma}^{x}_{i} &
\hat{\sigma}^{z}_{i} 
\end{pmatrix}
\hat{\mathfrak{t}}^{\dagger}_{\mathbb{Z}^{\,}_{2}}
=
\begin{pmatrix}
\hat{\sigma}^{x}_{i+1} &
\hat{\sigma}^{z}_{i+1} &
\hat{\tau}^{x}_{i} &
\hat{\tau}^{z}_{i} 
\end{pmatrix},
\label{eq:def Z2 half translation}
\end{equation}
This equivalence between the Hamiltonian \eqref{eq:def Ham gen S3} and \eqref{eq:def Ham gen Svee3}
is the $\mathbb{Z}^{\,}_{3}$ Kramers-Wannier (KW) duality due to gauging the
$\mathbb{Z}^{\,}_{3}$ subgroup of the $S^{\,}_{3}$ symmetry group. 
When $J^{\,}_{1}=J^{\,}_{2}$ and $J^{\,}_{5}=J^{\,}_{6}$, both
Hamiltonians \eqref{eq:def Ham gen S3} and \eqref{eq:def Ham gen Svee3}
become self-dual under the KW duality.
In this submanifold of parameter space, the KW duality becomes a genuine non-invertible
symmetry of the Hamiltonian. 
Without loss of generality, we 
focus on the dual Hamiltonian \eqref{eq:def Ham gen S3}.
The full KW duality operator\footnote{We should note that a closely-related duality defect operator was also formulated in terms of a
Temperley-Lieb algebra in \Rf{SYS231019703}.
} 
takes the form~\cite{CDZ230701267,SS230702534,MLSW240209520}
\begin{subequations}
\label{eq:def KW duality operator}
\begin{align}
\widehat{D}^{\,}_{\mathrm{KW}}
\=
\hat{\mathfrak{t}}^{\,}_{\mathbb{Z}^{\,}_{2}}\,
\widehat{P}^{U^{\,}_{r}=1}\,
\widehat{W}\,
\left(
\widehat{\mathfrak{H}}^{\dagger}_{1}\,
\widehat{CZ}^{\dagger}_{2,1}
\right)\,
\left(
\widehat{\mathfrak{H}}^{\dagger}_{2}\,
\widehat{CZ}^{\dagger}_{3,2}
\right)\,
\cdots
\left(
\widehat{\mathfrak{H}}^{\dagger}_{L-1}\,
\widehat{CZ}^{\dagger}_{L,L-1}
\right),
\label{eq:def KW duality operator a}
\end{align}
where (i) the unitary operator $\hat{t}^{\,}_{\mathbb{Z}^{\,}_{2}}$
is the half-translation operator defined in Eq.\ \eqref{eq:def Z2 half translation} that is necessary to preserve the form of the Hamiltonian \eqref{eq:def Ham gen S3},
(ii) the operator
\begin{align}
\widehat{P}^{U^{\,}_{r}=1}
\=
\frac{1}{3}
\sum_{\alpha=0}^{2}
\prod_{i=1}^{L}
\widehat{X}^{\alpha}_{i},
\end{align}
is the projector to the $\widehat{U}^{\,}_{r}=1$ subspace, 
(iii) the unitary operator 
\begin{align}
\widehat{W}
\=
\sum_{\alpha=0}^{2}
\widehat{Z}^{\alpha}_{L}\,
\widehat{P}^{Z^{\dagger}_{1}\,Z^{\,}_{L}=\omega^{\alpha}},
\quad
\widehat{W}\,
\widehat{X}^{\,}_{1}\,
\widehat{W}^{\dagger}
=
\widehat{Z}^{\dagger}_{L}\,
\widehat{X}^{\,}_{1},
\quad
\widehat{W}\,
\widehat{X}^{\,}_{L}\,
\widehat{W}^{\dagger}
=
\widehat{Z}^{\,}_{L}\,
\widehat{X}^{\,}_{L}\,
\widehat{Z}^{\,}_{L}\,
\widehat{Z}^{\dagger}_{1},
\end{align}
that contains the projector $\widehat{P}^{Z^{\dagger}_{1}\,Z^{\,}_{L}=\omega^{\alpha}}$ 
to the $\widehat{Z}^{\dagger}_{1}\,\widehat{Z}^{\,}_{L}=\omega^{\alpha}$ subspace
and acts nontrivially only on operators 
$\widehat{X}^{\,}_{1}$ and $\widehat{X}^{\,}_{L}$, 
and finally (iv) the unitary operators 
$\widehat{\mathfrak{H}}^{\dagger}_{i}$ and $\widehat{CZ}^{\dagger}_{i+1,i}$
are Hadamard and control Z operators with their only nontrivial actions
being
\begin{align}
\widehat{\mathfrak{H}}^{\dagger}_{i}\,
\begin{pmatrix}
\widehat{X}^{\,}_{i}\\
\widehat{Z}^{\,}_{i}
\end{pmatrix}
\widehat{\mathfrak{H}}^{\,}_{i}
=
\begin{pmatrix}
\widehat{Z}^{\,}_{i}\\
\widehat{X}^{\dagger}_{i}
\end{pmatrix},
\qquad
\widehat{CZ}^{\dagger}_{i+1,i}\,
\begin{pmatrix}
\widehat{X}^{\,}_{i+1}\\
\widehat{X}^{\,}_{i}
\end{pmatrix}
\widehat{CZ}^{\,}_{i+1,i}
=
\begin{pmatrix}
\widehat{Z}^{\dagger}_{i}\,
\widehat{X}^{\,}_{i+1}\\
\widehat{X}^{\,}_{i}\,
\widehat{Z}^{\dagger}_{i+1}
\end{pmatrix}.
\end{align}
\end{subequations}
As written in Eq.\ \Eq{eq:def KW duality operator a}, the KW duality operator 
can be thought as a sequential circuit~\cite{CDZ230701267} 
of control Z and Hadamard operators 
that are applied sequentially from site $L$ 
down to site $1$.  

The KW duality operator \eqref{eq:def KW duality operator a} is non-invertible
since it contains the projector $\widehat{P}^{U^{\,}_{r}=1}$. 
It becomes unitary in the subspace $\widehat{U}^{\,}_{r}=1$, where the
self-duality holds. 
Its action on the local operators 
can be read from the identities
\begin{equation}
\begin{split}
&
\widehat{D}^{\,}_{\mathrm{KW}}\,
\widehat{X}^{\,}_{i}
=
\widehat{Z}^{\,}_{i}\,
\widehat{Z}^{\dagger}_{i+1}\,
\widehat{D}^{\,}_{\mathrm{KW}},
\qquad\qquad\qquad\,\,\,
\widehat{D}^{\,}_{\mathrm{KW}}\,
\widehat{Z}^{\,}_{i}\,
\widehat{Z}^{\dagger}_{i+1}\,
=
\widehat{X}^{\,}_{i+1}\,
\widehat{D}^{\,}_{\mathrm{KW}},
\\
&
\widehat{D}^{\,}_{\mathrm{KW}}\,
\begin{pmatrix}
\hat{\sigma}^{x}_{i} &
\hat{\sigma}^{z}_{i}
\end{pmatrix}
=
\begin{pmatrix}
\hat{\tau}^{x}_{i} &
\hat{\tau}^{z}_{i}
\end{pmatrix}
\widehat{D}^{\,}_{\mathrm{KW}},
\qquad
\widehat{D}^{\,}_{\mathrm{KW}}\,
\begin{pmatrix}
\hat{\tau}^{x}_{i} &
\hat{\tau}^{z}_{i}
\end{pmatrix}
=
\begin{pmatrix}
\hat{\sigma}^{x}_{i+1} &
\hat{\sigma}^{z}_{i+1}
\end{pmatrix}
\widehat{D}^{\,}_{\mathrm{KW}}.
\end{split}
\end{equation}
In the parameter space where self-duality holds, 
the symmetry algebra is appended to 
\begin{align}
&
\widehat{D}^{\,}_{\mathrm{KW}}\,
\widehat{U}^{\,}_{r} 
= 
\widehat{U}^{\,}_{r}\,
\widehat{D}^{\,}_{\mathrm{KW}} 
=
\widehat{D}^{\,}_{\mathrm{KW}}, 
\label{eq: Z3 TY}
\\
&\widehat{D}^{\dagger}_{\mathrm{KW}}
=
\widehat{T}^{\dagger}\,
\widehat{D}^{\,}_{\mathrm{KW}},
\\
&
\widehat{D}^{2}_{\mathrm{KW}}
=
\widehat{P}^{U^{\,}_{r}=1}\,
\widehat{T},
\end{align}
where $\widehat{T}$ is the operator that implements translation by 
one lattice site for both qubits and qutrits.
Hence, action of the operator $\widehat{D}^{\,}_{\mathrm{KW}}$ can be thought of as a \emph{half-translation} operator in the subspace $\widehat{U}^{\,}_{r}=1$.
We note that the operator $\hat{\mathfrak{t}}^{\,}_{\mathbb{Z}^{\,}_{2}}$ in Eq.\ 
\eqref{eq:def Z2 half translation} implements this half-translation only for 
qubits. This operator exists owing to the fact that
each unit cell contains two flavors of qubits
for a single flavor of qutrit. Had we defined a 
6 dimensional local Hilbert space which supports single flavor of $\mathbb{Z}^{\,}_{2}$-
and $\mathbb{Z}^{\,}_{3}$-clock operators, KW self-duality would only hold
when the couplings $J^{\,}_{5}$ and $J^{\,}_{6}$ are zero, \ie when qubits and 
qutrits are decoupled.

The symmetry algebra above includes \eqn{eq: Z3 TY} as a lattice analogue of the fusion rules of 
the $ \Z^{\,}_3 $ Tambara-Yamagami fusion category symmetry. 
In the continuum limit, we expect that the $ \Z^{\,}_{3}$ symmetry generator 
$ \widehat{U}^{\,}_r $ flows to a $ \Z^{\,}_{3}$ topological line
$ \xi $, while both $\sqrt 3 \,\widehat{D}^{\,}_{\mathrm{KW}}$ and 
its Hermitian conjugate flow to the continuum 
duality topological line $ \cD $. They satisfy the fusion rules
\begin{equation}
\label{eq: continuum Z3 KW}
\xi \, \cD = \cD \, \xi = \cD\,, \quad \cD^2 = 1+\xi+\xi^2\,, \quad \xi^3 = 1\,,
\end{equation}
where the operator that implements single lattice site translation becomes an internal symmetry in the continuum limit. 
This interpretation follows the approach presented in \Rf{SS230702534}.

\subsection{Phase diagram}
\label{sec:phase transitions S3}
To discuss the phase diagram of the Hamiltonian \eqref{eq:def Ham gen S3}, we first reparameterize it as 
\begin{equation}\label{eq:Ham gen S3 reparameterized}
\begin{alignedat}{3}
\widehat{H}^{\,}_{S^{\,}_3}=
&
-
J^{\,}_{1} \cos \th\,
\sum_{i=1}^{L}
\left(
\widehat{Z}^{\,}_{i}\,
\widehat{Z}^{\dagger}_{i+1}
+
\widehat{Z}^{\dagger}_{i}\,
\widehat{Z}^{\,}_{i+1}
\right)
-
J^{\,}_{1} \sin \th\,
\sum_{i=1}^{L}
\mathrm{i}\,
\hat{\tau}^{z}_{i}
\left(
\widehat{Z}^{\,}_{i}\,
\widehat{Z}^{\dagger}_{i+1}
-
\widehat{Z}^{\dagger}_{i}\,
\widehat{Z}^{\,}_{i+1}
\right)
\\
&
-
J^{\,}_{2} \cos \th\,
\sum_{i=1}^{L}
\left(
\widehat{X}^{\,}_{i}
+
\widehat{X}^{\dagger}_{i}
\right)	
-
J^{\,}_{2} \sin \th\,
\sum_{i=1}^{L}
\mathrm{i}\,
\hat{\sigma}^{z}_{i}
\left(
\widehat{X}^{\,}_{i}
-
\widehat{X}^{\dagger}_{i}
\right)
\\
&	
-
J^{\,}_{3}\,
\sum_{i=1}^{L}
\left(
\hat{\sigma}^{z}_{i}\,
\hat{\tau}^{z}_{i}
+
\hat{\tau}^{z}_{i}\,
\hat{\sigma}^{z}_{i+1}
\right)
-
J^{\,}_{4}\,
\sum_{i=1}^{L}
\left(
\hat{\sigma}^{x}_{i}
+
\hat{\tau}^{x}_{i}
\right)	\,.
\end{alignedat}
\end{equation}
In what follows, we will explore the phase diagram of this Hamiltonian as a function of 
dimensionless ratios $J^{\,}_{1}/J^{\,}_{2}$ and $J^{\,}_{3}/J^{\,}_{4}$, 
for the cases of $\theta=0$,  non-zero but small $\theta\approx 0$, and large $\theta \sim 0.7$.

\subsubsection{Analytical arguments}
\begin{figure}[t]
	\centering
	\includegraphics[width=.4\textwidth]{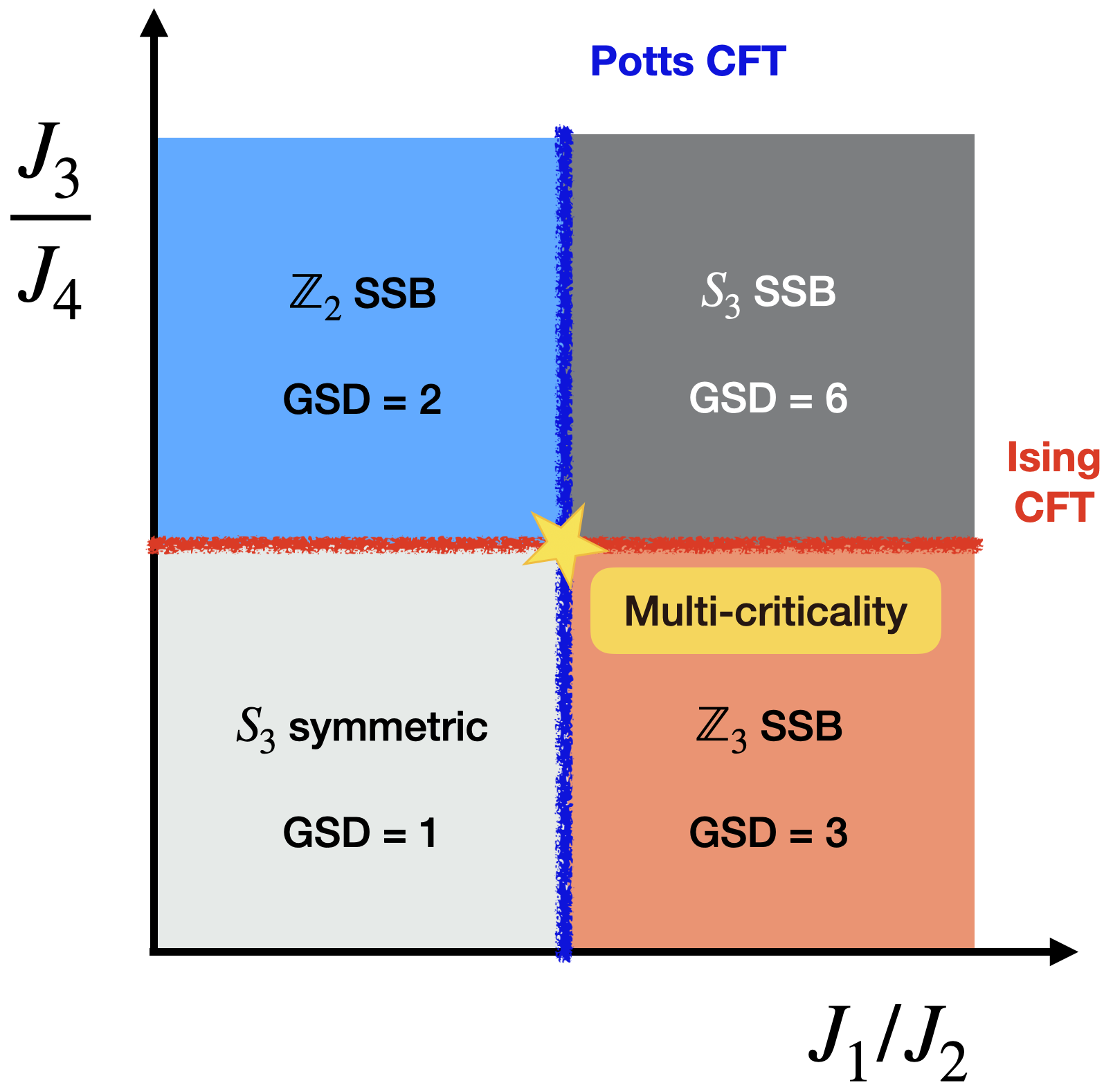}  
	\caption{Phase diagram of Hamiltonian 
		\eqref{eq:Ham gen S3 reparameterized} 
		based on analytical arguments for $ \th\approx 0 $. 
		The ground state degeneracy (GSD) for each of the 
		SSB phases are labeled. 
		The vertical and horizontal critical lines correspond to 
		the Potts (6,5) and Ising (4,3) minimal model CFTs, respectively. 
		They intersect at a multi-critical point belonging to the 3-state 
		Potts $ \boxtimes $ Ising universality class. }
	\label{fig:s3pd-analytical}
\end{figure}

When $\theta=0$, the Hamiltonian \eqref{eq:Ham gen S3 reparameterized} describes decoupled 
quantum Ising and 3-state Potts chains, for which the phase diagram is known.
There are four gapped phases which correspond to four different symmetry breaking 
patterns for $S^{\,}_{3}$. At four fixed-points, we can write the wave-functions exactly:
\begin{enumerate}[(i)]
\item 
When $J^{\,}_{1}=J^{\,}_{3}=0$, there is only a single ground state 
\begin{align}
\label{eq:GS S3}
\ket{\mathrm{GS}^{\,}_{S^{\,}_{3}}}\=
\bigotimes_{i=1}^{L}
\ket{\sigma^{x}_{i}=1,\,\tau^{x}_{i}=1,\,X^{\,}_{i}=1},
\end{align}
which describes the $S^{\,}_{3}$-disordered phase. 

\item 
When $J^{\,}_{1}=J^{\,}_{4}=0$, there are two degenerate ground states
\begin{equation}
\label{eq:GS Z3}
\ket{\mathrm{GS}^{\pm}_{\mathbb{Z}^{\,}_{3}}}\=
\bigotimes_{i=1}^{L}
\ket{\sigma^{z}_{i}=\pm 1,\,\tau^{z}_{i}=\pm 1,\,X^{\,}_{i}=1},
\end{equation}
that describe the phase where qubits
are ordered and $S^{\,}_{3}$ symmetry is broken down to $\mathbb{Z}^{\,}_{3}$.

\item 
When $J^{\,}_{2}=J^{\,}_{3}=0$, there are three degenerate ground states
\begin{equation}
\label{eq:GS Z2}
\ket{\mathrm{GS}^{\alpha}_{\mathbb{Z}^{\,}_{2}}}\=
\bigotimes_{i=1}^{L}
\ket{\sigma^{x}_{i}=+1,\,\tau^{x}_{i}=+1,\,Z^{\,}_{i}=\omega^{\alpha}},
\end{equation}
with $\alpha=0,1,2$, that 
describe the phase where qutrits
are ordered. One each ground state, $S^{\,}_{3}$ symmetry is broken down to 
a $\mathbb{Z}^{\,}_{2}$ subgroup.

\item 
When $J^{\,}_{2}=J^{\,}_{4}=0$, there are six degenerate ground states
\begin{equation}
\label{eq:GS Z1}
\ket{\mathrm{GS}^{\pm,\alpha}_{\mathbb{Z}^{\,}_{1}}}\=
\bigotimes_{i=1}^{L}
\ket{\sigma^{z}_{i}=\pm 1,\,\tau^{z}_{i}=\pm 1,\,Z^{\,}_{i}=\omega^{\alpha}},
\end{equation}
with $\alpha=0,1,2$ that describe the $S^{\,}_{3}$ ordered phase.
\end{enumerate}
See \scn{subsec:correlation funcs} for the discussion of expectation values
of the correlation functions and disorder operators in these ground states.

The lines $J^{\,}_{1}/J^{\,}_{2}=1$
and $J^{\,}_{3}/J^{\,}_{4}=1$ correspond to the transition points between the gapped phases 1 and 2, and 3 and 4, respectively.
They are described by the $3$-state Potts CFT and the Ising CFT, respectively. The 3-state Potts CFT is one of the $(6,5)$ minimal models
with $c=4/5$, while the Ising CFT is the $(4,3)$ minimal model wth $c=1/2$. At $J^{\,}_{1}/J^{\,}_{2}=1$ and
$J^{\,}_{3}/J^{\,}_{4}=1$, there is a multicritical 
point described by the stacking of the two 
CFTs, with total central charge $c=13/10$.

If we turn on small $\theta\neq 0$, the gapped phases 
are expected to remain unaffected by the virtue of finiteness of the gap (in the thermodynamic limit). 
However, one may wonder what the fate of the critical lines 
and the multicritical point is under these perturbations. 
To understand what happens to the critical lines, we first argue that
three out of the four critical lines are stable 
against small nonzero $\theta$ as follows.
Along the line $J^{\,}_{3}/J^{\,}_{4}=1$,
when $J^{\,}_{2}<J^{\,}_{1}$, qutrits are ordered and gapped. 
This means that both $\widehat{X}^{\,}_{i}-\widehat{X}^{\dagger}_{i}$ and 
$\widehat{Z}^{\,}_{i}\,\widehat{Z}^{\dagger}_{i+1}-\widehat{Z}^{\dagger}_{i}\,\widehat{Z}^{\,}_{i+1}$
vanish in the low-energy states below the qutrit excitation gap.
The same line of thought holds when $J^{\,}_{1}<J^{\,}_{2}$
for which qutrits are disordered and gapped. 
Similarly, along the line $J^{\,}_{1}/J^{\,}_{2}=1$,
when $J^{\,}_{3}<J^{\,}_{4}$, qubits are disordered and gapped. 
Both $\hat{\sigma}^{z}_{i}$ and $\hat{\tau}^{z}_{i}$ vanish in the low-energy states below the qubit excitation gap.

\begin{figure}[t!]
\centering
\begin{subfigure}{.4\textwidth}
\centering
\includegraphics[width=\linewidth]{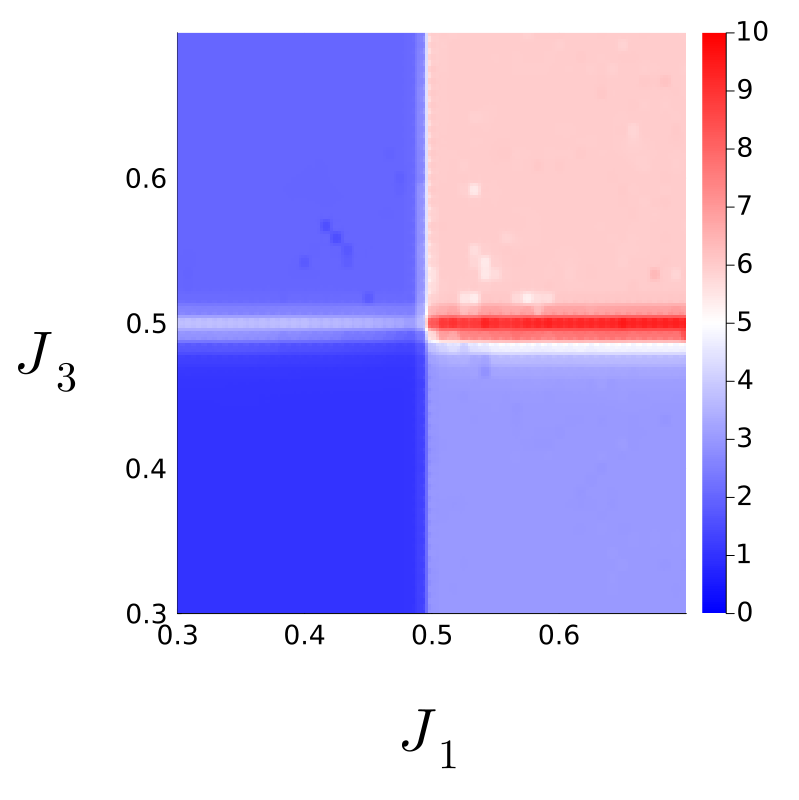}  
\caption{}
\label{fig:s3pd-numerics-gsd}
\end{subfigure}
\qquad
\begin{subfigure}{.4\textwidth}
\centering
\includegraphics[width=\linewidth]{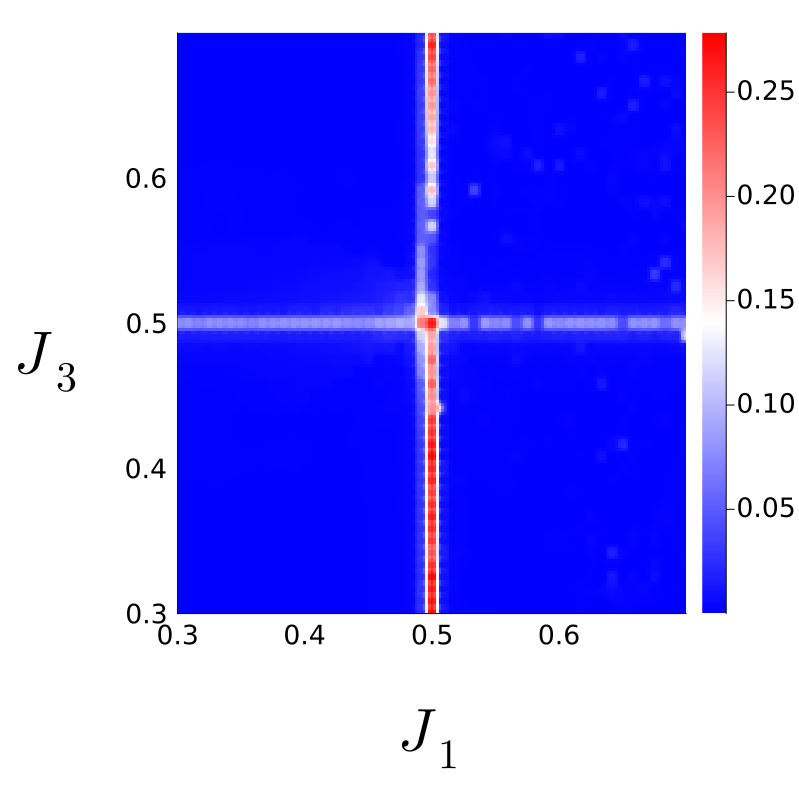} 
\caption{}
\label{fig:s3pd-numerics-c}
\end{subfigure}

\medskip

\begin{subfigure}{.35\textwidth}
\centering
\includegraphics[width=\linewidth]{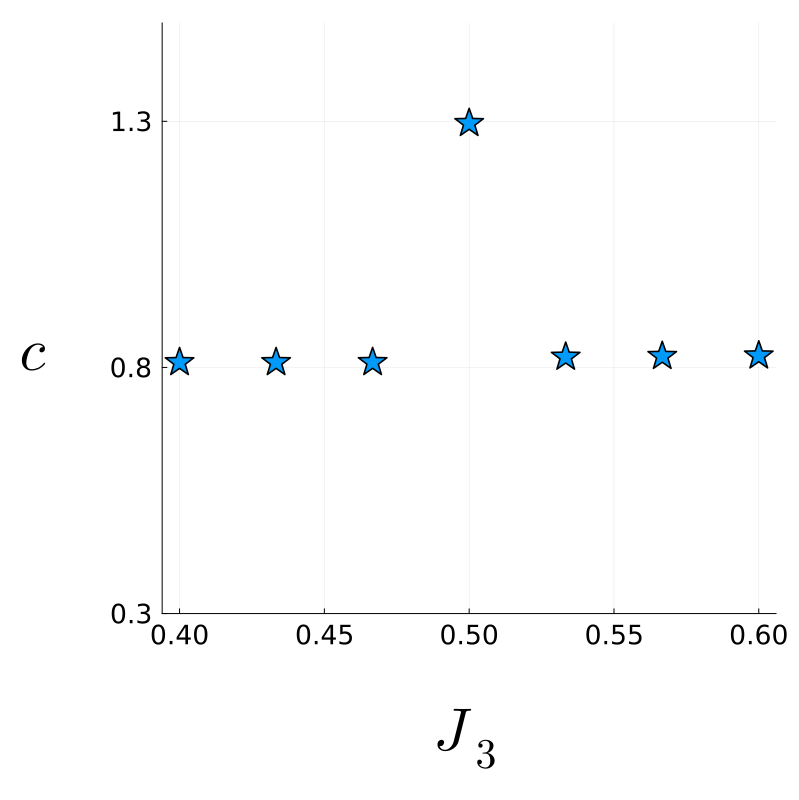}  
\caption{}
\label{fig:J1=0.5}
\end{subfigure}
\qquad
\begin{subfigure}{.35\textwidth}
\centering
\includegraphics[width=\linewidth]{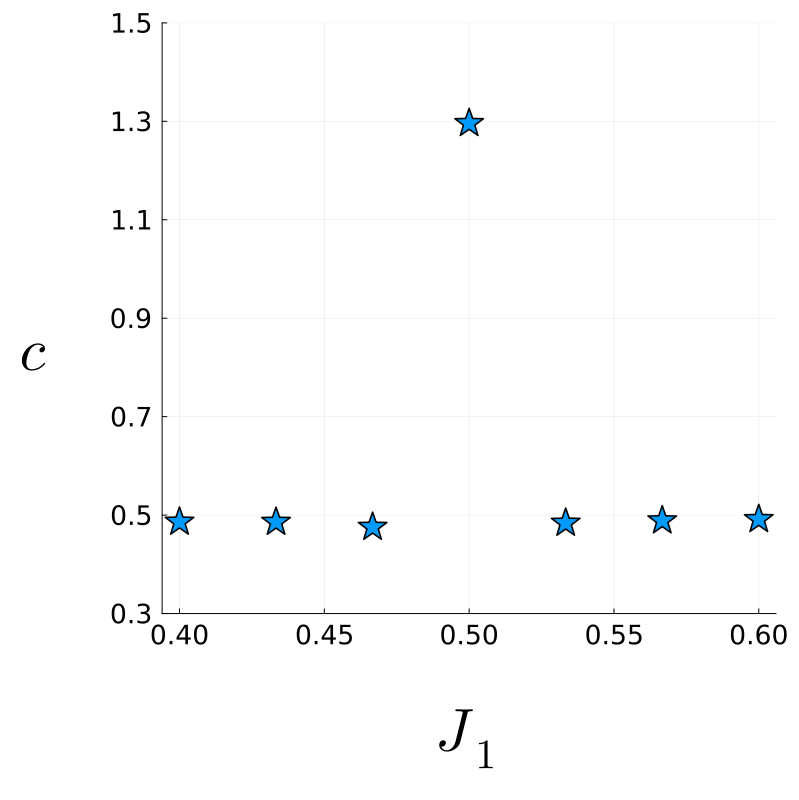}  
\caption{}
\label{fig:J3=0.5}
\end{subfigure}
\caption{
Numerical phase diagram obtained from the TEFR algorithm showing 
GSD (a) and central charge (b) 
as a function of $ J^{\,}_1 $ and $ J_3 $.  
The effective system size for these plots is $L=128$.  
Figs.~(c) and (d) show the 
central charges computed from bipartite entanglement entropy
scaling in the ground state obtained from DMRG, as discussed 
in the main text. 
We plot the central charges along a vertical ($J^{\,}_1=0.5$) and a 
horizontal ($J_3=0.5$) slice of the phase diagram shown in Fig.~(b), for
a chain of $L=100$ sites. 
We fix $\th=0.1$, with $ J^{\,}_2=1-J^{\,}_1 $, $ J_4=1-J_3 $ 
in all of these plots.
\label{fig:s3pd-numerics}
}
% \caption{
% Numerical phase diagram obtained from the TEFR algorithm showing 
% $1-\frac{1}{GSD}$ (a) and central charge (b) 
% as a function of $ J^{\,}_1 \in [0.4,0.6]$ (horizontal) and $ J^{\,}_{3} \in [0.4,0.6] $ (vertical).  
% The effective system sizes for these plots are 
% %$L=64$.  
% $L=256$ (blue),
% $L=128$ (green),
% $L=64$ (red).  
% Figs.~(c) and (d) show the 
% central charges computed from bipartite entanglement entropy
% scaling in the ground state obtained from DMRG, as discussed 
% in the main text. 
% We plot the central charges along a vertical ($J^{\,}_1=0.5$) and a 
% horizontal ($J^{\,}_{3}=0.5$) slice of the phase diagram shown in Fig.~(b), for
% a chain of $L=100$ sites. 
% We fix $\th=0.1$, with $ J^{\,}_2=1-J^{\,}_1 $, $ J_4=1-J^{\,}_{3} $ 
% in all of these plots.
% \label{fig:s3pd-numerics}
% }
\end{figure}

The situation is different when $J^{\,}_{1}/J^{\,}_{2}=1$ and
$J^{\,}_{4}< J^{\,}_{3}$ for which
quibts order, \ie the terms with $\sin\theta$ coefficient 
are not trivially vanishing.
On this line (excluding the multicritical point), the terms
with $\sin\theta$ coefficient must flow to a 
primary or descendant operator $\mathcal{O}$ in $3$-state Potts CFT that is 
odd under the charge conjugation symmetry. 
Using holographic modular bootstrap techniques~\cite{JW190513279,CW220506244},
we identify that 
(see Appendix \ref{app:potts cft partition function}) only possible relevant 
operators are those primaries with scaling dimension
$\Delta^{\,}_{\mathcal{O}}=9/5$ 
and conformal spin $\pm1$. Indeed, in Ref.\ \cite{MCF14060846},
the terms with coefficients $\sin\theta$ in Hamiltonian 
\eqref{eq:Ham gen S3 reparameterized} are identified 
with these primary operators. 
Even though these primaries are relevant, we expect that 
as long as the KW self-duality symmetry is preserved 
one cannot open a KW-symmetric gap owing to the fact that this non-invertible
symmetry is anomalous. This is consistent with the results in 
Ref.\ \cite{ZCTH150205049} where it was shown that  (see Fig.\ 2.(f) therein)
the critical line with $J^{\,}_{1}/J^{\,}_{2}=1$ and $J^{\,}_{4}< J^{\,}_{3}$ 
is stable for small $\theta$ and an incommensurate phase opens up 
for $\theta$ larger than a critical value. We confirm this numerically
for our model as we shall see in the next section. 
In summary,  we conclude that the phase diagram at perturbatively small $ \th $
has the same form as that for $ \th=0 $.
The phase diagram at $\th=0$ plane is shown in \fig{fig:s3pd-analytical}. 
We verify our predictions for non-zero and small $\theta$ numerically in \fig{fig:s3pd-numerics}.

\subsubsection{Numerical results}
\label{sec: s3 numerics}

We mapped out the phase diagram of Hamiltonian \eqref{eq:Ham gen S3 reparameterized} numerically, using 
the tensor entanglement filtering renormalization (TEFR) algorithm \cite{LN0701,GW0931}. The ground state 
degeneracies of each gapped phase and the central charges associated with continuous phase transitions 
were obtained using this algorithm. The results are shown in \fig{fig:s3pd-numerics} as a function of $ 
J^{\,}_1 $ and $ J^{\,}_{3} $ with $ \th=0.1 $. We chose a 2d slice in the full parameter space such that at 
every point of the phase diagram shown here, $ J^{\,}_2 = 1-J^{\,}_1 $ and $ J_4=1-J^{\,}_{3} $.

% \begin{figure}[t]
% \centering
% % fig 1a
% \begin{subfigure}{.31\textwidth}
% \centering
% \includegraphics[width=\linewidth]{FIGURES/cslice_dS3-selfdual_J1_10_J3_40_60_th10_Jperp0.png}  
% \caption{$J^{\,}_1=0.1$}
% \label{fig:J1=0.1}
% \end{subfigure}
% % fig 1b
% \begin{subfigure}{.31\textwidth}
% \centering
% \includegraphics[width=\linewidth]{FIGURES/cslice_dS3-selfdual_J1_90_J3_40_60_th10_Jperp0.png}  
% \caption{$J^{\,}_1=0.9$}
% \label{fig:J1=0.9}
% \end{subfigure}
% % fig 1c
% \begin{subfigure}{.31\textwidth}
% \centering
% \includegraphics[width=\linewidth]{example-image-a}  
% \caption{$J^{\,}_1=0.5$}
% \label{fig:J1=0.5}
% \end{subfigure}
% % fig 2a
% \begin{subfigure}{.31\textwidth}
% \centering
% \includegraphics[width=\linewidth]{example-image-a}  
% \caption{$J^{\,}_{3}=0.1$}
% \label{fig:J3=0.1}
% \end{subfigure}
% % fig 2b
% \begin{subfigure}{.31\textwidth}
% \centering
% \includegraphics[width=\linewidth]{example-image-a}  
% \caption{$J^{\,}_1=0.9$}
% \label{fig:J3=0.9}
% \end{subfigure}
% % fig 2c
% \begin{subfigure}{.31\textwidth}
% \centering
% \includegraphics[width=\linewidth]{example-image-a}  
% \caption{$J^{\,}_{3}=0.5$}
% \label{fig:J3=0.5}
% \end{subfigure}
% \caption{Central charge computed from entanglement entropy using the DMRG-based approach discussed in the main text. We plot the central charge along various slices of the phase diagrams shown in \fig{fig:s3pd-numerics}. We fix $\th=0.1$ in all of these plots.
% \label{fig:s3pd-slices}
% }
% \end{figure}

The TEFR algorithm does not give numerically precise values of 
central charge even though it is extremely precise at 
extracting ground state degeneracy.\footnote{
The interested reader is referred to Appendix \ref{app:numerics} 
for more details on this point.} 
In order to extract numerically precise central charges, 
we used the density matrix renormalization group (DMRG) 
algorithm from the iTensor library \cite{FWM200714822,itensor-r0.3}. 
On the phase diagram shown in \fig{fig:s3pd-numerics}, 
we make two cuts, one horizontal and one vertical, 
and compute the central charges using finite size bipartite 
entanglement entropy scaling that is computed using DMRG
calculation. Our results for finite chain of $L=100$ sites
are shown in Figs.\ 
\ref{fig:J1=0.5} and \ref{fig:J3=0.5}.

\begin{figure}[t!]
\centering
\begin{subfigure}{.4\textwidth}
\centering
\includegraphics[width=\linewidth]{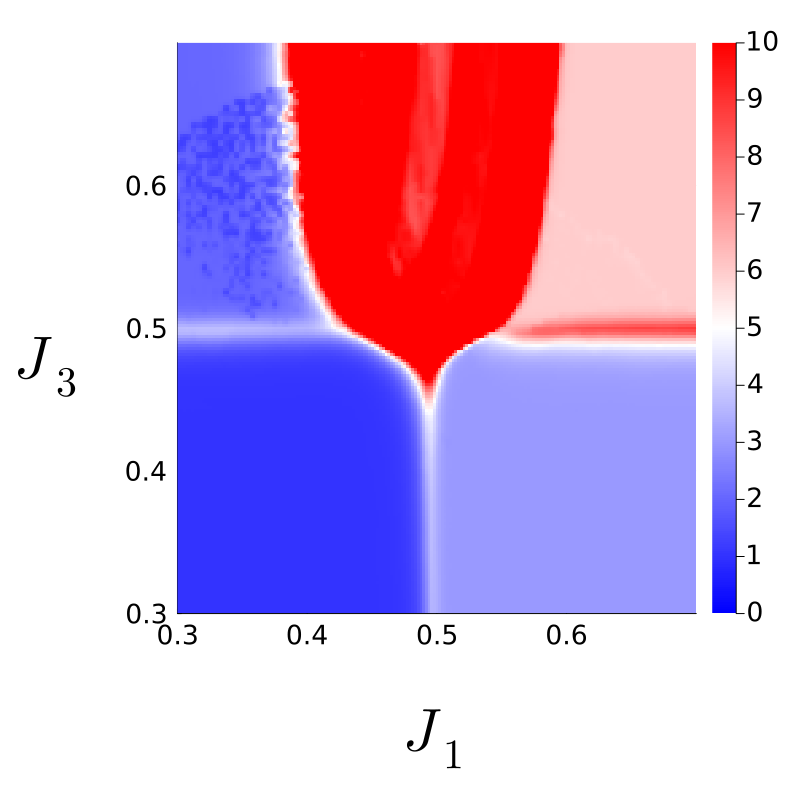} 
\caption{}
\label{fig:s3-incomm-gsd}
\end{subfigure}
%	\hfill
\begin{subfigure}{.4\textwidth}
\centering
\includegraphics[width=\linewidth]{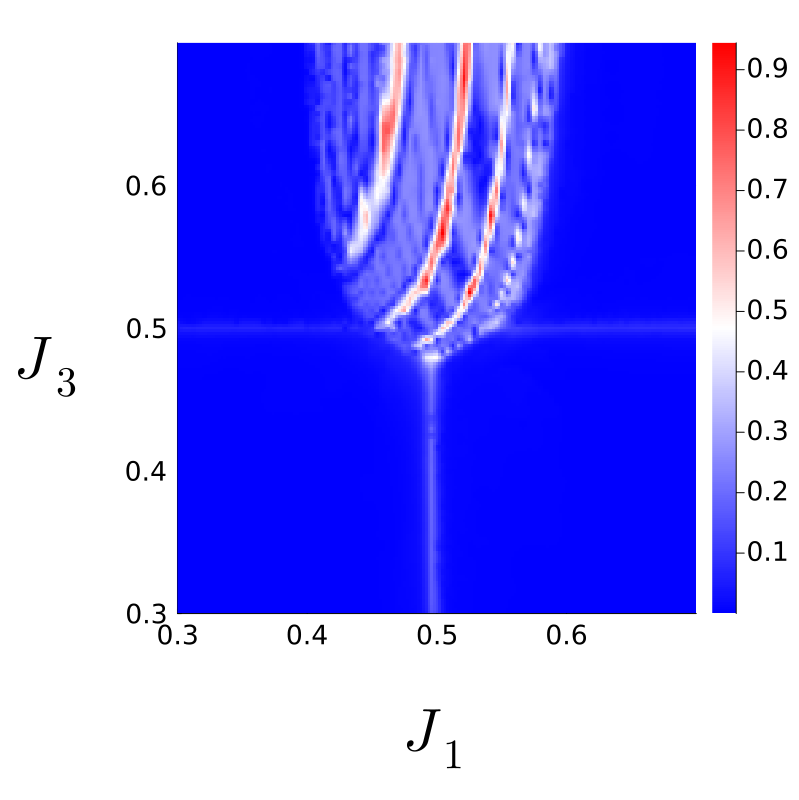}
\caption{}
\label{fig:s3-incomm-c}
\end{subfigure}

\medskip

\begin{subfigure}{.55\textwidth}
\centering
\includegraphics[width=\linewidth]{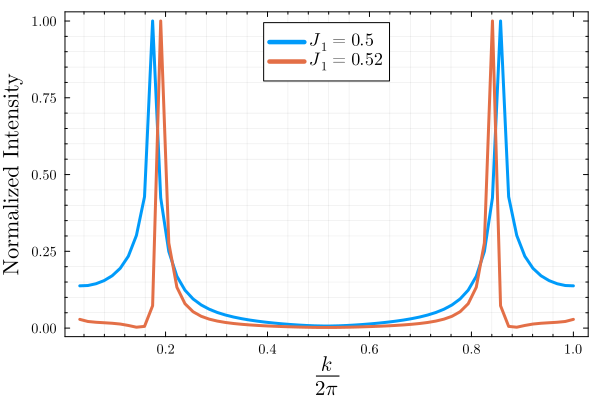}
\caption{}
\label{fig:s3-incomm-osc1}
\end{subfigure}
\caption{Numerical phase diagram obtained from the TEFR algorithm with fixed $ \th=0.7\approx\frac{2\pi}{9} $, showing GSD (Fig.~(a)), and central charge (Fig.~(b)) as a function of $ J^{\,}_1$ and $ J_3 $, setting $ J^{\,}_2=1-J^{\,}_1 $ and $ J_4=1-J_3 $ everywhere. 
The nominal GSD in this gapless region is much larger than that shown in Fig.~(a); we have capped the maximum allowed values to 10 so that these plots may be easily compared with the ones in \fig{fig:s3pd-numerics}. 
In these figures, the effective system size is $ L=128 $. 
Fig.~(c) shows the absolute value of the Fourier transform of ground state expectation value 
$\<\hat{\sigma}^{z}_{i}\>$ for $J^{\,}_{1}=0.5$ and $J^{\,}_{1}=0.52$ with fixed $J^{\,}_{3}=0.5$
and $L=101$ sites. 
\label{fig:s3-incomm}
}
% \caption{Numerical phase diagram obtained from the TEFR algorithm with fixed $ \th=0.7\approx\frac{2\pi}{9} $, showing $1-\frac{1}{GSD}$ (Fig.~(a)), and central charge (Fig.~(b)) as a function of $ J^{\,}_1 \in [0.4,0.6]$
% (horizontal) and $ J^{\,}_{3} \in [0.4,0,6] $ (vertical), setting $ J^{\,}_2=1-J^{\,}_1 $ and $ J_4=1-J^{\,}_{3} $ everywhere. 
% %The nominal GSD in this gapless region is much larger than that shown in Fig.~(a); we have capped the maximum allowed values to 10 so that these plots may be easily compared with the ones in \fig{fig:s3pd-numerics}. 
% In these figures, the effective system size is 
% %$ L=64 $. 
% $L=256$ (blue),
% $L=128$ (green),
% $L=64$ (red). 
% The stripe feature in $c\neq 0$ area suggests an incommensurate gapless phase.
% Fig.~(c) shows the absolute value of the Fourier transform of ground state expectation value 
% $\braket{\hat{\sigma}^{z}_{i}}$ for $J^{\,}_{1}=0.5$ and $J^{\,}_{1}=0.52$ with fixed $J^{\,}_{3}=0.5$
% and $L=101$ sites. 
% \label{fig:s3-incomm}
% }
\end{figure}

We find that the both $c=1/2$ and $c=4/5$ lines are stable against 
small values of $\th$, in agreement with our argument in the previous section.
For large values of $ \th $, a gapless region opens up in the Ising ordered regime, 
surrounding the $\Z^{\,}_3$ KW symmetric line $ J^{\,}_1=J^{\,}_2 $ 
in the parameter space of the Hamiltonian \eqref{eq:Ham gen S3 reparameterized}. 
From numerical estimation, we find the  critical value of $\th$
to be around $ \th_* \simeq \frac{\pi}{8}$. 
Heuristically, the gapless region appears first in the $\Z^{\,}_{2}$
ordered phase as the terms in Hamiltonian \eqref{eq:Ham gen Rep S3 reparameterized}
with $\sin\th$ are proportional to $\hat{\sigma}^{z}_{i}$ and $\hat{\tau}^{z}_{i}$
operators which have vanishing expectation values when the $\mathbb{Z}^{\,}_{2}$
subgroup of $S^{\,}_{3}$ is unbroken.
The multicritical point is engulfed by the gapless 
region beyond a certain value of $ \th \in (\pi/8,2\pi/9)$. 
Several comments are due: 
\begin{enumerate}[(i)]
\item 
Since the $\Z^{\,}_{3}$ KW duality symmetry is anomalous, the only 
compatible phase without symmetry breaking is gapless~\cite{TW191202817}.
This is compatible with the gapless region we numerically observe. Also, because of the
KW duality, the gapless phase must be symmetrically placed about the $ J^{\,}_1=0.5=J^{\,}_2 $ 
line of the phase diagram, consistent with the numerically obtained phase diagram in \fig{fig:s3-incomm}.

\item 
Extraction of the central charge in the gapless region is somewhat subtle. 
The phase diagram obtained from TEFR  (Fig.\ \ref{fig:s3-incomm-c}) shows fluctuations in $ c $
throughout the gapless region.
From DMRG calculations, we find that the ground state expectation value 
$\braket{\hat{\sigma}^{z}_i}$ shows oscillatory behavior around its mean value. 
In Fig.\ \ref{fig:s3-incomm-osc1}, we plot the absolute value of Fourier transform of 
this expectation value (minus its mean) for two points in the 
gapless region of the parameter space for $L=101$ sites. 
We find that the position of the peak
value changes for $J^{\,}_{1}=0.50$ and $J^{\,}_{1}=0.52$ for fixed $J^{\,}_{3}=0.5$,
with corresponding periods being $63/11$ and $63/12$ lattice constants, respectively.
This suggests a smooth variation of the oscillation period as a function of $J^{\,}_{1}$ and $ J^{\,}_{3} $ 
in the thermodynamic limit. 
We conjecture that in this gapless region, the system realizes a incommensurate phase
with central charge $c=1$. In other words,
the ground state contains low-energy states at a non-zero quasi-momentum which smoothly
varies as a function of the couplings $ J^{\,}_{1} $ and $ J^{\,}_{3} $. 
This echoes the behavior of the self-dual deformed Ising model, which is exactly solvable,
discussed in Sec.\ \ref{sec:incommensurate}.
 
\item
Results from the TEFR algorithm also show that the interface between 
the gapless region and the neighboring 
gapped phases has a vanishing central charge. Since the transition from a gapless phase 
to a gapped one is expected to not be a first-order transition, we conjecture that this 
must correspond to a continuous transition with a dynamical critical exponent $z>1$.
\end{enumerate}

\begin{figure}[t]
\centering
\begin{subfigure}{.7\textwidth}
\centering
\includegraphics[width=0.475\linewidth]{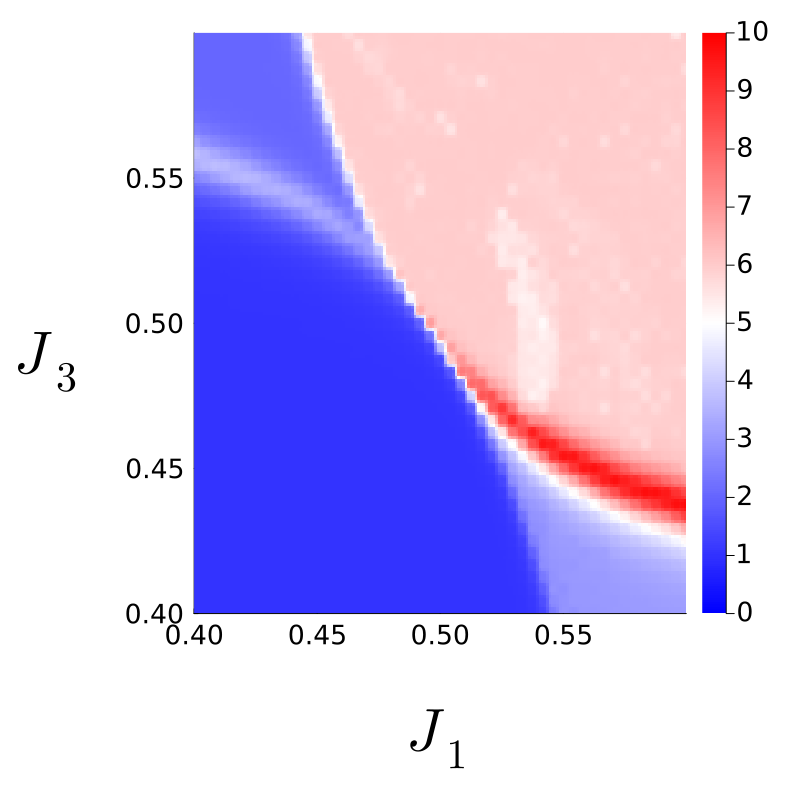}  
\quad
\includegraphics[width=0.475\linewidth]{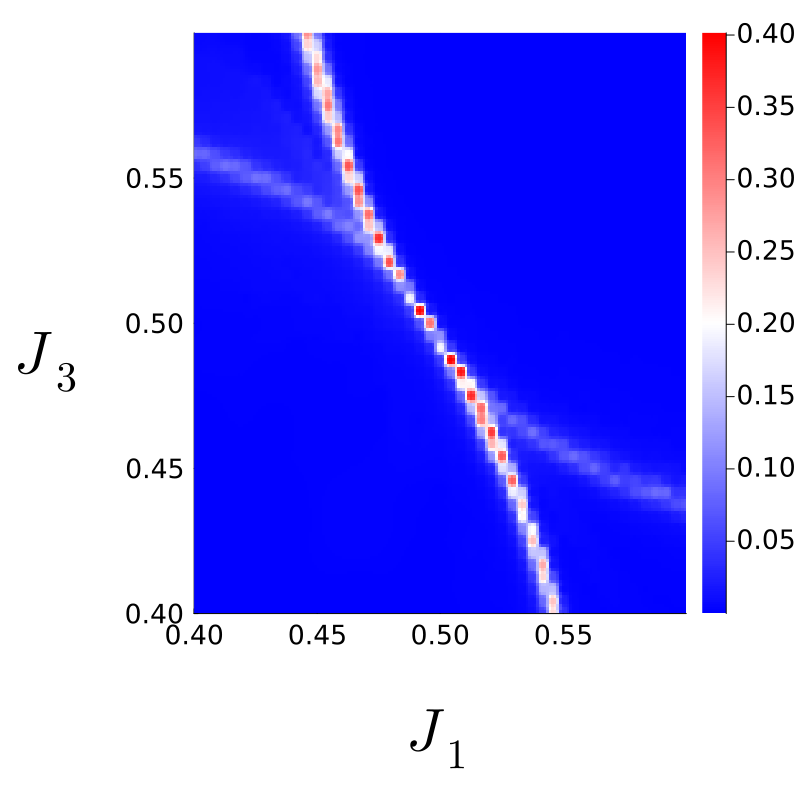} 
\vspace*{-8mm}
\caption{$J_{\perp}=0.05$
}
\label{fig:s3-jperp-positive}
\end{subfigure}

\medskip

\begin{subfigure}{.7\textwidth}
\centering
\includegraphics[width=0.475\linewidth]{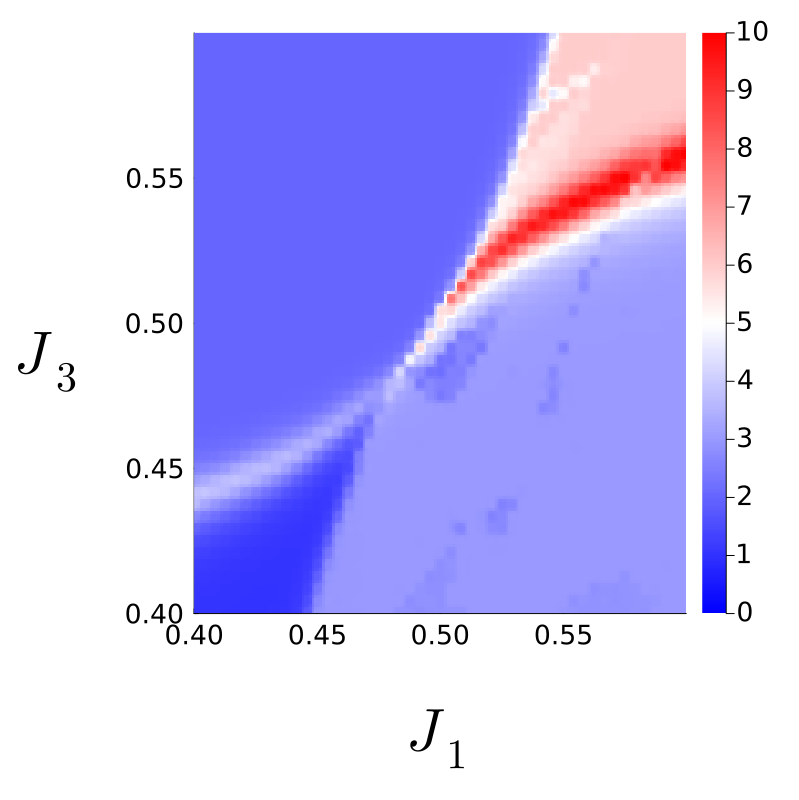}
\quad
\includegraphics[width=0.475\linewidth]{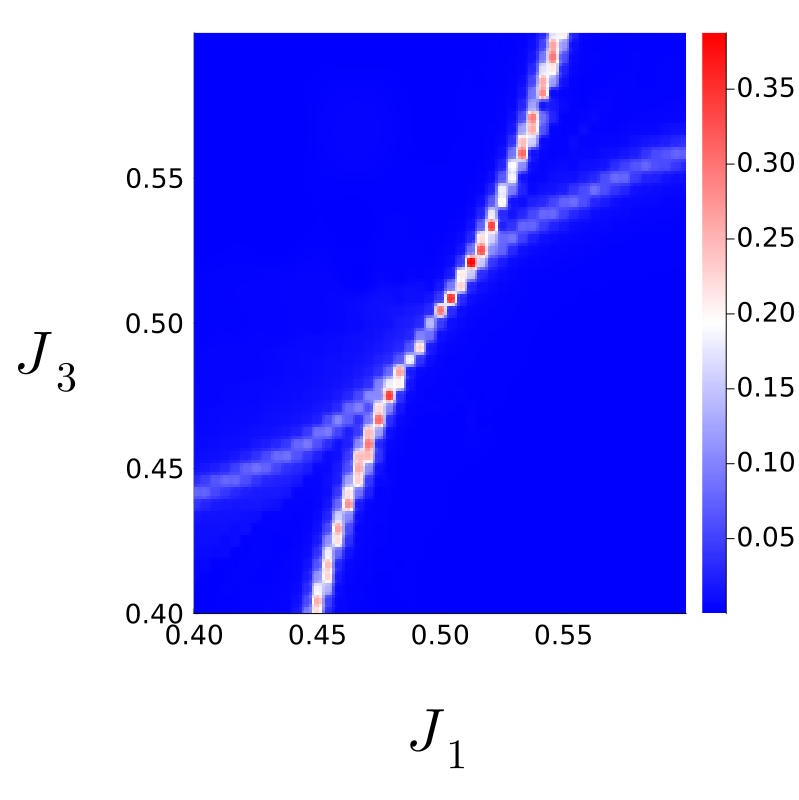} 
\vspace*{-8mm}
\caption{$J_{\perp}=-0.05$}
\label{fig:s3-jperp-negative}
\end{subfigure}
\caption{Numerical phase diagram from TEFR algorithm, with fixed $J_{\perp}$ and $ \th=0 $, showing GSD (left) and  central charge(right) as heatmaps, in the $ J^{\,}_1, J^{\,}_{3} $ plane (with $ J^{\,}_2=1-J^{\,}_1 $ and $ J_4=1-J^{\,}_{3} $ everywhere). The effective system size is $ L=128 $.  (a) For positive $J_{\perp}$, we find the multicritical point widens into a critical line between the $S_3$ symmetric and $S_3$ SSB phases. (b) For negative $J_{\perp}$, we find the multicritical point widens into a critical line between the phases which spontaneously break $S_3$ down to $\Z_3$ and $\Z_2$, instead.
\label{fig:s3-jperp}
}
\end{figure}

\subsubsection{Explicitly breaking KW self-duality symmetry}
\label{sec:third relevant perturbation S3}

Hamiltonian \eqref{eq:Ham gen S3 reparameterized} consists of terms such that
the lines $J^{\,}_{1}=J^{\,}_{2}$ is symmetric under the $\Z^{\,}_{3}$ 
KW self-duality symmetry generated by the operator \eqref{eq:def KW duality operator}.
As explained in the previous section,
the multicritical point, $J^{\,}_{1}=J^{\,}_{2}$ and $J^{\,}_{3}=J^{\,}_{4}$,
is described by $3$-state Potts $\boxtimes$ Ising CFT with central charge $c=13/10$.
This multicritical point has three relevant perturbations, see Appendix 
\ref{app:potts x ising cft partition function}. 
Two of these are generated when $J^{\,}_{1}-J^{\,}_{2}\neq 0$ 
and $J^{\,}_{3}-J^{\,}_{4}\neq 0$, which gap out the 
qutrits and qubits, respectively. 
In the continuum description, these perturbations correspond to the ``energy'' primaries
$\epsilon^{\,}_{\mathrm{I}}$ and $\epsilon^{\,}_{\mathrm{P}}$
of the Ising and 3-state Potts CFTs, respectively.
The former is self-dual under the $\mathbb{Z}^{\,}_{3}$
KW duality and gaps out the qubits.
In contrast, the latter breaks KW duality symmetry explicitly and gaps out 
the qutrits. 

The third and final relevant perturbation is given by the product of two 
energy primaries $\epsilon^{\,}_{\mathrm{I}}\,\epsilon^{\,}_{\mathrm{P}}$ 
which is odd under the $\Z^{\,}_{3}$ self-duality symmetry.
In the lattice model, this perturbation is generated by 
\begin{align}
\label{eq:def Hperp}
\widehat{H}^{\,}_{\perp}
:=
-
J^{\,}_{\perp}\,
\sum_{i=1}^{L}
\left\{
\left(
\hat{\tau}^{z}_{i}\,
\hat{\sigma}^{z}_{i+1}
-
\hat{\sigma}^{x}_{i+1}
\right)
\left(\widehat{Z}^{\,}_{i}\,\widehat{Z}^{\dagger}_{i+1}+
\text{H.c.}
\right)
-
\left(
\hat{\tau}^{z}_{i}\,
\hat{\sigma}^{z}_{i}
-
\hat{\tau}^{x}_{i}
\right)
\left(\widehat{X}^{\,}_{i}+\text{H.c.}\right)
\right\},
\end{align}
which is odd under the $\Z^{\,}_{3}$ KW 
duality symmetry \eqref{eq:def KW duality operator}.

From the TEFR algorithm, we find that adding this term has 
the effect of replacing the multicritical point 
by a critical line which separates the $ S^{\,}_3 $ symmetric 
and the $ S^{\,}_3 $ completely broken phases for positive $ J^{\,}_{\perp} $ 
and by a critical line which separates the $ \Z^{\,}_3 $ symmetric and $ \Z^{\,}_2 $ 
symmetric phases for negative $ J^{\,}_{\perp} $. 
This is shown in the GSD and central charge plots obtained from the TEFR algorithm in \fig{fig:s3-jperp}. 

While it is rather hard to precisely determine the points on the critical lines, 
we know that the point with $J^{\,}_{1}=J^{\,}_{2}=0.5$ and $J^{\,}_{3}=J^{\,}_{4}=0.5$
must be critical. This is because  when $\th=0$ the Hamiltonian \eqref{eq:def Ham gen S3}
has an additional non-invertible $\mathbb{Z}^{\,}_{2}$ KW self-duality symmetry along the 
$J{\,}_{3}=J^{\,}_{4}$ line. In the continuum limit, the perturbation \eqref{eq:def Hperp} 
flows to $\epsilon^{\,}_{\mathrm{I}}\epsilon^{\,}_{\mathrm{P}}$, \ie product of energy primaries, 
which is odd under both $\Z^{\,}_{2}$ and $Z^{\,}_{3}$ KW self-duality symmetries.\footnote{
At the lattice level, the perturbation \eqref{eq:def Hperp} is exactly odd only under
the $\mathbb{Z}^{\,}_{3}$ KW self-duality operator \eqref{eq:def KW duality operator}.
While it explicitly breaks the $\Z^{\,}_{2}$ KW self-duality symmetry too,
it does not go to minus itself under $\Z^{\,}_{2}$ KW self-duality transformation.} 
Under both of these dualities, the point with $J^{\,}_{1}=J^{\,}_{2}$ and  $J^{\,}_{3}=J^{\,}_{4}=0.5$
is invariant, and hence must be gapless in both phase diagrams with $J^{\,}_{\perp}>0$ 
and $J^{\,}_{\perp}<0$. At this point the DMRG results suggest a central charge of 
$ c \approx 1.2 $ which we believe to hold for the entire the critical line; 
a precise characterization of this transition in terms of an associated conformal 
field theory is beyond the scope of the present paper. 

We note that the continuous phase transition which separates the $ \Z^{\,}_3 $ symmetric and $ \Z^{\,}_2 $ 
symmetric phases for $ J^{\,}_{\perp} <0 $ is a 
beyond Landau-Ginzburg (LG) transition as it is between phases that break
different subgroups of the full symmetry group. 
Under the $\Z^{\,}_{3}$ KW-duality symmetry, \ie when $ J^{\,}_{\perp} $ 
becomes positive, this beyond-LG 
transition is mapped to the ordinary LG-type transition between the $ S^{\,}_3 $ 
symmetric and the $ S^{\,}_3 $ SSB phases.
\section{\texorpdfstring{$\cRep(S^{\,}_{3})$}{Rep(S3)}-symmetric spin chain}
\label{sec:RepS3model}

In Sec.\ \ref{sec:S3model}, we studied a spin chain with $S^{\,}_{3}$ symmetry.
We discussed how this $S^{\,}_{3}$ is enriched by non-invertible
$\mathbb{Z}^{\,}_{3}$ Kramers-Wannier self-duality symmetry, at special points in the parameter space.
Therein, the presence of KW duality symmetry ensures that the ground 
is either gapless or degenerate.\footnote{This is because this 
non-invertible symmetry is anomalous\cite{TW191202817,CRBSS230509713,ZC230401262}.} 
In this section, we are going to show that 
our $S^{\,}_{3}$-symmetric spin chain is dual to 
another model which has non-invertible symmetries in its entire parameter space. 
As we shall see in Sec.\ \ref{sec:gauging Z2 in S3}, this duality follows from 
gauging a $\mathbb{Z}^{\,}_{2}$ subgroup of $S^{\,}_{3}$, which delivers a dual model 
with $\cRep(S^{\,}_{3})$ fusion category symmetry. 
We discuss the gapped phases and phase transitions of this $\cRep(S^{\,}_{3})$-symmetric
model in Sec.\ \ref{sec:phase transitions repS3}.
As it was the case for Hamiltonian \eqref{eq:def Ham gen S3}, we will see in Sec.\ 
\ref{sec:gauging S3 in S3}, its dual with $\cRep(S^{\,}_{3})$ symmetry also has
an additional non-invertible symmetry at special points in the parameter space. 
We are going to describe how this additional symmetry is also associated with a 
gauging procedure which can be implemented through a sequential circuit.

\subsection{Gauging \texorpdfstring{$\Z^{\,}_2$}{Z2} subgroup: dual \texorpdfstring{$\cRep(S^{\,}_{3})$}{Rep(S3)} symmetry}
\label{sec:gauging Z2 in S3}

We shall gauge the $\mathbb{Z}^{\,}_{2}$ subgroup of $S^{\,}_{3}$.
We follow the same prescription as in Sec.\ \ref{sec:gauging Z3 in S3}. 
As we shall see, as opposed to gauging $\mathbb{Z}^{\,}_{3}$ subgroup,
the dual symmetry will be the category $\cRep(S^{\,}_{3})$, owing to the 
fact that $\mathbb{Z}^{\,}_{2}$ is not a normal subgroup of $S^{\,}_{3}$. 

In order to gauge the $\mathbb{Z}^{\,}_{2}$ symmetry, on 
each link between sites $i$ and $i+1$, we introduce 
$\mathbb{Z}^{\,}_{2}$ clock operators $\left\{\hat{\mu}^{x}_{i+1/2},\, 
\hat{\mu}^{z}_{i+1/2}\right\}$ that satisfy the algebra
\begin{equation}
\begin{split}
&
\hat{\mu}^{z}_{i+1/2}\,\hat{\mu}^{x}_{j+1/2}
=
(-1)^{\delta^{\,}_{ij}}
\hat{\mu}^{x}_{j+1/2}\,\hat{\mu}^{z}_{i+1/2},
\\
&
\left(\hat{\mu}^{z}_{i+1/2}\right)^{2}
=
\left(\hat{\mu}^{x}_{i+1/2}\right)^{2}
=
\hat{\mathbbm{1}},
\\
&
\hat{\mu}^{z}_{i+1/2+L}
=
\hat{\mu}^{z}_{i+1/2},
\quad
\hat{\mu}^{x}_{i+1/2+L}
=
\hat{\mu}^{x}_{i+1/2},
\end{split}
\end{equation}
where we have imposed periodic boundary conditions. 
Accordingly, we define the Gauss operator
\begin{align}
\widehat{G}^{\mathbb{Z}^{\,}_{2}}_{i}
\=
\hat{\mu}^{z}_{i-1/2}\,
\hat{\sigma}^{x}_{i}\,
\hat{\tau}^{x}_{i}\,
\widehat{C}^{\,}_{i}\,
\hat{\mu}^{z}_{i+1/2},
\qquad
\left[
\widehat{G}^{\mathbb{Z}^{\,}_{2}}_{i}
\right]^{2}
=
\hat{\mathbbm{1}}.
\label{eq:Gauss operator Z2 gauging}
\end{align}
The physical states are those in the subspace $\widehat{G}^{\mathbb{Z}^{\,}_{2}}_{i}=1$
for all $i=1,\cdots,L$. By way of minimally coupling the bond algebra 
\eqref{eq:def S3 algebra}, we obtain the gauge invariant bond algebra~
\footnote{Here, we introduce the short-hand notations 
\begin{align*}
\widehat{Z}^{\hat{\mu}^{x}_{i+1/2}}_{i}
\equiv
\frac{1+\hat{\mu}^{x}_{i+1/2}}{2}\,
\widehat{Z}^{\,}_{i}
+
\frac{1-\hat{\mu}^{x}_{i+1/2}}{2}\,
\widehat{Z}^{\dagger}_{i},
\qquad
\widehat{Z}^{-\hat{\mu}^{x}_{i+1/2}}_{i}
\equiv
\frac{1-\hat{\mu}^{x}_{i+1/2}}{2}\,
\widehat{Z}^{\,}_{i}
+
\frac{1+\hat{\mu}^{x}_{i+1/2}}{2}\,
\widehat{Z}^{\dagger}_{i}.
\end{align*}}
\small
\begin{align}
\mathfrak{B}^{\mathrm{mc}}_{S^{\,}_{3}/\Z^{\,}_{2}}
\=
\Big\langle
&
\hat{\sigma}^{z}_{i}\,
\hat{\tau}^{z}_{i},\,
\hat{\tau}^{z}_{i}\,
\hat{\mu}^{x}_{i+1/2}\,
\hat{\sigma}^{z}_{i+1},\,
\hat{\sigma}^{x}_{i},\,
\hat{\tau}^{x}_{i},\,
\left(\widehat{X}^{\,}_{i}+\widehat{X}^{\dagger}_{i}\right),\,
\left(\widehat{Z}^{\hat{\mu}^{x}_{i+1/2}}_{i}\,
\widehat{Z}^{\dagger}_{i+1}
+\widehat{Z}^{-\hat{\mu}^{x}_{i+1/2}}_{i}\,
\widehat{Z}^{\,}_{i+1}\right),
\nonumber\\
&
\hat{\sigma}^{z}_{i}\,\left(\widehat{X}^{\,}_{i}-\widehat{X}^{\dagger}_{i}\right),\,
\hat{\tau}^{z}_{i}\,
\hat{\mu}^{x}_{i+1/2}\,
\left(
\widehat{Z}^{\hat{\mu}^{x}_{i+1/2}}_{i}\,
\widehat{Z}^{\dagger}_{i+1}
-\widehat{Z}^{-\hat{\mu}^{x}_{i+1/2}}_{i}\,
\widehat{Z}^{\,}_{i+1}\right)
\Big\vert 
\,\,
\widehat{G}^{\mathbb{Z}^{\,}_{2}}_{i}=1,
\quad
i\in \Lambda
\Big\rangle.
\label{eq:def S3 Z2 minimally coupled algebra}
\end{align}
\normalsize
As it was in Sec.\ \ref{sec:gauging Z3 in S3}, we can simplify the 
Gauss constraint by applying the unitary transformation
\begin{equation}\label{eq:Z2 gauging unitary}
\begin{aligned}
&
\widehat{U}\,
\hat{\sigma}^{x}_{i}\,
\widehat{U}^{\dagger}
=
\hat{\mu}^{z}_{i-1/2}\,
\hat{\sigma}^{x}_{i}\,
\hat{\tau}^{x}_{i}\,
\widehat{C}^{\,}_{i}\,
\hat{\mu}^{z}_{i+1/2},
\qquad
&&
\widehat{U}\,
\hat{\sigma}^{z}_{i}\,
\widehat{U}^{\dagger}
=
\hat{\sigma}^{z}_{i},
\\
&
\widehat{U}\,
\hat{\tau}^{x}_{i}\,
\widehat{U}^{\dagger}
=
\hat{\tau}^{x}_{i},
\qquad
&&
\widehat{U}\,
\hat{\tau}^{z}_{i}\,
\widehat{U}^{\dagger}
=
\hat{\tau}^{z}_{i}\,
\hat{\sigma}^{z}_{i},
\\
&
\widehat{U}\,
\widehat{X}^{\,}_{i}\,
\widehat{U}^{\dagger}
=
\widehat{X}^{\hat{\sigma}^{z}_{i}}_{i},
\qquad
&&
\widehat{U}\,
\widehat{Z}^{\,}_{i}\,
\widehat{U}^{\dagger}
=
\widehat{Z}^{\hat{\sigma}^{z}_{i}}_{i},
\\
&
\widehat{U}\,
\hat{\mu}^{x}_{i+1/2}\,
\widehat{U}^{\dagger}
=
\hat{\sigma}^{z}_{i}\,
\hat{\mu}^{x}_{i+1/2}\,
\hat{\sigma}^{z}_{i+1},
\qquad
&&
\widehat{U}\,
\hat{\mu}^{z}_{i+1/2}\,
\widehat{U}^{\dagger}
=
\hat{\mu}^{z}_{i+1/2}.
\end{aligned}
\end{equation}
Under this unitary the Gauss operator simplifies 
$\widehat{U}\,\widehat{G}^{\mathbb{Z}^{\,}_{2}}_{i}\,\widehat{U}^{\dagger}
=\hat{\sigma}^{x}_{i}$. 
We apply this unitary to the minimally coupled algebra 
\eqref{eq:def S3 Z2 minimally coupled algebra} and project onto the 
$\hat{\sigma}^{x}_{i}=1$ sector. After shifting the link degrees of freedom
by $i+1/2 \mapsto i+1$, we obtain the dual algebra
\begin{align}
\mathfrak{B}^{\,}_{S^{\,}_{3}/\Z^{\,}_{2}}
&\=
\widehat{U}\,
\mathfrak{B}^{\mathrm{mc}}_{S^{\,}_{3}/\mathbb{Z}^{\,}_{2}}\,
\widehat{U}^{\dagger}\,
\Big\vert^{\,}_{\hat{\sigma}^{x}_{i}=1}
\nonumber\\
&=
\Big\langle
\hat{\tau}^{z}_{i},\,
\hat{\tau}^{z}_{i}\,
\hat{\mu}^{x}_{i+1},\,
\hat{\mu}^{z}_{i}\,
\hat{\tau}^{x}_{i}\,
\widehat{C}^{\,}_{i}\,
\hat{\mu}^{z}_{i+1},\,
\hat{\tau}^{x}_{i},\,
\left(\widehat{X}^{\,}_{i}+\widehat{X}^{\dagger}_{i}\right),\,
\left(\widehat{Z}^{\hat{\mu}^{x}_{i+1}}_{i}\,
\widehat{Z}^{\dagger}_{i+1}
+\widehat{Z}^{-\hat{\mu}^{x}_{i+1}}_{i}\,
\widehat{Z}^{\,}_{i+1}\right),
\nonumber\\
&\qquad
\left(\widehat{X}^{\,}_{i}-\widehat{X}^{\dagger}_{i}\right),\,
\hat{\tau}^{z}_{i}\,
\hat{\mu}^{x}_{i+1}\,
\left(
\widehat{Z}^{\hat{\mu}^{x}_{i+1}}_{i}\,
\widehat{Z}^{\dagger}_{i+1}
-\widehat{Z}^{-\hat{\mu}^{x}_{i+1}}_{i}\,
\widehat{Z}^{\,}_{i+1}\right)
\Big\vert 
\,\,
i\in \Lambda
\Big\rangle.
\label{eq:def S3/Z2 algebra}
\end{align}

What is the symmetry described by the dual algebra \eqref{eq:def S3/Z2 algebra}?
We claim that $\mathfrak{B}^{\,}_{S^{\,}_{3}/\mathbb{Z}^{\,}_{2}}$ 
is the algebra of $\cRep(S^{\,}_{3})$-symmetric operators.
The fusion category $\cRep(S^{\,}_{3})$ consists of three simple objects, $\one$, $\one'$, 
and $\two$, which are labeled by the three irreducible representations (irreps) of $S^{\,}_{3}$.
The object $\one$ is represented by the unitary identity operator $\widehat{W}^{\,}_{\one}
=\hat{\mathbbm{1}}$, while the object $\one'$ is represented by the unitary operator
\begin{align}
\widehat{W}^{\,}_{\one'}
\=
\prod_{i=1}^{L} 
\hat{\mu}^{x}_{i},
\end{align}
which is the generator of $\mathbb{Z}^{\,}_{2}$ dual symmetry associated with gauging the 
$\mathbb{Z}^{\,}_{2}$ subgroup of $S^{\,}_{3}$. 
Consistency in gauging with imposing periodic boundary conditions on 
the operator $\left\{\hat{\mu}^{x}_{i},\,\hat{\mu}^{z}_{i}\right\}$
and the operators $\left\{\hat{\sigma}^{x}_{i},\,\hat{\sigma}^{z}_{i}\right\}$
requires the conditions
\begin{subequations}
\label{eq:Z2 gauging dual subspaces}
\begin{align}
\widehat{U}^{\,}_{s} = 1,
\qquad
\widehat{W}^{\,}_{\one'}
=
1,
\label{eq:Z2 gauging consistency cond}
\end{align}
to be satisfied, respectively. 
In other words, the duality holds between the subalgebras 
\begin{align}
\mathfrak{B}^{\,}_{S^{\,}_{3}}\Big\vert^{\,}_{\widehat{U}^{\,}_{s}=1}
\cong 
\mathfrak{B}^{\,}_{S^{\,}_{3}/\mathbb{Z}^{\,}_{2}}\Big\vert^{\,}_{\widehat{W}^{\,}_{\one'}=1},
\end{align}
\end{subequations}
where conditions in Eq.\ \eqref{eq:Z2 gauging consistency cond} are satisfied.

Finally, we notice that gauging $\mathbb{Z}^{\,}_{2}$ subgroup
breaks the $\mathbb{Z}^{\,}_{3}$ symmetry, since the former is not a normal 
subgroup. Under conjugation by $\widehat{U}^{\,}_{r}$, the Gauss operator
\eqref{eq:Gauss operator Z2 gauging} transforms nontrivially
\begin{align}
\widehat{U}^{\,}_{r}\,
\widehat{G}^{\mathbb{Z}^{\,}_{2}}_{i}\,
\widehat{U}^{\dagger}_{r}
=
\widehat{G}^{\mathbb{Z}^{\,}_{2}}_{i}\,
\widehat{X}^{\,}_{i}.
\end{align}
Therefore, $\widehat{U}^{\,}_{r}$ cannot be made 
gauge invariant by coupling to the gauge fields $\hat{\mu}^{x}_{i+1/2}$.
However, the non-unitary and non-invertible operator 
\begin{align}
\widehat{U}^{\,}_{r\oplus r^{2}}
\=
\widehat{U}^{\,}_{r}\,
+
\widehat{U}^{\dagger}_{r}
=
\prod_{i=1}^{L}
\widehat{X}^{\,}_{i}
+
\prod_{i=1}^{L}
\widehat{X}^{\dagger}_{i}
\label{eq:r+r2 symmetry operator}
\end{align}
commutes with all the generators of the algebra \eqref{eq:def S3 algebra}
and the global symmetry operator $\widehat{U}^{\,}_{s}$
when periodic boundary conditions for all operators  in the algebra 
\eqref{eq:def S3 algebra}. 
This is the representation of direct sum $r\oplus r^{2}$ of simple objects $r$
and $r^{2}$ in the symmetry category $\cVec_{S^{\,}_{3}}$.
Since it commutes with $\widehat{U}^{\,}_{s}$, $\widehat{U}^{\,}_{r\oplus r^{2}}$
can be made gauge invariant. Minimally coupling the operator 
\eqref{eq:r+r2 symmetry operator}, and applying the unitary transformation 
\eqref{eq:Z2 gauging unitary} delivers the operator \footnote{
Each of the two terms in square brackets individually commutes with the Gauss operators 
associated with Gauss operators $\widehat{G}_2^{\Z^{\,}_2}$ through 
$\widehat{G}_L^{\Z^{\,}_2}$, while $\widehat{G}_1^{\Z^{\,}_2}$ simply 
exchanges them so that the sum is still gauge invariant.}
\begin{align}
\widehat{W}^{\,}_{\two}
\=
\frac{1}{2}
\left(
1
+
\prod_{i=1}^{L}
\hat{\mu}^{x}_{i}
\right)
\left[
\prod_{i=1}^{L}
\widehat{X}^{\prod_{k=2}^{i}\hat{\mu}^{x}_{k^{\,}_{\,}}}_{i}
+
\widehat{X}^{-\prod_{k=2}^{i}\hat{\mu}^{x}_{k^{\,}_{\,}}}_{i}
\right].
\label{eq:operator for element 2}
\end{align}
This is the representation of simple object $\two$ in the category 
$\cRep(S^{\,}_{3})$. 
It can be expressed as a matrix product operator (MPO) with a 2-dimensional
virtual index, as follows~\footnote{
See Appendix \ref{app:G and Rep G} for construction of
$\cRep(G)$ operators in terms of MPOs that results from gauging a finite 
$G$ symmetry.
}
\begin{equation}\label{eq:W2 MPO}
\widehat{W}^{\,}_{\two}
=
\Tr \left [\del_{\al_1,\al_{L+1}} \prod_{j=1}^{L} \widehat{M}^{(j)}_{\al_j \al_{j+1}}	\right ]
,
\quad 
\widehat{\bm M}^{(j)}=
	\begin{pmatrix}
		\widehat{X}_j \, P^{\, \hat{\mu}^x_{j+1}=1} 
		& 
		\widehat{X}_j^\dagger \, P^{\,\hat{\mu}^x_{j+1}=-1}
		\\
		\widehat{X}_j \, P^{\,\hat{\mu}^x_{j+1}=-1} 
		& 
		\widehat{X}_j^\dagger \, P^{\,\hat{,\mu}^x_{j+1}=1}
	\end{pmatrix}
\end{equation}
Together with $\widehat{W}^{\,}_{\one}$ and 
$\widehat{W}^{\,}_{\one'}$, 
they satisfy the fusion rules of $\cRep(S^{\,}_{3})$, \ie
\begin{align}
\widehat{W}^{\,}_{\one'}\,
\widehat{W}^{\,}_{\one'}
=
\widehat{W}^{\,}_{\one},
\qquad
\widehat{W}^{\,}_{\one'}\,
\widehat{W}^{\,}_{\two}
=
\widehat{W}^{\,}_{\two},
\qquad
\widehat{W}^{\,}_{\two}\,
\widehat{W}^{\,}_{\two}
=
\widehat{W}^{\,}_{\one}
+
\widehat{W}^{\,}_{\one'}
+
\widehat{W}^{\,}_{\two}.
\end{align}
We note that because of the projector, $\widehat{W}^{\,}_{\two}$ is non-invertible.
This projector to the $\prod_{i=1}^{L}\hat{\mu}^{x}_{i}=1$ subspace ensures 
that there is no $\mathbb{Z}^{\,}_{2}$ twist in the $S^{\,}_3$-symmetric algebra
\eqref{eq:def S3 algebra}. This is needed as $\widehat{U}^{\,}_{r\oplus r^{2}}$
is not a symmetry of the algebra \eqref{eq:def S3 algebra} when 
$\mathbb{Z}^{\,}_{2}$-twisted boundary conditions are imposed.\footnote{
A $g\in G$ twist reduces the full symmetry $G$ to the centralizer $\mathrm{C}^{\,}_{G}(g)$
of $g$. Since $S^{\,}_{3}$ is non-Abelian, imposing $\mathbb{Z}^{\,}_{3}$- 
and $\mathbb{Z}^{\,}_{2}$-twisted boundary conditions, 
reduce $S^{\,}_{3}$ down to $\mathbb{Z}^{\,}_{3}$ and 
$\mathbb{Z}^{\,}_{2}$, respectively.} In other words, in the presence of a 
$\mathbb{Z}^{\,}_{2}$ twist, one expect the dual symmetry to be $\mathbb{Z}^{\,}_{2}$
generated by $\widehat{W}^{\,}_{\one'}$ instead of the full $\cRep(S^{\,}_{3})$.

\begin{figure}[t!]
	\centering
	\includegraphics[width=0.7\linewidth]{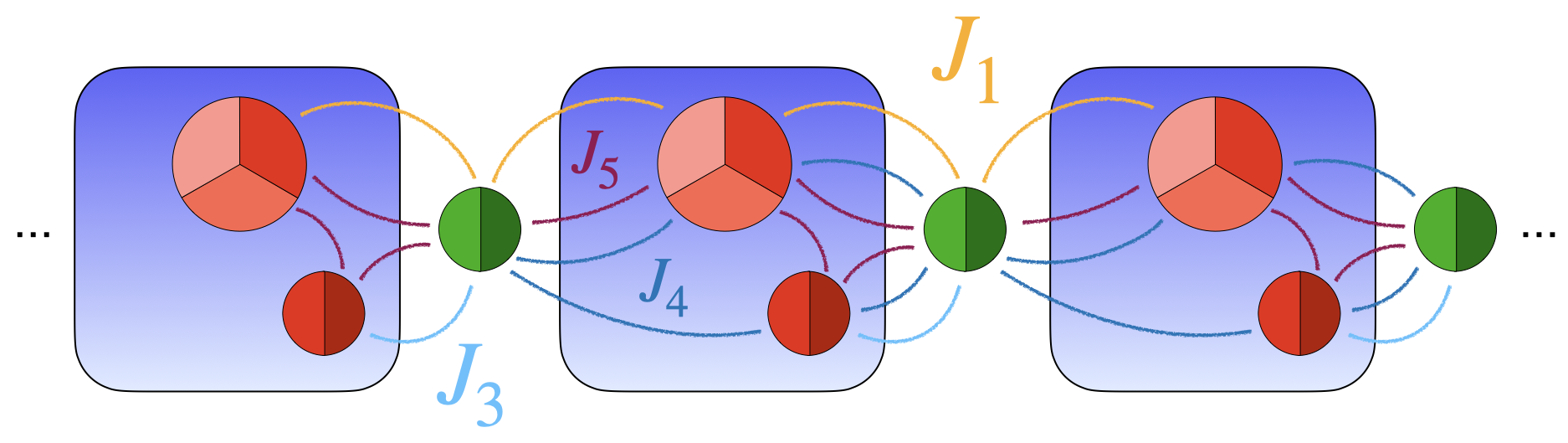}
	\caption{Schematic of the Hamiltonian \eqref{eq:def Ham gen RepS3} 
		showing the couplings between qutrit (depicted by a tripartitioned disk) 
		and qubit (depicted by a bipartitioned disk) degrees of freedom. 
		Single-body terms $ J^{\,}_2,J^{\,}_6 $ are suppressed.}
	\label{fig:reps3-chain-ham}
\end{figure}

We now use the duality mapping between the local operator algebras $\mathfrak{B}^{\,}_{S^{\,}_{3}}$ and $\mathfrak{B}^{\,}_{S^{\,}_{3}/\mathrm{Z}^{\,}_{2}}$ (recall Eqs.\ \eqref{eq:def S3 algebra}
and \eqref{eq:def S3/Z2 algebra}) to construct a local Hamiltonian with $\cRep(S^{\,}_{3})$
symmetry. Under this map, the image of Hamiltonian \eqref{eq:def Ham gen S3} is
\begin{equation}\label{eq:def Ham gen RepS3}
\begin{split}
\widehat{H}^{\,}_{\cRep(S^{\,}_{3})}
\=
&-
J^{\,}_{1}\,
\sum_{i=1}^{L}
\left(
\widehat{Z}^{\hat{\mu}^{x}_{i+1}}_{i}\,
\widehat{Z}^{\dagger}_{i+1}
+
\widehat{Z}^{-\hat{\mu}^{x}_{i+1}}_{i}\,
\widehat{Z}^{\,}_{i+1}
\right)
-
J^{\,}_{2}\,
\sum_{i=1}^{L}
\left(
\widehat{X}^{\,}_{i}
+
\widehat{X}^{\dagger}_{i}
\right)
\\
&
-
J^{\,}_{3}\,
\sum_{i=1}^{L}
\left(
\hat{\tau}^{z}_{i}
+
\hat{\tau}^{z}_{i}\,
\hat{\mu}^{x}_{i+1}
\right)
-
J^{\,}_{4}\,
\sum_{i=1}^{L}
\left(
\hat{\mu}^{z}_{i}\,
\hat{\tau}^{x}_{i}\,
\widehat{C}^{\,}_{i}\,
\hat{\mu}^{z}_{i+1}
+
\hat{\tau}^{x}_{i}
\right)
\\
&
-
J^{\,}_{5}\,
\sum_{i=1}^{L}
\mathrm{i}\,
\hat{\tau}^{z}_{i}\,
\hat{\mu}^{x}_{i+1}\,
\left(
\widehat{Z}^{\hat{\mu}^{x}_{i+1}}_{i}\,
\widehat{Z}^{\dagger}_{i+1}
-\widehat{Z}^{-\hat{\mu}^{x}_{i+1}}_{i}\,
\widehat{Z}^{\,}_{i+1}
\right)
-
J^{\,}_{6}\,
\sum_{i=1}^{L}
\mathrm{i}\,
\left(
\widehat{X}^{\,}_{i}
-
\widehat{X}^{\dagger}_{i}
\right).
\end{split}
\end{equation}
In what follows, we are going to study the phase diagram of this Hamiltonian
and identify spontaneous symmetry breaking patterns for $\cRep(S^{\,}_{3})$ 
symmetry  as well as the transitions between various ordered phases. 
Before doing so, we will briefly digress to discuss a duality 
that delivers a dual $\cRep(S^{\vee}_{3})$ symmetric bond algebra.\footnote{
We use the superscript $\vee$ to differentiate 
the $\cRep(S^{\,}_{3})$ symmetry of 
the bond algebra \eqref{eq:def S3/Z2 algebra} from the $\cRep(S^{\vee}_{3})$ 
symmetry of the bond algebra \eqref{eq:def S3/S3 algebra}.}

\subsection{Another non-invertible self-duality symmetry}
\label{sec:gauging S3 in S3}

As we discussed in Sec.\ \ref{sec:gauging Z3 in S3}, $S^{\,}_{3}$-symmetric
Hamiltonian \eqref{eq:def Ham gen S3} enjoys a self-duality when 
$J^{\,}_{1}=J^{\,}_{2}$ and $J^{\,}_{5}=J^{\,}_{6}$, which is induced by
gauging the $\mathbb{Z}^{\,}_{3}$ subgroup. We hence expect 
the same self-duality to hold for $\cRep(S^{\,}_{3})$-symmetric
Hamiltonian \eqref{eq:def Ham gen RepS3} too. 
To understand the self-duality of Hamiltonian \eqref{eq:def Ham gen RepS3}, 
we will show that gauging the entire
$S^{\,}_{3}$ symmetry delivers another bond algebra with $\cRep(S^{\vee}_{3})$
symmetry. This can be achieved by first gauging 
$\mathbb{Z}^{\,}_{3}$ and then $\mathbb{Z}^{\vee}_{2}$ 
symmetry of dual $S^{\vee}_{3}$ symmetry. 
Starting from the bond algebra \eqref{eq:def S3/Z3 algebra}
and gauging the $\mathbb{Z}^{\vee}_{2}$ symmetry generated by
$\widehat{U}^{\vee}_{s}$ defined in Eq.\ \eqref{eq:Svee3 representations}, 
we find the bond algebra
\begin{align}
\mathfrak{B}^{\,}_{S^{\,}_{3}/S^{\,}_{3}}
&\=
\Big\langle
\hat{\tau}^{z}_{i},\,
\hat{\tau}^{z}_{i}\,
\hat{\mu}^{x}_{i+1},\,
\hat{\mu}^{z}_{i}\,
\hat{\tau}^{x}_{i}\,
\hat{c}^{\,}_{i}\,
\hat{\mu}^{z}_{i+1},\,
\hat{\tau}^{x}_{i},\,
\left(\hat{z}^{\hat{\mu}^{x}_{i+1}}_{i}\,\hat{z}^{\dagger}_{i+1}
+
\hat{z}^{-\hat{\mu}^{x}_{i+1}}_{i}\,\hat{z}^{\,}_{i+1}\right),\,
\left(\hat{x}^{\,}_{i}+\hat{x}^{\dagger}_{i}\right),
\nonumber\\
&\qquad
\hat{\mu}^{x}_{i+1}\,
\left(
\hat{z}^{\hat{\mu}^{x}_{i+1}}_{i}\,
\hat{z}^{\dagger}_{i+1}
-\hat{z}^{-\hat{\mu}^{x}_{i+1}}_{i}\,
\hat{z}^{\,}_{i+1}\right),\,
\hat{\tau}^{z}_{i}\,
\hat{\mu}^{x}_{i+1}\,
\left(\hat{x}^{\,}_{i+1}-\hat{x}^{\dagger}_{i+1}\right)
\Big\vert 
\,\,
i\in\Lambda
\Big\rangle,
\label{eq:def S3/S3 algebra}
\end{align}
that is dual to the algebra \eqref{eq:def S3/Z3 algebra} 
under gauging the $\mathbb{Z}^{\vee}_{2}$ symmetry. 
We notice that this algebra has the same terms as algebra \eqref{eq:def S3/Z2 algebra}.
Therefore, its commutant algebra is that of the category $\cRep(S^{\vee}_{3})$.
The simple objects in $\cRep(S^{\vee}_{3})$ are represented by the operators
\begin{align}
\widehat{W}^{\vee}_{\one} \= \hat{\mathbbm{1}},
\quad
\widehat{W}^{\vee}_{\one'} \= \prod_{i=1}^{L} 
\hat{\mu}^{x}_{i},
\quad
\widehat{W}^{\vee}_{\two}
\=
\frac{1}{2}
\left(
1
+
\prod_{i=1}^{L}
\hat{\mu}^{x}_{i}
\right)
\left[
\prod_{i=1}^{L}
\hat{x}^{\prod_{k=2}^{i}\hat{\mu}^{x}_{k^{\,}_{\,}}}_{i}
+
\hat{x}^{-\prod_{k=2}^{i}\hat{\mu}^{x}_{k^{\,}_{\,}}}_{i}
\right].
\end{align}
To find in which subalgebra the duality holds, we
combine Eqs.\ \eqref{eq:Z3 gauging dual subspaces} and 
\eqref{eq:Z2 gauging dual subspaces} that describe the consistency conditions
imposed by gauging $\mathbb{Z}^{\,}_{3}$ and $\mathbb{Z}^{\,}_{2}$ subgroups, respectively.
We find that the duality induced by 
gauging the entire group $S^{\,}_{3}$ holds between
the subalgebras
\begin{align}
\mathfrak{B}^{\,}_{S^{\,}_{3}}\Big\vert^{\,}_{\widehat{U}^{\,}_{s}=\widehat{U}^{\,}_{r}=1}
\cong 
\mathfrak{B}^{\,}_{S^{\,}_{3}/S^{\,}_{3}}\Big\vert^{\,}_{\widehat{W}^{\vee}_{\one'}=1,\,
\widehat{W}^{\vee}_{\two}=2}\,.
\end{align}
This consistency condition says that the duality maps the $S^{\,}_{3}$-singlet
subalgebra of $\mathfrak{B}^{\,}_{S^{\,}_{3}}$ to the subalgebra of 
$\mathfrak{B}^{\,}_{S^{\,}_{3}/S^{\,}_{3}}$ where the representation of each 
simple object is equal to its quantum dimension.
The image of Hamiltonian \eqref{eq:def Ham gen S3} under gauging the entire 
$S^{\,}_{3}$ symmetry is 
\begin{equation}
\label{eq:def Ham gen RepS3vee}
\begin{split}
\widehat{H}_{\cRep(S^{\vee}_{3})}
\=
&-
J^{\,}_{1}\,
\sum_{i=1}^{L}
\left(
\hat{x}^{\,}_{i}
+
\hat{x}^{\dagger}_{i}
\right)
-
J^{\,}_{2}\,
\sum_{i=1}^{L}
\left(
\hat{z}^{\hat{\mu}^{x}_{i+1}}_{i}\,
\hat{z}^{\dagger}_{i+1}
+
\hat{z}^{-\hat{\mu}^{x}_{i+1}}_{i}\,
\hat{z}^{\,}_{i+1}
\right)
\\
&
-
J^{\,}_{3}\,
\sum_{i=1}^{L}
\left(
\hat{\tau}^{z}_{i}
+
\hat{\tau}^{z}_{i}\,
\hat{\mu}^{x}_{i+1}
\right)
-
J^{\,}_{4}\,
\sum_{i=1}^{L}
\left(
\hat{\mu}^{z}_{i}\,
\hat{\tau}^{x}_{i}\,
\widehat{C}^{\,}_{i}\,
\hat{\mu}^{z}_{i+1}
+
\hat{\tau}^{x}_{i}
\right)
\\
&
-
J^{\,}_{5}\,
\sum_{i=1}^{L}
\mathrm{i}\,
\hat{\tau}^{z}_{i}\,
\hat{\mu}^{x}_{i+1}\,
\left(\hat{x}^{\,}_{i+1}-\hat{x}^{\dagger}_{i+1}\right)
-
J^{\,}_{6}\,
\sum_{i=1}^{L}
\mathrm{i}\,
\hat{\mu}^{x}_{i+1}\,
\left(
\hat{z}^{\hat{\mu}^{x}_{i+1}}_{i}\,
\hat{z}^{\dagger}_{i+1}
-\hat{z}^{-\hat{\mu}^{x}_{i+1}}_{i}\,
\hat{z}^{\,}_{i+1}\right).
\end{split}
\end{equation}
This Hamiltonian is unitarily equivalent to the Hamiltonian \eqref{eq:def Ham gen RepS3} 
under the interchange $(J^{\,}_{1},J^{\,}_{5}) \leftrightarrow (J^{\,}_{2}, J^{\,}_{6})$.
The unitary transformation connecting the two Hamiltonians is 
the unitary $ \hat{\mathfrak{t}}^{\,}_{\cRep(S^{\,}_{3})} $, whose definition and action on the $ \cRep(S^{\,}_3) $-symmetric bond algebra generators are described in \app{app:RepS3 duality}.

In \scn{sec:gauging Z3 in S3}, we gave the explicit form of an operator that performs the $ \Z^{\,}_3 $ Kramers-Wannier duality transformation 
for the $S^{\,}_{3}$-symmetric Hamiltonian \eqref{eq:def Ham gen S3}. 
Gauging the $ \Z^{\,}_2 $ subgroup of the original $ S^{\,}_{3} $ symmetry leads to the
$\cRep(S^{\,}_{3}) $ symmetry. 
So we would like to apply the same $ \Z^{\,}_2 $ gauging map to 
$\widehat{D}_{\mathrm{KW}} $ to obtain the sequential quantum circuit 
that implements the duality under the 
$(J^{\,}_1,J^{\,}_5)\leftrightarrow(J^{\,}_2,J^{\,}_6) $ exchange.
To that end, we follow how the individual operators 
(or, gates in the quantum circuit language) in \eqref{eq:def KW duality operator} 
transform under this gauging map.
The full gauged operator has the form
\begin{equation}\label{eq: def DRepS3}
\widehat{D}_{\cRep(S^{\,}_3)} 
\= 
\hat{\mathfrak{t}}^{\,}_{\cRep(S^{\,}_{3})} \
\widehat{P}^{\,}_{\mathrm{reg}}\ \widehat{D}^{\,}_0\,,  
\end{equation}
where (i) the unitary $ \widehat{D}^{\,}_0  $ is defined as
\begin{equation}
\widehat{D}^{\,}_0 \=
\left (\sum_{\al=0}^{2} 
\widehat{Z}_L^{\, \al} \ 
\widehat{P}_{Z_1^{\mu^{x}_{1}} Z_L=\om^{ \al} } \right )
\left(
\prod_{j=1}^{L-1} 
\widehat{\mathfrak H}_j^\dagger \ \widehat{\mathrm{CZ}}_{j+1,j}^{-\hat{\mu}^x_{j+1}} \right )\,,
\end{equation}
(ii) the projector
$ \widehat{P}^{\,}_{\mathrm{reg}} \= \frac{1}{6}\widehat{W}_\text{reg}$ is defined in terms of the operator $\widehat{W}_\text{reg}$, corresponding to the (non-simple) regular representation object $\text{reg}=\mathbf{1} \oplus \one'\oplus 2 \ \mathbf{2}$ of the $\cRep(S^{\,}_{3})$ fusion category, \ie
\begin{equation*}
\widehat{W}_\text{reg}
\=
\widehat{W}_{\mathbf{1}} + \widehat{W}_{\one'} + 2 \widehat{W}_{\mathbf{2}}\,,
\end{equation*} 
and (iii) the unitary $ \hat{\mathfrak{t}}^{\,}_{\cRep(S^{\,}_{3})} $ is the operator obtained under the action of the $ \Z^{\,}_2 $-gauging map on the \emph{half-translation} operator $ \hat{\mathfrak{t}}^{\,}_{\mathbb{Z}^{\,}_{2}} $ defined in Eq.\ \eqref{eq:def Z2 half translation}. The explicit details of this operator are provided in \app{app:RepS3 duality}.
The projector
$\widehat{P}^{\,}_{\mathrm{reg}}$ 
annihilates  any state that is not $ \cRep(S^{\,}_{3})$ symmetric
since for any irrep $R=\one,\,\one',\, \two$ 
of $S^{\,}_{3}$, the identities
\begin{equation*}
\widehat{P}^{\,}_{\mathrm{reg}}  
\, 
\frac{1}{d^{\,}_R}\widehat{W}^{\,}_{R} 
= \widehat{P}^{\,}_{\mathrm{reg}},
\qquad
\widehat{P}^{\,}_{\mathrm{reg}} 
\left(
\hat{1}- \frac{1}{d^{\,}_R}\widehat{W}^{\,}_{R} 
\right) 
= 
0,
\end{equation*}
hold. The duality operator $\widehat{D}_{\cRep(S^{\,}_3)}$ acts on the 
$\cRep(S^{\,}_3)$-symmetric bond algebra (see \app{app:RepS3 duality} for details) as
\begin{equation}
\widehat{D}_{\cRep(S^{\,}_3)}
\,
\begin{pmatrix}
\widehat{X}_j + \text{H.c.} \\[0.4em]
\widehat{Z}_{j}^{\hat{\mu}^x_{j+1}} 
\widehat{Z}_{j+1}^\dagger  +\text{H.c.} \\[0.4em]
\widehat{X}_j - \text{H.c.} \\[0.5em]
\hat{\tau}^{z}_{i}\,\hat{\mu}^{x}_{i+1}
\left(
\widehat{Z}^{\hat{\mu}^{x}_{i+1}}_{i}\,\widehat{Z}^{\dagger}_{i+1}
-\text{H.c.}
\right) 
\end{pmatrix}
=
\begin{pmatrix}
\widehat{Z}_{j}^{\hat{\mu}^x_{j+1}} 
\widehat{Z}_{j+1}^\dagger + \text{H.c.} \\[0.4em]
\widehat{X}_{j+1}+\text{H.c.}\\[0.4em]
\hat{\tau}^{z}_{j}
\hat{\mu}^x_{j+1}
\left (
\widehat{Z}_{j}^{\hat{\mu}^x_{j+1}} 
\widehat{Z}_{j+1}^\dagger 
-
\text{H.c.}
\right )\\[0.5em]
\widehat{X}_{j+1}
- \text{H.c.}
\end{pmatrix}
\,\widehat{D}_{\cRep(S^{\,}_3)},
\end{equation}
which implements the self-duality transformation $ (J^{\,}_1,J^{\,}_5) \leftrightarrow (J^{\,}_2 , J^{\,}_6 )$ of Eq.\ \eqref{eq:def Ham gen RepS3}.
The operator $\widehat{D}_{\cRep(S^{\,}_3)}$ is non-invertible 
since it contains the projector $\widehat{P}^{\,}_{\mathrm{reg}}$. 
The operator $\widehat{D}_{\cRep(S^{\,}_3)}$ obeys the algebraic relations~\footnote{
We use the fact that $ \hat{\mathfrak{t}}^{\,}_{\cRep(S^{\,}_{3})}  $ and $ \widehat{D}_{0} $ commute with $ \widehat{P}^{\,}_{\mathrm{reg}} $.}
\begin{subequations}
\begin{align}
\widehat{D}_{\cRep(S^{\,}_3)}\,
\widehat{W}_{R} 
&= 
\widehat{W}_{R}\,
\widehat{D}_{\cRep(S^{\,}_3)}
=	
d^{\,}_R \ \widehat{D}_{\cRep(S^{\,}_3)}, 
\\
\left (\widehat{D}_{\cRep(S^{\,}_3)}\right )^{2} 
&= 
\widehat{P}^{\,}_{\mathrm{reg}} 
\left (
\hat{\mathfrak{t}}^{\,}_{\cRep(S^{\,}_{3})}
\widehat{D}_{0}
\right ) ^{2}
=
\widehat{P}^{\,}_{\mathrm{reg}} 
\widehat{T}
%=
%\widehat{T}
%\widehat{P}^{\,}_{\mathrm{reg}} 
,
\\
\left (\widehat{D}_{\cRep(S^{\,}_3)}\right )^\dagger 
&=  
\widehat{P}^{\,}_{\mathrm{reg}}  
\left (\hat{\mathfrak{t}}^{\,}_{\cRep(S^{\,}_{3})} \widehat{D}_{0}\right )^{\dagger} 
=
\widehat{T}^\dagger
\widehat{D}_{\cRep(S^{\,}_3)},
\end{align}
\end{subequations}
where $ d^{\,}_R $ is the dimension of the irreducible representation $R$, and 
$\widehat{T}$ is the operator translating both qubits and 
qutrits by one lattice site. 
Let us note that, the second line in the above set of equations implies
\begin{equation}\label{eq: Dreps3 qdim}
\left (	
\sqrt{6}  \widehat{D}_{\cRep(S^{\,}_3)} 
\right )^2 
= 
\left (
\widehat{W}_{\mathbf{1}} + \widehat{W}_{\one'} + 2 \widehat{W}_{\mathbf{2}}
\right )
\widehat{T}\,.
\end{equation}
Following the discussion for Eq.\ \eqref{eq: continuum Z3 KW}, an
analogous calculation of the quantum dimension suggests
\begin{equation}\label{eq:qdim Dreps3}
d_{\widehat{D}^{\,}_{\cRep(S^{\,}_3)}}^{2} 
= 
1+1+2\cdot 2 = 6 
\implies 
d^{\,}_{\widehat{D}^{\,}_{\cRep(S^{\,}_3)} } 
= 
\sqrt{6}
\end{equation}
The quantum dimension calculated above as well as the fusion rule in
Eq.\ \eqref{eq: Dreps3 qdim} are in tension with the category theoretic result
\cite{DLW231117044} which suggests that a duality defect symmetry 
$ \cD $ arising from ``half-gauging" by an algebra object $ A $ must satisfy 
\begin{equation}\label{eq:cat duality}
\cD^2 = A, \quad d^{\,}_{\cD} = \sqrt{\<A\>}.
\end{equation}
In our case, the self-duality symmetry generated by
$ \widehat{D}_{\cRep(S^{\,}_3)}  $
corresponds to a duality defect associated with gauging by the algebra 
object $ A= \one \oplus \two $, as we argue in 
\scn{sec:discussion gauging dualities}.
Therefore we should expect the quantum dimension to be $ \sqrt{3} $.
This highlights an important subtlety in calculating quantum dimension of
self-duality symmetries on the lattice when considering 
self-duality symmetries associated with gauging of non-invertible
symmetries by general algebra objects.
A more careful way to compute the quantum dimension, as well as
the fusion rules, involves unitary operators that move
non-invertible symmetry defects in a lattice Hamiltonian, as is done 
in Ref.\ \cite{SSS240112281}.

\subsection{Phase diagram}
\label{sec:phase transitions repS3}

To discuss the phase diagram of the Hamiltonian \eqref{eq:def Ham gen RepS3}, 
we first reparameterize it as 
\begin{equation}
\label{eq:Ham gen Rep S3 reparameterized}
\begin{alignedat}{3}
\widehat{H}^{\,}_{\cRep(S^{\,}_{3})}
=
&
-J^{\,}_{1}\cos \th\,
\sum_{i=1}^{L}
\left(
\widehat{Z}^{\hat{\mu}^{x}_{i+1}}_{i}\,
\widehat{Z}^{\dagger}_{i+1}
+
\text{H.c.}
\right) 
- J^{\,}_{1} \sin \th\,
\sum_{i=1}^{L}
\mathrm{i}\,
\hat{\tau}^{z}_{i}\,
\hat{\mu}^{x}_{i+1}
\left(
\widehat{Z}^{\hat{\mu}^{x}_{i+1}}_{i}\,
\widehat{Z}^{\dagger}_{i+1}
- \text{H.c.}
\right)
\\
&
-
J^{\,}_{2} \cos \th\,
\sum_{i=1}^{L}
\left(
\widehat{X}^{\,}_{i}
+
\widehat{X}^{\dagger}_{i}
\right)
-
J^{\,}_{2} \sin \th\,
\sum_{i=1}^{L}
\mathrm{i}
\left(
\widehat{X}^{\,}_{i}
-
\widehat{X}^{\dagger}_{i}
\right)
\\
%		\ +\
&- J^{\,}_{3}\,
\sum_{i=1}^{L}
\left(
\hat{\tau}^{z}_{i}
+
\hat{\tau}^{z}_{i}\,
\hat{\mu}^{x}_{i+1}
\right)
-
J^{\,}_{4}\,
\sum_{i=1}^{L}
\left(
\hat{\mu}^{z}_{i}\,
\hat{\tau}^{x}_{i}\,
\widehat{C}^{\,}_{i}\,
\hat{\mu}^{z}_{i+1}
+
\hat{\tau}^{x}_{i}
\right) \,.
\end{alignedat}
\end{equation}
As in Sec.\ \ref{sec:phase transitions S3}, 
we will explore the phase diagram of this Hamiltonian as a function of 
dimensionless ratios $J^{\,}_{1}/J^{\,}_{2}$ and $J^{\,}_{3}/J^{\,}_{4}$, 
for the cases of $\theta=0$,  non-zero but small $\theta\approx 0$, and large $\theta \sim 0.7$.

\subsubsection{Analytical arguments}

When studying the phase diagram of \eqref{eq:Ham gen Rep S3 reparameterized},
we can utilize its duality to the Hamiltonian \eqref{eq:Ham gen S3 reparameterized}, under the $\Z^{\,}_2$-gauging map.
When $\th=0$, we again identify four gapped phases that correspond to 
four distinct symmetry breaking patterns for $\cRep(S^{\,}_{3})$ as follows.

\begin{enumerate}[(i)]
\item
When $J^{\,}_{1}=J^{\,}_{4}=0$, 
Hamiltonian \eqref{eq:Ham gen Rep S3 reparameterized} becomes
\begin{equation}
\label{eq:HRepS3 J2, J3}
\widehat{H}^{\,}_{\cRep(S^{\,}_{3});2,3}
\= 	
-
\sum_{i=1}^{L}
\left[
J^{\,}_{2}
\left(
\widehat{X}^{\,}_{i} + 
\widehat{X}^{\dagger}_{i}
\right)
+
J^{\,}_{3}\,
\left (
\hat{\tau}^{z}_{i}\,
+
\hat{\tau}^{z}_{i}\,
\hat{\mu}^{x}_{i+1}
\right )
\right].
\end{equation}
The qubits and qutrits are decoupled
and all terms in the Hamiltonian pairwise commute. There is a single nondegenerate gapped ground state
\begin{align}
\label{eq:GS repS3}
\ket{\mathrm{GS}^{\,}_{\cRep(S^{\,}_{3})}}
\=
\bigotimes_{i=1}^{L}
\ket{\tau^{z}_{i}=1,\,\mu^{x}_{i}=1,\, X^{\,}_{i}=1},
\end{align}
which is symmetric under the entire 
$\cRep(S^{\,}_{3})$ symmetry. 
At this point, it is instructive to note that 
there is a subtle distinction between states symmetric 
under invertible and non-invertible symmetries that is 
implicitly used in the above discussion. 
Namely, the state $\ket{\mathrm{GS}^{\,}_{\cRep(S^{\,}_{3})}}$
transforms as
\begin{equation}
\label{eq:symmstate}
\widehat{W}^{\,}_{\one'} \ket{\mathrm{GS}^{\,}_{\cRep(S^{\,}_{3})}} 
= 
\ket{\mathrm{GS}^{\,}_{\cRep(S^{\,}_{3})}}, 
\quad 
\widehat{W}^{\,}_{{\mathbf{2}}} 
\ket{\mathrm{GS}^{\,}_{\cRep(S^{\,}_{3})}}= 2\,\ket{\mathrm{GS}^{\,}_{\cRep(S^{\,}_{3})}}
\end{equation}
where the factor of 2 reflects the quantum dimension of the non-invertible symmetry 
$ \widehat{W}^{\,}_{\mathbf{2}}$.\footnote{
To ensure the consistency with the fusion rules of $ \cRep(S^{\,}_{3}) $, 
the numerical pre-factor in Eq.\ \eqref{eq:symmstate} is essential. 
On the one hand,
\begin{equation*}
\widehat{W}^{\,2}_{{\mathbf{2}}} \ket{\mathrm{GS}^{\,}_{\cRep(S^{\,}_{3})}} 
=\widehat{W}^{\,}_{{\mathbf{2}}} \left (\widehat{W}^{\,}_{{\mathbf{2}}}
\ket{\mathrm{GS}^{\,}_{\cRep(S^{\,}_{3})}} \right )
= 4 \ket{\mathrm{GS}^{\,}_{\cRep(S^{\,}_{3})}} \, ,
\end{equation*}
while on the other,
\begin{equation*}
\widehat{W}^{\,2}_{{\mathbf{2}}}
\ket{\mathrm{GS}^{\,}_{\cRep(S^{\,}_{3})}} 
= \left (
\widehat{W}^{\,}_{\mathbf{1}} + 
\widehat{W}^{\,}_{\one'}+ 
\widehat{W}^{\,}_{\mathbf{2}}
\right ) 
\ket{\mathrm{GS}^{\,}_{\cRep(S^{\,}_{3})}}  
= 
(1+1+2)
\ket{\mathrm{GS}^{\,}_{\cRep(S^{\,}_{3})}}  
= 
4
\ket{\mathrm{GS}^{\,}_{\cRep(S^{\,}_{3})}}\, .
\end{equation*}
For a general non-invertible symmetry $\si$, its 
eigenvalue corresponding to a symmetric state must match its quantum dimension $ d_{\si} $.}
This is the lattice analogue of the field theory result that 
the vacuum expectation value (vev) of a non-invertible 
topological line defect is its quantum dimension. 
More generally, we say that a
state spontaneously breaks a non-invertible symmetry if
its expectation value is vanishing.

The $S^{\,}_{3}$-symmetric Hamiltonian \eqref{eq:Ham gen S3 reparameterized}
has twofold degenerate ground states \eqref{eq:GS Z3} when $J^{\,}_{1}=J^{\,}_{4}$ and $\th=0$.
Under the duality mapping, the unique ground state \eqref{eq:GS repS3} is the image of the 
symmetric linear combination of these ground states, i.e., 
$\ket{\mathrm{GS}^{+}_{\mathbb{Z}^{\,}_{2}}} + \ket{\mathrm{GS}^{-}_{\mathbb{Z}^{\,}_{2}}}$.
This is because the duality only holds in the subspace specified in Eq.\ 
\eqref{eq:Z2 gauging dual subspaces}.

\begin{figure}[t]
	\centering
	\includegraphics[width=.4\linewidth]{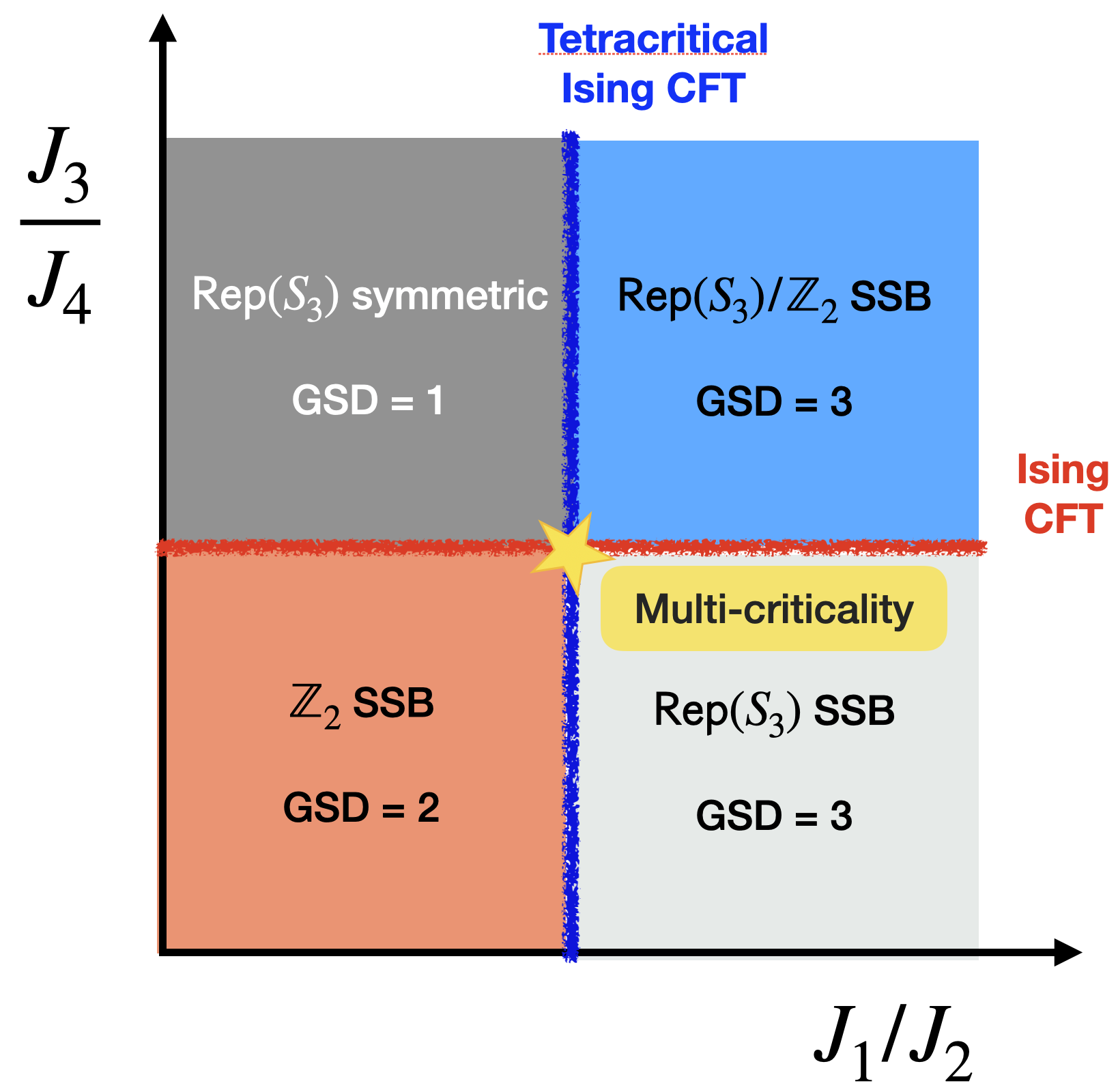}  
	\caption{Phase diagram of Hamiltonian \eqref{eq:Ham gen Rep S3 reparameterized} based on 
		analytical arguments at $ \th= 0 $. The critical points are guesses based on duality arguments
		and various simple limits.}
	\label{fig:reps3pd-analytical}
\end{figure}

\item 
When $J^{\,}_{2}=J^{\,}_{4}=0$, 
Hamiltonian \eqref{eq:Ham gen Rep S3 reparameterized} becomes
\begin{equation}
\label{eq:HRepS3 J1, J3}
\widehat{H}^{\,}_{\cRep(S^{\,}_{3});1,3}
\= 	
-
\sum_{i=1}^{L}
\left[
J^{\,}_{1}
\left(
\widehat{Z}^{\hat{\mu}^{x}_{i+1}}_{i}\,
\widehat{Z}^{\dagger}_{i+1}
+
\text{H.c.}
\right)
+
J^{\,}_{3}\,
\left (
\hat{\tau}^{z}_{i}\,
+
\hat{\tau}^{z}_{i}\,
\hat{\mu}^{x}_{i+1}
\right )
\right].
\end{equation}
Qubits are again in the disordered phase which pins
their value to $\hat{\tau}^{z}_{i}=1$ and $\hat{\mu}^{x}_{i}=1$ subspace. 
This means that the $J^{\,}_{1}$ term simply reduces to the classical 3-state Potts model.
There are three degenerate ground states 
\begin{align}
\label{eq:GS W1}
\ket{\mathrm{GS}^{\alpha}_{\one'}}
\=
\bigotimes_{i=1}^{L}
\ket{\tau^{z}_{i}=1,\,\mu^{x}_{i}=1,\, Z^{\,}_{i}=\omega^{\alpha}}.
\end{align}
These ground states preserve the $\mathbb{Z}^{\,}_{2}$ subgroup generated by 
$\widehat{W}^{\,}_{\one'}$ while they break the non-invertible
$\widehat{W}^{\,}_{\mathbf{2}}$ symmetry.
Under the latter each ground state is mapped to equal superposition
of the the other two, i.e.,
\begin{equation}
\begin{split}
&
\widehat{W}^{\,}_{\mathbf{2}}\,
\ket{\mathrm{GS}^{1}_{\one'}}
=
\ket{\mathrm{GS}^{2}_{\one'}}
+
\ket{\mathrm{GS}^{3}_{\one'}},
\\
&
\widehat{W}^{\,}_{\mathbf{2}}\,
\ket{\mathrm{GS}^{2}_{\one'}}
=
\ket{\mathrm{GS}^{3}_{\one'}}
+
\ket{\mathrm{GS}^{1}_{\one'}},
\\
&
\widehat{W}^{\,}_{\mathbf{2}}\,
\ket{\mathrm{GS}^{3}_{\one'}}
=
\ket{\mathrm{GS}^{1}_{\one'}}
+
\ket{\mathrm{GS}^{2}_{\one'}}.
\end{split}
\end{equation}
Note that the expectation value of $\widehat{W}^{\,}_{\mathbf{2}}$
is zero in any of these ground states. 
We call this phase $\cRep(S^{\,}_{3})/\mathbb{Z}^{\,}_{2}$ SSB phase. 

The $S^{\,}_{3}$-symmetric Hamiltonian \eqref{eq:Ham gen S3 reparameterized}
has sixfold degenerate ground states 
\eqref{eq:GS Z1} when $J^{\,}_{2}=J^{\,}_{4}$ and $\th=0$.
Under the duality mapping, each ground state $\ket{\mathrm{GS}^{\alpha}_{\one'}}$
is the image of the linear combinations  
$\ket{\mathrm{GS}^{+,\alpha}_{\mathbb{Z}^{\,}_{1}}}+\ket{\mathrm{GS}^{-,\alpha}_{\mathbb{Z}^{\,}_{1}}}$
that are in the subspace \eqref{eq:Z2 gauging dual subspaces}.

\item 
When $J^{\,}_{1}=J^{\,}_{3}=0$, 
the Hamiltonian \eqref{eq:Ham gen Rep S3 reparameterized} becomes
\begin{equation}
\label{eq:HRepS3 J2, J4}
\widehat{H}^{\,}_{\cRep(S^{\,}_{3});2,4}
\= 	
-
\sum_{i=1}^{L}
\left[
J^{\,}_{2}
\left(
\widehat{X}^{\,}_{i} + 
\widehat{X}^{\dagger}_{i}
\right)
+
J^{\,}_{4}\,
\left (
\hat{\mu}^{z}_{i}\,
\hat{\tau}^{x}_{i}\,
\widehat{C}^{\,}_{i}\,
\hat{\mu}^{z}_{i+1}
+
\hat{\tau}^{x}_{i}
\right )
\right].
\end{equation}
First, we note that all terms in the Hamiltonian pairwise commute.  
Therefore we can set $\hat{\tau}^{x}_i=1$. 
Second, we can minimize the $J^{\,}_{2}$ term by setting $\widehat{X}^{\,}_{i}=1$,
for which $\widehat{C}^{\,}_{i}=1$ too. 
This leaves us with the $ J^{\,}_4$
that is reduced to $ J^{\,}_{4}\,
\hat{\mu}^{z}_{i}\,
\hat{\mu}^{z}_{i+1}$. This term favors twofold degenerate ground states 
for $\hat{\mu}$ degrees of freedom, i.e.,
\begin{align}
\label{eq:GS W2}
\ket{\mathrm{GS}^{\pm}_{\mathbf{2}}}
\=
\bigotimes_{i=1}^L
\ket{\tau^{x}_{i}=1,\,\mu^{z}_{i}=\pm 1,\, X^{\,}_{i}=1}.
\end{align}
These ground states break the entire $\cRep(S^{\,}_{3})$ symmetry. 
First, the two ground states are mapped to each other under the
$\mathbb{Z}^{\,}_{2}$ symmetry
generated by $\widehat{W}^{\,}_{\one'}$.
Second, one verifies
\begin{equation}
\widehat{W}^{\,}_{\mathbf{2}} 
\ket{\mathrm{GS}^{\pm}_{\mathbf{2}}}
= 
\ket{\mathrm{GS}^{+}_{\mathbf{2}}}
+
\ket{\mathrm{GS}^{-}_{\mathbf{2}}},
\end{equation}
i.e., both ground states are mapped to the same linear combination under $\widehat{W}^{\,}_{\mathbf{2}}$.
This is to say that the vev of $\widehat{W}^{\,}_{2}$ is $1$ 
in both of the ground states. While this does not 
match the quantum dimension of by $\widehat{W}^{\,}_{\mathbf{2}}$,
we say that the non-invertible $\widehat{W}^{\,}_{\mathbf{2}}$
symmetry is not spontaneously broken. 
For this reason we call this phase $\mathbb{Z}^{\,}_{2}$
SSB phase.\footnote{See also  Ref.\ \cite{BBS231003784} where 
the same terminology is used.}
We provide further motivation for this interpretation in Sec.\ 
\ref{subsec:correlation funcs}
where we computed expectation values of order and disorder operators
in ground states \eqref{eq:GS W2}.

The $S^{\,}_{3}$-symmetric Hamiltonian \eqref{eq:Ham gen S3 reparameterized}
has a non-degenerate ground state \eqref{eq:GS S3} when $J^{\,}_{1}=J^{\,}_{3}$ and $\th=0$.
Under the duality mapping, the \emph{cat state}, $\ket{\mathrm{GS}^{+}_{\mathbf{2}}}+ 
\ket{\mathrm{GS}^{-}_{\mathbf{2}}}$ is the image of this non-degenerate ground state 
$\ket{\mathrm{GS}^{\,}_{S^{\,}_{3}}}$.

\item
When $J^{\,}_{2}=J^{\,}_{3}=0$, 
the Hamiltonian \eqref{eq:Ham gen Rep S3 reparameterized} becomes
\begin{equation}
\label{eq:HRepS3 J1, J4}
\widehat{H}^{\,}_{\cRep(S^{\,}_{3});1,4} 
:= 
-
\sum_{i=1}^L
\left \{
J^{\,}_{1}
\left(
\widehat{Z}^{\hat{\mu}^{x}_{i+1}}_{i}\,
\widehat{Z}^{\dagger}_{i+1}
+
\text{H.c.}
\right)
+
J^{\,}_{4}\,
\hat{\mu}^{z}_{i}\,
\widehat{C}^{\,}_{i}\,
\hat{\mu}^{z}_{i+1}
\right \}	+\text{const}\,.
\end{equation}
Again, the two set of operators commute, so we can simultaneously 
diagonalize the operators and minimize their eigenvalues.
There are three degenerate ground states. 
Two of them are quite simple, because they are obtained by setting 
$\widehat{Z}^{\,}_{i}=1 $ for all sites. 
As in the discussion of Eq.\ \eqref{eq:HRepS3 J2, J4}, 
such states are eigenvalue $+1$ eigenstates of the charge conjugation operators $ \widehat{C}_i $. 
The second term of Eq.\ \eqref{eq:HRepS3 J1, J4} simply becomes an Ising-like term for the 
qubits, which favors a twofold degenerate ground state manifold spanned by
\begin{subequations}
\label{eq:GS 1}
\begin{equation}
\label{eq:gs12}
\Ket{\text{GS}^{\pm}_{\mathbf{1}}} 
\= 
\bigotimes_{i=1}^{L} \Ket{\tau_i^x=1,\mu^z_{i} =\pm1, Z^{\,}_i=1}.
\end{equation}
The third degenerate ground state is 
\begin{equation}
\label{eq:gs3}
\Ket{\text{GS}^{3}_{\mathbf{1}}} 
\= 
\frac{1}{2^{L/2}}
\sum_{\{s_i = \pm 1 \}} 
\bigotimes_{i=1}^{L} 
\Ket{\tau_i^x=1, 
\mu^x_{i}=  s^{\,}_{i}\,s^{\,}_{i-1},
 Z^{\,}_i=\om^{s_i} }.
\end{equation}
\end{subequations}
The assignment of the $ \hat{\mu}^x $ eigenvalues ensures 
that the $J^{\,}_1$ term of Eq.\ \eqref{eq:HRepS3 J1, J4} is minimized in 
each summand of \eqref{eq:gs3}, while the superposition 
over different $\{s_i\}$ configurations ensures that the $J^{\,}_4$ 
term is minimized.\footnote{
Even though it is not immediately obvious that
$\Ket{\text{GS}^{3}_{\mathbf{1}}}$ 
is a short-range entangled state, in fact,
it is related to a product state by 
the action of a finite depth local unitary circuit,
$\Ket{\text{GS}^{3}_{\mathbf{1}}} = 
\prod_{j=1}^{L} \left ( \widehat{C\mu^z}^{\,}_{j,j} \widehat{C\mu^z}^{\,}_{j,j+1} \right )
\bigotimes_{i=1}^L 
\left [
\Ket{\tau^x_i = 1, \mu^x_{i} = 1}
\otimes \frac{1}{\sqrt 2} \left (\Ket{ Z^{\,}_i = \om} + \Ket{ Z^{\,}_i = \om^*} \right )
\right ] 
$, where $ \widehat{C\mu^z}_{i,j} $ is a kind of CZ operator that acts as the identity operator if 
$ Z^{\,}_i = \om $ and as $ \hat{\mu}^z_j $ if $ Z^{\,}_i=\om^* $.
}
All three of these states have the minimum possible energy 
associated with minimizing the eigenvalue of 
each of the two set of commuting terms. 
Under the action of $\cRep(S^{\,}_{3})$
symmetry these ground states transform as 
\begin{subequations}
\begin{equation}
\widehat{W}^{\,}_{\one'} \Ket{\text{GS}^{+}_{\mathbf{1}}}  = \Ket{\text{GS}^{-}_{\mathbf{1}}},
\quad
\widehat{W}^{\,}_{\one'} \Ket{\text{GS}^{-}_{\mathbf{1}}}  = \Ket{\text{GS}^{+}_{\mathbf{1}}},
\quad
\widehat{W}^{\,}_{\one'} \Ket{\text{GS}^{3}_{\mathbf{1}}}  = \Ket{\text{GS}^{3}_{\mathbf{1}}} ,
\end{equation}
and
\begin{equation}
\widehat{W}^{\,}_{\mathbf{2}} 
\Ket{\text{GS}^{3}_{\mathbf{1}}}  
=
\Ket{\text{GS}^{+}_{\mathbf{1}}}  + \Ket{\text{GS}^{-}_{\mathbf{1}}} +\Ket{\text{GS}^{3}_{\mathbf{1}}},
\quad
\widehat{W}^{\,}_{\mathbf{2}} 
\left (
\Ket{\text{GS}^{+}_{\mathbf{1}}}  + \Ket{\text{GS}^{-}_{\mathbf{1}}} 
\right )
=
2\Ket{\text{GS}^{3}_{\mathbf{1}}}.
\end{equation}
\end{subequations}
We interpret this as the $\cRep(S^{\,}_{3})$ SSB pattern
as the vev $\widehat{W}^{\,}_{\mathbf{2}}$ is vanishing in 
ground states $\ket{\mathrm{GS}^{\pm}_{\mathbf{1}}}$.
We provide further motivation for this interpretation in Sec.\ 
\ref{subsec:correlation funcs}
where we computed expectation values of order and disorder operators
in ground states \eqref{eq:GS 1}.

The $S^{\,}_{3}$-symmetric Hamiltonian \eqref{eq:Ham gen S3 reparameterized}
has a threefold degenerate ground states \eqref{eq:GS Z2} when $J^{\,}_{2}=J^{\,}_{3}$ and $\th=0$.
Under the duality mapping, the linear combination $\Ket{\text{GS}^{+}_{\mathbf{1}}} + 
\Ket{\text{GS}^{-}_{\mathbf{1}}}$ is the image of the linear combination $\ket{\mathrm{GS}^{1}_{\mathbb{Z}^{\,}_{2}}}+ \ket{\mathrm{GS}^{2}_{\mathbb{Z}^{\,}_{2}}}$
while the ground state $\Ket{\text{GS}^{3}_{\mathbf{1}}}$ is the image of 
$\ket{\mathrm{GS}^{0}_{\mathbb{Z}^{\,}_{2}}}$.
\end{enumerate}

\begin{figure}[t]
\centering
\begin{subfigure}{.35\textwidth}
\centering
\includegraphics[width=\linewidth]{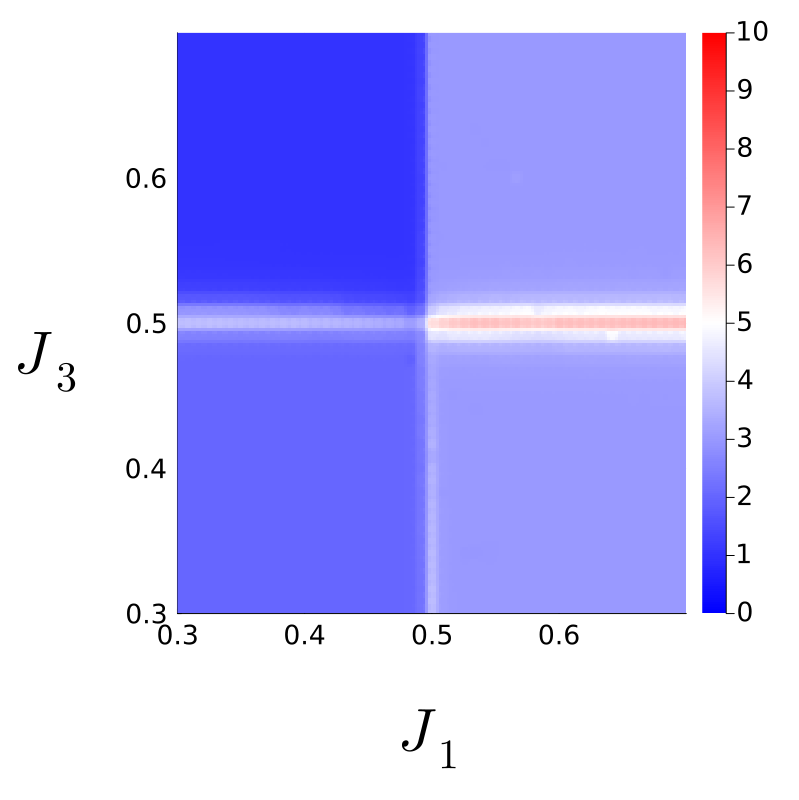}  
\caption{}
\label{fig:reps3pd-numerics-gsd}
\end{subfigure}
\begin{subfigure}{.35\textwidth}
\includegraphics[width=\linewidth]{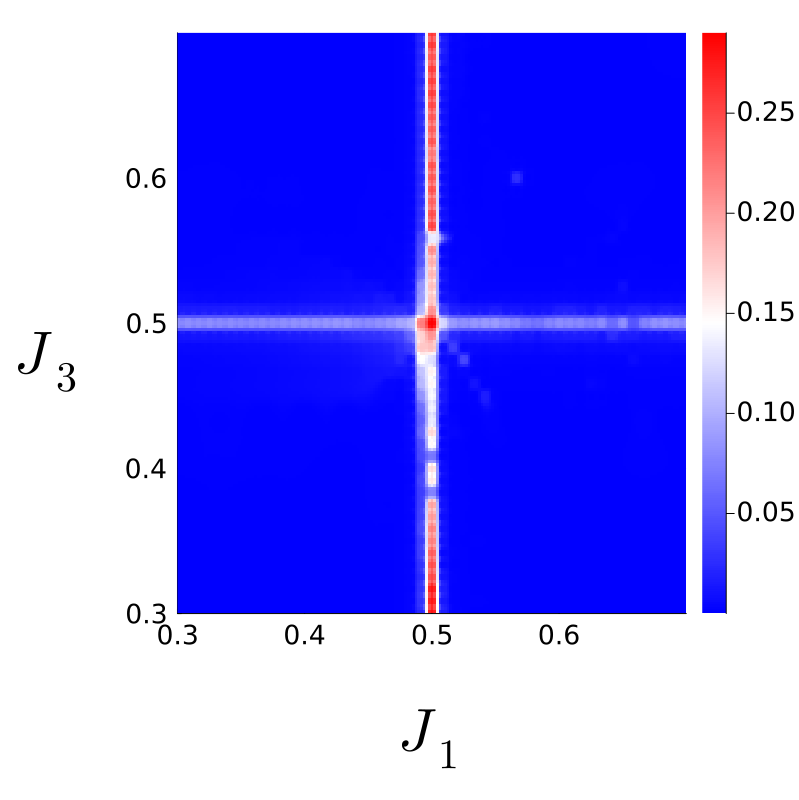} 
\caption{}
\label{fig:reps3pd-numerics-c}
\end{subfigure}
\caption{Numerical phase diagram showing GSD (a) and central charge (b) as heatmaps, as a function of $ J^{\,}_1 $ and $ J^{\,}_3 $, with $ J^{\,}_2=1-J^{\,}_1 $, $ J^{\,}_4=1-J^{\,}_3 $ everywhere and with fixed $ \th=0.1$. 
The effective system size is $L=128$.
\label{fig:reps3pd-numerics}}
\end{figure}

In conclusion, we have identified 4 fixed-point gapped ground states of the 
$\cRep(S^{\,}_3) $-symmetric Hamiltonian \eqref{eq:Ham gen Rep S3 reparameterized}. 
On general grounds (see \scn{sec:disc}), this symmetry category 
is indeed expected to have 4 gapped phases. 
Therefore, we find consistency between our lattice model 
and general category theoretic arguments. Again, for the gapped phases, 
turning on small non-zero $\th$ makes no difference.

The continuous phase transitions between these gapped phase can also be obtained from 
those between the gapped phases of $S^{\,}_{3}$-symmetric Hamiltonian 
\eqref{eq:Ham gen S3 reparameterized}. More precisely, in the language of conformal field theory, the gauging
procedure we performed in Sec.\ \ref{sec:gauging Z2 in S3} corresponds to 
the orbifold construction. 
Namely, for the Ising and 3-state Potts CFTs, gauging the $\mathbb{Z}^{\,}_{2}$ subgroup 
of $S^{\,}_{3}$ can be achieved by orbifolding the Ising symmetry and 
charge conjugation symmetry of these CFTs, respectively. 
Under orbifolding the central charge of the CFT does not change while the 
local operator content does~\cite{CFT12,G1988,DVV1989,MS1989,CRTR170600557}. 
In particular, under the $\Z_2$ orbifold operation, the Ising CFT is isomorphic to itself, while the Potts CFT is mapped to tetracritical Ising (TCI).\footnote{It is known that the TCI CFT has $\cRep(S_3)$ topological defect lines~\cite{CLY180204445}.}
As a result, we expect the same reasoning behind 
the stability of the critical lines and multicritical point to small non-zero 
$\th$ to hold 
for the $\cRep(S^{\,}_3)$-symmetric 
Hamiltonian \eqref{eq:Ham gen Rep S3 reparameterized} as well.
In the following section, we verify these expectations 
by providing numerical evidence obtained through the TEFR and DMRG algorithms.

\subsubsection{Numerical results}

As we did in \scn{sec: s3 numerics}, 
we implement the TEFR algorithm to obtain 
the phase diagram of the Hamiltonian \eqref{eq:Ham gen Rep S3 reparameterized}. 
We extract the ground state degeneracies and the central charges 
using the approach described in \scn{sec: s3 numerics}. 
For simplicity, we only focus on the case of small $\th\approx 0$ limit.
The results are shown in \fig{fig:reps3pd-numerics}. 
We find, as expected, four gapped phases of the $\cRep(S^{\,}_3)$-symmetric 
Hamiltonian \eqref{eq:Ham gen Rep S3 reparameterized} 
along with continuous phase transitions separating them from each other. 
We confirm using DMRG that the central charges at the continuous 
transition lines matches those for the phase diagram of 
$S^{\,}_{3}$-symmetric Hamiltonian~\eqref{eq:def Ham gen S3}.
In contrast, the ground state degeneracies of four gapped phases differ
as they follow the $\cRep(S^{\,}_{3})$ SSB patterns. 
The duality between the gapped ground states then holds only in the 
symmetric subspaces~\eqref{eq:Z2 gauging dual subspaces}. 

In fact, the above reasoning also holds in the large $\th$ limit of Hamiltonian
\eqref{eq:def Ham gen RepS3}. Just as it was the case for the Hamiltonian 
\eqref{eq:def Ham gen S3} with $S^{\,}_{3}$ symmetry, around $\th^{\,}_{*}\sim 
\frac{\pi}{8}$ an extended gapless region opens up in the phase diagram. 
Similarly, we can add a term that breaks the non-invertible self-duality symmetry 
implemented by $\widehat{D}^{\,}_{\cRep(S^{\,}_{3})}$ (recall Eq.\ 
\eqref{eq: def DRepS3}). This can be achieved by dualizing the perturbation
\eqref{eq:def Hperp}. Under gauging the $\mathbb{Z}^{\,}_{2}$ subgroup of $S^{\,}_{3}$
perturbation \eqref{eq:def Hperp} is mapped to
\begin{equation}
\label{eq:def Hperp RepS3}
\begin{split}
\widehat{H}^{\,}_{\perp}
:=
&-
J^{\,}_{\perp}\,
\sum_{i=1}^{L}
\left(
\hat{\tau}^{z}_{i}\,
\hat{\mu}^{x}_{i+1}
-
\hat{\mu}^{z}_{i+1}\,
\hat{\tau}^{x}_{i+1}\,
\widehat{C}^{\,}_{i+1}\,
\hat{\mu}^{z}_{i+2}
\right)
\left(\widehat{Z}^{\hat{\mu}^{x}_{i+1}}_{i}\,
\widehat{Z}^{\dagger}_{i+1}
+\widehat{Z}^{-\hat{\mu}^{x}_{i+1}}_{i}\,
\widehat{Z}^{\,}_{i+1}\right)
\\
&
+
J^{\,}_{\perp}\,
\sum_{i=1}^{L}
\left(
\hat{\tau}^{z}_{i}
-
\hat{\tau}^{x}_{i}
\right)
\left(\widehat{X}^{\,}_{i}+\widehat{X}^{\dagger}_{i}\right),
\end{split}
\end{equation}
which is odd under the non-invertible $\widehat{D}^{\,}_{\cRep(S^{\,}_{3})}$
symmetry. When this term is added to the $\cRep(S^{\,}_{3})$-symmetric
Hamiltonian \eqref{eq:def Ham gen RepS3}, depending on the sign of $J^{\,}_{\perp}$,
shape of the phase diagram matches either that in Fig.\ \ref{fig:s3-jperp-negative}
or Fig.\ \ref{fig:s3-jperp-positive}. This allows the direct continuous phase 
transitions between $\cRep(S^{\,}_{3})$-symmetric and $\cRep(S^{\,}_{3})$
SSB phases or between $\cRep(S^{\,}_{3})/\mathbb{Z}^{\,}_{2}$
SSB and $\mathbb{Z}^{\,}_{2}$ SSB phases.

\section{Self-dual spin chains and their SymTO description } 
\label{sec:symTO}

Emergent symmetries are an important characteristic feature of so-called gapless 
liquid states. These symmetries often take the form of generalized symmetries \eg 
higher-form, higher-group, non-invertible; each of these types 
may also be ('t Hooft) anomalous.  
A complete understanding of gapless phases, therefore, 
requires a general theory of emergent generalized symmetries.
The SymTO framework~\footnote{See Appendix
\ref{app:SymTO} for a review of SymTO.} is a proposal for such a theory; it attempts to classify gapless states
in terms of topological orders in one higher dimension.  

In Secs.\ \ref{sec:S3model} and \ref{sec:RepS3model}, we introduced spin chains which respect non-invertible self-duality symmetries. 
To understand how these non-invertible symmetries
constrain the low energy dynamics of the lattice model, we need to use the
symmetry-topological-order correspondence, 
and find out which SymTOs describe them.
In what follows, we use this correspondence to first understand $S^{\,}_{3}$
and $\cRep(S^{\,}_{3})$ symmetries. We will then obtain the SymTO that corresponds 
to the enhancement of these symmetries by the non-invertible self-duality symmetries.

\begin{table}[t!]
\begin{subtable}[l]{0.3\textwidth}
\small
\centering
\begin{tabular}{|c|c|c|}
\hline
$\eD(S^{\,}_{3})$ & $s$ & $d$ 
\\ 
\hline
$\one$ & 0 & 1 
\\ 
\hline
$\one'$ & 0 & 1
\\ 
\hline
$\two$ & 0 & 2
\\ 
\hline
$r$ & 0 & 2
\\ 
\hline
$r_1$ & $\frac{1}{3}$ & 2
\\ 
\hline
$r_2$ &  $\frac{2}{3}$ & 2
\\ 
\hline
$s$ & 0 & 3
\\ 
\hline
$s_1$ & $\frac{1}{2}$ & 3
\\ 
\hline
\end{tabular}
\newline\newline
(a)\hspace{0.15\textwidth}
\end{subtable}
\hfill
\begin{subtable}[r]{0.6\textwidth}
\small
\centering
\begin{align*}
S= 
\begin{pmatrix}
1
& 1
& 2
& 2
& 2
& 2
& 3
& 3 \\ 
1
& 1
& 2
& 2
& 2
& 2
& -3
& -3 \\ 
2
& 2
& 4
& -2
& -2
& -2
& 0
& 0 \\ 
2
& 2
& -2
& 4
& -2
& -2
& 0
& 0 \\ 
2
& 2
& -2
& -2
& -2
& 4
& 0
& 0 \\ 
2
& 2
& -2
& -2
& 4
& -2
& 0
& 0 \\ 
3
& -3
& 0
& 0
& 0
& 0
& 3
& -3 \\ 
3
& -3
& 0
& 0
& 0
& 0
& -3
& 3 \\ 
\end{pmatrix}
\end{align*}
\\
\vspace{0.75mm}
(b)
\end{subtable}
\\\\
\begin{subtable}[b]{0.9\textwidth}
\centering\small
\begin{tabular}{ |c||c|c|c|c|c|c|c|c|}
 \hline 
 & $\one$  & $\one'$  & $\two$  & $r$  & $r_1$  & $r_2$  & $s$  & $s_1$ \\ 
\hline 
 \hline 
$\one$  & $ \one$  & $ \one'$  & $ \two$  & $ r$  & $ r_1$  & $ r_2$  & $ s$  & $ s_1$  \\ 
 \hline 
$\one'$  & $ \one'$  & $ \one$  & $ \two$  & $ r$  & $ r_1$  & $ r_2$  & $ s_1$  & $ s$  \\ 
 \hline 
$\two$  & $ \two$  & $ \two$  & $ \one \oplus \one' \oplus \two$  & $ r_1 \oplus r_2$  & $ r \oplus r_2$  & $ r \oplus r_1$  & $ s \oplus s_1$  & $ s \oplus s_1$  \\ 
 \hline 
$r$  & $ r$  & $ r$  & $ r_1 \oplus r_2$  & $ \one \oplus \one' \oplus r$  & $ \two \oplus r_2$  & $ \two \oplus r_1$  & $ s \oplus s_1$  & $ s \oplus s_1$  \\ 
 \hline 
$r_1$  & $ r_1$  & $ r_1$  & $ r \oplus r_2$  & $ \two \oplus r_2$  & $ \one \oplus \one' \oplus r_1$  & $ \two \oplus r$  & $ s \oplus s_1$  & $ s \oplus s_1$  \\ 
 \hline 
$r_2$  & $ r_2$  & $ r_2$  & $ r \oplus r_1$  & $ \two \oplus r_1$  & $ \two \oplus r$  & $ \one \oplus \one' \oplus r_2$  & $ s \oplus s_1$  & $ s \oplus s_1$  \\ 
 \hline 
$s$  & $ s$  & $ s_1$  & $ s \oplus s_1$  & $ s \oplus s_1$  & $ s \oplus s_1$  & $ s \oplus s_1$  & $ \one \oplus \two \oplus r \oplus r_1 \oplus r_2$  & $ \one' \oplus \two \oplus r \oplus r_1 \oplus r_2$  \\ 
 \hline 
$s_1$  & $ s_1$  & $ s$  & $ s \oplus s_1$  & $ s \oplus s_1$  & $ s \oplus s_1$  & $ s \oplus s_1$  & $ \one' \oplus \two \oplus r \oplus r_1 \oplus r_2$  & $ \one \oplus \two \oplus r \oplus r_1 \oplus r_2$  \\ 
 \hline 
\end{tabular}
\newline\newline
(c)
\end{subtable}
\caption{(a) Topological spin $s$ and quantum dimension $d$ of
the eight anyons of SymTO $\eD(S^{\,}_{3})$. (b) $S$-matrix of 
these excitations which encodes mutual braiding statistics. (c) Fusion rules of
these excitations.
}
\label{tab:DS3 data}
\end{table}

\subsection{SymTO of \texorpdfstring{$S^{\,}_3$}{S^{\,}_3} symmetry}

The symmetry data of a 1+1d bosonic system with $ S^{\,}_{3} $ symmetry can be
encapsulated completely in its SymTO~\cite{JW191213492},
which is the $S^{\,}_{3}$ quantum double $\eD(S^{\,}_{3})$, 
\ie $S^{\,}_3$ topological order in 2+1d. From the SymTO point of view, it
has been argued~\cite{KZ200514178,CW220506244} that the gapped phases allowed by 
$S^{\,}_{3} $ symmetry are in one-to-one correspondence with the gapped boundaries
of the SymTO $\eD(S^{\,}_{3})$.\footnote{Similar statements have also appeared
elsewhere in the literature.\cite{MMT220710712,ZC230401262,BBS231003784}}

The $\eD(S^{\,}_{3})$ SymTO has eight anyons whose
topological spins $s$, quantum dimensions $d$, $S$-matrix, and fusion rules are
given in Table \ref{tab:DS3 data}. 
The anyons labeled $s$ and $r$ carry the  gauge fluxes for their associated conjugacy class.\footnote{See \app{app:S3} for notations.} The
anyon $\one'$ carries the 1-dimensional irrep of $ S^{\,}_{3}$ and
can be viewed as a $ \Z^{\,}_{2}$ charge.  From the $S$ matrix, we see that 
$\one'$ and $s$ have mutual $\pi$ statistics, consistent with $s$ being the $ \Z^{\,}_{2}$ flux.

There are four Lagrangian condensable algebras of $ \eD(S^{\,}_{3}) $, which
correspond to four maximal subsets of bosonic anyons in $\eD(S^{\,}_{3})$ with
trivial mutual statistics between them.  We can condense all the anyons in a
Lagrangian condensable algebra on 1+1d boundary of the 2+1d SymTO, which 
gives rise to a gapped boundary~\cite{KK11045047}. In turn, such condensable algebras 
correspond to gapped phases for systems with $S^{\,}_3$ symmetry.  
The four Lagrangian condensable algebras of $\eD(S^{\,}_{3})$
are denoted as follows: 
\begin{enumerate}[(i)]
\item 
$\cA^{\,}_1 = \one \oplus \one' \oplus 2\,\two $ corresponds to an $S^{\,}_{3}$ SSB phase, 
\ie $S^{\,}_{3}$ ferromagnet

\item 
$\cA^{\,}_2 = \one \oplus \one' \oplus 2\,r$ corresponds to a $\Z^{\,}_2$ SSB phase

\item 
$\cA^{\,}_3 = \one \oplus \two \oplus s $ corresponds to a $\Z^{\,}_{3}$ SSB phase

\item 

$\cA^{\,}_4 = \one \oplus r \oplus s $ corresponds to an $S^{\,}_{3}$-symmetric phase,
\ie $S^{\,}_{3}$ paramagnet
\end{enumerate}

\subsection{SymTO of $\cRep(S^{\,}_3)$ symmetry}

As we exemplified in Sec.\ \ref{sec:RepS3model}, the Morita equivalent
$S^{\,}_{3}$ and $ \cRep(S^{\,}_{3}) $ symmetries have the property that Hamiltonians
with these symmetries have identical spectra when restricted to the
respective symmetric sub-Hilbert space; this latter aspect was emphasized as ``holo-equivalence" in \Rf{KZ200514178}. As a result, $S^{\,}_{3}$-symmetric
models and $ \cRep(S^{\,}_{3}) $-symmetric models have identical phase diagrams. 
However, the corresponding phases and phase
transitions may be given different names due to the difference in symmetry labels. 
The Morita equivalence between $S^{\,}_{3}$ and $ \cRep(S^{\,}_{3})
$ symmetries follows from the fact that they have the same SymTO.  
The SymTO can be calculated by computing the algebra of the associated patch operators; some
related examples were discussed in \Rfs{CW220303596,IW231005790}. 

The four Lagrangian condensable algebras of $\eD(S^{\,}_{3})$ also give
rise to four gapped phases for $\cRep(S^{\,}_3)$ symmetric models. 
In terms of the Morita equivalent $\cRep(S^{\,}_{3})$ symmetry, $\cA^{\,}_{1}$ 
and $\cA^{\,}_{4}$ correspond to $\cRep(S^{\,}_{3})$-symmetric and $\cRep(S^{\,}_{3})$
SSB phases, respectively. The Lagrangian algebras $\cA^{\,}_{2}$
and $\cA^{\,}_{3}$
are more subtle; guided by the phase diagram of the lattice models introduced in \scn{sec:RepS3model},
we find that they correspond to $\cRep(S^{\,}_{3})/\mathbb{Z}^{\,}_{2}$ and 
$\mathbb{Z}^{\,}_{2}$ SSB phases, respectively.\footnote{Here, we match the 
Lagrangian algebras with the gapped 
phases of the Hamiltonian \eqref{eq:def Ham gen RepS3vee} with $\cRep(S^{\vee}_{3})$
symmetry.}

\subsection{SymTO of the self-duality symmetry}

In \Rf{CW220506244}, the authors also highlighted the importance of an
automorphism of $\eD(S^{\,}_{3})$ associated with the permutation of the anyons $\two$ and $r$. This automorphism of the SymTO suggests
a non-invertible self-duality symmetry of the boundary theory.

Here, we would like stress an important difference between $\eD(S^{\,}_{3})$ SymTO
and $\eD(S^{\,}_{3})$ SymTFT.  In $\eD(S^{\,}_{3})$ SymTFT, the automorphism
$\two \leftrightarrow r$ implies a symmetry of SymTFT.  In contrast,
$\eD(S^{\,}_{3})$ SymTO is just an 2+1d $S^{\,}_3$ lattice gauge theory with matter.
Thus in general, the SymTO (\ie the lattice gauge theory) does not have
any symmetry.  This corresponds to the fact that our $S^{\,}_3$ and
$\cRep(S^{\,}_{3})$ lattice models, in general, do not have the self-dual
symmetry, and their SymTO is the $\eD(S^{\,}_{3})$ SymTO, \emph{without the
$\two \leftrightarrow r$ automorphism symmetry 
$\Z^{\two \leftrightarrow r}_2$}.

The presence of $\two \leftrightarrow r$ automorphism in the $\eD(S^{\,}_{3})$ SymTFT implies that we
can fine tune the  $\eD(S^{\,}_{3})$ SymTO so that it has the automorphism
symmetry $\Z^{\two \leftrightarrow r}_2$. This, in turn, implies that we can
fine tune the  $S^{\,}_3$ and $\cRep(S^{\,}_{3})$ symmetric lattice models, so that
these fine-tuned models have an additional self-duality symmetry.  Such an
existence of lattice self-duality symmetry was assumed in \Rf{CJW221214432}; 
in Secs.\ \ref{sec:S3model} and \ref{sec:RepS3model},
we explicitly constructed this lattice self-duality
symmetry, and confirmed this conjecture.

\begin{table}[t!]
\begin{subtable}{\textwidth}
\small
\centering
\hspace{0.14\textwidth}
\begin{tabular}{|c|c|c|}
\hline
$\mathrm{SU}(2)^{\,}_{4}$ & $s$ & $d$
\\ 
\hline
$\one$ & $0$ & $1$
\\ 
\hline
$e$ & $0$ & $1$
\\ 
\hline
$m$ & $\frac{1}{8}$ & $\sqrt{3}$
\\ 
\hline
$m_1$ & $\frac{5}{8}$ & $\sqrt{3}$
\\ 
\hline
$q$ & $\frac{1}{3}$ & $2$
\\ 
\hline
\end{tabular}
\qquad
\begin{tabular}{|c|c|c|}
\hline
$\mathrm{JK}^{\,}_{4}$ & $s$ & $d$
\\ 
\hline
$\one$ & $0$ & $1$
\\ 
\hline
$e$ & $0$ & $1$
\\ 
\hline
$m$ & $\frac{1}{8}$ & $\sqrt{3}$
\\ 
\hline
$m_1$ & $\frac{5}{8}$ & $\sqrt{3}$
\\ 
\hline
$q$ & $\frac{2}{3}$ & $2$
\\ 
\hline
\end{tabular}
\newline\newline
(a)
\end{subtable}
\newline
\begin{subtable}{\textwidth}
\small
\centering
\begin{align*}
S^{\,}_{\mathrm{SU(2)}^{\,}_{4}}
=
\begin{pmatrix}
1
& 1
& \sqrt{3}
& \sqrt{3}
& 2 \\ 
1
& 1
& -\sqrt{3}
& -\sqrt{3}
& 2 \\ 
\sqrt{3}
& -\sqrt{3}
& \sqrt{3}
& -\sqrt{3}
& 0 \\ 
\sqrt{3}
& -\sqrt{3}
& -\sqrt{3}
& \sqrt{3}
& 0 \\ 
2
& 2
& 0
& 0
& -2 \\ 
\end{pmatrix}
\qquad
S^{\,}_{\mathrm{JK}^{\,}_{4}}
= 
\begin{pmatrix}
1
& 1
& \sqrt{3}
& \sqrt{3}
& 2 \\ 
1
& 1
& -\sqrt{3}
& -\sqrt{3}
& 2 \\ 
\sqrt{3}
& -\sqrt{3}
& -\sqrt{3}
& \sqrt{3}
& 0 \\ 
\sqrt{3}
& -\sqrt{3}
& \sqrt{3}
& -\sqrt{3}
& 0 \\ 
2
& 2
& 0
& 0
& -2 \\ 
\end{pmatrix}
\end{align*}
\\
\vspace{0.75mm}
(b)
\end{subtable}
\\\\
\begin{subtable}[b]{\textwidth}
\centering\small
\hspace{0.14\textwidth}
\begin{tabular}{ |c||c|c|c|c|c|}
\hline 
& $\one$  & $e$  & $m$  & $m_1$  & $q$ \\ 
\hline 
 \hline 
$\one$  & $ \one$  & $ e$  & $ m$  & $ m_1$  & $ q$  \\ 
 \hline 
$e$  & $ e$  & $ \one$  & $ m_1$  & $ m$  & $ q$  \\ 
 \hline 
$m$  & $ m$  & $ m_1$  & $ \one \oplus q$  & $ e \oplus q$  & $ m \oplus m_1$  \\ 
\hline 
$m_1$  & $ m_1$  & $ m$  & $ e \oplus q$  & $ \one \oplus q$  & $ m \oplus m_1$  \\ 
\hline 
$q$  & $ q$  & $ q$  & $ m \oplus m_1$  & $ m \oplus m_1$  & $ \one \oplus e \oplus q$  \\ 
\hline 
\end{tabular}
\newline\newline
(c)
\end{subtable}
\caption{(a) Topological spin $s$ and quantum dimension $d$ of
the five anyons of $\mathrm{SU(2)}^{\,}_{4}$ and $\mathrm{JK}^{\,}_{4}$ topological orders. 
(b) $S$-matrices of these excitations for the two topological 
orders which encodes mutual braiding statistics. 
(c) Fusion rules of these five anyons which is identical for $\mathrm{SU(2)}^{\,}_{4}$ and 
$\mathrm{JK}^{\,}_{4}$ topological orders.
}
\label{tab:SU(2)4 and JK4 Data}
\end{table}

The fine-tuned self-dual lattice models have a larger symmetry which include
both the self-duality symmetry and either the $S^{\,}_3$ or the $\cRep(S^{\,}_{3})$
symmetries. Thus, the self-dual lattice models must have a larger SymTO.  Such a
larger SymTO can be obtained by gauging the $ \Z_{2}^{\two \leftrightarrow r}$
automorphism symmetry in $\eD(S^{\,}_{3})$ SymTO. In \Rf{BZ14104540}, this
guaging procedure was carried out and 
the larger SymTO is found to be either 
$\mathrm{SU}(2)^{\,}_{4}\boxtimes\overline{\mathrm{SU}(2)}^{\,}_{4}$ 
or $\mathrm{JK}^{\,}_{4}\boxtimes \overline{\mathrm{JK}}^{\,}_{4}$ topological order. Note
that there still remains a two-fold ambiguity, coming from the possibility
of the stacking a $ \Z_{2}^{\two \leftrightarrow r}$ SPT state to the SymTO,
before the $ \Z_{2}^{\two \leftrightarrow r}$ gauging.  
The anyon data for the $\mathrm{SU}(2)^{\,}_{4}$ and
$\mathrm{JK}^{\,}_{4}$ topological orders are shown in Table \ref{tab:SU(2)4 and JK4 Data}.
From the fusion rule $e\otimes e = \one$, we see that $e$ carries a $ \Z^{\,}_{2}$ gauge charge.
From the $S$-matrix, we see that $m$ (and $m_1$) carries the corresponding
$ \Z^{\,}_{2}$ gauge flux. On an appropriate boundary of the SymTO, $e$ would reduce to a 
$ \Z^{\,}_{2}$ symmetry charge while $m$ (and $m_1$) would reduce to the corresponding domain walls.

Later, we will show that the generalized symmetries in our self-dual
$S^{\,}_3$-symmetric and  self-dual $\cRep(S^{\,}_3)$-symmetric models are both
described by the $\mathrm{JK}^{\,}_{4}\boxtimes \overline{\mathrm{JK}}^{\,}_{4}$ SymTO. 
Therefore, we momentarily concentrate on 
$\mathrm{JK}^{\,}_{4}\boxtimes \overline{\mathrm{JK}}^{\,}_{4}$ SymTO.
The  $\mathrm{JK}^{\,}_{4}\boxtimes \overline{\mathrm{JK}}^{\,}_{4}$ 
and $\eD(S^{\,}_{3})$ SymTOs can
have a gapped domain wall between them, which describes the breaking of the
$ \Z_{2}^{\two \leftrightarrow r}$ self-duality symmetry. 
This reduces the $\mathrm{JK}^{\,}_{4}\boxtimes \overline{\mathrm{JK}}^{\,}_{4}$ SymTO to 
$\eD(S^{\,}_{3})$ SymTO.  Such a domain wall is 
created by the condensation of $e \bar e$ in $\mathrm{JK}^{\,}_{4}\boxtimes
\overline{\mathrm{JK}}^{\,}_{4}$ SymTO 
(and no  condensation in the $\eD(S^{\,}_{3})$ SymTO). 
More precisely, the  domain wall is described by the following
condensable algebra 
\begin{align}
\cA &=  (\one,\one,\one)
\oplus (e,\bar e,\one)
\oplus (\one,\bar e,\one')
\oplus (e,\one,\one')
\oplus (q,\bar q,\two)
\oplus (q,\bar q,r)
\oplus (q,\one,r_1)
\oplus (q,\bar e,r_1)
\nonumber\\ &\ \ \ \ 
\oplus (\one,\bar q,r_2)
\oplus (e,\bar q,r_2)
\oplus (m,\bar m,s)
\oplus (m_1,\bar m_1,s)
\oplus (m,\bar m_1,s_1)
\oplus (m_1,\bar m,s_1)
%\nonumber\\
%\cA &=  % for SU(2)
%(\one,\one,\one)
%\oplus (e,\bar e,\one)
%\oplus (\one,\bar e,\one')
%\oplus (e,\one,\one')
%\oplus (q,\bar q,\two)
%\oplus (q,\bar q,r)
%\oplus (\one,\bar q,r_1)
%\oplus (e,\bar q,r_1)
%\nonumber\\ &\ \ \ \
%\oplus (q,\one,r_2)
%\oplus (q,\bar e,r_2)
%\oplus (m,\bar m,s)
%\oplus (m_1,\bar m_1,s)
%\oplus (m,\bar m_1,s_1)
%\oplus (m_1,\bar m,s_1),
\end{align}
in the topological order
$
\mathrm{JK}^{\,}_{4}\boxtimes \overline{\mathrm{JK}}^{\,}_{4} \boxtimes \overline{\eD}(S^{\,}_3)
= \mathrm{JK}^{\,}_{4}\boxtimes \overline{\mathrm{JK}}^{\,}_{4} \boxtimes \eD(S^{\,}_3)
$.
This condensable algebra allows us to relate the anyons in 
$\eD(S^{\,}_{3})$ SymTO to the anyons in 
$\mathrm{JK}^{\,}_{4}\boxtimes \overline{\mathrm{JK}}^{\,}_{4}$ SymTO.  
The term  $(\one,\bar
e,\one')$ in $\cA$ means that the anyon $\one'$ in $\eD(S^{\,}_{3})$ SymTO and
the anyon $\one\otimes \bar e=\bar e$ in 
$\mathrm{JK}^{\,}_{4}\boxtimes\overline{\mathrm{JK}}^{\,}_{4}$ SymTO can
condense on the domain wall.  
In other words, after going through 
the domain wall, the $\bar e$ anyon  in 
$\mathrm{JK}^{\,}_{4}\boxtimes \overline{\mathrm{JK}}^{\,}_{4}$ SymTO turns into the
$\one'$ anyon in $\eD(S^{\,}_{3})$ SymTO.  
Similarly, the term $(e,\one,\one')$ in $\cA$ means that,  
after going through the domain wall, 
the $e$ anyon in $\mathrm{JK}^{\,}_{4}\boxtimes \overline{\mathrm{JK}}^{\,}_{4}$ 
SymTO turns into the $\one'$ anyon in $\eD(S^{\,}_{3})$ SymTO. 
Thus the $e$ anyon and  the $\bar e$ anyon  in
$\mathrm{JK}^{\,}_{4}\boxtimes \overline{\mathrm{JK}}^{\,}_{4}$ 
SymTO carries the $ \Z^{\,}_{2}$-charge of the
$S^{\,}_3= \Z^{\,}_{3}\rtimes  \Z^{\,}_{2}$ symmetry.  
The corresponding $ \Z^{\,}_{2}$-flux is carried by
$m\bar m$ anyon in $\mathrm{JK}^{\,}_{4}\boxtimes \overline{\mathrm{JK}}^{\,}_{4}$ SymTO, 
which has a $\pi$-mutual statistics with both $e$ and $\bar e$ anyons. 
This is also consistent with the term $(m,\bar m,s) $ in $\cA$, 
which implies that after going through
the domain wall, the $m\bar m$ anyon 
in $\mathrm{JK}^{\,}_{4}\boxtimes \overline{\mathrm{JK}}^{\,}_{4}$ SymTO
turns into the $s$ anyon (the $ \Z^{\,}_{2}$-flux) in $\eD(S^{\,}_{3})$ SymTO.  Thus,
the string operator that creates a pair of  $m\bar m$-anyons generates the
$ \Z^{\,}_{2}$ (of $S^{\,}_3$) transformation.

The Abelian anyon $e\bar e$ 
in $\mathrm{JK}^{\,}_{4}\boxtimes \overline{\mathrm{JK}}^{\,}_{4}$ SymTO does not
carry the $ \Z^{\,}_{2}$ charge of the $S^{\,}_3$.  But it carries the 
$\Z_{2}^{\two\leftrightarrow r}$ charge of the self-duality symmetry.  This is consistent with
the fact that the condensation $\cA= \one\oplus e\bar e$ breaks the self-dual
symmetry and reduces the $\mathrm{JK}^{\,}_{4}\boxtimes \overline{\mathrm{JK}}^{\,}_{4}$ SymTO to
$\eD(S^{\,}_{3})$ SymTO~\cite{CW220506244,CJW221214432,BBS231217322}
\begin{align}
 (\mathrm{JK}^{\,}_{4}\boxtimes \overline{\mathrm{JK}}^{\,}_{4})_{/\one\oplus e\bar e}=
\eD(S^{\,}_{3}).
\end{align}
To summarize, the
anyons in $\mathrm{JK}^{\,}_{4}\boxtimes \overline{\mathrm{JK}}^{\,}_{4}$ and $\eD(S^{\,}_{3})$
SymTOs have the following identification under the condensation of $\one \oplus e\bar e$
\begin{align}
\label{SymTOSymTO}
 e\bar e \to \one,\ \ \
 e \to \one',\ \ \
 \bar e \to \one',\ \ \
 m\bar m \to s,\ \ \
 q\bar q \to \two \oplus r,\ \ \
 q \to r_2,\ \ \
 \bar q \to r_1 .
\end{align}
Since $e\bar e$ has $\pi$-mutual statistics with $m$ and $\bar m$ in
$\mathrm{JK}^{\,}_{4}\boxtimes \overline{\mathrm{JK}}^{\,}_{4}$ SymTO, 
anyons $m$ and $\bar m$ are $ \Z_{2}^{\two\leftrightarrow r}$-flux 
for the self-duality symmetry. In other words, the
string operator that creates a pair of $m$-anyons generates the self-duality
transformation. Similarly, the string operator that creates a pair of  $\bar
m$-anyons also generates the self-duality transformation. The two
transformations differ by a $ \Z^{\,}_{2}$ transformation. Because $m$ has quantum
dimension $\sqrt 3$, the transformation generated by the string operator of
$m$ is an intrinsic non-invertible symmetry.\footnote{A (generalized) symmetry
is called intrinsically non-invertible \cite{KZZ220501104,KOZ220911062} if all its Morita equivalent symmetries are
non-invertible. This is to say that 
a symmetry with non-integral quantum dimension cannot be Morita equivalent to
(sum of) simple objects with integral quantum dimension.} The non-integral quantum dimension also implies that the symmetry
is anomalous,\footnote{By definition, a (generalized) symmetry is
anomaly-free if it allows gapped non-degenerate ground state on all closed
space.} since the anomaly-free generalized symmetry 
always have integral quantum dimensions~\cite{TW191202817,KZ200308898,KZ200514178}.

\begin{figure}[t!]
\centering
\includegraphics[width=0.4\linewidth]{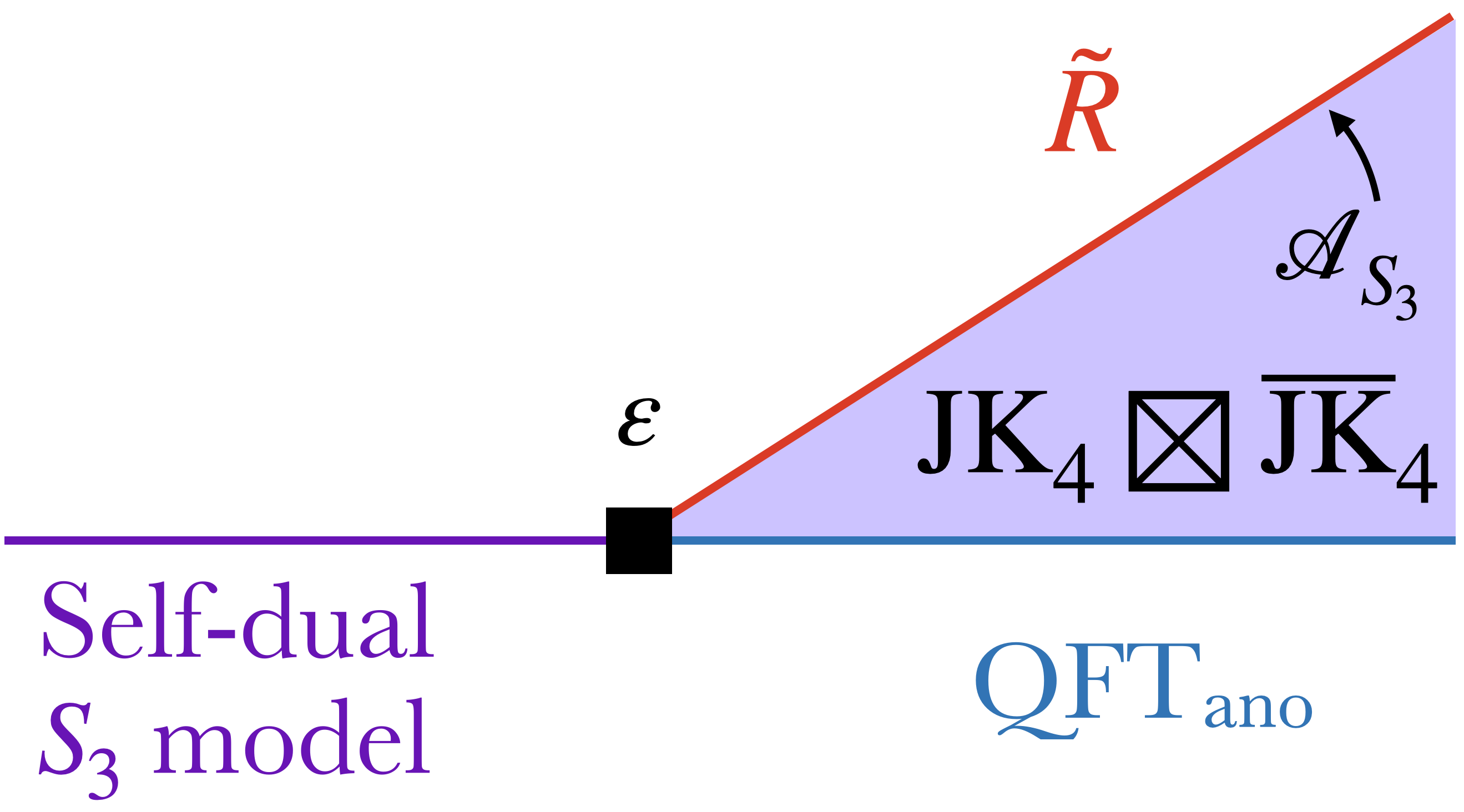}
% {FIGURES/QFTS3}
\ \ \ \ \
\includegraphics[width=0.4\linewidth]
{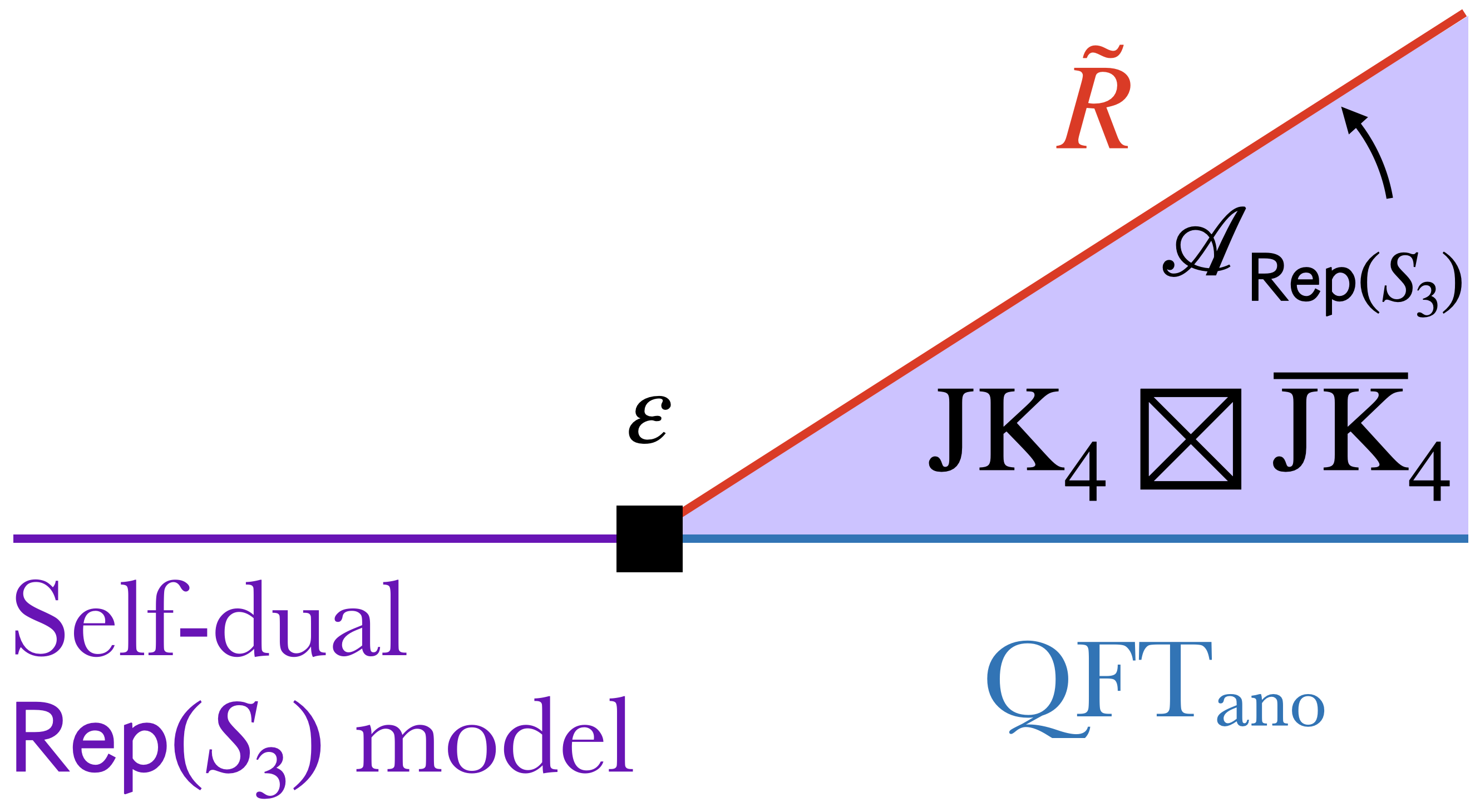}
% {FIGURES/QFTRepS3}
\caption{The isomorphic holographic decomposition reveals the generalized symmetry in self-dual
$S_3$-symmetric models and in self-dual $\cRep(S_3)$-symmetric models, \ie reveal the SymTO 
$\eM=  \mathrm{JK}^{\,}_{4}\boxtimes \overline{\mathrm{JK}}^{\,}_{4}$ and the fusion category $\t\cR$ for the symmetry defects: 
(Left) The bulk $\mathrm{JK}^{\,}_{4}\boxtimes \overline{\mathrm{JK}}^{\,}_{4}$  SymTO and the
$\cA_{S^{\,}_3}$-condensation induced topological boundary $\t\cR$ describes the
$S^{\,}_3$-symmetry plus the self-duality symmetry in the self-dual $S^{\,}_3$-symmetric
models.  (Right) The bulk $\mathrm{JK}^{\,}_{4}\boxtimes \overline{\mathrm{JK}}^{\,}_{4}$  SymTO and the
$\cA_{\cRep(S^{\,}_3)}$-condensation induced topological boundary $\t\cR$ describes
the $\cRep(S^{\,}_3)$-symmetry plus the self-duality symmetry in the self-dual
$\cRep(S^{\,}_3)$-symmetric models.  } \label{fig:QFTS3}
\end{figure}

In Refs.\ \cite{KZ150201690,KZ200308898,KZ200514178}, an isomorphic holographic
decomposition $(\eps, \t \cR)$ of a model is introduced to expose the symmetry
and the SymTO in the model (see Fig. \ref{fig:QFTS3}):
\begin{align}
\label{decomp}
 \text{model}
\stackrel{(\eps, \t \cR)}{\cong} \t \cR \boxtimes_{\mathrm{JK}^{\,}_{4}\boxtimes
\overline{\mathrm{JK}}^{\,}_{4}} \mathrm{QFT}_\text{ano},
\end{align}
where the boundary $\t\cR$ and the bulk $\mathrm{JK}^{\,}_{4}\boxtimes \overline{\mathrm{JK}}^{\,}_{4}$ 
are assumed to have infinite energy gap.  
A similar picture was also obtained later in Refs.\ \cite{ABS211202092,FT220907471}; 
also see Ref.\ \cite{CJW221214432} and Appendix \ref{app:SymTO} for a short review.  
The isomorphic holographic decomposition \eqref{decomp} has the following
physical meaning. The model is exactly simulated by the composite system 
$ \t\cR \boxtimes_{\mathrm{JK}^{\,}_{4}\boxtimes
\overline{\mathrm{JK}}^{\,}_{4}} \mathrm{QFT}^{\,}_\text{ano}$.
For example, the model and the composite system have the identical energy spectrum below the
energy gaps of the boundary $\t\cR$ and the bulk.
The local low-energy properties of the model are captured by a quantum field theory
$\mathrm{QFT}^{\,}_\text{ano}$~\footnote{This is referred to as ``physical boundary" in the SymTFT literature.} 
(which has a gravitational anomaly characterized by the $\mathrm{JK}^{\,}_{4}\boxtimes 
\overline{\mathrm{JK}}^{\,}_{4}$ SymTO), while the fully gapped boundary $\t\cR$ and the
bulk $\mathrm{JK}^{\,}_{4}\boxtimes \overline{\mathrm{JK}}^{\,}_{4}$ SymTO cover the global properties of the
model (such as ground state degeneracy).

Using the SymTO correspondence described above, 
we find that the $\mathrm{JK}^{\,}_{4}\boxtimes \overline{\mathrm{JK}}^{\,}_{4}$ SymTO 
has only two Morita equivalent symmetries, characterized by two different choices of the gapped boundary 
$\t\cR$ in Fig.\ \ref{fig:QFTS3}. 
The two Lagrangian condensable algebras that give rise to these two choices of $\t\cR$ are:
\begin{equation}
\label{LcAlg}
\begin{split}
\cA_{S^{\,}_3} &=  (\one,\one)
\oplus (\one,\bar e)
\oplus (e,\one)
\oplus (e,\bar e)
\oplus 2(q,\bar q),
\\
\cA_{\cRep(S^{\,}_{3})} &=  (\one,\one)
\oplus (e,\bar e)
\oplus (m,\bar m)
\oplus (m_1,\bar m_1)
\oplus (q,\bar q).
\end{split}
\end{equation}
From Eq.\ \eqref{SymTOSymTO}, we see that the
$\cA_{S^{\,}_3}$ condensation condenses the $S^{\,}_3$-charges
$\one'$ and $\two$, as well as the $ \Z_{2}^{\two \leftrightarrow
r}$-charge of the self-duality symmetry. This suggests that the
$\cA_{S^{\,}_3}$ condensation in the $\t\cR$ boundary leads to the $S^{\,}_3$ symmetry together with
the self-duality symmetry of $\mathrm{QFT}^{\,}_\text{ano}$, \ie 
the symmetry of the self-dual $S^{\,}_3$ symmetric model studied in Section
\ref{sec:S3model} (see Fig. \ref{fig:QFTS3} (left)).
Following the same logic, the $\cA_{\cRep(S^{\,}_3)}$ condensation condenses the
$S^{\,}_3$ fluxes $s$ and $r$, as well as the $ \Z_{2}^{\two
\leftrightarrow r}$-charge of the self-duality symmetry.  This suggests that the
$\cA_{\cRep(S^{\,}_3)}$ condensation in the $\t\cR$ boundary 
leads to the $\cRep(S^{\,}_3)$ symmetry 
together with the self-duality symmetry
of $\mathrm{QFT}^{\,}_\text{ano}$,
\ie the symmetry of the self-dual $\cRep(S^{\,}_3)$
symmetric model studied in Section \ref{sec:RepS3model} (see Fig. \ref{fig:QFTS3}(right)).

The SymTO identified here classifies the gapped
phases of the self-dual $S^{\,}_3$ symmetric model
in terms of possible gapped boundaries $\mathrm{QFT}^{\,}_\text{ano}$, induced by
Lagrangian condensable algebras in \eqref{LcAlg}.  The self-dual $S^{\,}_3$
symmetric model has only two possible gapped phases.  
The ground state degeneracy of a gapped phase is given by the inner product of the integer vectors associated with the Lagrangian algebras~\footnote{This integer vector $\vec{n}$ 
for a particular Lagrangian algebra $\cA$ is given by its decomposition into simple anyon types as 
$\cA = \bigoplus_{a\in \eM} n_a\,  a$. In our context, $\eM$ is the SymTO. We will use the notation $(\cA_1,\cA_2)$ to denote the inner product $(\vec{n}_1,\vec{n}_2)$ of the integer vectors associated with the Lagrangian algebras $\cA_1$ and $\cA_2$.} giving rise to the $\t \cR$ and $\mathrm{QFT}^{\,}_\text{ano}$ boundaries~\cite{CJW221214432}. So if we choose the $\cA_{\cRep(S^{\,}_3)}$ condensation to describe the gapped $\mathrm{QFT}^{\,}_\text{ano}$ 
boundary, the associated ground state degeneracy is
\begin{align}
\mathrm{GSD} = (\cA_{S^{\,}_3},\cA_{\cRep(S^{\,}_3)}) = 4\, .
\end{align}
This phase carries the degeneracies of $\mathbb{Z}^{\,}_{3}$ SSB phase (with GSD = 3)
and the $S^{\,}_{3}$-symmetric phase (with GSD= 1). This gapped phase describes a first-order 
quantum phase transition between these two phases.
The second gapped
phase of the self-dual $S^{\,}_3$ symmetric model corresponds to a $\cA_{S^{\,}_3}$-condensed
$\mathrm{QFT}^{\,}_\text{ano}$ boundary.
The corresponding ground state degeneracy is 
\begin{align}
 \mathrm{GSD} = (\cA_{S^{\,}_3},\cA_{S^{\,}_3}) = 8.
\end{align}
This phase carries the degeneracies of $\mathbb{Z}^{\,}_{2}$ SSB phase (with GSD = 2)
and the $S^{\,}_{3}$ SSB phase (with GSD= 6). It describes a first-order 
quantum phase transition between these two phases.
We note that the self-dual $S^{\,}_3$ symmetric model does not have any gapped phase
with a non-degenerate ground state. This is consistent with the fact that it is anomalous.

We can repeat the same analysis for the 
self-dual $\cRep(S^{\,}_3)$ symmetric model that is given by 
Fig.\ \ref{fig:QFTS3}(right). Here, the $\t\cR$ boundary is given by
$\cA_{\cRep(S^{\,}_3)}$ condensation. 
The  self-dual $\cRep(S^{\,}_3)$ symmetric model also has only two gapped
phases, with ground state degeneracies
\begin{align}
\mathrm{GSD} = (\cA_{S^{\,}_3},\cA_{\cRep(S^{\,}_3)}) = 4,\ \ \ \ \
\mathrm{GSD}= (\cA_{\cRep(S^{\,}_3)},\cA_{\cRep(S^{\,}_3)}) = 5.
\end{align}
These gapped phases describe the first-order transitions between 
$\cRep(S^{\,}_{3})$-symmetric phase (with GSD=1) and $\cRep(S^{\,}_{3})/\mathbb{Z}^{\,}_{2}$
SSB phase (with GSD = 3), and that between 
$\mathbb{Z}^{\,}_{2}$ SSB phase (with GSD=2) and $\cRep(S^{\,}_{3})$
SSB phase (with GSD = 3), respectively.
The self-dual $\cRep(S^{\,}_3)$ symmetric model also does not have any gapped
phases with non-degenerate ground state, since the non-inverible self-duality symmetry is anomalous.

So far, we have used the Lagrangian algebras of the SymTO to classify all
possible gapped phases for systems with the corresponding symmetry.  
These classify the ways in which the SymTO can be ``maximally broken".  
Non-Lagrangian condensable algebras lead to a non-trivial \emph{unbroken SymTO}, and the associated phase via SymTO correspondence must be gapless. 
Such gapless states are described by the $\one$-condensed boundaries of the unbroken SymTOs~\cite{CW220506244},
\ie the boundaries induced by the minimal condensable algebra $\cA =
\one$.  
We find that two of the $\one$-condensed boundaries of the $\mathrm{JK}^{\,}_{4}\boxtimes
\overline{\mathrm{JK}}^{\,}_{4}$ SymTO are described by the $(6,5)$ minimal model.
The first one is described by the
multi-component partition function~\cite{JW191213492,CW220506244}
\begingroup
\allowdisplaybreaks
\begin{align}
\label{ZJK1}
 Z_{\mathbf{1}; \mathbf{1}}^{\mathrm{JK}^{\,}_{4} \boxtimes \overline{\mathrm{JK}}^{\,}_{4}} &= \chi^{m6 \times \overline{m6}}_{1,0; 1,0} +  \chi^{m6 \times \overline{m6}}_{10,\frac{7}{5}; 10,-\frac{7}{5}},
 & 
Z_{\mathbf{1}; \bar e}^{\mathrm{JK}^{\,}_{4} \boxtimes \overline{\mathrm{JK}}^{\,}_{4}} &= \chi^{m6 \times \overline{m6}}_{1,0; 5,-3} +  \chi^{m6 \times \overline{m6}}_{10,\frac{7}{5}; 6,-\frac{2}{5}},
 \nonumber \\ 
Z_{\mathbf{1}; \bar m}^{\mathrm{JK}^{\,}_{4} \boxtimes \overline{\mathrm{JK}}^{\,}_{4}} &= \chi^{m6 \times \overline{m6}}_{1,0; 2,-\frac{1}{8}} +  \chi^{m6 \times \overline{m6}}_{10,\frac{7}{5}; 9,-\frac{21}{40}},
 & 
Z_{\mathbf{1}; \bar m_1}^{\mathrm{JK}^{\,}_{4} \boxtimes \overline{\mathrm{JK}}^{\,}_{4}} &= \chi^{m6 \times \overline{m6}}_{1,0; 4,-\frac{13}{8}} +  \chi^{m6 \times \overline{m6}}_{10,\frac{7}{5}; 7,-\frac{1}{40}}, 
 \nonumber \\ 
Z_{\mathbf{1}; \bar q}^{\mathrm{JK}^{\,}_{4} \boxtimes \overline{\mathrm{JK}}^{\,}_{4}} &= \chi^{m6 \times \overline{m6}}_{1,0; 3,-\frac{2}{3}} +  \chi^{m6 \times \overline{m6}}_{10,\frac{7}{5}; 8,-\frac{1}{15}}, 
 & 
Z_{e; \mathbf{1}}^{\mathrm{JK}^{\,}_{4} \boxtimes \overline{\mathrm{JK}}^{\,}_{4}} &= \chi^{m6 \times \overline{m6}}_{5,3; 1,0} +  \chi^{m6 \times \overline{m6}}_{6,\frac{2}{5}; 10,-\frac{7}{5}}, 
 \nonumber \\ 
Z_{e; \bar e}^{\mathrm{JK}^{\,}_{4} \boxtimes \overline{\mathrm{JK}}^{\,}_{4}} &= \chi^{m6 \times \overline{m6}}_{5,3; 5,-3} +  \chi^{m6 \times \overline{m6}}_{6,\frac{2}{5}; 6,-\frac{2}{5}}, 
 & 
Z_{e; \bar m}^{\mathrm{JK}^{\,}_{4} \boxtimes \overline{\mathrm{JK}}^{\,}_{4}} &= \chi^{m6 \times \overline{m6}}_{5,3; 2,-\frac{1}{8}} +  \chi^{m6 \times \overline{m6}}_{6,\frac{2}{5}; 9,-\frac{21}{40}}, 
 \nonumber \\ 
Z_{e; \bar m_1}^{\mathrm{JK}^{\,}_{4} \boxtimes \overline{\mathrm{JK}}^{\,}_{4}} &= \chi^{m6 \times \overline{m6}}_{5,3; 4,-\frac{13}{8}} +  \chi^{m6 \times \overline{m6}}_{6,\frac{2}{5}; 7,-\frac{1}{40}}, 
 & 
Z_{e; \bar q}^{\mathrm{JK}^{\,}_{4} \boxtimes \overline{\mathrm{JK}}^{\,}_{4}} &= \chi^{m6 \times \overline{m6}}_{5,3; 3,-\frac{2}{3}} +  \chi^{m6 \times \overline{m6}}_{6,\frac{2}{5}; 8,-\frac{1}{15}}, 
 \nonumber \\ 
Z_{m; \mathbf{1}}^{\mathrm{JK}^{\,}_{4} \boxtimes \overline{\mathrm{JK}}^{\,}_{4}} &= \chi^{m6 \times \overline{m6}}_{2,\frac{1}{8}; 1,0} +  \chi^{m6 \times \overline{m6}}_{9,\frac{21}{40}; 10,-\frac{7}{5}}, 
 & 
Z_{m; \bar e}^{\mathrm{JK}^{\,}_{4} \boxtimes \overline{\mathrm{JK}}^{\,}_{4}} &= \chi^{m6 \times \overline{m6}}_{2,\frac{1}{8}; 5,-3} +  \chi^{m6 \times \overline{m6}}_{9,\frac{21}{40}; 6,-\frac{2}{5}}, 
 \nonumber \\ 
Z_{m; \bar m}^{\mathrm{JK}^{\,}_{4} \boxtimes \overline{\mathrm{JK}}^{\,}_{4}} &= \chi^{m6 \times \overline{m6}}_{2,\frac{1}{8}; 2,-\frac{1}{8}} +  \chi^{m6 \times \overline{m6}}_{9,\frac{21}{40}; 9,-\frac{21}{40}}, 
 & 
Z_{m; \bar m_1}^{\mathrm{JK}^{\,}_{4} \boxtimes \overline{\mathrm{JK}}^{\,}_{4}} &= \chi^{m6 \times \overline{m6}}_{2,\frac{1}{8}; 4,-\frac{13}{8}} +  \chi^{m6 \times \overline{m6}}_{9,\frac{21}{40}; 7,-\frac{1}{40}}, 
 \nonumber \\ 
Z_{m; \bar q}^{\mathrm{JK}^{\,}_{4} \boxtimes \overline{\mathrm{JK}}^{\,}_{4}} &= \chi^{m6 \times \overline{m6}}_{2,\frac{1}{8}; 3,-\frac{2}{3}} +  \chi^{m6 \times \overline{m6}}_{9,\frac{21}{40}; 8,-\frac{1}{15}}, 
 & 
Z_{m_1; \mathbf{1}}^{\mathrm{JK}^{\,}_{4} \boxtimes \overline{\mathrm{JK}}^{\,}_{4}} &= \chi^{m6 \times \overline{m6}}_{4,\frac{13}{8}; 1,0} +  \chi^{m6 \times \overline{m6}}_{7,\frac{1}{40}; 10,-\frac{7}{5}}, 
 \nonumber \\ 
Z_{m_1; \bar e}^{\mathrm{JK}^{\,}_{4} \boxtimes \overline{\mathrm{JK}}^{\,}_{4}} &= \chi^{m6 \times \overline{m6}}_{4,\frac{13}{8}; 5,-3} +  \chi^{m6 \times \overline{m6}}_{7,\frac{1}{40}; 6,-\frac{2}{5}}, 
 & 
Z_{m_1; \bar m}^{\mathrm{JK}^{\,}_{4} \boxtimes \overline{\mathrm{JK}}^{\,}_{4}} &= \chi^{m6 \times \overline{m6}}_{4,\frac{13}{8}; 2,-\frac{1}{8}} +  \chi^{m6 \times \overline{m6}}_{7,\frac{1}{40}; 9,-\frac{21}{40}}, 
 \nonumber \\ 
Z_{m_1; \bar m_1}^{\mathrm{JK}^{\,}_{4} \boxtimes \overline{\mathrm{JK}}^{\,}_{4}} &= \chi^{m6 \times \overline{m6}}_{4,\frac{13}{8}; 4,-\frac{13}{8}} +  \chi^{m6 \times \overline{m6}}_{7,\frac{1}{40}; 7,-\frac{1}{40}},
 & 
Z_{m_1; \bar q}^{\mathrm{JK}^{\,}_{4} \boxtimes \overline{\mathrm{JK}}^{\,}_{4}} &= \chi^{m6 \times \overline{m6}}_{4,\frac{13}{8}; 3,-\frac{2}{3}} +  \chi^{m6 \times \overline{m6}}_{7,\frac{1}{40}; 8,-\frac{1}{15}},
 \nonumber \\ 
Z_{q; \mathbf{1}}^{\mathrm{JK}^{\,}_{4} \boxtimes \overline{\mathrm{JK}}^{\,}_{4}} &= \chi^{m6 \times \overline{m6}}_{3,\frac{2}{3}; 1,0} +  \chi^{m6 \times \overline{m6}}_{8,\frac{1}{15}; 10,-\frac{7}{5}}, 
 & 
Z_{q; \bar e}^{\mathrm{JK}^{\,}_{4} \boxtimes \overline{\mathrm{JK}}^{\,}_{4}} &= \chi^{m6 \times \overline{m6}}_{3,\frac{2}{3}; 5,-3} +  \chi^{m6 \times \overline{m6}}_{8,\frac{1}{15}; 6,-\frac{2}{5}}, 
 \nonumber \\ 
Z_{q; \bar m}^{\mathrm{JK}^{\,}_{4} \boxtimes \overline{\mathrm{JK}}^{\,}_{4}} &= \chi^{m6 \times \overline{m6}}_{3,\frac{2}{3}; 2,-\frac{1}{8}} +  \chi^{m6 \times \overline{m6}}_{8,\frac{1}{15}; 9,-\frac{21}{40}}, 
 & 
Z_{q; \bar m_1}^{\mathrm{JK}^{\,}_{4} \boxtimes \overline{\mathrm{JK}}^{\,}_{4}} &= \chi^{m6 \times \overline{m6}}_{3,\frac{2}{3}; 4,-\frac{13}{8}} +  \chi^{m6 \times \overline{m6}}_{8,\frac{1}{15}; 7,-\frac{1}{40}},
 \nonumber \\ 
Z_{q; \bar q}^{\mathrm{JK}^{\,}_{4} \boxtimes \overline{\mathrm{JK}}^{\,}_{4}} &= \chi^{m6 \times \overline{m6}}_{3,\frac{2}{3}; 3,-\frac{2}{3}} +  \chi^{m6 \times \overline{m6}}_{8,\frac{1}{15}; 8,-\frac{1}{15}}, 
 \end{align}
while the second one is described by
\begin{align}
\label{ZJK2}
 Z_{\mathbf{1}; \mathbf{1}}^{\mathrm{JK}^{\,}_{4} \boxtimes \overline{\mathrm{JK}}^{\,}_{4}} &= \chi^{m6 \times \overline{m6}}_{1,0; 1,0} +  \chi^{m6 \times \overline{m6}}_{10,\frac{7}{5}; 10,-\frac{7}{5}},
 & 
Z_{\mathbf{1}; \bar e}^{\mathrm{JK}^{\,}_{4} \boxtimes \overline{\mathrm{JK}}^{\,}_{4}} &= \chi^{m6 \times \overline{m6}}_{1,0; 5,-3} +  \chi^{m6 \times \overline{m6}}_{10,\frac{7}{5}; 6,-\frac{2}{5}}, 
 \nonumber \\ 
Z_{\mathbf{1}; \bar m}^{\mathrm{JK}^{\,}_{4} \boxtimes \overline{\mathrm{JK}}^{\,}_{4}} &= \chi^{m6 \times \overline{m6}}_{5,3; 2,-\frac{1}{8}} +  \chi^{m6 \times \overline{m6}}_{6,\frac{2}{5}; 9,-\frac{21}{40}}, 
 & 
Z_{\mathbf{1}; \bar m_1}^{\mathrm{JK}^{\,}_{4} \boxtimes \overline{\mathrm{JK}}^{\,}_{4}} &= \chi^{m6 \times \overline{m6}}_{5,3; 4,-\frac{13}{8}} +  \chi^{m6 \times \overline{m6}}_{6,\frac{2}{5}; 7,-\frac{1}{40}},
 \nonumber \\ 
Z_{\mathbf{1}; \bar q}^{\mathrm{JK}^{\,}_{4} \boxtimes \overline{\mathrm{JK}}^{\,}_{4}} &= \chi^{m6 \times \overline{m6}}_{1,0; 3,-\frac{2}{3}} +  \chi^{m6 \times \overline{m6}}_{10,\frac{7}{5}; 8,-\frac{1}{15}}, 
 & 
Z_{e; \mathbf{1}}^{\mathrm{JK}^{\,}_{4} \boxtimes \overline{\mathrm{JK}}^{\,}_{4}} &= \chi^{m6 \times \overline{m6}}_{5,3; 1,0} +  \chi^{m6 \times \overline{m6}}_{6,\frac{2}{5}; 10,-\frac{7}{5}},
 \nonumber \\ 
Z_{e; \bar e}^{\mathrm{JK}^{\,}_{4} \boxtimes \overline{\mathrm{JK}}^{\,}_{4}} &= \chi^{m6 \times \overline{m6}}_{5,3; 5,-3} +  \chi^{m6 \times \overline{m6}}_{6,\frac{2}{5}; 6,-\frac{2}{5}},
 & 
Z_{e; \bar m}^{\mathrm{JK}^{\,}_{4} \boxtimes \overline{\mathrm{JK}}^{\,}_{4}} &= \chi^{m6 \times \overline{m6}}_{1,0; 2,-\frac{1}{8}} +  \chi^{m6 \times \overline{m6}}_{10,\frac{7}{5}; 9,-\frac{21}{40}}, 
 \nonumber \\ 
Z_{e; \bar m_1}^{\mathrm{JK}^{\,}_{4} \boxtimes \overline{\mathrm{JK}}^{\,}_{4}} &= \chi^{m6 \times \overline{m6}}_{1,0; 4,-\frac{13}{8}} +  \chi^{m6 \times \overline{m6}}_{10,\frac{7}{5}; 7,-\frac{1}{40}}, 
 & 
Z_{e; \bar q}^{\mathrm{JK}^{\,}_{4} \boxtimes \overline{\mathrm{JK}}^{\,}_{4}} &= \chi^{m6 \times \overline{m6}}_{5,3; 3,-\frac{2}{3}} +  \chi^{m6 \times \overline{m6}}_{6,\frac{2}{5}; 8,-\frac{1}{15}}, 
 \nonumber \\ 
Z_{m; \mathbf{1}}^{\mathrm{JK}^{\,}_{4} \boxtimes \overline{\mathrm{JK}}^{\,}_{4}} &= \chi^{m6 \times \overline{m6}}_{2,\frac{1}{8}; 5,-3} +  \chi^{m6 \times \overline{m6}}_{9,\frac{21}{40}; 6,-\frac{2}{5}}, 
 & 
Z_{m; \bar e}^{\mathrm{JK}^{\,}_{4} \boxtimes \overline{\mathrm{JK}}^{\,}_{4}} &= \chi^{m6 \times \overline{m6}}_{2,\frac{1}{8}; 1,0} +  \chi^{m6 \times \overline{m6}}_{9,\frac{21}{40}; 10,-\frac{7}{5}},
 \nonumber \\ 
Z_{m; \bar m}^{\mathrm{JK}^{\,}_{4} \boxtimes \overline{\mathrm{JK}}^{\,}_{4}} &= \chi^{m6 \times \overline{m6}}_{4,\frac{13}{8}; 4,-\frac{13}{8}} +  \chi^{m6 \times \overline{m6}}_{7,\frac{1}{40}; 7,-\frac{1}{40}},
 & 
Z_{m; \bar m_1}^{\mathrm{JK}^{\,}_{4} \boxtimes \overline{\mathrm{JK}}^{\,}_{4}} &= \chi^{m6 \times \overline{m6}}_{4,\frac{13}{8}; 2,-\frac{1}{8}} +  \chi^{m6 \times \overline{m6}}_{7,\frac{1}{40}; 9,-\frac{21}{40}},
 \nonumber \\ 
Z_{m; \bar q}^{\mathrm{JK}^{\,}_{4} \boxtimes \overline{\mathrm{JK}}^{\,}_{4}} &= \chi^{m6 \times \overline{m6}}_{2,\frac{1}{8}; 3,-\frac{2}{3}} +  \chi^{m6 \times \overline{m6}}_{9,\frac{21}{40}; 8,-\frac{1}{15}},
 & 
Z_{m_1; \mathbf{1}}^{\mathrm{JK}^{\,}_{4} \boxtimes \overline{\mathrm{JK}}^{\,}_{4}} &= \chi^{m6 \times \overline{m6}}_{4,\frac{13}{8}; 5,-3} +  \chi^{m6 \times \overline{m6}}_{7,\frac{1}{40}; 6,-\frac{2}{5}}, 
 \nonumber \\ 
Z_{m_1; \bar e}^{\mathrm{JK}^{\,}_{4} \boxtimes \overline{\mathrm{JK}}^{\,}_{4}} &= \chi^{m6 \times \overline{m6}}_{4,\frac{13}{8}; 1,0} +  \chi^{m6 \times \overline{m6}}_{7,\frac{1}{40}; 10,-\frac{7}{5}},
 & 
Z_{m_1; \bar m}^{\mathrm{JK}^{\,}_{4} \boxtimes \overline{\mathrm{JK}}^{\,}_{4}} &= \chi^{m6 \times \overline{m6}}_{2,\frac{1}{8}; 4,-\frac{13}{8}} +  \chi^{m6 \times \overline{m6}}_{9,\frac{21}{40}; 7,-\frac{1}{40}},
 \nonumber \\ 
Z_{m_1; \bar m_1}^{\mathrm{JK}^{\,}_{4} \boxtimes \overline{\mathrm{JK}}^{\,}_{4}} &= \chi^{m6 \times \overline{m6}}_{2,\frac{1}{8}; 2,-\frac{1}{8}} +  \chi^{m6 \times \overline{m6}}_{9,\frac{21}{40}; 9,-\frac{21}{40}},
 & 
Z_{m_1; \bar q}^{\mathrm{JK}^{\,}_{4} \boxtimes \overline{\mathrm{JK}}^{\,}_{4}} &= \chi^{m6 \times \overline{m6}}_{4,\frac{13}{8}; 3,-\frac{2}{3}} +  \chi^{m6 \times \overline{m6}}_{7,\frac{1}{40}; 8,-\frac{1}{15}},
 \nonumber \\ 
Z_{q; \mathbf{1}}^{\mathrm{JK}^{\,}_{4} \boxtimes \overline{\mathrm{JK}}^{\,}_{4}} &= \chi^{m6 \times \overline{m6}}_{3,\frac{2}{3}; 1,0} +  \chi^{m6 \times \overline{m6}}_{8,\frac{1}{15}; 10,-\frac{7}{5}}, 
 & 
Z_{q; \bar e}^{\mathrm{JK}^{\,}_{4} \boxtimes \overline{\mathrm{JK}}^{\,}_{4}} &= \chi^{m6 \times \overline{m6}}_{3,\frac{2}{3}; 5,-3} +  \chi^{m6 \times \overline{m6}}_{8,\frac{1}{15}; 6,-\frac{2}{5}}, 
 \nonumber \\ 
Z_{q; \bar m}^{\mathrm{JK}^{\,}_{4} \boxtimes \overline{\mathrm{JK}}^{\,}_{4}} &= \chi^{m6 \times \overline{m6}}_{3,\frac{2}{3}; 2,-\frac{1}{8}} +  \chi^{m6 \times \overline{m6}}_{8,\frac{1}{15}; 9,-\frac{21}{40}},
 & 
Z_{q; \bar m_1}^{\mathrm{JK}^{\,}_{4} \boxtimes \overline{\mathrm{JK}}^{\,}_{4}} &= \chi^{m6 \times \overline{m6}}_{3,\frac{2}{3}; 4,-\frac{13}{8}} +  \chi^{m6 \times \overline{m6}}_{8,\frac{1}{15}; 7,-\frac{1}{40}},
 \nonumber \\ 
Z_{q; \bar q}^{\mathrm{JK}^{\,}_{4} \boxtimes \overline{\mathrm{JK}}^{\,}_{4}} &= \chi^{m6 \times \overline{m6}}_{3,\frac{2}{3}; 3,-\frac{2}{3}} +  \chi^{m6 \times \overline{m6}}_{8,\frac{1}{15}; 8,-\frac{1}{15}}.
 \end{align}
\endgroup
The various terms in each component of the partition function are conformal
characters of the (6,5) minimal model. The expression $\chi^{m6\times
\overline{m6}}_{a,h_a;\, b, -h_b}$ is a short-hand notation for the product of
the left moving chiral conformal character associated with the primary operator
labeled by $a$ (set by an arbitrary indexing convention) with conformal weight
$(h_a,0)$, and the right moving chiral conformal character associated with the
primary operator labeled by $b$ with conformal weight $(0,h_b)$. The superscript
$m6 \times \overline{m6}$ indicates that both the left and right moving chiral
conformal characters are picked from the same (6,5) minimal model.

Note that in the above multi-component ``SymTO-resolved" partition function,
the $\one$ sector contains all the primary operators that respect the symmetry.
We see from the term $\chi^{m6 \times \overline{m6}}_{10,\frac{7}{5}; 10,-\frac{7}{5}} $ in $Z_{\mathbf{1};
\mathbf{1}}^{\mathrm{JK}^{\,}_{4} \boxtimes \overline{\mathrm{JK}}^{\,}_{4}}$
that the scaling dimensions of the symmetric operators to be $7/5 + 7/5 + 2n> 2$ with a
non-negative integer $n$ for both gapless states. 
Such symmetric operators are then irrelevant.  Therefore, both of the above two gapless states are
in fact gapless phases with no relevant perturbation that respects
the symmetries dictated by $\mathrm{JK}^{\,}_{4} \boxtimes \overline{\mathrm{JK}}^{\,}_{4}$ SymTO.

We remark that the above calculation was also performed for the other candidate
$\mathrm{SU}(2)^{\,}_{4} \boxtimes \overline{\mathrm{SU}(2)}^{\,}_{4}$ SymTO.  
We find that for systems with $\mathrm{SU}(2)^{\,}_{4} \boxtimes \overline{\mathrm{SU}(2)}^{\,}_{4}$ SymTO, all gapless states that are described by
(6,5) minimal model contain at least one relevant perturbation that respects
the $\mathrm{SU}(2)^{\,}_{4} \boxtimes \overline{\mathrm{SU}(2)}^{\,}_{4}$ SymTO.  
This contradicts our
numerical calculations in Secs.\ \ref{sec:S3model} and \ref{sec:RepS3model}
where we found a stable gapless phase described by (6,5)
minimal model in the presence of $S^{\,}_3$ (respectively, $\cRep(S_3)$) and self-duality symmetry.  
We conclude that the SymTO in our self-dual $S^{\,}_3$-symmetric model and self-dual
$\cRep(S^{\,}_3)$-symmetric model is given by $\mathrm{JK}^{\,}_{4} \boxtimes \overline{\mathrm{JK}}^{\,}_{4}$ and not by
$\mathrm{SU}(2)^{\,}_{4} \boxtimes \overline{\mathrm{SU}(2)}^{\,}_{4}$.

The operators that break the self-duality symmetry live in the $e\bar e$
sector. From the partition function 
$Z_{e; \bar e}^{\mathrm{JK}^{\,}_{4} \boxtimes \overline{\mathrm{JK}}^{\,}_{4}}$, we find that
the scaling dimensions of the operators breaking the self-duality symmetry to be
$2/5 +2/5 + 2n$ or $3 +3 + 2n$ (with a non-negative integer $n$), for both gapless
states. Thus, the two gapless states have only one relevant operator that 
break the self-duality symmetry but not the $S^{\,}_3$ symmetry.  
They can be identified with the upper and lower vertical lines that meet 
at the multi-critical point in phase diagrams in Figs.\
\ref{fig:s3pd-numerics} and \ref{fig:reps3pd-numerics}
which indeed have only one such relevant perturbation.
In fact, SSB of the
self-duality symmetry due to the $e\bar e$ condensation can be seen from
the following relation between 
$\mathrm{JK}^{\,}_{4}\boxtimes \overline{\mathrm{JK}}^{\,}_{4}$-SymTO-resolved
partition functions \eqref{ZJK1} and \eqref{ZJK2}, and
$\eD(S^{\,}_3)$-SymTO-resolved partition function \eqref{eq:potts mod bootstrap}:
\begin{align}
 Z_{\one}^{\eD(S^{\,}_3)} 
 &= Z_{\one;\one}^{\mathrm{JK}^{\,}_{4} \boxtimes \overline{\mathrm{JK}}^{\,}_{4}} 
+ Z_{e; \bar e}^{\mathrm{JK}^{\,}_{4} \boxtimes \overline{\mathrm{JK}}^{\,}_{4}},
&
Z_{\one'}^{\eD(S^{\,}_3)} &= Z_{e;\one}^{\mathrm{JK}^{\,}_{4} \boxtimes \overline{\mathrm{JK}}^{\,}_{4}} 
+ Z_{\one; \bar e}^{\mathrm{JK}^{\,}_{4} \boxtimes \overline{\mathrm{JK}}^{\,}_{4}}, 
\nonumber\\
Z_{s}^{\eD(S^{\,}_3)} &= Z_{m;\bar m}^{\mathrm{JK}^{\,}_{4} \boxtimes \overline{\mathrm{JK}}^{\,}_{4}} 
+ Z_{m_1; \bar m_1}^{\mathrm{JK}^{\,}_{4} \boxtimes \overline{\mathrm{JK}}^{\,}_{4}},
&
 &\cdots
\end{align}
where the sectors in $\mathrm{JK}^{\,}_{4}\boxtimes \overline{\mathrm{JK}}^{\,}_{4}$ SymTO connected by $e\bar e$
are combined into a single sector in $\eD(S^{\,}_3)$ SymTO.

The above two gapless states with self-duality symmetry are very similar. 
The only difference is that $Z_{s}^{\eD(S^{\,}_3)}$ splits differently when we add the
self-duality symmetry as
\begin{align}
Z^{\eD(S^{\,}_3)}_{s} &=
\underbrace{\chi^{m6 \times \overline{m6}}_{2,\frac{1}{8}; 2,-\frac{1}{8}} 
+  \chi^{m6 \times \overline{m6}}_{9,\frac{21}{40}; 9,-\frac{21}{40}} 
}_{Z_{m;\bar m}^{\mathrm{JK}^{\,}_{4} \boxtimes \overline{\mathrm{JK}}^{\,}_{4}}}
+  
\underbrace{\chi^{m6 \times \overline{m6}}_{4,\frac{13}{8}; 4,-\frac{13}{8}} 
+  \chi^{m6 \times \overline{m6}}_{7,\frac{1}{40}; 7,-\frac{1}{40}} 
}_{Z_{m_1;\bar m_1}^{\mathrm{JK}^{\,}_{4} \boxtimes \overline{\mathrm{JK}}^{\,}_{4}}},
\nonumber\\
Z^{\eD(S^{\,}_3)}_{s} &=
\underbrace{\chi^{m6 \times \overline{m6}}_{4,\frac{13}{8}; 4,-\frac{13}{8}}
+  \chi^{m6 \times \overline{m6}}_{7,\frac{1}{40}; 7,-\frac{1}{40}} 
}_{Z_{m;\bar m}^{\mathrm{JK}^{\,}_{4} \boxtimes \overline{\mathrm{JK}}^{\,}_{4}}}
+  
\underbrace{\chi^{m6 \times \overline{m6}}_{2,\frac{1}{8}; 2,-\frac{1}{8}} 
+  \chi^{m6 \times \overline{m6}}_{9,\frac{21}{40}; 9,-\frac{21}{40}} 
}_{Z_{m_1;\bar m_1}^{\mathrm{JK}^{\,}_{4} \boxtimes \overline{\mathrm{JK}}^{\,}_{4}}}.
\end{align}

In addition to the two Lagrangian condensable algebras \eqref{LcAlg},
$\eM=\mathrm{JK}^{\,}_{4}\boxtimes \overline{\mathrm{JK}}^{\,}_{4}$ SymTO also has six
non-Lagrangian condensable algebras, listed below. 
The condensation of these six algebras
reduces the SymTO $\eM=\mathrm{JK}^{\,}_{4}\boxtimes \overline{\mathrm{JK}}^{\,}_{4}$ to
a smaller SymTO $\eM_{/\cA}$.
The reduced, unbroken SymTO can be identified from its total quantum dimension  
$D_{\eM_{/\cA}}= \sqrt{\sum_i d_i^2}$ 
and topological spins of the anyons, as well as confirmed by the presence of a domain wall between
$\eM$ and $\eM_{/\cA}$ that has no condensation of $\eM_{/\cA}$ anyons:
\begin{enumerate}[(i)]

\item
Condensable algebra
$\cA_1 = $ $(\one,\one)$
$\oplus$ $(e,\bar e)$
$\oplus$ $(q,\bar q)$:
\\
SymTO 
$\eM_{/\cA_1}$ contains topological spins 
$( 0 ) $
and has
$D_{\eM_{/\cA_1}} = 2$.
\\ 
We conclude that $\eM_{/\cA_1} = \eD( \Z^{\,}_{2})$.

\item
Condensable algebra
$\cA_2 = $ $(\one,\one)$
$\oplus$ $(\one,\bar e)$
$\oplus$ $(e,\one)$
$\oplus$ $(e,\bar e)$:
\\
SymTO
$\eM_{/\cA_2}$ contains topological spins 
$\left( 0,
\frac{1}{3},
\frac{2}{3} \right)
$
and has  
$D_{\eM_{/\cA_2}} = 3$.
\\ 
We conclude that $\eM_{/\cA_2} = \eD( \Z^{\,}_{3})$.

\item
Condensable algebra
$\cA_3 = $ $(\one,\one)$
$\oplus$ $(e,\bar e)$:
\\
SymTO
$\eM_{/\cA_3}$ contains topological spins 
$\left( 0,
\frac{1}{3},
\frac{1}{2},
\frac{2}{3} \right)
$
and has
$D_{\eM_{/\cA_3}} = 6$.
\\
We conclude that $\eM_{/\cA_3} = \eD(S^{\,}_3)$.

\item
Condensable algebra
$\cA_4 = $ $(\one,\one)$
$\oplus$ $(e,\one)$:
\\
SymTO
$\eM_{/\cA_4}$ contains topological spins 
$\left( 0,
\frac{1}{24},
\frac{1}{3},
\frac{3}{8},
\frac{13}{24},
\frac{2}{3},
\frac{7}{8} \right)
$
and has
 $D_{\eM_{/\cA_4}} = 6$.
% 5_2^13 x 3_6^8
% JK_-4 x K(-3)
\\
We conclude that 
$\eM_{/\cA_4} = 
\overline{\eZ}_3 
\boxtimes 
\overline{\mathrm{JK}}^{\,}_{4}$.

\item
Condensable algebra
$\cA_5 = $ $(\one,\one)$
$\oplus$ $(\one,\bar e)$:
\\
SymTO
$\eM_{/\cA_5}$ contains topological spins 
$\left( 0,
\frac{1}{8},
\frac{1}{3},
\frac{11}{24},
\frac{5}{8},
\frac{2}{3},
\frac{23}{24} \right)
$
and has
 $D_{\eM_{/\cA_5}} = 6$.
\\ 
We conclude that
$\eM_{/\cA_5} = \mathrm{JK}^{\,}_{4}\boxtimes \eZ_3 $.

\item 
Condensable algebra
$\cA_6 = $ $(\one,\one)$:
\\
$\eM_{/\cA_6}= \mathrm{JK}^{\,}_{4}\boxtimes \overline{\mathrm{JK}}^{\,}_{4}$.
\end{enumerate}
Here $\eZ_3$ is the Abelian topological order described by the $K$-matrix
$\begin{pmatrix}
 2 & 1\\
 1 & 2\\
\end{pmatrix}$, where the fusion of the anyons form a $\Z^{\,}_{3}$ group.
These  condensable algebras describe the possible spontaneous symmetry breaking
patterns of the $\mathrm{JK}^{\,}_{4}\boxtimes \overline{\mathrm{JK}}^{\,}_{4}$ SymTO, where the unbroken SymTO is
given by $\eM_{/\cA}=(\mathrm{JK}^{\,}_{4}\boxtimes \overline{\mathrm{JK}}^{\,}_{4})_{/\cA}$.  Because the
unbroken SymTO is non-trivial, those non-maximal SymTO broken states are
gapless \cite{KZ190504924,CW220506244}, and are given by the $\one$-condensed boundaries of
$\eM_{/\cA}=(\mathrm{JK}^{\,}_{4}\boxtimes \overline{\mathrm{JK}}^{\,}_{4})_{/\cA}$.  
In turn, such $\one$-condensed
boundaries are some possible gapless states of our self-dual $S^{\,}_3$-symmetric or
$\cRep(S^{\,}_3)$-symmetric models.

\begin{figure}[tb]
\centerline{
\begin{minipage}{2in}
\includegraphics[width=2in]{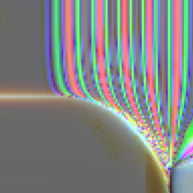}
\\[-1.7in]
  {\color{white}\large
\hspace*{4mm}$A$\hfill Incommensurate\ \ \ \\[10mm]
\hspace*{4mm}$C$ \hfill.\\[10mm]
\hspace*{4mm}$B$ \hfill.\\

\ 

}
\end{minipage}
\ \ \ 
\ \ \ 
\begin{minipage}{2in}
\includegraphics[width=2in]{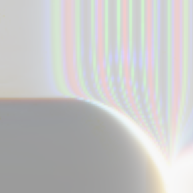}
\\[-1.7in]
{
\large
\hspace*{4mm}$A$\hfill Incommensurate\ \ \ \\[10mm]
\hspace*{4mm}$C$ \hfill.\\[10mm]
\hspace*{4mm}$B$ \hfill.\\

\ 

}
\end{minipage}
}
\caption{(Left) The central charge $c$ and (Right) GSD from tensor network
numerical calculation, for the self-dual $S_3$-symmetric model with $J_1=J_2 =
1/2$ and $J_4=1-J_3$.  The horizontal axis is $\th \in [0,\pi/2]$.  The
vertical axis is $J_3 \in [0,1]$.  Here GSD is obtained from the partition
function: $\mathrm{GSD}\equiv Z^2(L, L)/Z(L, 2L)$ where $Z(L_x, L_t)$ is the partition
function for spacetime of size $L_x\times L_t$.  For gapped quantum systems,
GSD happen to be the ground state degeneracy.  The central charge $c$ is also
obtained from the partition function, which has a form 
$Z(L, L_\infty) = \ee^{- L_\infty[ \eps L -\frac{2\pi c v }{24 L} +o(L^{-1})]}$ 
(which determines $cv$)
and
$Z(L_\infty,L) = \ee^{- L_\infty[ \eps L -\frac{2\pi c/v  }{24 L} +o(L^{-1})]}$
(which determines $c/v$),
 when $L_\infty \gg L$.
This way,  the central charge $c$ is defined even for non-critical states.  The
red-channel of the colored image is for lattice of size $64$,
green-channel for size $128$, and blue-channel for size $256$.  (Left)
The color intensity $[0,1]$ corresponds to central charge $c \in [0,2]$.
(Right) The color intensity $[0,1]$ corresponds to $1-\frac{1}{\mathrm{GSD}}$.  } 
\label{S3SDc} 
\end{figure}

To see the actual gapless states of our self-dual $S^{\,}_3$-symmetric model,
we have calculated the phase diagram for the model
\eqref{eq:Ham gen S3 reparameterized} (see Fig. \ref{S3SDc}). We find gapless phases
A and B (as indicated by the non-zero central charge $c$) and
gapless incommensurate phase (as indicated by the ``striped'' non-zero central charge $c$), as well as continuous phase transitions between them.
To understand such a phase diagram,
in the following, we list the gapless states that we found using holographic modular bootstrap \cite{CZ190312334,JW191213492,CW220506244} for the self-dual
$S_3$-symmetric model \eqref{eq:Ham gen S3 reparameterized}.  We only list gapless phases (which have no
symmetric relevant operator), and gapless critical points at continuous
phase transitions (which have only one symmetric relevant operator).  We group
those gapless states by the condensable algebras $\cA_i$, whose condensation
lead to the corresponding gapless states.
Those gapless states have a reduced
%, as well as reduces the SymTO $\eM =JK_4\boxtimes \overline{JK}_4$ to a small 
SymTO $\eM_{/\cA_i}$ as discussed above:
\\[3mm]
$\cA_1 = (\one,\one) \oplus(e,\bar e) \oplus(q,\bar q),\ \ \eM_{/\cA_1}= \eD(\Z_2)$: \\
Gapless phase: 
$m4\times m4\times \overline{U1} $ (see \eqref{m4m4U1} for the SymTO-resolved partition function)  \\
Critical point:
 $m4\times \overline{m4}$ (see \eqref{m4-eeqq})
\\[3mm]
$\cA_2 = (\one,\one) \oplus(\one,\bar e) \oplus(e,\one) \oplus(e,\bar e),\ \ \eM_{/\cA_2}= \eD(\Z_3) $: \\
Critical point:
  $m6\times \overline{m6}$
(see \eqref{m6-ebe})
\\[3mm]
$\cA_3 = (\one,\one) \oplus(e,\bar e),\ \ \eM_{/\cA_3}= \eD(S_3) $: \\
Critical point:
  $m6\times \overline{m6}$ (see \eqref{m6-ee}).
\\[3mm]
$\cA_4 = (\one,\one) \oplus(\one,\bar e),\ \ \eM_{/\cA_4} = \overline{\eZ}_3  \boxtimes  \overline{\mathrm{JK}}^{\,}_{4} $: \\
Gapless phase: 
 ${m6\times \overline{m6}}$
(see \eqref{m6-1be})
\\
Critical point:
$m7\times \overline{m7}$ 
(see \eqref{m7-1be})
\\[3mm]
$\cA_5 = (\one,\one) \oplus(e,\one),\ \ \eM_{/\cA_5} = \mathrm{JK}^{\,}_{4}\boxtimes \eZ_3 $: \\
Gapless phase: 
 ${m6\times \overline{m6}}$
(see \eqref{m6-1e})
\\
Critical point:
$m7\times \overline{m7}$
(see \eqref{m7-1e})
\\[3mm]
$\cA_6 = (\one,\one),\ \ \eM_{/\cA_6}= \mathrm{JK}^{\,}_{4}\boxtimes \overline{\mathrm{JK}}^{\,}_{4}$:\\
Gapless phase:  
 $m6\times \overline{m6}_1$ (see \eqref{ZJK1}),
 $m6\times \overline{m6}_2$ (see \eqref{ZJK2}) 
\\
Critical point:
$m7\times \overline{m7}_1$ (see \eqref{m7-82}),
$m7\times \overline{m7}_2$ (see \eqref{m7-85}),
$m4\times m6\times \overline{m4}\times \overline{m6}_1$ (see \eqref{m4m6-1112}),
$m4\times m6\times \overline{m4}\times \overline{m6}_2$ (see \eqref{m4m6-1512}),
$m4\times m6\times \overline{m4}\times \overline{m6}_3$ (see \eqref{m4m6-1534}),
$m4\times m6\times \overline{m4}\times \overline{m6}_4$ (see \eqref{m4m6-1134}),

Here $m4$, $m6$, $m7$ represent $(4,3)$, $(6,5)$, $(7,6)$ chiral minimal model
CFT.  $U1$ represents chiral compact boson CFT.
For example, $m4\times m4\times \overline{U1} $ is a gapless state
whose right-movers are described by $m4\times m4$ CFT and
whose left-movers are described by $\overline{U1}$ CFT.

The SymTO-resolved partition function $\mathbf{Z}$ for $m4\times m4\times \overline{U1} $
state is given by \eqref{m4m4U1}.
From the $(\one,\one)$ component of the partition function
\begin{align}
  Z_{\mathbf{1}; \mathbf{1}}^{JK_4 \boxtimes \overline{JK}_4} &={\color{blue} \chi^{m4 \times m4 \times \overline{U1}_4}_{1,0; 1,0; 1,0}} + {\color{red} \chi^{m4 \times m4 \times \overline{U1}_4}_{1,0; 3,\frac{1}{2}; 3,-\frac{1}{2}}} + {\color{red} \chi^{m4 \times m4 \times \overline{U1}_4}_{3,\frac{1}{2}; 1,0; 3,-\frac{1}{2}}} +  \chi^{m4 \times m4 \times \overline{U1}_4}_{3,\frac{1}{2}; 3,\frac{1}{2}; 1,0} 
\end{align}
we conclude that the $m4\times m4\times \overline{U1} $ state has two $JK_4
\boxtimes \overline{JK}_4$ symmetric relevant perturbations (highlighted in red).  However, if we want
to use $m4\times m4\times \overline{U1} $ state to describe the incommensurate
state in the phase diagram Fig. \ref{S3SDc}, we also need to include the lattice translation
symmetry.  In this case, the crystal momentum of a many-body state become the
$U(1)$ charge of the $\overline{U1}$ CFT.  The primary field for the conformal
character $\chi^{m4 \times m4 \times \overline{U1}_4}_{1,0; 3,\frac{1}{2};
3,-\frac{1}{2}}$ carries a non-zero  $U(1)$ charge, \ie a non-zero  crystal
momentum.  Such a primary field is not symmetric under the translation
symmetry.  As a result, the $m4\times m4\times \overline{U1} $ state  has no
$JK_4 \boxtimes \overline{JK}_4$ symmetric and translation symmetric relevant
perturbations.  So it is a stable gapless phase.

Let us point out that the $m4\times m4\times \overline{U1} $ state is
nothing but the incommensurate state discussed in Section
\ref{sec:incommensurate}. Both states have the same unbroken SymTO
$\eM_{/\cA}=\eD(\Z_2)$, and both states are described by the same CFTs.

In the phase diagram shown in Fig. \ref{S3SDc}, we identify the gapless phase-B
as the $m6\times \overline{m6}_1$ phase (see \eqref{ZJK1}) or $m6\times
\overline{m6}_2$ phase (see \eqref{ZJK2}).  Both phases have the full SymTO
$JK_4 \boxtimes \overline{JK}_4$.  
Such phases for the self-dual $S_3$ symmetric model
have the following modular invariant partition function, which is obtained via the inner product
of the vector-valued partition function $\mathbf{Z}$ and the Lagrangian condensable Algebra $\cA_{S_3}$
\begin{align}
 Z &= (\mathbf{Z}, \cA_{S_3}) =
 Z_{\mathbf{1}; \mathbf{1}}^{JK_4 \boxtimes \overline{JK}_4}
+Z_{\mathbf{1}; \bar e}^{JK_4 \boxtimes \overline{JK}_4}
+Z_{e; \mathbf{1}}^{JK_4 \boxtimes \overline{JK}_4}
+Z_{e; \bar e}^{JK_4 \boxtimes \overline{JK}_4}
+2Z_{q; \bar q}^{JK_4 \boxtimes \overline{JK}_4}
\nonumber\\
&=
% Z_{\mathbf{1}; \mathbf{1}}^{JK_4 \boxtimes \overline{JK}_4} &=
{\color{blue} \chi^{m6 \times \overline{m6}}_{1,0; 1,0}} +  \chi^{m6 \times \overline{m6}}_{10,\frac{7}{5}; 10,-\frac{7}{5}} 
%Z_{\mathbf{1}; \bar e}^{JK_4 \boxtimes \overline{JK}_4} &= 
+\chi^{m6 \times \overline{m6}}_{1,0; 5,-3} +  \chi^{m6 \times \overline{m6}}_{10,\frac{7}{5}; 6,-\frac{2}{5}} 
%Z_{e; \mathbf{1}}^{JK_4 \boxtimes \overline{JK}_4} &= 
+\chi^{m6 \times \overline{m6}}_{5,3; 1,0} +  \chi^{m6 \times \overline{m6}}_{6,\frac{2}{5}; 10,-\frac{7}{5}} 
%Z_{e; \bar e}^{JK_4 \boxtimes \overline{JK}_4} &= 
\nonumber\\
&\ \ \ \
+\chi^{m6 \times \overline{m6}}_{5,3; 5,-3} +  \chi^{m6 \times \overline{m6}}_{6,\frac{2}{5}; 6,-\frac{2}{5}} 
%Z_{q; \bar q}^{JK_4 \boxtimes \overline{JK}_4} &= 
+2(\chi^{m6 \times \overline{m6}}_{3,\frac{2}{3}; 3,-\frac{2}{3}} +  \chi^{m6 \times \overline{m6}}_{8,\frac{1}{15}; 8,-\frac{1}{15}} 
)
\end{align}

We also identify the gapless phase-A as
${m6\times \overline{m6}}$ phase with $\cA_4$-condensation (see \eqref{m6-1be})
or ${m6\times \overline{m6}}$ phase with $\cA_5$-condensation (see
\eqref{m6-1e}).
The two phases for the self-dual $S_3$ symmetric model
have the following modular invariant partition function
\begin{align}
 Z &= (\mathbf{Z}, \cA_{S_3}) =
 Z_{\mathbf{1}; \mathbf{1}}^{JK_4 \boxtimes \overline{JK}_4}
+Z_{\mathbf{1}; \bar e}^{JK_4 \boxtimes \overline{JK}_4}
+Z_{e; \mathbf{1}}^{JK_4 \boxtimes \overline{JK}_4}
+Z_{e; \bar e}^{JK_4 \boxtimes \overline{JK}_4}
+2Z_{q; \bar q}^{JK_4 \boxtimes \overline{JK}_4}
\nonumber\\
&=
% Z_{\mathbf{1}; \mathbf{1}}^{JK_4 \boxtimes \overline{JK}_4} &=
2({\color{blue} \chi^{m6 \times \overline{m6}}_{1,0; 1,0}} +  \chi^{m6 \times \overline{m6}}_{5,3; 1,0} +  \chi^{m6 \times \overline{m6}}_{6,\frac{2}{5}; 10,-\frac{7}{5}} +  \chi^{m6 \times \overline{m6}}_{10,\frac{7}{5}; 10,-\frac{7}{5}} 
%Z_{\mathbf{1}; \bar e}^{JK_4 \boxtimes \overline{JK}_4} &= 
+ \chi^{m6 \times \overline{m6}}_{1,0; 5,-3} +  \chi^{m6 \times \overline{m6}}_{5,3; 5,-3} 
\nonumber\\ &\ \ \ \
+  \chi^{m6 \times \overline{m6}}_{6,\frac{2}{5}; 6,-\frac{2}{5}} +  \chi^{m6 \times \overline{m6}}_{10,\frac{7}{5}; 6,-\frac{2}{5}} 
%Z_{e; \mathbf{1}}^{JK_4 \boxtimes \overline{JK}_4} &=
%\nonumber\\
%&\ \ \ \
%+ {\color{blue} \chi^{m6 \times \overline{m6}}_{1,0; 1,0}} +  \chi^{m6 \times \overline{m6}}_{5,3; 1,0} +  \chi^{m6 \times \overline{m6}}_{6,\frac{2}{5}; 10,-\frac{7}{5}} +  \chi^{m6 \times \overline{m6}}_{10,\frac{7}{5}; 10,-\frac{7}{5}} 
%%Z_{e; \bar e}^{JK_4 \boxtimes \overline{JK}_4} &=
%+ \chi^{m6 \times \overline{m6}}_{1,0; 5,-3} +  \chi^{m6 \times \overline{m6}}_{5,3; 5,-3} +  \chi^{m6 \times \overline{m6}}_{6,\frac{2}{5}; 6,-\frac{2}{5}} +  \chi^{m6 \times \overline{m6}}_{10,\frac{7}{5}; 6,-\frac{2}{5}} 
%%Z_{q; \bar q}^{JK_4 \boxtimes \overline{JK}_4} &= 
+2\chi^{m6 \times \overline{m6}}_{3,\frac{2}{3}; 3,-\frac{2}{3}} +  2\chi^{m6 \times \overline{m6}}_{8,\frac{1}{15}; 8,-\frac{1}{15}} 
)
\end{align}
Since the conformal character for the identity operator $ \chi^{m6 \times
\overline{m6}}_{1,0; 1,0}$ appears twice, the ${m6\times \overline{m6}}$ phase
with $\cA_4$ or $\cA_5$ condensation has 2-fold degenerate ground states on a
ring, as one can see from Fig. \ref{S3SDc}(Right).  In fact, such phases
spontaneously breaks the $\Z_2$ symmetry, due to the $e$ or $\bar e$
condensation, which leads to the  2-fold degenerate ground states. 
Let us note here that such phases were termed ``gapless SSB" phases in \Rf{BPW240300905}.

We see that the gapless phase-A has a $\Z_2$ SSB while the
gapless phase-B has no symmetry breaking, despite both phases being described by
$m6\times \overline{m6}$ CFT.  The phase transition between phase-A and phase-B
is a $\Z_2$ symmetry breaking transition.  Thus the critical point of the
transition is described by $m4 \times \overline{m4}$ Ising CFT on top of $m6\times \overline{m6}$ CFT, which is one of the four gapless critical states: $m4\times
m6\times \overline{m4}\times \overline{m6}_i$, $i=1,2,3,4$, listed above.

\section{Discussion}
\label{sec:disc}

Let us recap and give a detailed discussion of the main lessons from Secs.\
\ref{sec:S3model} and \ref{sec:RepS3model}.  We collect our key results under
three directions.  Sec.\ \ref{sec:discussion gauging dualities} reviews the web
of dualities we have obtained by gauging various subgroups of $S^{\,}_{3}$.  In
Sec.\ \ref{subsec:correlation funcs}, we describe symmetry-breaking patterns in
terms of patch operators, and compute their expectation values in the gapped
fixed-point ground states. We argue that these can be used to detect  ordered
and disordered phases of models with general fusion category symmetries.
Finally, in Sec.~\ref{sec:incommensurate} we study a Hamiltonian of spin-1/2 degrees
of freedom that is exactly solvable, has an exact non-invertible self-duality symmetry 
in a certain parameter regime, and supports a gapless incommensurate phase
in its phase diagram. We use this model to 
draw analogies with the gapless regions in the 
phase diagrams of $S^{\,}_{3}$- and $\cRep(S^{\,}_{3})$-symmetric 
models and better understand the latter two.

\subsection{Gauging-induced dualities}
\label{sec:discussion gauging dualities}

\begin{figure}[t] \centering
\includegraphics[width=0.6\linewidth]{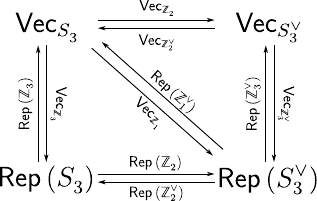} \caption{The web
of dualities obtained induced by gauging.  The four categories $\cVec_{S_{3}}$,
$\cVec_{S^{\vee}_{3}}$, $\cRep(S^{\,}_{3})$, and $\cRep(S^{\vee}_{3})$ at the
corners of the diagram denote the fusion category symmetry of the bond algebras
\eqref{eq:def S3 algebra}, \eqref{eq:def S3/Z3 algebra}, \eqref{eq:def S3/Z2
algebra}, and  \eqref{eq:def S3/S3 algebra}, respectively.  Each directed arrow
between these fusion categories denote a duality induced by gauging. The label
on each arrow denotes the corresponding module category over the category at
the source of the arrow. } \label{fig:duality web} \end{figure}

In Secs.\ \ref{sec:S3model} and \ref{sec:RepS3model}, we presented several
dualities that are induced by gauging subgroups of $S^{\,}_{3}$ symmetry.  In
Fig.\ \ref{fig:duality web}, we summarize the corresponding web of dualities.
The corners of the diagram label the symmetry categories of the dual bond
algebras while the each arrow implies a duality map induced by gauging. It has
been shown in Ref.\ \cite{BT170402330}, distinct gaugings of a symmetry
category $\mathcal{C}$ are in one-to-one correspondence with the left (or
right) module categories over $\mathcal{C}$. Accordingly, we label each arrow
in Fig.~\ref{fig:duality web} by the choice of the corresponding module
category.\footnote{Module categories over $\cVec_G$ ($\cRep(G)$) are given by
the categories $\cVec_H$ ($\cRep(H)$) where $H \leqslant G$, \ie $H$ is a
subgroup of G~\cite{EGNV16}.} Alternatively, each left (or right) module
category over $\mathcal{C}$ is equivalent to the category of right (or left)
$A$-modules over $\mathcal{C}$ where $A\in \mathcal{C}$ is some algebra object
(see Appendix A of Ref.\ \cite{CRBSS230509713}).  This correspondence forms the
connection between the module categories over $\mathcal{C}$ and the perspective
on gauging as summing over symmetry defect insertions in two-dimensional
spacetime.  In the context of fusion category symmetries, gauging can then be
understood as summing over all insertions of $A$-defects in the partition
function in two-dimensional spacetime.

In \fig{fig:duality web}, for the fusion category
$\cVec_{S^{\,}_{3}}$,\footnote{For any group $G$, the fusion category $\cVec_G$
consists of simple objects that can be thought of as $G$-graded vector spaces,
which fuse according to group multiplication law. The morphisms of this
category are graded $G$-graded linear maps. In this subsection, we will refer
to $S_3$ symmetry as $\cVec_{S^{\,}_{3}}$ to emphasize the general language of
fusion category symmetry.} the module categories label the symmetry category of
the subgroups that are not gauged. Note that the module categories are in
one-to-one correspondence with (conjugacy classes) of subgroups of $S^{\,}_3$.
In particular, the module categories $\cVec_{S^{\,}_{3}}$ and
$\cVec_{\Z^{\,}_2}$ of $\cVec_{S^{\,}_{3}}$ correspond to gauging by the
algebra objects $e$ (the trivial algebra object, implementing trivial gauging)
and $e\oplus r\oplus r^2$, respectively. These gauging maps give back the same
symmetry category $\cVec_{S^{\,}_{3}}$.  The latter, in particular, is what we
would ordinarily describe as the $\Z^{\,}_3$ Kramers-Wannier duality. We
provided a recipe for implementing this gauging map in \scn{sec:gauging Z3 in
S3}. On the other hand, the module categories $\cVec_{\Z^{\,}_3}$ and
$\cVec_{\Z_1}$, where $\Z_1$ is the trivial group, correspond to gauging
$\cVec_{S^{\,}_{3}}$ by the algebra objects $e\oplus s$ and $e\oplus r\oplus
r^2\oplus s \oplus sr \oplus sr^2$, respectively. These gauging maps lead to
the dual $\cRep(S^{\,}_{3})$ symmetry category. The first of these is exactly
the gauging map we used to construct the $\cRep(S^{\,}_{3})$ spin chain in
\scn{sec:RepS3model}, starting from the $S^{\,}_{3}$ spin chain of
\scn{sec:S3model}. The second one of these can be implemented by first gauging
$\Z^{\,}_3$ in $S^{\,}_{3}$ and then gauging the $\Z^{\,}_2$ subgroup of the
dual $S^{\vee}_3$, as discussed in \scn{sec:gauging S3 in S3}.

For the fusion category $\cRep(S^{\,}_{3})$, we find that the algebra objects
corresponding to three distinct gaugings of $\cRep(S^{\,}_{3})$ labeled by
module categories $\cRep(\mathbb{Z}^{\,}_{3})$, $\cRep(\mathbb{Z}^{\,}_{2})$,
and $\cRep(\mathbb{Z}^{\,}_{1})$ to be $\bm{1}\oplus\one'$,
$\bm{1}\oplus\bm{2}$, and $\bm{1}\oplus\one'\oplus2\,\bm{2}$,
respectively.\footnote{These were derived using the internal Hom construction
outlined in Appendix A.3 of \Rf{CRBSS230509713}.} Gauging by certain algebra
objects can give back the same symmetry category. For other algebra objects,
one gets a new (dual) symmetry category.  This is a generalization of
Kramers-Wannier duality to arbitrary fusion category symmetries.  In our
example with $\cRep(S^{\,}_{3})$ symmetry, we find that gauging by either of
the algebra objects $\bm{1}\oplus\one'$ or
$\bm{1}\oplus\one'\oplus2\,\bm{2}$ (\ie the regular representation
object) of $\cRep(S^{\,}_3)$ gives rise to $S^{\,}_3$ symmetry.  On the other
hand, gauging by either the trivial algebra object $\bm 1$ (\ie implementing
trivial gauging) or the algebra object $\bm{1}\oplus\bm{2}$ gives back a
``dual" $\cRep(S^\vee_3)$ symmetry. Like the familiar KW duality associated
with gauging Abelian groups, the gauging by $\bm{1}\oplus\bm{2}$ implements a
duality transformation that exchanges pairs of gapped phases of the system. In
our analysis, we did not provide an explicit description of how gauging by
algebra objects, via the insertion of defects approach put forward in
\Rf{BT170402330}, works at the level of microscopic Hamiltonians. However, we
note that gauging by the $\bm{1}\oplus\one'$ algebra object should
proceed very identically to gauging of an ordinary $\Z^{\,}_2$ symmetry since
$\widehat{W}_{\bar{\bm 1}}$ indeed generates a $\Z^{\,}_2$ sub-symmetry of
$\cRep(S^{\,}_{3})$, and gauging by the regular representation algebra object
should be identical to first gauging by $\bm{1}\oplus\one'$ and then
gauging the $\Z^{\,}_3$ subgroup of the resulting $S^{\,}_{3}$ symmetry.
Finally, we identify a sequential quantum circuit \eqref{eq: def DRepS3}, that
implements a duality transformation of our $\cRep(S^{\,}_{3})$ spin chain,
which therefore must correspond to the remaining option of gauging by the
$\bm{1}\oplus\bm{2}$ algebra object.

\subsection{SSB patterns and order/disorder operators}
\label{subsec:correlation funcs}

\setlength{\tabcolsep}{4pt} \renewcommand{\arraystretch}{1.5} \begin{table}
\caption{ \label{tab:S3 order and disorder ops GS} Expectation values of
$S^{\,}_{3}$ order and disorder operators defined in Eq.\ \eqref{eq:S3 order
disorder ops} in the fixed point ground state wavefunctions defined in Eqs.\
\eqref{eq:GS S3}, \eqref{eq:GS Z3}, \eqref{eq:GS Z2}, and \eqref{eq:GS Z1},
respectively.  The non-zero (zero) expectation values of order (disorder)
operators detect the spontaneous symmetry breaking and long-range order in the
ground states.  } \centering \begin{tabular}{l|cccc} \hline \hline $\text{GS}$
& $\widehat{C}^{\,}_{\Z^{\,}_2}(j,\ell)$ & $\widehat{U}^{\,}_{[s]}(j,\ell)$ &
$\widehat{C}^{\,}_{\Z^{\,}_3}(j,\ell)$ & $\widehat{U}^{\,}_{[r]}(j,\ell)$ \\
\hline $\ket{\mathrm{GS}^{\,}_{S^{\,}_{3}}}$ & $0$ & $+3$ & $0$ & $+2$ \\
$\ket{\mathrm{GS}^{\pm}_{\Z^{\,}_{3}}}$ & $+1$ & $0$ & $0$ & $+2$ \\
$\ket{\mathrm{GS}^{\alpha}_{\Z^{\,}_{2}}}$ & $0$ & +1 & $+2$ & $0$ \\
$\ket{\mathrm{GS}^{\pm,\alpha}_{\Z^{\,}_{1}}}$ & $+1$ & $0$ & $+2$ & $0$ \\
\hline \hline \end{tabular} \end{table} \renewcommand{\arraystretch}{1}

Ordered phases in which symmetries are spontaneously broken can be detected by
non-zero values of appropriate correlation functions of order operators.  In
contrast, disordered phases can be detected by non-zero values of appropriate
correlation functions of disorder operators.  The expectation values of
correlation functions of order and disorder operators, considered together,
have been found to be a tool that can detect gaplessness \Rf{L190309028}. The
idea of order and disorder operators can be generalized to non-invertible
symmetries in the form of patch operators
\cite{JW191213492,CW220303596,IW231005790}.  Depending on which gapped phase
the system is in, different patch operators will get a non-zero expectation
value in the ground state(s).  This gives a way to detect symmetry-breaking
even if we restrict to the symmetric sub-Hilbert space so that ground state
degeneracy is no longer a reliable tool.

For the $S^{\,}_{3}$ symmetry, we have the patch operators~\footnote{ We
choose, without loss of generality, $\hat{\si}^z_j \,\hat{\si}^z_{j+\ell}$ to
be the $\Z^{\,}_{2}$ order operator. Alternatively, we could have chosen
$\hat{\si}^z_j \,\hat{\tau}^z_{j+\ell}$, $\hat{\tau}^z_j
\,\hat{\si}^z_{j+\ell}$, or $\hat{\tau}^z_j \,\hat{\tau}^z_{j+\ell}$ as well.
Any of these choices for the order operators produce the same expectation
values in the ground states of gapped fixed-points of the Hamiltonian
\eqref{eq:Ham gen S3 reparameterized}.  } \begin{subequations} \label{eq:S3
order disorder ops} \begin{align} \widehat{C}^{\,}_{\Z^{\,}_2}(j,\ell) \=
\hat{\si}^z_j \,\hat{\si}^z_{j+\ell}, \qquad
\widehat{C}^{\,}_{\Z^{\,}_3}(j,\ell) \= \widehat{Z}^{\,}_j \
\widehat{Z}^{\dagger}_{j+\ell} + \widehat{Z}^{\dagger}_j \
\widehat{Z}^{\,}_{j+\ell}, \end{align} that are associated with $\Z^{\,}_{2}$
and $\Z^{\,}_{3}$ order operators, respectively, and the patch operators
\begin{align} \widehat{U}^{\,}_{[s]}(j,\ell) \= \sum_{\alpha=0}^{2}
\prod_{k=j}^{j+\ell}
\hat{\si}^x_k\,\hat{\tau}^x_k\,\widehat{C}^{\,}_k\,\widehat{X}^{\alpha}_{k},
\qquad \widehat{U}^{\,}_{[r]}(j,\ell) \= \prod_{k=j}^{j+\ell}
\widehat{X}^{\,}_k + \prod_{k=j}^{j+\ell}  \widehat{X}^{\dagger}_k, \end{align}
\end{subequations} that are associated with $\Z^{\,}_{2}$ and $\Z^{\,}_{3}$
disorder operators, respectively.  We note that both of these classes of patch
operators are symmetric under the entire $S^{\,}_{3}$ group, i.e., they are
constructed out of the generators of $S^{\,}_{3}$-symmetric bond algebra
\eqref{eq:def S3 algebra}, while the disorder operators are closed under the
action of $S^{\,}_{3}$.  Both order and disorder operators can be thought of as
\textit{transparent} patch operators \cite{CW220303596} in the sense that they
commute all the terms in the $S^{\,}_{3}$ Hamiltonian \eqref{eq:Ham gen S3
reparameterized} that are supported between sites $j+1$ and $j+\ell-1$.  On the
Hamiltonian their nontrivial actions only appear at their boundaries.

The expectation values attained by the patch operators in the gapped
fixed-point  ground states \eqref{eq:GS S3}, \eqref{eq:GS Z3}, \eqref{eq:GS
Z2}, and \eqref{eq:GS Z1} are given in Table \ref{tab:S3 order and disorder ops
GS}.  In the fixed-point ground states, non-zero expectation values of order
operators accompany the vanishing expectation values of disorder operators and
detect spontaneous symmetry breaking in the ground states.  We note that when
the $S^{\,}_{3}$ symmetry is broken down to $\mathbb{Z}^{\,}_{2}$, each of the
threefold degenerate ground states preserve a different $\mathbb{Z}^{\,}_{2}$
subgroup which reflected in the non-vanishing expectation value of
$\mathbb{Z}^{\,}_{2}$ disorder operators on only one of the degenerate ground
states. Away from the fixed-points the zero expectation values are expected to
be replaced by an exponential decay $\propto e^{-|j-\ell|/\xi}$ with a finite
non-zero correlation length $\xi$ (with gapped fixed-points corresponding to
$\xi\to 0$ limit).

\setlength{\tabcolsep}{4pt} \renewcommand{\arraystretch}{1.5} \begin{table}
\caption{ \label{tab:repS3 order and disorder ops GS} Expectation values of
$\cRep(S^{\,}_{3})$ order and disorder operators defined in Eq.\
\eqref{eq:repS3 order disorder ops} in the fixed point ground state
wavefunctions defined in Eqs.\ \eqref{eq:GS repS3}, \eqref{eq:GS W1},
\eqref{eq:GS W2}, \eqref{eq:gs12}, and \eqref{eq:gs3}, respectively.  The
non-zero (zero) expectation values of order (disorder) operators detect the
spontaneous symmetry breaking and long-range order in the ground states.  }
\centering \begin{tabular}{l|cccc} \hline \hline $\text{GS}$ &
$\widehat{C}^{\,}_{\one'}(j,\ell)$ &
$\widehat{W}^{\,}_{\one'}(j,\ell)$ &
$\widehat{C}^{\,}_{\mathbf{2}}(j,\ell)$ &
$\widehat{W}^{\,}_{\mathbf{2}}(j,\ell)$ \\ \hline
$\ket{\mathrm{GS}^{\,}_{\cRep(S^{\,}_{3})}}$ & $0$ & $+1$ & $0$ & $+2$ \\
$\ket{\mathrm{GS}^{\alpha}_{\one'}}$ & $0$ & $+1$ & $+2$ & $0$ \\
$\ket{\mathrm{GS}^{\pm}_{\mathbf{2}}}$ & $+3$ & $0$ & $0$ & $+2$ \\
$\ket{\mathrm{GS}^{\pm}_{\mathbf{1}}}$ & $+1$ & $0$ & $+2$ & $0$ \\
$\ket{\mathrm{GS}^{3}_{\mathbf{1}}}$ & $+1$ & $0$ & $+2$ & $0$ \\ \hline \hline
\end{tabular} \end{table} \renewcommand{\arraystretch}{1}

For the $\cRep(S^{\,}_{3})$ symmetry, we can apply the $\mathbb{Z}^{\,}_{2}$
gauging map, derived in Sec.\ \ref{sec:RepS3model}, to obtain the
$\cRep(S^{\,}_{3})$ patch operators.  Since we are gauging $\Z^{\,}_{2}$
subgroup of $S^{\,}_{3}$, we expect the $\mathbb{Z}^{\,}_{2}$ order and
disorder operators to be mapped to disorder and order operators of the dual
$\Z^{\,}_{2}$ symmetry generated by $\widehat{W}^{\,}_{\one'}$
operator.  Accordingly, we identify the patch operators \begin{subequations}
\label{eq:repS3 order disorder ops} \begin{align} &
\widehat{C}^{\,}_{\one'}(j,\ell) \= \hat{\mu}^{z}_{j}\,
\sum_{\al=0}^2 \left( \prod_{k=j}^{j+\ell}
\widehat{X}^{\prod_{q=j+1}^{k}\hat{\mu}^{x}_{q}}_{k} \right)^\al
\hat{\mu}^{z}_{j+\ell},
% \hat{\mu}^{z}_{j}\, \left( 1 + \prod_{k=j}^{j+\ell}
% \widehat{X}^{\prod_{q=j+1}^{k}\hat{\mu}^{x}_{q}}_{k} + \prod_{k=j}^{j+\ell}
% \widehat{X}^{-\prod_{q=j+1}^{k}\hat{\mu}^{x}_{q}}_{k} \right)
% \hat{\mu}^{z}_{j+\ell},
\\ & \widehat{W}^{\,}_{\one'}(j,\ell) \= \prod_{k=j+1}^{j+\ell}
\hat{\mu}^{x}_{k}, \end{align} that correspond to
$\widehat{W}^{\,}_{\one'}$ order and disorder operators,
respectively.  In contrast to this, under gauging $\mathbb{Z}^{\,}_{2}$
subgroup of $S^{\,}_{3}$ symmetry, the $\Z^{\,}_{3}$ order and disorder
operators are mapped to \begin{align} & \widehat{C}^{\,}_{\mathbf{2}}(j,\ell)
\= \widehat{Z}^{\prod_{k=j+1}^{j+\ell}\hat{\mu}^{x}_{k}}_{j}\,
\widehat{Z}^{\dagger}_{j+\ell} +
\widehat{Z}^{-\prod_{k=j+1}^{j+\ell}\hat{\mu}^{x}_{k}}_{j}\,
\widehat{Z}^{\,}_{j+\ell}, \\ & \widehat{W}^{\,}_{\mathbf{2}}(j,\ell) \=
\prod_{k=j}^{j+\ell} \widehat{X}^{\prod_{q=j+1}^{k}\hat{\mu}^{x}_{q}}_{k} +
\prod_{k=j}^{j+\ell} \widehat{X}^{-\prod_{q=j+1}^{k}\hat{\mu}^{x}_{q}}_{k},
\end{align} \end{subequations} which are the order and disorder operators,
respectively, associated with the non-invertible symmetry operator
$\widehat{W}^{\,}_{\mathbf{2}}$.  We note that for $\cRep(S^{\,}_{3})$
symmetry, order operators are non-local string-like objects, as opposed to the
case of $S^{\,}_{3}$ symmetry for which order operators are bilocal, i.e.,
products of two local operators. In other words, the spontaneous breaking of
non-invertible $\cRep(S^{\,}_{3})$ is detected by non-local string order
parameters.  The expectation values attained by the patch operators in the
gapped fixed-point ground states \eqref{eq:GS repS3}, \eqref{eq:GS W1},
\eqref{eq:GS W2}, and \eqref{eq:GS 1} are given in Table~\ref{tab:repS3 order
and disorder ops GS}.  We see that the expectation values of operators
\eqref{eq:repS3 order disorder ops} in these ground states are consistent with
interpreting the corresponding phases as $\cRep(S^{\,}_{3})$-symmetric,
$\cRep(S^{\,}_{3})/\mathbb{Z}^{\,}_{2}$ SSB, $\mathbb{Z}^{\,}_{2}$ SSB, and
$\cRep(S^{\,}_{3})$ SSB phases, respectively.

\subsection{Incommensurate phase in a self-dual spin-1/2 chain}
% anomalous $\mathrm{U}(1)$ symmetry, and \texorpdfstring{$z>1$}{z>1} continuous transition }
\label{sec:incommensurate}

In the phase diagram of the $S_3$-symmetric model shown in Fig.\ \ref{fig:s3-incomm}, 
we observed an extended gapless phase that is centered around the self-dual line.
We have argued that such a gapless phase is an incommensurate phase.  
We define an incommensurate state as a gapless state that has gapless excitations 
carrying quasi-momentum that is incommensurate with the size of the Brillouin zone.  
As a result, an incommensurate state contain gapless excitations whose quasi-momenta form a
dense set that covers the whole Brillouin zone.  
To better understand such an incommensurate phase, we will first consider the following spin-1/2 chain
\begin{align}
\widehat{H}
\=
-
\sum_{j=1}^{L}
\left\{
J\,
\hat{\sigma}^{z}_{j}\,
\hat{\sigma}^{z}_{j+1}
+
h\,\hat{\sigma}^{x}_{j}
+
\lambda
(
\hat{\sigma}^{y}_{j}\,
\hat{\sigma}^{z}_{j+1}
-
\hat{\sigma}^{z}_{j}\,
\hat{\sigma}^{y}_{j+1}
)
\right\},
\label{eq:Ham gen Ising selfdual}
\end{align}
which also has an incommensurate phase. 
Hamiltonian \eqref{eq:Ham gen Ising selfdual} describes the transverse field Ising model perturbed
by a Dzyaloshinskii-Moriya-type interaction. 
The latter is one of the simplest
two-body interactions that is symmetric under the (non-invertible)
$\Z^{\,}_2$ KW self-duality symmetry of the critical point of the Ising model ($h=J, \la=0$). 
The Hamiltonian \eqref{eq:Ham gen Ising selfdual} then has the self-duality symmetry when $h=J$.

\begin{comment}
As opposed to the $S^{\,}_{3}$-symmetric
Hamiltonian \eqref{eq:def Ham gen S3}, this Hamiltonian is exactly solvable
by applying JW transformation. This maps the Hamiltonian 
to a fermionic one with only
quadratic interactions~\cite{SCL09051849,OF171206662}.  
After the JW
transformation, one can diagonalize the fermionic Hamiltonian via a Bogoliubov
transformation which yields
\begin{equation}\label{eq:Ham fermion}
\widehat{H} 
= 
\sum_{-\pi< k\leq \pi} 
E^{\,}_k \, 
\hat{\psi}_k^{\dagger} \hat{\psi}^{\,}_k + E^{\,}_0
\end{equation}
where
\begin{equation}\label{eq:Ek fermion}
E^{\,}_k = 
4\lambda \sin k 
\pm 
2\sqrt{J^2\sin^2 k+(h-J\cos k)^2}
\end{equation}
Here, we note that electrons at quasi-momentum $ k $ are identified with holes
at quasi-momentum $ -k $.  We can either choose to only look at both electrons
and holes for quasi-momenta in $ [0,\pi] $ or only the electrons (or,
equivalently only the holes) for all $ k\in (-\pi,\pi] $.  We take the latter
point of view.
\end{comment}

Unlike the $S^{\,}_{3}$-symmetric
Hamiltonian \eqref{eq:def Ham gen S3}, this Hamiltonian is exactly solvable
by applying Jordan-Wigner transformation. This maps the Hamiltonian 
to a fermionic one with only
quadratic interactions~\cite{SCL09051849}. Closely related results have also been obtained with other self-dual deformations of the critical Ising model~\cite{RZA150405192,RZA150503966,OF171206662}. 
After the JW
transformation, one can diagonalize the fermionic Hamiltonian via a Bogoliubov
transformation which yields
\begin{equation}\label{eq:Ham fermion}
\widehat{H} 
= 
\sum_{-\pi< k\leq \pi} 
E^{\,}_k \, 
\widehat{\psi}_k^{\, \dagger} \widehat{\psi}^{\,}_k + E^{\,}_0
\end{equation}
where
\begin{equation}\label{eq:Ek fermion}
E^{\,}_k = 
4\lambda \sin k 
+
2\sqrt{J^2\sin^2 k+(h-J\cos k)^2}
\end{equation}
In the ground state of this Hamiltonian, the $k$ modes with negative energy are filled in the ground state. 
The excitations are hole-like/particle-like for the negative/positive $E_k$ modes.
\begin{figure}[t!]
\centering
\begin{subfigure}{0.24\textwidth}
\centering
\includegraphics[width=\linewidth]{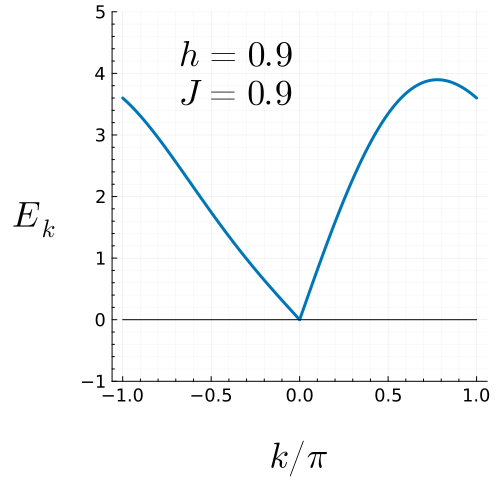}
\caption{}
\end{subfigure}
\begin{subfigure}{0.24\textwidth}
\centering
\includegraphics[width=\linewidth]{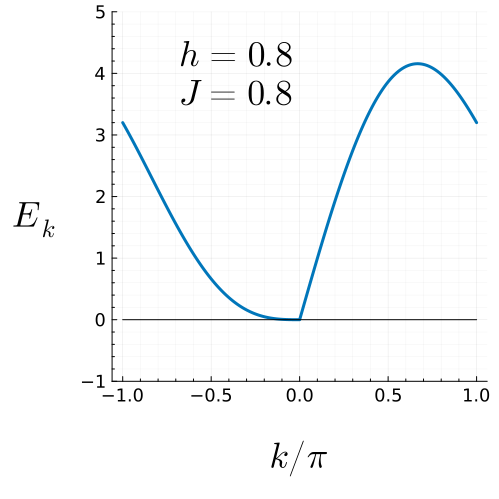}
\caption{}
\end{subfigure}
\begin{subfigure}{.24\textwidth}
\centering
\includegraphics[width=\linewidth]{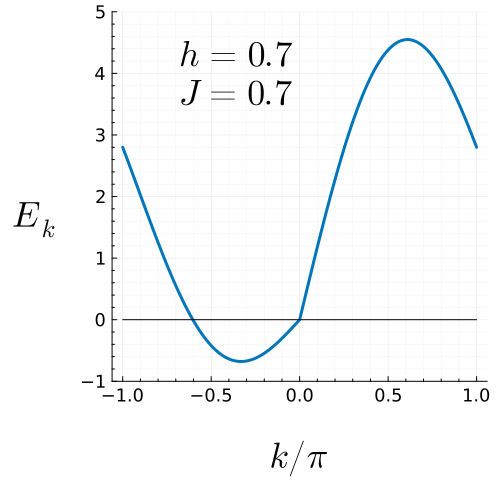}
\caption{}
\end{subfigure}
\begin{subfigure}{.24\textwidth}
\centering
\includegraphics[width=\linewidth]{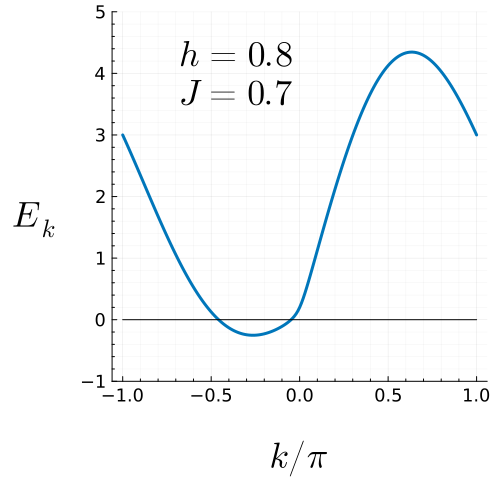}
\caption{}
\end{subfigure}
\caption{
The fermion dispersion \eqref{eq:Ek fermion} for
$\la = 2-h-J$ and
(a) $h=0.9,\ J=0.9$ (the $c=\frac12$ self-dual critical point),
(b) $h=0.8,\ J=0.8$ (the $z=3$ self-dual multi-critical point),
(c) $h=0.7,\ J=0.7$ (the $c=1$ self-dual gapless phase),
(d) $h=0.8,\ J=0.7$ (the $c=1$ gapless phase)}
\label{fig:IsingEk}
\end{figure}

\begin{figure}[t!]
\centering
\includegraphics[height=2.1in]{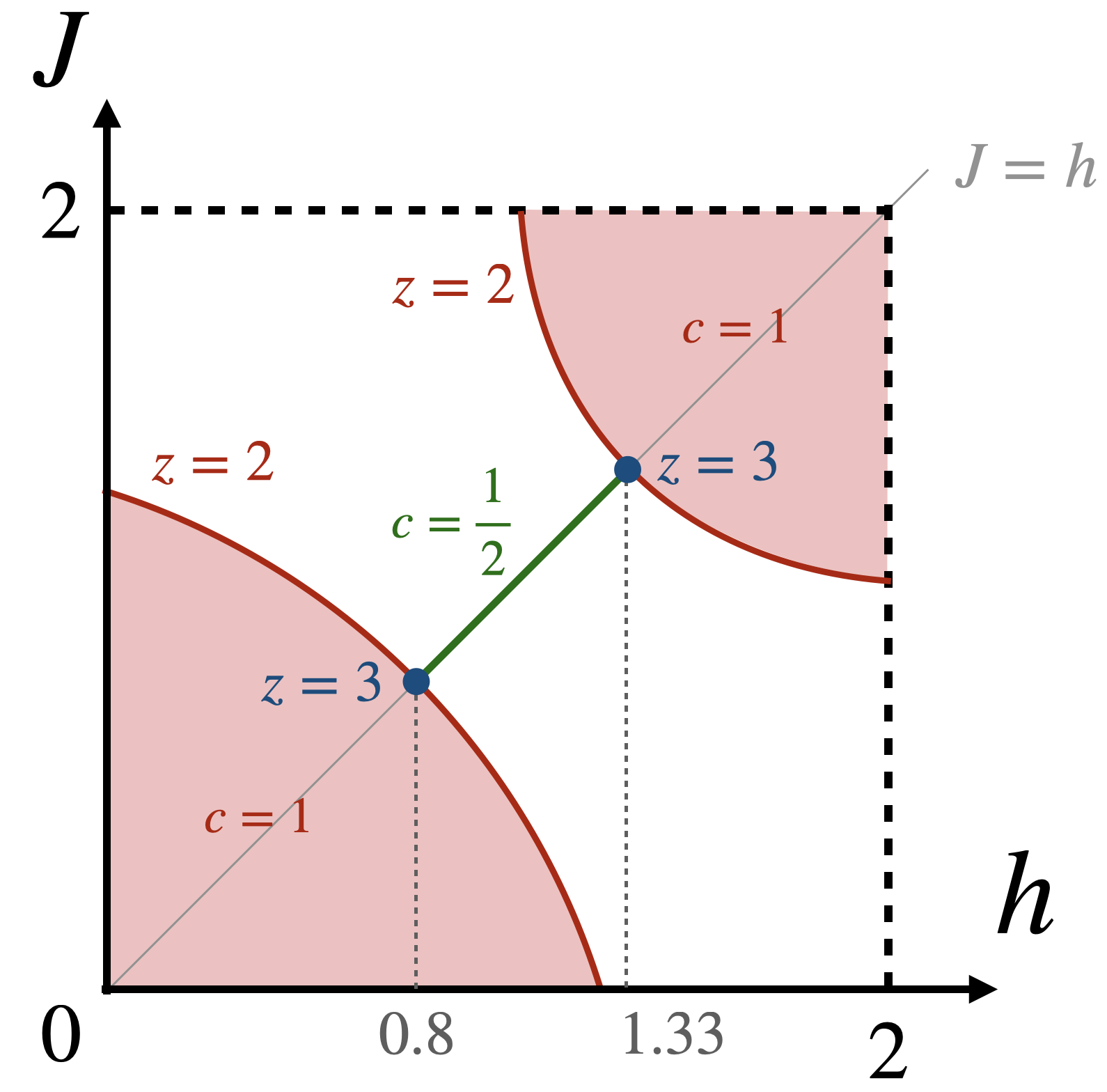}
\caption{Phase diagram of Hamiltonian \eqref{eq:Ham gen Ising selfdual} in the
$ (h,J) $ plane with $ \la=2-J-h $ at each point. The unshaded regions are
gapped.  The $h=J$ line has the self-duality symmetry.
} \label{fig:ising-self-dual-phase-diagram}
\end{figure}

The fermion dispersion  \eqref{eq:Ek fermion} is plotted in Fig.\  
\ref{fig:IsingEk} for some values of $h,J,\la$.  From this spectrum, we
can identify the low-energy degrees of freedom and whether the ground state is
gapped or not.  
Along the self-dual line, the central charge computed from DMRG
evaluates to $ c=1/2 $ for $\la\neq 0$. On the other hand, in the large $|\la|$ limit, we find $c=1$. In fact there are extended gapless phases with $c=1$ even away from the self-dual line in the large $|\la|$ limit.
The full phase diagram is shown in \fig{fig:ising-self-dual-phase-diagram}, with the gapped phases in white and two gapless phases shaded in red. The critical point corresponding to the Ising CFT lies at the $J=h$, $\la=0$ point. 

% Whenever there are linear dispersing modes at zero energy with
% left and right moving partners, we interpret the low energy effective field
% theory as a $ c=1/2 $ Majorana CFT.  Whenever there are two pairs of such
% linear dispersing modes, we identify the low energy effective field theory to
% be a $ c=1 $ compact boson theory.  The phase diagram we obtain from this
% analysis is shown in \fig{fig:ising-self-dual-phase-diagram}.

The extended gapless regions labeled $ c=1 $ have gapless linearly dispersing
modes near a ``Fermi momentum" that smoothly varies throughout the regions.  
This
behavior is seen only when $ \la $ is larger than $ 0.4 $ or smaller
than $  -0.66 $.  
Moreover, we see that the value of $c=1/2$ is stable
to a finite non-zero value of $ \la $. In other words, the $c=1/2$ critical point becomes a stable gapless phase in the presence of the self-duality symmetry on the $h=J$ line. This is consistent with the fact that the $\lambda$ term in \eqref{eq:Ham gen Ising selfdual} corresponds to an exactly marginal chiral operator $\propto T-\bar{T}$ in the Ising CFT~\cite{ZMV190106439}.

Let us make a few comments:
\begin{enumerate}[(i)]
\item
Time reversal and reflection symmetries are explicitly broken by the $\lambda$ term in \eqref{eq:Ham gen Ising selfdual}.

\item
% The self-duality symmetry pins a pair of left- or right-moving Majorana fermions
% to $k=0$ point.
The self-duality symmetry pins the right-moving particle- and hole-like excitations at the $k=0$ point.

\item
% In the $c=1$ gapless phase, there is at least one pair of left- and
% right-moving Majorana fermions that has non-zero Fermi momenta $k^{\,}_{\mathrm{F}}$.
% The smoothly varying non-zero Fermi momenta $k^{\,}_{\mathrm{F}}$ implies that
% the $c=1$ gapless phase is incommensurate.
In each $c=1$ gapless region, there is a chiral complex fermion mode that has non-zero Fermi momentum $k^{\,}_{\mathrm{F}}$.
This $k^{\,}_{\mathrm{F}}$ varies smoothly throughout the region, hence we refer to it as an incommensurate gapless phase.

\item
% Without the self-duality symmetry,
% there are two pairs of left- and
% right-moving Majorana fermions that have non-zero Fermi momenta $k^{\,}_{\mathrm{F}}$.
Without the self-duality symmetry,
there are both left- and right-moving complex fermion modes with non-zero Fermi momenta $k^{\,}_{\mathrm{F}}$.

\item
% The $h=J$ line is a continuous transition line even within the
% $c=1$ incommensurate phase. This is because the fermion dispersion has a singular change as we go across the $h=J$ line in the
% $c=1$ incommensurate phase.
The fermion dispersion has a singular change as we go across the $h=J$ line in the
gapless incommensurate phase. 
This suggests that this is a continuous phase transition within the
incommensurate phase. 

\end{enumerate}

In the incommensurate gapless phase for large $\la$, a low-energy chiral (complex) fermion mode near Fermi momentum $k_F$ furnishes a representation of a $\mathrm{U}(1)$ symmetry.
This U(1) symmetry emanates from the lattice translation symmetry. For instance when $\la>0.4$, there are left-moving fermion modes satisfying 
\begin{equation}
\widehat T^{\,}_{\text{lat}} \widehat\psi_{L,q} \widehat T_{\text{lat}}^{\dagger} 
= e^{ik_F} e^{iq}\widehat \psi_{L,q}
\, ,
\end{equation}
while the right-moving fermion modes satisfy
\begin{equation}
\widehat{T}^{\,}_{\text{lat}} \widehat\psi_{R,q} 
\widehat{T}_{\text{lat}}^{\dagger} 
= e^{iq} \widehat\psi_{R,q} \, .
\end{equation}
So this chiral U(1) symmetry acts as $e^{ik_F \widehat N_L}$ where $\widehat N_L$ is the number operator in the left-moving sector of the IR theory. 
When $k_F$ is not a rational multiple of $\pi$, the set $\{e^{ik_F \widehat N_L|N_L\in \Z_{\geq 0}}\}$ is dense in U(1). 
However when $k_F=2\pi p/q$, with $\gcd(p,q)=1$, the emanent symmetry becomes a finite cyclic group of order $q$. 
We still refer to the emanant symmetry as $U(1)$ since the finite-order case happens only for a measure 0 subset of values of $k_F\in (-\pi,\pi]$.

The presence of U(1)
symmetries emanating from lattice translation may be a general feature 
of the low energy effective field theory of an incommensurate gapless state. 
Note that the right-moving particles and holes (for $\la>0.4$) carry zero $\mathrm{U}(1)$ charge (see Fig. \ref{fig:IsingEk}c). 
The emanant chiral $\mathrm{U}(1)$ symmetry is anomalous. This anomaly is reflected in the UV by the fact that the ground state carries momentum 
\begin{equation}
k_{0} = \sum_{k\in(-\pi,\pi]} k \, \Theta(-E_k) \stackrel{L\to \infty}{\approx} \frac{k_F^2}{2} \, .
\end{equation}
So the UV translation symmetry acts projectively on the low-energy states.
This has close parallels with the examples discussed in \Rf{CS221112543}. 
In addition to the appearance of anomalous chiral $\mathrm{U}(1)$ symmetry, 
the notion of incommensurate state also allows us to make
the following conjecture. Some continuous transitions, \eg Mott insulator to superfluid transition,
involve the introduction of gapless modes. If such gapless modes are incommensurate,
\ie if the transition is between commensurate and incommensurate phases, then
the dynamical exponent must satisfy $z > 1$.  The transition along the self-dual $h=J$ line
at $h=J=0.8$ in Fig. \ref{fig:ising-self-dual-phase-diagram} is an example of
such a transition, with $z=3$.

In the $c=1$ incommensurate phase, away from the $h=J$ line,
there are two chiral complex fermion modes, carrying different non-zero $\mathrm{U}(1)$ charges 
(\ie quasi-momenta).
On the $h=J$ line, the $\mathrm{U}(1)$ charge of
one set of chiral complex fermion modes vanishes, and the associated particle and hole branches acquire unequal velocities.
This causes a singularity in the single-particle fermion spectrum, signaling a possible continuous phase transition. 
This is an unusual kind of phase transition, since it does not have more gapless modes compared to the phases on the two sides of the transition.

\begin{figure}[t!]
\centering
\includegraphics[height=1.5in]{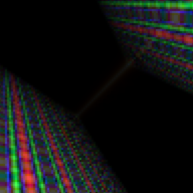}\ \
\includegraphics[height=1.5in]{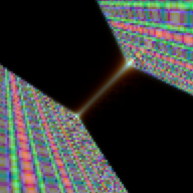}\ \
\includegraphics[height=1.5in]{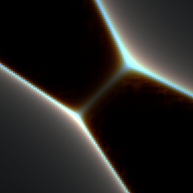}
\caption{
(Middle) Central charge $c$ computed via tensor network numerical approach as a
function of $h \in [0,2]$ (horizontal) and $J \in [0,2]$ (vertical) with $
\la=2-J-h $, for the Ising chain \eqref{eq:Ham gen Ising selfdual} of sizes $L=256$ (blue), $L=128$ (green),
$L=64$ (red).  The range of the color intensity [0,1] corresponds to the range
of the central charge $c\in [0,2]$.  The stripe pattern in the gapless $c\neq
0$ area comes from incommensurate nature of the gapless phase.  (Left) A plot
of $vc$ for the same range of $h$ and $J$. (Right) A plot of $c/v$ for the same
range of $h$ and $J$.  Here $v$ is the velocity of the gapless mode.
} \label{fig:Isingc}
\end{figure}

\begin{figure}[t!]
\centering
\includegraphics[height=1.5in]{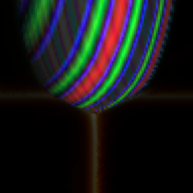}\ \
\includegraphics[height=1.5in]{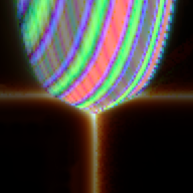}\ \
\includegraphics[height=1.5in]{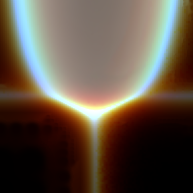}
\caption{
(Middle) Central charge $c$ computed via tensor network numerical approach as a
function of $ J^{\,}_1 \in [0.4,0.6]$ (horizontal) and $ J^{\,}_{3} \in
[0.4,0,6] $ (vertical) for the $S_3$-symmetric model \eqref{eq:Ham gen S3
reparameterized} of sizes $L=256$ (blue), $L=128$ (green), $L=64$ (red).  We
have chosen $ J^{\,}_2=1-J^{\,}_1 $, $ J_4=1-J^{\,}_{3} $ and
$\th=0.7\approx\frac{2\pi}{9} $. 
The $J_1=J_2=0.5$ line is the self-dual line.
The range of the color intensity [0,1] corresponds to the range
of the central charge $c\in [0,2]$.  The stripe pattern in the gapless $c\neq
0$ area comes from incommensurate nature of the gapless phase.  (Left) A plot
of $vc$ for the same range of $h$ and $J$. (Right) A plot of $c/v$ for the same
range of $h$ and $J$.  Here $v$ is the velocity of the gapless mode.
} \label{fig:S3c}
\end{figure}

In Fig.\ \ref{fig:Isingc}, we present a tensor network calculation 
of the central charge for the Hamiltonian 
\eqref{eq:Ham gen Ising selfdual}. In the gapless incommensurate phase we find a stripe pattern
in central charge $c$.  To understand the appearance of stripes, 
we note that the central charge is contained in the $1/L$
term in the ground state energy:
\begin{align}
 E^{\,}_\text{grnd} = \eps L - \frac{cv}{24}\frac{2\pi}{L}  + \cO\left(\frac 1 {L^2}\right).
\end{align}
By computing the ground state energy at size $L$ and $L/2$, we can extract the
$1/L$ and obtain the value of $vc$. 
In particular, we note that
\begin{equation}\label{eq:cv}
    \frac{4L}{\pi}\left[E_{\text{grnd}}(L) - 2E_{\text{grnd}}(L/2)\right] =  cv + \cO\left(\frac 1 {L}\right)
\end{equation}
so that in the large $L$ limit, the quantity on the l.h.s.\  approaches $cv$.
By doing a similar calculation with
space and time exchanged, we can compute $c/v$.  Then the central charge is the
geometric mean of $vc$ and $c/v$. Certainly, 
this approach works only when all the gapless modes have the same velocity $v$.
This is how we obtained the Fig.\ \ref{fig:Isingc}.
We see that $vc$ is small at the boundary of
the incommensurate phase, while $c/v$ is large at the boundary.  This is
consistent with our exact result, which implies that $v=0$ at the boundary.
As we change parameters, the $\mathrm{U}(1)$ charge (\ie the quasi-momentum) of the
ground state may jump, which cause a change of ground state energy of order $1/L$. This leads to the stripe pattern in the calculated central charge in the incommensurate phase.
We contrast this with the case  when $\eps$ is $L$-independent, for which the analysis in \eqref{eq:cv} gives a constant $L$-independent $vc$.

Armed with this understanding, we compute the central charge $c$, as well as
$vc$ and $c/v$, for the $ S^{\,}_{3} $ symmetric Hamiltonians (see Fig.
\ref{fig:S3c}).  The similarity between Figs.\ \ref{fig:S3c} and 
\ref{fig:Isingc} suggests that the gapless phase in Fig.\ \ref{fig:S3c} is
also incommensurate.  
The continuous transition between commensurate and
incommensurate phases has a dynamical exponent $z> 1$. 
Also, the gapless incommensurate phase of our $ S^{\,}_{3} $-symmetric model has a local low energy 
effective field theory with  internal $\mathrm{U}(1)$ symmetry 
(coming from the lattice translation symmetry). Since $S_3$ and $\cRep(S_3)$ symmetries
are Morita equivalent as fusion categories,
the same results also apply to our $ \cRep(S^{\,}_{3}) $-symmetric model.

The parameter $ \th $ in our $ S^{\,}_{3} $ and $ \cRep(S^{\,}_{3}) $ symmetric
models, \eqref{eq:Ham gen S3 reparameterized} and \eqref{eq:Ham gen Rep S3
reparameterized}, plays a very similar role to that of $ \la $ in Hamiltonian 
\eqref{eq:Ham gen Ising selfdual}.
For small $ \th $ and $ \la $ along the self-dual line, the (multi-)critical point remains gapless and 
behaves like stable gapless phase protected by the non-invertible self-duality symmetry.
At large enough values of $ \th $ and $ \la $ in respective
models, this gapless phase turns into an incommensurate phase via a $z>1$ continuous transition. 
\section{Conclusion}
\label{sec:conclusion}

In this paper, we have studied the consequences of non-invertible symmetries
that are realized as genuine UV symmetries (\ie not emergent IR symmetries) 
of 1+1d Hamiltonian lattice models defined on Hilbert spaces having tensor
product decompositions with finite-dimensional on-site Hilbert spaces; we refer
to them as \emph{spin chains} in short. We constructed a spin chains with $\cRep(S^{\,}_3)$ symmetry by gauging the non-normal 
$\Z_2$ subgroup symmetry of an $S^{\,}_{3}$-symmetric spin chain.
This provides an explicit microscopic verification of an instance of the fact 
that gauging non-normal subgroups in a theory with an ordinary group-like 
symmetry leads to a dual theory with a non-invertible symmetry, which was known 
using more abstract methods in the literature.\cite{BT170402330,TW210612577}
We explored SSB phases of the non-invertible $\cRep(S^{\,}_3)$ symmetry as well as continuous phase transitions between them.
Both the $S^{\,}_{3}$ and $\cRep(S^{\,}_{3})$ spin chains demonstrate intrinsically non-invertible self-duality symmetries in special subspaces of the full parameter space. 
We identify the SymTO description of this enhanced symmetry and use it to obtain model-independent constraints on the phase diagrams with this symmetry.

There are various directions for future work. It would be interesting to study 
how gauging by algebra objects may be implemented at the lattice level. This 
requires a deeper understanding of how symmetry twists of non-invertible 
symmetries are implemented in the context of spin chain Hamiltonians. Such an 
understanding would also be extremely useful in making statements about the 
gauging-related dualities in the context of non-invertible symmetries, including 
those discussed in this paper, more precise. We note that similar questions have 
been addressed in the context of lattice models built on Hilbert spaces that do 
not necessarily have a tensor product decomposition, such as in 
\Rfs{I211012882,LV211209091,LDV221103777}. We believe that our treatment of 
these questions is complementary to these previous works. In fact, it is not 
clear that all fusion category symmetries can be realized in spin chain 
Hamiltonians as strictly internal symmetries (\emph{cf.}~footnote 32 of 
\Rf{SSS240112281}). 
Using our lattice model, we also explored KW-symmetric perturbations to the 
analytically tractable limit of the $S^{\,}_{3}$ spin chain
and uncovered a stable gapless phase. 
We find evidence that the numerically computed central charge vanishes at the 
phase boundaries of this gapless region, which is consistent with a
dynamical critical exponent $z>1$. It would be interesting to characterize this 
gapless phase and transitions out of it, on or away from the KW-symmetric line, 
from a low-energy field theory perspective.
Lastly, it is possible to gauge the $\Z_2$ sub-symmetry of the $ S^{\,}_{3} $ 
symmetric spin chain using fermionic $ \Z_2 $ variables instead of bosonic ones. 
This leads to a theory of fermions coupled to spin variables with a fermionic $ 
\mathsf{SRep}(S^{\,}_{3}) $ symmetry. We discuss this in Appendix 
\ref{app:fermion} as a straightforward generalization of the story presented in 
\scn{sec:gauging Z2 in S3} to fermionic lattice models with non-invertible 
symmetries. The fermion parity symmetry $ \Z_2^{\text{F}} $ becomes a part of 
the symmetry category in the gauged model. We leave a discussion of more non-
trivial examples for future work.

%----------------------------------------

\section*{Acknowledgements}
We thank Liang Kong, Sal Pace, Shu-Heng Shao for helpful comments on the manuscript, and
Alex Avdoshkin, Clement Delcamp, Zhihuan Dong, Andreas L\"auchli, Frederic Mila, Christopher Mudry, Seth Musser, Sal Pace, Rahul Sahay, Sahand Seifnashri, Tomo Soejima, Ophelia Sommer, Saran Prembabu, Nat Tantivasadakarn, and Carolyn Zhang for many helpful discussions. 
We thank Apoorv Tiwari for helpful comments on Appendix \ref{app:fermion}.

\paragraph{Funding information}
This work is partially supported by NSF DMR-2022428 and by the Simons 
Collaboration on Ultra-Quantum Matter, which is a grant from the Simons 
Foundation (651446, XGW). \"O. M.\ A.\ is also 
supported by Swiss National Science Foundation (SNSF)
under Grant No.\ P500PT-214429.

\begin{appendix}
\section{Group \texorpdfstring{$S^{\,}_{3}$}{S3}}
\label{app:S3}

The group $S^{\,}_{3}$ has cardinality $|S^{\,}_{3}|=6$ and 
generated by two elements $s$ and $r$ such that $s^{2}=r^{3}=e$, $s\,r = r^{2}\,s$, 
where $e$ is the identity element of the group. 
It has four non-trivial proper subgroups that we denote by
\begin{align}
\mathbb{Z}^{r}_{3}
\=
\left\{e,\,r,\,r^{2}\right\},
\quad
\mathbb{Z}^{s}_{2}
\=
\left\{e,\,s\right\},
\quad
\mathbb{Z}^{sr}_{2}
\=
\left\{e,\,s\,r\right\},
\quad
\mathbb{Z}^{sr^{2}}_{2}
\=
\left\{e,\,s\,r^{2}\right\},
\end{align}
respectively. 
From now on, we will choose $\mathbb{Z}^{s}_{2}$ as the 
$\mathbb{Z}^{\,}_{2}$ subgroup and drop the superscripts $r$ and $s$ when 
referring to the $\mathbb{Z}^{\,}_{3}$ and $\mathbb{Z}^{\,}_{2}$ subgroups of 
$S^{\,}_{3}$.

There are 3 irreducible representations (irreps) of the group $S^{\,}_3$. The trivial one, with all group elements represented by the number 1, is denoted $ \one$,
\begin{equation}
U_{\one}(g) = 1\, , \quad \forall g\in S_3.
\end{equation}
There is a second one-dimensional irrep, denoted $\one'$, with
\begin{equation}
U_{\one'}(e) = U_{\one'}(r) = U_{\one'}(r^2) = 1\, , \qquad U_{\one'}(s) = U_{\one'}(sr) = U_{\one'}(sr^2) = -1.
\end{equation}
The third irrep is a two-dimensional one, denoted $\two$, with
\begin{alignat*}{3}
& U_{\two}(e)
=
\begin{pmatrix}
1 & 0\\
0 & 1
\end{pmatrix},
\quad 
&& U_{\two}(r)
=
\begin{pmatrix}
\ee^{\ii \frac{2\pi}{3}} & 0\\
0 & \ee^{-\ii \frac{2\pi}{3}}
\end{pmatrix},
\quad
&& U_{\two}(r^2)
=
\begin{pmatrix}
\ee^{-\ii \frac{2\pi}{3}} & 0\\
0 & \ee^{\ii \frac{2\pi}{3}}
\end{pmatrix},
\\
& U_{\two}(s)
=
\begin{pmatrix}
0 & 1\\
1 & 0
\end{pmatrix},
\quad
&& U_{\two}(sr)
=
\begin{pmatrix}
0 & \ee^{-\ii \frac{2\pi}{3}}\\
\ee^{\ii \frac{2\pi}{3}} & 0
\end{pmatrix},
\quad
&& U_{\two}(sr^2)
=
\begin{pmatrix}
0 & \ee^{\ii \frac{2\pi}{3}}\\
\ee^{-\ii \frac{2\pi}{3}} & 0
\end{pmatrix}.	
\end{alignat*}
The irrep $\two$ is the only faithful irrep of $S^{\,}_3$. The tensor product of the irreps forms a so-called fusion ring with $\one$ as the identity of the ring. The product operation (or, fusion) is commutative with the following non-trivial fusion rules:
\begin{equation}
\one' \otimes \one' =\one,
\quad 
\one' \otimes \two = \two, 
\quad 
\two\otimes\two  = \one \oplus \one' \oplus \two
\end{equation}
The group $S^{\,}_3$ has three conjugacy classes
\begin{subequations}
\begin{align}
[e]\=\{e\},
\qquad
[s]\=\{s,\,s\,r,\,s\,r^{2}\},
\qquad
[r]\=\{r,\,r^{2}\},
\end{align}
labeled by a representative element.
For each of these conjugacy classes, the centralizer of the representative are 
\begin{align}
\mathrm{C}^{\,}_{S^{\,}_{3}}(e) = S^{\,}_{3},
\qquad
\mathrm{C}^{\,}_{S^{\,}_{3}}(s) = \mathbb{Z}^{s}_{2},
\qquad
\mathrm{C}^{\,}_{S^{\,}_{3}}(r) = \mathbb{Z}^{r}_{3},
\end{align}
respectively. These centralizers have the irreps
\begin{align}
&
\pi^{\,}_{e}\= \one,\,\one',\,\two,
\\
&
\pi^{\,}_{s}\= \one,\,\one',
\\
&
\pi^{\,}_{r}\= \one,\,\one_{\omega},\,\one_{\omega^{*}},
\end{align}
\end{subequations}
respectively. Here, all irreps are one-dimensional except 
$\two$. $\one'$ denotes the non-trivial 
one-dimensional irrep of $\mathbb{Z}^{\,}_{2}$, while
$\one_{\omega}$ and $\one_{\omega^{*}}$ 
are the non-trivial one-dimensional irreps of $\mathbb{Z}^{r}_{3}$
where $\omega=\mathrm{exp}\{\mathrm{i}2\pi/3\}$.

\section{Brief review of SymTO}
\label{app:SymTO}

The wide variety of (finite) generalized symmetries considered in the context
of quantum field theories and quantum many body physics can be provided a
unified description in the language of topological order in one higher
dimension. This general philosophy was put forward and discussed in
\Rfs{JW191213492,JW190513279,KZ200308898,KZ200514178,KZ201102859,KZ210703858,KZ220105726,CW220303596,FT220907471},
while a related connection with noninvertible gravitational anomaly was
explored in \Rf{KW1458,KZ150201690}. The correspondence between finite
symmetries in $ d $ spacetime dimensions and topological order in $ d+1 $
spacetime dimensions was referred to as Symmetry/Topological Order
correspondence in older work of two of the present authors \cite{CJW221214432}.
Closely related constructions have been referred to by various other names in
the generalized symmetries literature -- SymTFT, topological holography,
categorical symmetry, topological symmetry \etc  Similar ideas were discussed
for specialized situations, including for 1+1d systems, for rational conformal
field theories, or in the context of duality and gauging in
\Rfs{FSh0204148,FSh0607247,FS09095013,YS13094596,HV14116932,KZ170501087,FT180600008,KZ190504924,KZ191201760,CZ190312334,TW191202817,JW191209391,LB200304328,GK200805960}.
In this appendix, we summarize the aspects of the Sym/TO correspondence that
are relevant for the present paper.

\subsection{Algebra of local symmetric operators}

The most general way to define generalized symmetry is to start with a subset
of local operators, that is closed under addition and multiplication, \ie to
start with a sub-algebra of the  local operator algebra.  We define the
operators in sub-algebra as the symmetric operators of a yet-to-be-determined
symmetry.  The symmetry transformations are defined as the commutant of the
algebra of the local symmetric operators.  The symmetry defined this way is
very general, which include anomalous, higher-form, higher-group, and/or
non-invertible symmetries.  What is the mathematical frame work that can
describe and classify the generalized symmetry defined this way?

To reveal the underlying  mathematical structure of the algebra of the local
symmetric operators, \Rfs{JW191213492,CW220303596} introduced the notion of
transparent patch operators to capture the essence of isomorphic algebras of
the local symmetric operators.
Even though the  algebra is generated by local symmetric operators, the algebra
must contain extended operators.  Patch operators are a type of extended
operators, that have an \emph{extended} spatial support and are created by a
combination of a \emph{large} number of so-called local symmetric
operators.\footnote{The italicized terms can be made more precise. For a system
with linear size $ L $, the patch operators have a support on a number of sites
that is somewhere between $ \cO(1) $ and $ \cO(L) $, say $ \cO(\sqrt{L}) $.}
A patch operator has the transparent property if it commutes with
all local symmetric operators far away from the patch boundary. 

So the bulk of the transparent patch operators is invisible, and can be ignored
physically.  The boundaries of the transparent patch operators correspond to
fractionalized topological excitations, or the super selection sectors of the
corresponding generalized symmetry.  With this understanding, now it is easy to
see that operator algebra of transparent patch operators is equivalent to the braiding
and fusion of topological excitations (which is encoded in the algebra of
transparent string, membrane, \etc  operators).  This gives a correspondence
between the boundaries of transparent patch operators in $ d $ dimensional
quantum systems and topological excitations of the $ d+1 $ dimensional
topological order.

Through such a consideration, \Rfs{JW191213492,CW220303596} reveals a close
connection between isomorphic algebras of the local symmetric operators and
topological orders in one higher dimension.  Such a topological order in one
higher dimension is called symmetry topological order (SymTO).\footnote{In
\Rfs{JW191213492,KZ200514178,CW220303596} SymTO was referred to as
``categorical symmetry''.  
The name is motivated by the following
consideration: ``categorical symmetry'' contains conservation of both symmetry
charges and symmetry defects, plus the additional braiding structure of the
those symmetry objects.  Conservation corresponds to ``symmetry'' and the
additional braiding structure  corresponds to ``categorical'' in the name.}

Note that different set of local symmetric operators may generate isomorphic operator algebras. In this case, the corresponding symmetries are called holo-equivalent~\cite{KZ200514178}. The holo-equivalent symmetries give rise to the same local low energy properties and have the same SymTO.
The holo-equivalent symmetries are known in the math literature as Morita
equivalent symmetries.
Many examples of holo-equivalent symmetries are given in \Rf{CW220303596}.

The algebra of local symmetric operators and their description by transparent patch operators give rise to derivation\cite{CW220303596} of the topological holographic principle: \emph{boundary determines bulk} \cite{KW1458,KZ150201690}, but bulk
does not determine boundary.  The algebra of local
symmetric operators correspond to the ``boundary'', and the obtained topological
order in one higher dimension is the ``bulk''.

\subsection{Phases and phase transitions from SymTO}\label{app:symto-phases}

Through the Sym/TO correspondence, the gapped boundaries of the SymTO can be
mapped to gapped phases of the symmetric systems.  The gapped boundaries of
topological orders were classified by Lagrangian condensible algebras in
\Rf{KK11045047}, and thus the  gapped phases of the symmetric systems can be
classified by Lagrangian condensible algebras of the corresponding SymTO.

More generally, \Rf{CW220506244} argued that the non-Lagrangian condensable
algebras correspond to gapless phases or critical points of the system -- see
also \Rf{BBS231217322}, for a closely-related discussion. Each non-Lagrangian
algebra, in turn, has an associated \emph{reduced SymTO} which constrains the
CFT that can describe the corresponding gapless state. In more concrete terms,
gapped phases described by Lagrangian condensable algebras $ \cA_1 $ and $
\cA_2 $ have a phase transition corresponding to the non-Lagrangian algebra $
\cA_{12}= \cA_1 \cap \cA_2 $. Anyon permutation symmetries that preserve $
\cA_{12} $ are associated with emergent symmetries of the IR theory that
describes the corresponding phase transition. In \Rf{CW220506244}, the
algebraic structure of the condensable algebras did not play any explicit role.
There is a suggestion that the algebraic properties of the order parameters for
various gapped phases allowed by the SymTO may be encoded in the algebra
product of the corresponding Lagrangian algebra. This connection has not been
explored in the literature yet.

\subsection{Holo-equivalence and gauging}

Symmetries whose SymTOs are identical were referred to as ``holo-equivalent" in
previous literature \cite{KZ200514178}. This is a much more general statement
than the Morita equivalence of symmetry (fusion) categories in 1+1d, but they
coincide in the latter case \cite{N12080840}. In 1+1d, for instance, the
symmetry categories $ \cVec_G $ and $ \cRep(G) $ are related under gauging.
More specifically, gauging the entire group $ G $ in $ \cVec_G $, which can be
achieved by gauging by the algebra object $ \cA_{G} = \sum_g a_g \in \cVec_G$,
gives the dual symmetry category $ \cRep(G) $. Here, by $ a_g $ we refer to the
simple object of $ \cVec_G $ labeled by the group element $ g \in G$. On the
other hand, gauging by the regular representation algebra object $
\cA_{\text{reg}} = \sum_R d_R a_R \in \cRep(G)$, where $ d_R  $ is the
dimension of the irreducible representation $ R $, gives the dual symmetry
category $ \cVec_G $. For more details on gauging by algebra objects, the
reader is encouraged to refer to \Rf{BT170402330}. 

Following the discussion in \scn{app:symto-phases}, we conclude that the phases
of systems with Morita equivalent symmetries should have a one-to-one
correspondence in their local, low energy properties. One way to see this is
that since these Morita equivalent symmetries are related to each other by
gauging, only certain global features of the corresponding phase diagrams
should be altered. In particular, we expect the same CFTs (up to global
modifications, \eg orbifolding) to describe the phase transitions related under
this correspondence.

\section{Details of duality transformations}
\label{app:Duality ops}

\subsection{\texorpdfstring{$ \Z^{\,}_{3}$}{Z3} Kramers-Wannier duality}
\label{app:Z3 KW}
The duality transformation of the model $ \widehat{H}_{S^{\,}_{3}} $ is implemented by the operator
\begin{align}
\widehat{D}_{\mathrm{KW}}
\=
\hat{\mathsf{t}}^{\,}_{\mathbb{Z}^{\,}_{2}}\,
\widehat{P}^{U_{r}=1}\,
\widehat{W}\,
\left(
\widehat{\mathfrak{H}}^{\dagger}_{1}\,
\widehat{\mathrm{CZ}}^{\dagger}_{2,1}
\right)\,
\left(
\widehat{\mathfrak{H}}^{\dagger}_{2}\,
\widehat{\mathrm{CZ}}^{\dagger}_{3,2}
\right)\,
\cdots
\left(
\widehat{\mathfrak{H}}^{\dagger}_{L-1}\,
\widehat{\mathrm{CZ}}^{\dagger}_{L,L-1}
\right),
\label{eq:Z3KW duality operator app}
\end{align}
as discussed in the main text. 
As written in Eq.\ \eqref{eq:Z3KW duality operator app}, 
it has the form of a sequential circuit where each operator 
can be thought of as a unitary quantum gate acting on ket states
sequentially starting from the rightmost operator. 
The unitary operators in $\widehat{D}^{\,}_{\mathrm{KW}}$
are defined as follows:
\begin{enumerate}[(i)]
\begin{subequations}
\item
We denote by $\widehat{\mathrm{CZ}}^{\dagger}_{i,j}$ the controlled-Z operator
\begin{align}
\widehat{\mathrm{CZ}}^{\dagger}_{i,j}
\=
\sum_{\alpha=0}^{2}
\widehat{Z}^{-\alpha}_{j}\,
\widehat{P}^{Z=\omega^{\alpha}}_{i},
\end{align}
where $\widehat{P}^{Z=\omega^{\alpha}}_{i}$ is the projector
onto the $\widehat{Z}^{\,}_{i}=\omega^{\alpha}$ subspace.

\item 
We denote by $\widehat{\mathfrak{H}}^{\dagger}_{i}$ the Hadamard operator
\begin{align}
\widehat{\mathfrak{H}}^{\dagger}_{i}
\=
\frac{1}{\sqrt{3}}
\sum_{\alpha,\beta=0}^{2}
\omega^{\alpha\,\beta}\,
\widehat{X}^{\alpha-\beta}_{i}\,
\widehat{P}^{Z=\omega^{\beta}}_{i}.
\end{align}

\item 
We denote by $\widehat{W}$ the unitary operator
\begin{align}
\widehat{W}
\=
\sum_{\alpha=0}^{2}
\widehat{Z}^{\alpha}_{L}\,
\widehat{P}^{Z^{\dagger}_{1}\,\widehat{Z}^{\,}_{L}=\omega^{\alpha}},
\end{align}
where $\widehat{P}^{Z^{\dagger}_{1}\,\widehat{Z}^{\,}_{L}=\omega^{\alpha}}$ is the projector 
onto $\widehat{Z}^{\dagger}_{1}\,\widehat{Z}^{\,}_{L}=\omega^{\alpha}$ subspace.
This unitary acts non-trivially only at sites $1$ and $L$. 

\item 
We denote by $\widehat{P}^{U_{r}=1}$ the projector
\begin{align}
\widehat{P}^{U_{r}=1}
\=
\frac{1}{3}
\sum_{\alpha=0}^{2}
\prod_{i=1}^{L}
\widehat{X}^{\alpha}_{i},
\end{align}
which projects onto the $\widehat{U}^{\,}_{r}\equiv\prod_{i}\widehat{X}^{\,}_{i}=1$
subspace.

\item 
Finally, we denote by $\hat{\mathsf{t}}^{\,}_{\mathbb{Z}^{\,}_{2}}$ the unitary operator
\begin{align}
\hat{\mathfrak{t}}^{\,}_{\mathbb{Z}^{\,}_{2}}
\=
\frac{1 + \bm{\sigma}^{\,}_{1}\cdot\bm{\tau}^{\,}_{1}}{2}\,
\prod_{i=1}^{L-1}
\frac{1 + \bm{\tau}^{\,}_{i}\cdot\bm{\sigma}^{\,}_{i+1}}{2}\,
\frac{1 + \bm{\sigma}^{\,}_{i+1}\cdot\bm{\tau}^{\,}_{i+1}}{2},
\end{align}
\end{subequations}
which implements a ``half-translation'' of qubits. We note that as written $\hat{\mathsf{t}}^{\,}_{\mathbb{Z}^{\,}_{2}}$
also has the form of a sequential quantum circuit.
\end{enumerate}
In what follows, we list the non-trivial action of
the $\mathbb{Z}^{\,}_{3}$ KW duality operator \eqref{eq:Z3KW duality operator app}
on generators of the bond algebra \eqref{eq:def S3 algebra}
at each step of the sequential circuit.
\begin{enumerate}[(i)]
\item 
Step 1: The only non-trivial action of the operator $\widehat{\mathrm{CZ}}_{L,L-1}^\dagger$ 
by conjugation is
\begin{equation}
\begin{split}
\widehat{X}^{\,}_{L} &\mapsto \widehat{Z}^{\dagger}_{L-1}\,\widehat{X}^{\,}_{L},
\\
\widehat{X}^{\,}_{L-1} &\mapsto \widehat{X}^{\,}_{L-1}\widehat{Z}^{\dagger}_{L}.
\end{split}
\end{equation}
Note that the controlled-Z operators commute with all $\widehat{Z}^{\,}_{i}$ operators.
\item 
Step 2: The only non-trivial action of the operator
$\widehat{\mathfrak H}^{\dagger}_{L-1}$ by conjugation is
\begin{equation}
\begin{split}
\widehat{Z}_{L-1}^\dagger \widehat{X}_L &\mapsto \widehat{X}_{L-1} \widehat{X}_L\, , \\
\widehat{X}_{L-1}\widehat{Z}_L^\dagger &\mapsto  \widehat{Z}_{L-1}\widehat{Z}_L^\dagger\, , \\
\widehat{Z}_{L-1}\widehat{Z}_{L}^\dagger &\mapsto \widehat{X}_{L-1}^\dagger\widehat{Z}_{L}^\dagger\, , \\
\widehat{Z}_{L-2}\widehat{Z}_{L-1}^\dagger &\mapsto \widehat{Z}_{L-2}\widehat{X}_{L-1}.
\end{split}
\end{equation}
\item 
Step 3: 
The only non-trivial action of the operator $\widehat{\mathrm{CZ}}_{L-1,L-2}^\dagger$ 
by conjugation is
\begin{equation}
\begin{split}
\widehat{X}_{L-1} \widehat{X}_L  &\mapsto \widehat{Z}_{L-2}^\dagger\widehat{X}_{L-1} \widehat{X}_L\, , \\	
\widehat{X}_{L-1}^\dagger\widehat{Z}_{L}^\dagger &\mapsto \widehat{Z}_{L-2} \widehat{X}_{L-1}^\dagger\widehat{Z}_{L}^\dagger\, , \\
\widehat{Z}_{L-2}\widehat{X}_{L-1} &\mapsto \widehat{X}_{L-1}\, , \\
\widehat{X}_{L-2} &\mapsto \widehat{X}_{L-2}	\widehat{Z}_{L-1}^\dagger.
\end{split}
\end{equation}
\item Step 4: 
The only non-trivial action of the operator
$\widehat{\mathfrak H}_{L-2}^\dagger $ 
by conjugation is given by 
\begin{equation}
\begin{split}
\widehat{Z}_{L-2}^\dagger\widehat{X}_{L-1} \widehat{X}_L 
&\mapsto \widehat{X}_{L-2}\widehat{X}_{L-1} \widehat{X}_L\, , \\
\widehat{Z}_{L-2} \widehat{X}_{L-1}^\dagger\widehat{Z}_{L}^\dagger 
&\mapsto \widehat{X}_{L-2}^\dagger \widehat{X}_{L-1}^\dagger\widehat{Z}_{L}^\dagger\, , \\
\widehat{X}_{L-2}\widehat{Z}_{L-1}^\dagger &\mapsto 	\widehat{Z}_{L-2}\widehat{Z}_{L-1}^\dagger\, , \\
\widehat{Z}_{L-3}\widehat{Z}_{L-2}^\dagger &\mapsto  \widehat{Z}_{L-3}\widehat{X}_{L-2}.
\end{split}
\end{equation}
\end{enumerate}
Following this pattern for $L-1$ steps maps $\widehat{X}_{j} $ to $\widehat{Z}_{j}\widehat{Z}_{j+1}^\dagger$ for $j\in\{1,2,\dots,L-1\}$, and $\widehat{Z}_{j}\widehat{Z}_{j+1}^\dagger$ to $\widehat{X}_{j+1} $ for $j\in\{1,2,\dots,L-2\}$. This matches the Kramers-Wannier transformation that we set out to achieve but only for all but a few terms around the sites $ j=1 $ and $ j=L $.
These terms are
\begin{equation}\label{eq:remaining}
\begin{aligned}
&\widehat{X}_L &&\mapsto \widehat{X}_1 \widehat{X}_2\dots \widehat{X}_L\, \, , \\
&\widehat{Z}_{L-1}\widehat{Z}_{L}^\dagger &&\mapsto \widehat{X}_1^\dagger 
\widehat{X}_2^\dagger \dots \widehat{X}_{L-1}^\dagger\widehat{Z}_{L}^\dagger\, , \\
&\widehat{Z}_{L}\widehat{Z}_{1}^\dagger  &&\mapsto \widehat{Z}_{L}\widehat{X}_{1}. 
\end{aligned}
\end{equation}
We now conjugate by the operator $\widehat{W}$, which
acts non-trivially on $\widehat{X}_1$ and $\widehat{X}_L$, and trivially on all other $\widehat{X}_j$. 
Its action on $\widehat{X}^{\,}_1$ and $\widehat{X}^{\,}_L$ delivers
\begin{equation}
\begin{aligned}
&\widehat{X}_1 && \mapsto  \widehat{Z}_L^\dagger \widehat{X}_1\, , \\
&\widehat{X}_L && \mapsto \widehat{Z}_L \widehat{X}_L \ \widehat{Z}_L 
\widehat{Z}_1^\dagger
\end{aligned}
\end{equation}
Given this, we find that the 
three operators on the right hand side of Eq.\ \eqref{eq:remaining}
becomes
\begin{equation}
\begin{aligned}
&\widehat{X}_1 \widehat{X}_2\dots \widehat{X}_L &&\mapsto 
\widehat{Z}_L^\dagger \widehat{X}_1 \widehat{X}_2\dots \widehat{X}_{L-1} 
= 
\left (\prod_{j=1}^{L}\widehat{X}_{j} \right ) \widehat{Z}_L\widehat{Z}_1^\dagger,  \\
&\widehat{X}_1^\dagger \widehat{X}_2^\dagger \dots \widehat{X}_{L-1}^\dagger\widehat{Z}_{L}^\dagger &&\mapsto
\widehat{Z}_{L}\widehat{X}_1^\dagger \widehat{X}_2^\dagger \dots \widehat{X}_{L-1}^\dagger\widehat{Z}_{L}^\dagger 
= 
\left (\prod_{j=1}^{L}\widehat{X}_{j}^\dagger \right ) \widehat{X}_{L}\, , \\
&\widehat{Z}_{L}\widehat{X}_{1} &&\mapsto \widehat{X}_{1}. 
\end{aligned}
\end{equation}
In summary, up to the projector $\widehat{P}^{U^{\,}_{r}=1}$, conjugation by the 
unitary operators produce
\begin{equation}
\begin{aligned}
& \widehat{X}_1 &&\mapsto \widehat{Z}_1 \widehat{Z}_2^\dagger\, , \qquad 
&& \widehat{Z}_1 \widehat{Z}_2^\dagger &&\mapsto \widehat{X}_{2},
\\
& && \vdots \qquad  && &&  \vdots
\\
& \widehat{X}_{L-2} &&\mapsto \widehat{Z}_{L-2} \widehat{Z}_{L-1}^\dagger, 
\qquad 
&& \widehat{Z}_{L-2} \widehat{Z}_{L-1}^\dagger &&\mapsto \widehat{X}_{L-1}\, , \\
& \widehat{X}_{L-1} &&\mapsto \widehat{Z}_{L-1} \widehat{Z}_{L}^\dagger, 
\qquad 
&& \widehat{Z}_{L-1} \widehat{Z}_{L}^\dagger &&\mapsto
\left (\prod_{j=1}^{L}\widehat{X}_{j}^\dagger \right )  \widehat{X}_{L},
\\
& \widehat{X}_L &&\mapsto \left (\prod_{j=1}^{L}\widehat{X}_{j} \right ) \widehat{Z}_L\widehat{Z}_1^\dagger,  
\qquad && \widehat{Z}_L\widehat{Z}_1^\dagger &&\mapsto \widehat{X}_{1}.
\end{aligned}
\end{equation}
Our sequential circuit achieves what one expects from the Kramers-Wannier transformation 
on the qutrits if
if we restrict ourselves to the $\mathbb{Z}^{\,}_{3}$ symmetric sector,
in which $\prod_{j=1}^{L}\widehat{X}_{j} =1$. 
This achieved by the inclusion of the projector $\widehat{P}^{U^{\,}_{r}=1}$ in operator
\eqref{eq:Z3KW duality operator app}. Notice that this transformation is not yet a
symmetry of the Hamiltonian \eqref{eq:def Ham gen S3} when $J^{\,}_{1}=J^{\,}_{2}$
and $J^{\,}_{5}=J^{\,}_{6}\neq 0$. This is because so far all the operators we considered act
on the qutrits, which leads to a relative half-translation of 
$\mathbb{Z}^{\,}_{3}$ and $\widehat{Z}^{\,}_{2}$. We correct for this by the final unitary operator
$\hat{\mathfrak{t}}^{\,}_{\mathbb{Z}^{\,}_{2}}$ which implements the transformation
\begin{equation}
\hat{\mathfrak{t}}^{\,}_{\mathbb{Z}^{\,}_{2}}
\begin{pmatrix}
\hat{\tau}^{x}_{i} &
\hat{\tau}^{z}_{i} &
\hat{\sigma}^{x}_{i} &
\hat{\sigma}^{z}_{i} 
\end{pmatrix}
\hat{\mathfrak{t}}^{\dagger}_{\mathbb{Z}^{\,}_{2}}
=
\begin{pmatrix}
\hat{\sigma}^{x}_{i+1} &
\hat{\sigma}^{z}_{i+1} &
\hat{\tau}^{x}_{i} &
\hat{\tau}^{z}_{i} 
\end{pmatrix}.
\end{equation}
This completes our proof that the operator \eqref{eq:Z3KW duality operator app}
is indeed the correct Kramers-Wannier duality operator. 
This operator, as written in Eq.\ \eqref{eq:Z3KW duality operator app}, represents a sequential 
quantum circuit of depth $4L-2$.\footnote{We note that one can 
also apply each unitary operator in $\hat{\mathfrak{t}}^{\,}_{\mathbb{Z}^{\,}_{2}}$ after 
applying one cycle of controlled-Z and Hadamard operators rendering the sequential circuit 
of depth $2L-1$.}
We note that this can be straightforwardly generalized to any finite Abelian group. 
Our sequential circuit is closely related to one that is given in App.\ A of \Rf{CDZ230701267} 
for general finite groups. One important difference is that our circuit involves gates (operators) 
that are not all $\mathbb{Z}^{\,}_{3}$-symmetric, while the full circuit is so.

\subsubsection*{Algebra of \texorpdfstring{$\widehat{D}^{\,}_{\mathrm{KW}}$}{D-KW} and other symmetry operators:} 
Let us note a few more algebraic properties of the duality operator above. 
First, we find that the Hermitian conjugate gives
\begin{align}
\widehat{D}^{\dagger}_{\mathrm{KW}} 
&= 
\left (	
\prod_{j=L-1}^{1} 
\widehat{\mathrm{CZ}}^{\,}_{j+1,j}\,
\widehat{\mathfrak H}^{\,}_j 
\right )
\widehat{W}^\dagger\,
\hat{\mathfrak{t}}^{\dagger}_{\mathbb{Z}^{\,}_{2}}\,
\widehat{P}^{U_{r}=1}
\nonumber\\
&=
\widehat{P}^{U_{r}=1}
\left (	
\prod_{j=L-1}^{1}
\widehat{\mathrm{CZ}}^{\,}_{j+1,j}\,
\widehat{\mathfrak H}^{\,}_j
\right )
\widehat{W}^\dagger \,
\hat{\mathfrak{t}}^{\dagger}_{\mathbb{Z}^{\,}_{2}}
= 	
\widehat{P}^{U_{r}=1} 
\left(
\hat{\mathfrak{t}}^{\,}_{\mathbb{Z}^{\,}_{2}}\,
\widehat{W}\,
\prod_{j=1}^{L-1} 
\widehat{\mathfrak H}_j^\dagger\,
\widehat{\mathrm{CZ}}_{j+1,j}^\dagger
\right )^\dagger,
\end{align}
which is also a sequential circuit. 
The $\Z^{\,}_3$ symmetric local operators are mapped by 
the unitary part of $ \widehat{D}^{\dagger}_{\mathrm{KW}} $ in exactly the 
inverse manner as by that of $\widehat{D}^{\,}_{\mathrm{KW}}$, \ie
\begin{equation}
\begin{aligned}
& \widehat{Z}^{\,}_1 \widehat{Z}_2^\dagger 
&&\mapsto  \widehat{X}^{\,}_1,
\qquad 
&& \widehat{X}^{\,}_{2}  
&&\mapsto \widehat{Z}^{\,}_1\,\widehat{Z}_2^\dagger,
\\
& && \vdots \qquad  && &&  \vdots
\\
& \widehat{Z}^{\,}_{L-2}\, \widehat{Z}_{L-1}^\dagger  
&&\mapsto  \widehat{X}^{\,}_{L-2},
\qquad 
&& \widehat{X}^{\,}_{L-1}  
&&\mapsto  \widehat{Z}^{\,}_{L-2}\,\widehat{Z}_{L-1}^\dagger,
\\
&  \widehat{Z}^{\,}_{L-1}\,\widehat{Z}_{L}^\dagger 
&&\mapsto \widehat{X}^{\,}_{L-1},
\qquad
&& 
\widehat{X}^{\,}_{L} 
&&\mapsto 
\left (\prod_{j=1}^{L}\widehat{X}_{j}^\dagger \right )  
\widehat{Z}^{\,}_{L-1}\,\widehat{Z}_{L}^\dagger,
\\
& 
\widehat{Z}^{\,}_L\,\widehat{Z}_1^\dagger 
&&\mapsto 
\left (\prod_{j=1}^{L}\widehat{X}^{\,}_{j} \right ) 
\widehat{X}^{\,}_L ,
\qquad 
&& \widehat{X}^{\,}_{1} &&\mapsto  \widehat{Z}^{\,}_L\,\widehat{Z}_1^\dagger.
\end{aligned}
\end{equation}
Now applying the projector $\widehat{P}^{U_{r}=1}$ and 
the half-translation operator $\hat{\mathfrak{t}}^{\,}_{\mathbb{Z}^{\,}_{2}}$
simply produces the transformation
\begin{equation}
\begin{split}
&
\widehat{D}^{\dagger}_{\mathrm{KW}} \widehat{X}_j = \widehat{Z}_{j-1} \widehat{Z}_{j}^\dagger \widehat{D}^{\dagger}_{\mathrm{KW}},  
\quad \widehat{D}^{\dagger}_{\mathrm{KW}} \widehat{Z}_{j} \widehat{Z}_{j+1}^\dagger =  \widehat{X}_j \widehat{D}^{\dagger}_{\mathrm{KW}}\,
\\
&
\widehat{D}^{\dagger}_{\mathrm{KW}} 
\begin{pmatrix}
\hat{\tau}^{x}_{i} &
\hat{\tau}^{z}_{i} &
\hat{\sigma}^{x}_{i} &
\hat{\sigma}^{z}_{i} 
\end{pmatrix}
=
\begin{pmatrix}
\hat{\sigma}^{x}_{i} &
\hat{\sigma}^{z}_{i} &
\hat{\tau}^{x}_{i-1} &
\hat{\tau}^{z}_{i-1} 
\end{pmatrix}
\widehat{D}^{\dagger}_{\mathrm{KW}},  
\end{split}
\end{equation}
from which we observe the relation
\begin{equation}
\widehat{D}^{\dagger}_{\mathrm{KW}} = \widehat{T}^\dagger\widehat{D}^{\,}_{\mathrm{KW}}.
\end{equation}
Here, the operator $ \widehat{T} $ implements
the single lattice site translation of both qubits and qutrits, \ie
\begin{equation}
\widehat{T}\,
\begin{pmatrix}
\widehat{X}^{\,}_{i} & 
\widehat{Z}^{\,}_{i} &
\hat{\tau}^{x}_{i} &
\hat{\tau}^{z}_{i} &
\hat{\sigma}^{x}_{i} &
\hat{\sigma}^{z}_{i} 
\end{pmatrix}
\widehat{T}^{\dagger}
=
\begin{pmatrix}
\widehat{X}^{\,}_{i+1} & 
\widehat{Z}^{\,}_{i+1} &
\hat{\tau}^{x}_{i+1} &
\hat{\tau}^{z}_{i+1} &
\hat{\sigma}^{x}_{i+1} &
\hat{\sigma}^{z}_{i+1} 
\end{pmatrix}.
\end{equation}
Combining the above results, we obtain the fusion rules
\begin{subequations}
\begin{equation}
\widehat{D}^{\dagger}_{\mathrm{KW}} \widehat{D}^{\,}_{\mathrm{KW}} =\widehat{P}^{U_{r}=1},
\end{equation}
and
\begin{equation}
\widehat{D}^{\,}_{\mathrm{KW}} \prod_j \widehat{X}_{j} = \widehat{D}^{\,}_{\mathrm{KW}}
= \left (\prod_j \widehat{X}_{j} \right ) \widehat{D}^{\,}_{\mathrm{KW}}.
\end{equation}
\end{subequations}

\subsection{\texorpdfstring{$ \cRep(S^{\,}_{3}) $}{Rep(S3)} self-duality}
\label{app:RepS3 duality}
We found that our $ S^{\,}_{3} $ symmetric Hamiltonian \eqref{eq:H s3} has a 
self-duality symmetry in some sub-manifold in the parameter space. We discussed a 
sequential circuit that performs this transformation above. Recall that gauging a 
$\Z^{\,}_2 $ subgroup of $ S^{\,}_{3} $ led us to 
the Hamiltonian \eqref{eq:def Ham gen RepS3} 
with $ \cRep(S^{\,}_{3})$ symmetry.
We now want to know what symmetry, if 
any, the above self-duality symmetry gets mapped to. To that end, we would 
like to follow how each operator in $\widehat{D}^{\,}_{\mathrm{KW}}$ (recall 
Eq.\ \eqref{eq:Z3KW duality operator app}) 
change under the gauging map. 
\begin{enumerate}[(i)]
\item 
We note that acting by $ \Z^{\,}_2 $ on half of the system, \ie on all degrees
of freedom to the right of particular site, say $i$, will leave every CZ
operator except $\widehat{\mathrm{CZ}}^{\,}_{i+1,i} $ unaffected.  This
particular operator gets transformed as
\begin{equation}
\begin{split}
\widehat{\mathrm{CZ}}_{i+1,i} = \sum_{\al=0}^{2} \widehat{Z}_i^{\al} 
\widehat{P}_{Z_{i+1} = \om^{\al} } 
\mapsto  
&\sum_{\al=0}^{2} \widehat{Z}_i^{\al} 
\widehat{P}_{Z_{i+1} = \om^{-\al}} 
\\
&= \sum_{\al=0}^{2}
\widehat{Z}_i^{-\al} 
\widehat{P}_{Z^{\,}_{i+1} = \om^\al} 
=  \widehat{\mathrm{CZ}}_{i+1,i} ^\dagger.
\end{split}
\end{equation}
Therefore, if we consider arbitrary $\Z^{\,}_2$ gauge field configurations, 
we obtain the minimally coupled CZ operators as
\begin{equation}
\widehat{\mathrm{CZ}}_{i+1,i}  \mapsto 
\widehat{\mathrm{CZ}}_{i+1,i}^{\,\hat{\mu}^x_{i+1}} 
\equiv
\sum_{\al=0}^{2}
\widehat{Z}_i^{\al \hat{\mu}^x_{i+1}} 
\widehat{P}_{Z^{\,}_{i+1} = \om^\al },
\end{equation}
where we shifted the subscript of the link degrees of freedom by $1/2$, as done in the main text.
This minimally coupled operator is the image of 
$\widehat{\mathrm{CZ}}^{\,}_{i+1,i}$ under duality map 
as it is unchanged after the unitary transformation
\eqref{eq:Z2 gauging unitary} and the projection in Eq.\ \eqref{eq:Z2 gauging consistency cond}.

\item 
The Hadamard operator commutes with the charge 
conjugation operator
\begin{equation}\label{eq:CHC}
	\begin{split}
		\widehat{C}^{\,}_{i}\,
		\widehat{\mathfrak{H}}^{\,}_{i}\,
		\widehat{C}^{\dagger}_{i} 
		&= 
		\frac{1}{\sqrt{3}}
		\sum_{\alpha,\beta=0}^{2}
		\omega^{\alpha\,\beta}\,
		\widehat{X}^{\beta-\alpha}_{i}\,
		\widehat{P}^{Z=\omega^{-\beta}}_{i}  
		\\
		&=
		\frac{1}{\sqrt{3}}
		\sum_{\alpha,\beta=0}^{2}
		\omega^{\alpha\,\beta}\,
		\widehat{X}^{\alpha-\beta}_{i}\,
		\widehat{P}^{Z=\omega^{\beta}}_{i}
		=
		\widehat{\mathfrak{H}}^{\,}_{i}.
	\end{split}
\end{equation}
Hence, the Hadamard operators in $\widehat{D}^{\,}_{\mathrm{KW}}$ 
are gauge invariant. 
Similarly, the Hadamard operator commutes with the 
unitary transformation \eqref{eq:Z2 gauging unitary}
and is mapped to itself under $\mathbb{Z}^{\,}_{2}$ gauging.

\item 
As it was the case for controlled-Z operator 
$\widehat{\mathrm{CZ}}^{\,}_{i+1,i}$, the unitary $\widehat{W}$
is not gauge invariant. We find its minimally coupled version
\begin{align}
\widehat{W} 
=
\sum_{\al=0}^{2}
\widehat{Z}^{\alpha}_L\,
\widehat{P}^{Z_1^{\dagger} Z^{\,}_L=\om^{\alpha}} 
\mapsto 
\widehat{W}_{\text{mc}}
=
\sum_{\al=0}^{2}
\widehat{Z}^{\alpha}_L\,
\widehat{P}^{Z_1^{-\mu^{x}_{1}} Z^{\,}_L=\om^{\alpha}},
\end{align}
which is the image of the operator $\widehat{W}$ under $\mathbb{Z}^{\,}_{2}$
gauging. Note that similar to the discussion of the minimally coupled CZ operators above, $ \widehat{W}_{\text{mc}} $ is unchanged by the unitary transformation
\eqref{eq:Z2 gauging unitary} and the projection in Eq.\ \eqref{eq:Z2 gauging consistency cond}.

\item The projector $\widehat{P}^{U_{r}=1}$ in the definition of $ \widehat{D}^{\,}_{\mathrm{KW}} $ \eqref{eq:Z3KW duality operator app}, under minimal coupling 
takes the following form: \footnote{Note that the choice to start the string of $\mu^z_{k+1/2}$'s at $k=1$ is completely arbitrary and unphysical. We could just as well put this ``branch cut" anywhere else in the periodic chain.}
\begin{equation*}
	\widehat{P}^{U_{r}=1}  
= 
\frac{1}{3}\sum_{\al=0}^{2} \prod_{j=1}^{L}\widehat{X}_j ^{\al}
\mapsto 
\widehat{P}_{\text{mc}}
\=
\frac{1}{3}\sum_{\al=0}^{2}  
\prod_{j=1}^{L}\widehat{X}_j^{\,\al\prod_{k=1}^{j-1}\hat{\mu}^x_{k+1}}
\end{equation*}
%We can rewrite this as follows
%\begin{equation*}
%	\frac{1}{3}
%	\left( 
%	\one
%	+
%	\prod_{j=1}^{L}\widehat{X}_j^{\,\mu^z_{j-1/2}\mu^z_{j+1/2}\dots\mu^z_{L-1/2}}
%	+ 
%	\prod_{j=1}^{L}\widehat{X}_j^{\,-\mu^z_{j-1/2}\mu^z_{j+1/2}\dots\mu^z_{L-1/2}}
%	\right ) 
%	=
%	\frac{1}{3}
%	\left( 
%	\one
%	+
%	\prod_{j=1}^{L}\widehat{X}_j^{\,\mu^z_{j-1/2}\mu^z_{j+1/2}\dots\mu^z_{L-1/2}}
%	+ 
%	\prod_{j=1}^{L}\widehat{X}_j^{\,-\mu^z_{j-1/2}\mu^z_{j+1/2}\dots\mu^z_{L-1/2}}
%	\right ) 
%\end{equation*}
The unitary transformation, \eqref{eq:Z2 gauging unitary}
leaves the operator $ \widehat{P}_{\text{mc}} $ unchanged,
\begin{align*}
\widehat{P}_{\text{mc}} & \mapsto \frac{1}{3}\sum_{\al=0}^{2}  
\prod_{j=1}^{L}\widehat{X}_j^{\,\al \,\hat{\si}^z_{1} \prod_{k=1}^{j-1}\hat{\mu}^x_{k+1}}
\\
&= \frac{1}{3}\sum_{\al=0}^{2} 
\left (\frac{\hat{1}+\hat{\si}^z_{1}}{2}\right )
\prod_{j=1}^{L}\widehat{X}_j^{\,\al \prod_{k=1}^{j-1}\hat{\mu}^x_{k+1}}
+  \frac{1}{3}\sum_{\al=0}^{2}  \left (\frac{\hat{1}-\hat{\si}^z_{1}}{2}\right )
\prod_{j=1}^{L}\widehat{X}_j^{\,-\al \prod_{k=1}^{j-1}\hat{\mu}^x_{k+1}}
\\
&= \frac{1}{3}\sum_{\al=0}^{2} 
\left (\frac{\hat{1}+\hat{\si}^z_{1}}{2}\right )
\prod_{j=1}^{L}\widehat{X}_j^{\,\al \prod_{k=1}^{j-1}\hat{\mu}^x_{k+1}}
+  \frac{1}{3}\sum_{\al'=0}^{2}  \left (\frac{\hat{1}-\hat{\si}^z_{1}}{2}\right )
\prod_{j=1}^{L}\widehat{X}_j^{\,\al' \prod_{k=1}^{j-1}\hat{\mu}^x_{k+1}}
\\
&= 
\frac{1}{3}\sum_{\al=0}^{2} 
\prod_{j=1}^{L}\widehat{X}_j^{\,\al\prod_{k=1}^{j-1}\hat{\mu}^x_{k+1}} = \widehat{P}_{\text{mc}} \,.
\end{align*}
Here we used the periodic boundary condition on the $ \hat{\si}^z $ degrees of 
freedom, \ie $ \hat{\si}^z_0 \equiv \hat{\si}^z_L $. 

\item Finally, the qubit ``half-translation" operator $ \hat{\mathfrak{t}}^{\,}_{\Z^{\,}_2} $ after the whole gauging procedure takes the following form:
\begin{equation}
\begin{split}
&
\hat{\mathfrak{t}}^{\,}_{\cRep(S^{\,}_{3})}
\=
\widehat{A}^{\,}_{1}\,
\prod_{i=1}^{L-1}
\widehat{B}^{\,}_{i}\,
\widehat{A}^{\,}_{i+1},
\\
&
\widehat{A}^{\,}_{i}
\=
\frac{1+\hat{\tau}^{z}_{i}}{2}
+ 
\frac{1-\hat{\tau}^{z}_{i}}{2}\,
\widehat{C}^{\,}_{i}\,\hat{\mu}^{z}_{i}\,\hat{\mu}^{z}_{i+1},
\\
&
\widehat{B}^{\,}_{i}
\=
\frac{1+\hat{\tau}^{z}_{i}\,\hat{\mu}^{x}_{i+1}}{2}
+
\frac{1-\hat{\tau}^{z}_{i}\,\hat{\mu}^{x}_{i+1}}{2}\,
\hat{\tau}^{x}_{i}\,
\hat{\tau}^{x}_{i+1}\,
\widehat{C}^{\,}_{i+1}\,
\hat{\mu}^{z}_{i+1}\,
\hat{\mu}^{z}_{i+2}.
\end{split}
\end{equation}
The action of this operator on the generators of the $\cRep(S^{\,}_{3})$-symmetric
bond algebra \eqref{eq:def S3/Z2 algebra} can be deduced from the 
action of $\hat{t}^{\,}_{\mathbb{Z}^{\,}_{2}}$ on the generators
of $S^{\,}_{3}$-symmetric bond algebra \eqref{eq:def S3 algebra}.
Namely, we find the transformation rules
\begin{equation}\label{eq: def tRepS3}
\hat{\mathfrak{t}}^{\,}_{\cRep(S^{\,}_{3})}
\colon
\begin{pmatrix}
\hat{\tau}^{z}_{i}\\[0.4em]
\hat{\tau}^{z}_{i}\,\hat{\mu}^{x}_{i+1} \\[0.4em]
\hat{\mu}^{z}_{i}\,\hat{\tau}^{x}_{i}\,\widehat{C}^{\,}_{i}\,\hat{\mu}^{z}_{i+1} \\[0.4em]
\hat{\tau}^{x}_{i}\\[0.4em]
\widehat{X}^{\,}_{i}+\widehat{X}^{\dagger}_{i}\\[0.4em]
\widehat{Z}^{\hat{\mu}^{x}_{i+1}}_{i}\,\widehat{Z}^{\dagger}_{i+1}
+\mathrm{H.c.} \\[0.4em]
\widehat{X}^{\,}_{i}-\widehat{X}^{\dagger}_{i}\\[0.4em]
\hat{\tau}^{z}_{i}\,\hat{\mu}^{x}_{i+1}
\left(
\widehat{Z}^{\hat{\mu}^{x}_{i+1}}_{i}\,\widehat{Z}^{\dagger}_{i+1}
-\mathrm{H.c.}
\right)
\end{pmatrix}
\longmapsto
\begin{pmatrix}
\hat{\tau}^{z}_{i}\,\hat{\mu}^{x}_{i+1} \\[0.4em]
\hat{\tau}^{z}_{i+1}\\[0.4em]
\hat{\tau}^{x}_{i} \\[0.4em]
\hat{\mu}^{z}_{i+1}\,\hat{\tau}^{x}_{i+1}\,\widehat{C}^{\,}_{i+1}\,\hat{\mu}^{z}_{i+2} \\[0.4em]
\widehat{X}^{\,}_{i}+\widehat{X}^{\dagger}_{i}\\[0.4em]
\widehat{Z}^{\hat{\mu}^{x}_{i+1}}_{i}\,\widehat{Z}^{\dagger}_{i+1}
+\mathrm{H.c.} \\[0.4em]
\hat{\tau}^{z}_{i}\,
\left(\widehat{X}^{\,}_{i}-\widehat{X}^{\dagger}_{i}\right)\\[0.4em]
\widehat{Z}^{\hat{\mu}^{x}_{i+1}}_{i}\,\widehat{Z}^{\dagger}_{i+1}
-\mathrm{H.c.}
\end{pmatrix},
\end{equation}
on the generators of $\cRep(S^{\,}_{3})$-symmetric bond algebra \eqref{eq:def S3/Z2 algebra}.
\end{enumerate}

Since we gauged the original $ \Z^{\,}_2 $ symmetry with 
periodic boundary conditions, we will end up in the symmetric sector of the dual 
$\Z^{\,}_2 $ symmetry, generated by $ \prod_j \hat{\mu}^x_{j} $. 
Therefore, we must include a projector to this symmetric sector in the gauged duality 
operator.\footnote{In fact, without this additional projector, the gauged duality operator would not actually commute with the Hamiltonian \eqref{eq:def Ham gen RepS3} on the self-dual manifold of parameters described by $ J_1=J_2, J_5=J_6 $.} So the full gauged operator has the form~\footnote{Note that $ \hat{\mathfrak{t}}^{\,}_{\cRep(S^{\,}_{3})} $ commutes with the projector $ \frac{\hat{1}+ \prod_j \hat{\mu}^x_{j}  }{2}  $.}
\begin{align*}
\widehat{D}_{\cRep(S_3)} &\=
\hat{\mathfrak{t}}^{\,}_{\cRep(S^{\,}_{3})}
\frac{\hat{1}+ \prod_j \hat{\mu}^x_{j}  }{2} 
\ \widehat{P}_{\text{mc}} \ \widehat{D}_0\, , \quad \text{where} \\
\widehat{D}_0 &\=
\left (\sum_{\al=0}^{2} 
\widehat{Z}_L^{\, \al} \ 
\widehat{P}_{Z_1^{\mu^{x}_{1}} Z_L=\om^{ \al} } \right )
\left(
\prod_{j=1}^{L-1} 
\widehat{\mathfrak H}_j^\dagger \ \widehat{\mathrm{CZ}}_{j+1,j}^{-\hat{\mu}^x_{j+1}} \right )
\end{align*}
We can further simplify $ \widehat{D}_{\cRep(S_3)} $ since
\begin{equation*}
\frac{\hat{1}+ \prod_j \hat{\mu}^x_{j} }{2} 
\ \widehat{P}_{\text{mc}} = \frac{\hat{1}+ \prod_j \hat{\mu}^x_{j} }{2} 
\
\frac{\hat{1}+\widehat{W}_{\two}}{3}
= 
\frac{1}{6}\left (\hat{1}+\widehat{W}_{\one' }\right ) 
\left (\hat{1}+\widehat{W}_{\two}\right ) 
= 
\frac{1}{6}
\left (\widehat{W}_{\one} + 
\widehat{W}_{\one'}+ 
2\widehat{W}_{\two}\right )
\end{equation*}
Hence, we have $ \widehat{D}_{\cRep(S_3)} = \frac{1}{6} \left (\widehat{W}_{\one} + 
\widehat{W}_{\one'}+ 
2\widehat{W}_{\two}\right )  	\widehat{D}_0 $, which we denote in 
short as $ \widehat{D}_{\cRep(S_3)} =
\widehat{P}_{\text{reg}}\widehat{D}_0 $, where $ \widehat{P}_{\text{reg}} \= \frac{1}{6}\widehat{W}_\text{reg} $ is the projector to the $ \cRep(S_3) $-symmetric sector, as discussed in the main text.

Under the action of $\widehat{D}_0$ by conjugation, going through calculations 
similar to those in \scn{app:Z3 KW}, we find the following operator maps:
\begin{alignat*}{4}
& \widehat{X}_1 &&\mapsto \widehat{Z}_1 
\widehat{Z}_2^{-\hat{\mu}^x_{2}} \qquad&& 
\widehat{Z}_1^{-\hat{\mu}^x_{2}} \widehat{Z}_2 &&\mapsto 
\widehat{X}_{2}^\dagger\\
& && \vdots \qquad  && &&  \vdots\\
& \widehat{X}_{L-2} &&\mapsto \widehat{Z}_{L-2} 
\widehat{Z}_{L-1}^{-\hat{\mu}^x_{L-1}} \qquad && 
\widehat{Z}_{L-2}^{-\hat{\mu}^x_{L-1}} \widehat{Z}_{L-1}  
&&\mapsto \widehat{X}_{L-1}^\dagger\\
& \widehat{X}_{L-1} &&\mapsto \widehat{Z}_{L-1} 
\widehat{Z}_{L}^{-\hat{\mu}^x_{L}} \qquad && 
\widehat{Z}_L^{-\hat{\mu}^x_{1}} \widehat{Z}_1  &&\mapsto 
\widehat{X}_{1}^\dagger
\end{alignat*}
and
\begin{align*}
\widehat{Z}_{L-1}^{-\hat{\mu}^x_{L}} \widehat{Z}_{L} 
&\mapsto 
\widehat{Z}_L^{-\prod_j \hat{\mu}^x_{j}}
\left (\prod_{j=1}^{L-1}\widehat{X}_{j}^{\prod_{k=j}^{L-1} 
\hat{\mu}^x_{k+1}} 
\right )
\widehat{Z}_L
\\
\widehat{X}_L 
&\mapsto 
\widehat{Z}_L^{-\prod_j \hat{\mu}^x_{j}} 
\left (\prod_{j=1}^{L-1}\widehat{X}_{j}^{\prod_{k=j}^{L-1} 
\hat{\mu}^x_{k+1}} 
\right )
\
\widehat{Z}_L
\widehat{X}_L
\  
\widehat{Z}_L \widehat{Z}_1^{-\hat{\mu}^x_{1}}
\end{align*}
In the above, we used
$\widehat{W}_{\text{mc}} \widehat{X}_1^\dagger \widehat{W}_{\text{mc}}^{\dagger}= \widehat{X}_1^\dagger \widehat{Z}_L^{\hat{\mu}^x_{1}}$
and
$\quad
\widehat{W}_{\text{mc}}\widehat{X}_L \widehat{W}_{\text{mc}}^\dagger 
=
\widehat{Z}_L \widehat{X}_L \widehat{Z}_L \widehat{Z}_1^{-\hat{\mu}^{x}_{1}} \ $.
The last two transformations can be re-written as
\begin{align}
\widehat{Z}_{L-1}^{-\hat{\mu}^x_{L}} \widehat{Z}_{L} 
&\mapsto 
\widehat{Z}_L^{-\prod_j \hat{\mu}^x_{j}}
\left (\prod_{j=1}^{L-1}\widehat{X}_{j}^{\prod_{k=0}^{j-1} 
\hat{\mu}^x_{k+1}} 
\right )
^{\prod_j \hat{\mu}^x_{j}}
\widehat{Z}_L
\label{eq:ZZL}
\\
\widehat{X}_L 
&\mapsto 
\widehat{Z}_L^{-\prod_j \hat{\mu}^x_{j}} 
\left (\prod_{j=1}^{L-1}\widehat{X}_{j}^{\prod_{k=0}^{j-1} 
\hat{\mu}^x_{k+1}} 
\right )
^{\prod_j \hat{\mu}^x_{j}}
\
\widehat{Z}_L
\widehat{X}_L
\  
\widehat{Z}_L \widehat{Z}_1^{-\hat{\mu}^x_{1}}
\label{eq:XL}
\end{align}

% -------- SOME DETAILS THAT MAY OR MAY NOT BE INCLUDED ------
%The action of $\widehat{P}_{\text{mc}}$ on r.h.s. of Eq.\ \eqref{eq:ZZL} is as follows:
%\begin{align*}
%\widehat{P}_{\text{mc}}
%(\dots)
%&=
%\frac{1}{3}\sum_{k\in\Z^{\,}_3}
%\left(
%\prod_{j=1}^{L}\widehat{X}_j^{\,\prod_{\ell=1}^{j-1}\hat{\mu}^z_{\ell+1/2}}
%\right )^{k} 
%\widehat{Z}_L^{-\prod_j \hat{\mu}^x_{j}}
%\left (\prod_{j=1}^{L-1}\widehat{X}_{j}^{\prod_{k=0}^{j-1} 
%\hat{\mu}^x_{k+1}} 
%\right )
%^{\prod_j \hat{\mu}^x_{j}}
%\widehat{Z}_L
%\\
%&=
%\frac{1}{3}\sum_{k\in\Z^{\,}_3}
%\widehat{Z}_L^{-\prod_j \hat{\mu}^x_{j}}
%\om^{k \hat{\mu}^x_{1}}
%\left(
%\prod_{j=1}^{L}\widehat{X}_j^{\,\prod_{\ell=1}^{j-1}\hat{\mu}^z_{\ell+1/2}}
%\right )^{k} 	
%\left (\prod_{j=1}^{L-1}\widehat{X}_{j}^{\prod_{k=0}^{j-1} 
%\hat{\mu}^x_{k+1}} 
%\right )
%^{\prod_j \hat{\mu}^x_{j}}
%\widehat{Z}_L
%\end{align*}
%Without the additional projector in the definition of $ \widehat{D} $, we won't be able to recover a locality-preserving action on the operators.

%If we include a projection to the $\t \Z^{\,}_2$ symmetric sector where $\prod_j \hat{\mu}^x_{j} =1$, the transformations above simplify to
%\begin{equation*}
%	\widehat{X}_{j} \mapsto \widehat{Z}_{j} \widehat{Z}_{j+1}^{-\hat{\mu}^z_{j+1/2}},\qquad
%	\widehat{Z}_{j}^{-\hat{\mu}^z_{j+1/2}} \widehat{Z}_{j+1}  \mapsto \widehat{X}_{j+1}^\dagger
%\end{equation*}

Next, note that $ \widehat{P}_{\text{reg}} $ commutes with the first set of 
operators above, but for the last two we have
\begin{align*}
\widehat{P}_{\text{reg}}\ 
\widehat{Z}_L^{-\prod_j \hat{\mu}^x_{j}}
&
\left (\prod_{j=1}^{L-1}\widehat{X}_{j}^{\prod_{k=0}^{j-1} 
\hat{\mu}^x_{k+1}} 
\right )^{\prod_j \hat{\mu}^x_{j}} 
\widehat{Z}_L
=
\widehat{P}_{\text{reg}}\ 
\left (\prod_{j=1}^{L-1}\widehat{X}_{j}^{\prod_{k=0}^{j-1} 
\hat{\mu}^x_{k+1}} 
\right )
\\
&=
\widehat{P}_{\text{reg}}\ 
\left (\prod_{j=1}^{L-1}\widehat{X}_{j}^{\prod_{k=0}^{j-1} 
\hat{\mu}^x_{k+1}} 
\right ) \widehat{X}_L \widehat{X}_L^\dagger
\\
&=
\frac{1+\prod_j \hat{\mu}^x_{j}}{2}\
\frac{1}{3}
\sum_{\al=0}^2
\prod_{j=1}^{L}\widehat{X}_j^{\,\al \prod_{\ell=1}^{j-1}\hat{\mu}^x_{\ell+1}}
\ 
\left (\prod_{j=1}^{L-1}\widehat{X}_{j}^{\prod_{k=0}^{j-1} 
\hat{\mu}^x_{k+1}} 
\right ) 
\widehat{X}_L^{\prod_j   \hat{\mu}^x_{j}} \widehat{X}_L^\dagger
\\
&=
\frac{1+\prod_j \hat{\mu}^x_{j+1}}{2}\ 
\frac{1}{3}
\sum_{\al=0}^2
\prod_{j=1}^{L}
\widehat{X}_j^{\,\left (\al+\hat{\mu}^x_{1}\right )\prod_{\ell=1}^{j-1}\hat{\mu}^x_{\ell+1}} 
\widehat{X}_L^\dagger
= 
\widehat{X}_L^\dagger 
\ \widehat{P}_{\text{reg}} 
\end{align*}
and
\begin{align*}
\widehat{P}_{\text{reg}} 
\
\widehat{Z}_L^{-\prod_j \hat{\mu}^x_{j}} 
&
\left (
\prod_{j=1}^{L-1}\widehat{X}_{j}^{\prod_{k=0}^{j-1} 
\hat{\mu}^x_{k+1}} 
\right )^{\prod_j \hat{\mu}^x_{j}} 
\widehat{Z}_L
\widehat{X}_L
\  
\widehat{Z}_L \widehat{Z}_1^{-\hat{\mu}^x_{1}}
\\
&=
\widehat{P}_{\text{reg}} 
\
\left (
\prod_{j=1}^{L-1}\widehat{X}_{j}^{\prod_{k=0}^{j-1} 
\hat{\mu}^x_{k+1}} 
\right )
\widehat{X}_L^{\prod_j \hat{\mu}^x_{j}}
\  
\widehat{Z}_L \widehat{Z}_1^{-\hat{\mu}^x_{1}}
\\
&=
\frac{1+\prod_j \hat{\mu}^x_{j}}{2}\ 
\frac{1}{3}
\sum_{\al=0}^2
\left (
\prod_{j=1}^{L}\widehat{X}_j^{\,\left (\al+\hat{\mu}^x_{1}\right )\prod_{\ell=1}^{j-1}\hat{\mu}^z_{\ell+1/2}} 
\right )
\  
\widehat{Z}_L \widehat{Z}_1^{-\hat{\mu}^x_{1}}
= 
\widehat{Z}_L 
\widehat{Z}_1^{-\hat{\mu}^x_{1}}  
\ \widehat{P}_{\text{reg}} 
\end{align*}

Finally, we act with the unitary $ \hat{\mathfrak{t}}^{\,}_{\cRep(S^{\,}_{3})} $, whose action on the $ \cRep(S_3) $-symmetric bond algebra is outlined in Eq.\ \eqref{eq: def tRepS3}. In all, we have the following transformations of the operators appearing in the $ \cRep(S_3) $-symmetric Hamiltonian \eqref{eq:def Ham gen RepS3}:
\begin{enumerate}[(i)]
\item For operators $ \widehat{X}_j + \text{H.c.}$,
	\begin{align*}
	\widehat{D}_{\cRep(S_3)}
	\left (\widehat{X}_j+ \text{H.c.}\right )
	 &=
	 \hat{\mathfrak{t}}^{\,}_{\cRep(S^{\,}_{3})}
	 \left (
	 \widehat{Z}_{j} 
	 \widehat{Z}_{j+1}^{-\hat{\mu}^x_{j+1}} 
	 + 
	 \widehat{Z}_{j}^\dagger 
	 \widehat{Z}_{j+1}^{\hat{\mu}^x_{j+1}} 
	 \right )
	 \widehat{P}_{reg}
	 \widehat{D}_{0} \\
	 &=
	 \hat{\mathfrak{t}}^{\,}_{\cRep(S^{\,}_{3})}
	 \left (
	 \widehat{Z}_{j}^{\hat{\mu}^x_{j+1}} 
	 \widehat{Z}_{j+1}^\dagger + \text{H.c.}
	 \right )
	 \widehat{P}_{reg}
	 \widehat{D}_{0} 
	 =
	  \left (
	 \widehat{Z}_{j}^{\hat{\mu}^x_{j+1}} 
	 \widehat{Z}_{j+1}^\dagger + \text{H.c.}
	 \right )
	 \widehat{D}_{\cRep(S_3)}\,.
	\end{align*}
\item For operators $  \widehat{Z}_{j}^{\hat{\mu}^x_{j+1}} 
\widehat{Z}_{j+1}^\dagger  +\text{H.c.} $,
	\begin{align*}
		\widehat{D}_{\cRep(S_3)}
		\left (
		\widehat{Z}_{j}^{\hat{\mu}^x_{j+1}} 
		\widehat{Z}_{j+1}^\dagger  +\text{H.c.}
		\right )
		&=
		\hat{\mathfrak{t}}^{\,}_{\cRep(S^{\,}_{3})}
		\left (
		\widehat{X}_{j+1}+\text{H.c.}
		\right )
		\widehat{P}_{reg}
		\widehat{D}_{0} 
		= 
		\left (
		\widehat{X}_{j+1}+\text{H.c.}
		\right )
		\widehat{D}_{\cRep(S_3)}\,.
	\end{align*}
\item For operators $ \widehat{X}_j - \text{H.c.} $,
	\begin{align*}
		\widehat{D}_{\cRep(S_3)}
		\left (
		 \widehat{X}_j - \text{H.c.}
		\right )
		&=
		\hat{\mathfrak{t}}^{\,}_{\cRep(S^{\,}_{3})}
		\left (
		\widehat{Z}_{j} 
		\widehat{Z}_{j+1}^{-\hat{\mu}^x_{j+1}} 
		-
		 \widehat{Z}_{j}^\dagger 
		\widehat{Z}_{j+1}^{\hat{\mu}^x_{j+1}}
		\right )
		\widehat{P}_{reg}
		\widehat{D}_{0} \\
		&=
		\hat{\mathfrak{t}}^{\,}_{\cRep(S^{\,}_{3})}
		\hat{\mu}^x_{j+1}
		\left (
		\widehat{Z}_{j}^{\hat{\mu}^x_{j+1}} 
		\widehat{Z}_{j+1}^\dagger 
		-
		\widehat{Z}_{j}^{-\hat{\mu}^x_{j+1}}
		\widehat{Z}_{j+1}
		\right )
		\widehat{P}_{reg}
		\widehat{D}_{0}
		\\
		&=
		\hat{\tau}^{z}_{j}
		\hat{\mu}^x_{j+1}
		\left (
		\widehat{Z}_{j}^{\hat{\mu}^x_{j+1}} 
		\widehat{Z}_{j+1}^\dagger 
		-
		\text{H.c.}
		\right )
		\widehat{D}_{\cRep(S_3)}\,.
	\end{align*}
\item For operators $ \hat{\tau}^{z}_{i}\,\hat{\mu}^{x}_{i+1}
\left(
\widehat{Z}^{\hat{\mu}^{x}_{i+1}}_{i}\,\widehat{Z}^{\dagger}_{i+1}
-\text{H.c.}
\right) $,
	\begin{align*}
		\widehat{D}_{\cRep(S_3)}
		\hat{\tau}^{z}_{i}\,\hat{\mu}^{x}_{i+1}
		\left(
		\widehat{Z}^{\hat{\mu}^{x}_{i+1}}_{i}\,\widehat{Z}^{\dagger}_{i+1}
		-\text{H.c.}
		\right)
		&=
		\hat{\mathfrak{t}}^{\,}_{\cRep(S^{\,}_{3})}
		\hat{\tau}^{z}_{i}\,\hat{\mu}^{x}_{i+1}
		\left(
		\widehat{X}_{j+1}
		-
		\widehat{X}_{j+1}^\dagger
		\right)
		\widehat{P}_{reg}
		\widehat{D}_{0}\\
		&=
		\left(
		\widehat{X}_{j+1}
		-
		\widehat{X}_{j+1}^\dagger
		\right)
		\widehat{D}_{\cRep(S_3)}\,.
	\end{align*}
\item The $ J_3 $ and $ J_4 $ terms in the Hamiltonian \eqref{eq:def Ham gen RepS3} are left invariant, even though some indices get shuffled.
\end{enumerate}

%\begin{align}
%&\widehat{D}_{\cRep(S_3)}\left (\widehat{X}_j + \widehat{X}_j^\dagger\right ) = \left 
%(\widehat{Z}_{j} 
%\widehat{Z}_{j+1}^{-\hat{\mu}^x_{j+1}} + \widehat{Z}_{j}^\dagger 
%\widehat{Z}_{j+1}^{\hat{\mu}^x_{j+1}} \right ) \widehat{D}_{\cRep(S_3)} 
%=\left ( 
%\widehat{Z}_{j}^{-\hat{\mu}^x_{j+1}} 
%\widehat{Z}_{j+1} + \widehat{Z}_{j}^{\hat{\mu}^x_{j+1}} 
%\widehat{Z}_{j+1}^\dagger   \right )  \widehat{D}_{\cRep(S_3)} \\
%&\widehat{D}_{\cRep(S_3)}\left ( \widehat{Z}_{j}^{-\hat{\mu}^x_{j+1}} 
%\widehat{Z}_{j+1} + \widehat{Z}_{j}^{\hat{\mu}^x_{j+1}} 
%\widehat{Z}_{j+1}^\dagger   \right ) = \left (
%\widehat{X}_{j+1}^\dagger + \widehat{X}_{j+1}
%\right ) \widehat{D}_{\cRep(S_3)}
%\end{align}

\section{Details about numerical methods}
\label{app:numerics}

Let us briefly comment on how the TEFR algorithm works. We Trotter-ize the imaginary time path integral (or, partition function) associated with the Hamiltonian to obtain a rank-4 tensor $ T $. Since the lattice models have discrete space but continuous time, we implement a renormalization transformation along the time direction to obtain an ``isotropic" partition function tensor $ T_{\text{iso}} $. The full partition function can then be expressed as a network of these tensors $ T_{\text{iso}} $. Applying the TEFR algorithm \cite{LN0701,GW0931} entails multiple iterations of singular value decomposition and tensor contractions; finally, we reach a fixed point tensor $ T_{\text{iso}}^{\,*} $, whose largest eigenvalue has the form
\begin{equation}\label{eq: T* eigval}
\la^* = \text{GSD}\cdot  \exp\left\{T\left (-E_0 +\frac{2\pi v}{24L}\, c +\cO\left (\frac{1}{L^2}\right ) \right )\right\}
\end{equation}
where $ T $ is the total length of the compactified imaginary time direction, GSD is the ground state degeneracy, $ E_0 $ is the ground state energy, and $ c $ is the central charge which is only non-zero when the system is in a gapless phase described by a CFT at low energies and $ v $ is the ``velocity" of the linear-dispersing mode of this CFT. Since $ v $ in general depends on details of the microscopic Hamiltonian, the precise numerical value of the central charge is difficult to extract. However, we should note that the algorithm is quite efficient at distinguishing gapless regions of the phase diagram, which have non-zero $ c $, from gapped regions where $ c =0$ (within pre-set limits of precision). By benchmarking various known limits, we also find that relative values of $ c $ extracted using this approach are in practice reflective of the true central charges of the corresponding CFTs. 

\begin{figure}[t!]
\centering
\begin{subfigure}{.48\textwidth}
\centering
\includegraphics[width=.9\linewidth]{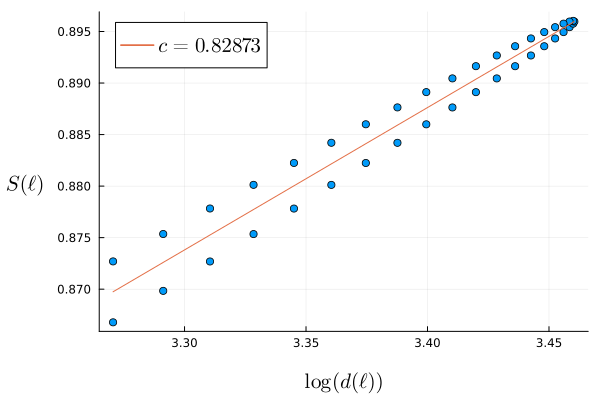}  
\caption{$ J_1 = 0.5, J_3=0.6 $}
\label{fig:J1=0.5,J3=0.6}
\end{subfigure}
\begin{subfigure}{.48\textwidth}
\centering
\includegraphics[width=.9\linewidth]{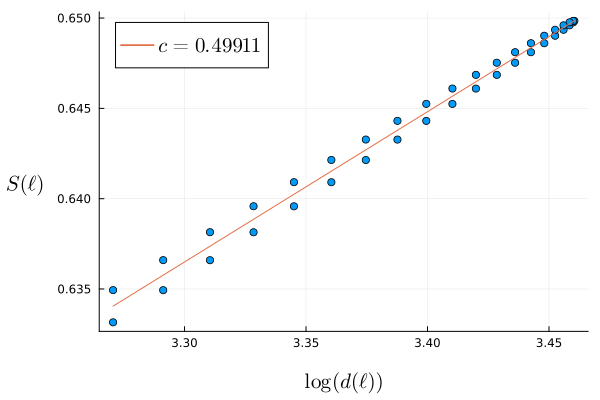}  
\caption{$ J_1 = 0.6, J_3=0.5 $}
\label{fig:J3=0.5,J1=0.6}
\end{subfigure}

\medskip

\begin{subfigure}{.48\textwidth}
\centering
\includegraphics[width=.9\linewidth]{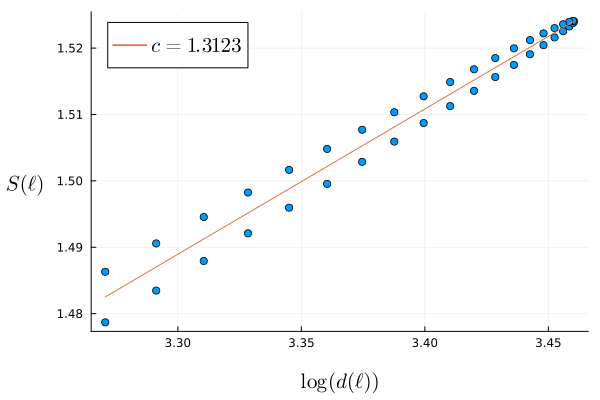}  
\caption{$ J_1 = 0.5, J_3=0.5 $}
\label{fig:J3=0.5,J1=0.5}
\end{subfigure}
\caption{Central charges extracted from fits to the Calabrese-Cardy function, as discussed in the main text. 
Figure (a) is consistent with Potts criticality, Figure (b) is consistent with Ising criticality, and Figure (c) is consistent
with a stacking of these two CFTs. 
The departures from exact values of $ c=0.8, 0.5, 1.3 $, respectively, are due to the
effects of finite system size. 
For these plots, we set the system size as $ L=100 $ and performed 
the central charge fits on the bulk sites by removing $ 30 $ sites from each boundary.
\label{fig:ccfits}
}
\end{figure}

In order to extract numerically precise central charges, we used the density matrix renormalization group (DMRG) algorithm from the iTensor library \cite{FWM200714822,itensor-r0.3}. Specifically, we used the Calabrese-Cardy formula \cite{CC0905} for the entanglement entropy of 1+1d CFTs,
\begin{equation}\label{eq: CCformula}
S(\ell) = \frac{c}{6}\log \left (\frac{2L}{\pi a} \sin \frac{\pi \ell}{L} \right )+c_1 \,,
\end{equation}
where $ \ell $ is the size of the subsystem, $ L $ is the size of the full system, $ S(\ell) $ is the entanglement entropy of the bipartition of the full system into two pieces of size $ \ell $ and $ L-\ell $, $ a $ is the lattice constant of the lattice model described by a CFT at low energies, and $ c $ is its central charge. By computing $ S(\ell) $ in the ground state for $ \ell =1,\dots L-1 $ and fitting it to the above functional form, we can quite reliably extract the central charge. We show examples of this method for two points in the phase diagram \fig{fig:s3pd-numerics} of our $ S_3 $ spin chain \eqref{eq:Ham gen S3 reparameterized} in \fig{fig:ccfits}, to demonstrate the quality of the fits.

%-------------------------

\section{Gapless boundaries of \texorpdfstring{$S_3$}{S3} SymTO}

\subsection{3-state Potts CFT boundary}
\label{app:potts cft partition function}
The 3-state Potts CFT is a $(6,5)$ minimal model. This CFT can be realized as a so-called $\one$-condensed gapless boundary of the 2+1d $S^{\,}_3$ topological order. Using the SymTO point of view, this gives us a way to identify operators of the CFT that carry various symmetry charges and symmetry twists. The vacuum sector $Z_\one$ contains contributions from local operators that are uncharged under the full $S^{\,}_3$ symmetry. The $\one'$ sector contains local operators that carry the sign representation of $S^{\,}_3$, and hence are charged under the $Z_2$ subgroups of $S^{\,}_3$ but uncharged under the $Z_3$ subgroup. The $\two$ sector contains local operators that carry the 2d irrep of $S^{\,}_3$. The sectors labeled by $b$ and $c$ contain (uncharged) operators that live on the boundaries of symmetry defects, whereas the sectors labeled by $r_1,r_2,s_1$ contain charged operators of the corresponding symmetry-twisted sectors. 

From modular bootstrap calculations, we find the following as the only $\one$-condensed boundary of $\eD(S_3)$ constructed out of the conformal characters of the Potts minimal model:
\begin{equation}
\label{eq:potts mod bootstrap}
\begin{split}
 Z^{\eD(S_3)}_{\one} &= \chi^{m6 \times \overline{m6}}_{1,0; 1,0} +  \chi^{m6 \times \overline{m6}}_{5,3; 5,-3} +  \chi^{m6 \times \overline{m6}}_{6,\frac{2}{5}; 6,-\frac{2}{5}} +  \chi^{m6 \times \overline{m6}}_{10,\frac{7}{5}; 10,-\frac{7}{5}} 
 \\ 
Z^{\eD(S_3)}_{\one'} &= \chi^{m6 \times \overline{m6}}_{1,0; 5,-3} +  \chi^{m6 \times \overline{m6}}_{5,3; 1,0} +  \chi^{m6 \times \overline{m6}}_{6,\frac{2}{5}; 10,-\frac{7}{5}} +  \chi^{m6 \times \overline{m6}}_{10,\frac{7}{5}; 6,-\frac{2}{5}} 
\\ 
Z^{\eD(S_3)}_{\two} &= \chi^{m6 \times \overline{m6}}_{3,\frac{2}{3}; 3,-\frac{2}{3}} +  \chi^{m6 \times \overline{m6}}_{8,\frac{1}{15}; 8,-\frac{1}{15}} 
\\ 
Z^{\eD(S_3)}_{r} &= \chi^{m6 \times \overline{m6}}_{3,\frac{2}{3}; 3,-\frac{2}{3}} +  \chi^{m6 \times \overline{m6}}_{8,\frac{1}{15}; 8,-\frac{1}{15}} 
 \\ 
Z^{\eD(S_3)}_{r_1} &= \chi^{m6 \times \overline{m6}}_{1,0; 3,-\frac{2}{3}} +  \chi^{m6 \times \overline{m6}}_{5,3; 3,-\frac{2}{3}} +  \chi^{m6 \times \overline{m6}}_{6,\frac{2}{5}; 8,-\frac{1}{15}} +  \chi^{m6 \times \overline{m6}}_{10,\frac{7}{5}; 8,-\frac{1}{15}} 
 \\ 
Z^{\eD(S_3)}_{r_2} &= \chi^{m6 \times \overline{m6}}_{3,\frac{2}{3}; 1,0} +  \chi^{m6 \times \overline{m6}}_{3,\frac{2}{3}; 5,-3} +  \chi^{m6 \times \overline{m6}}_{8,\frac{1}{15}; 6,-\frac{2}{5}} +  \chi^{m6 \times \overline{m6}}_{8,\frac{1}{15}; 10,-\frac{7}{5}} 
 \\ 
Z^{\eD(S_3)}_{s} &= \chi^{m6 \times \overline{m6}}_{2,\frac{1}{8}; 2,-\frac{1}{8}} +  \chi^{m6 \times \overline{m6}}_{4,\frac{13}{8}; 4,-\frac{13}{8}} +  \chi^{m6 \times \overline{m6}}_{7,\frac{1}{40}; 7,-\frac{1}{40}} +  \chi^{m6 \times \overline{m6}}_{9,\frac{21}{40}; 9,-\frac{21}{40}} 
 \\ 
Z^{\eD(S_3)}_{s_1} &= \chi^{m6 \times \overline{m6}}_{2,\frac{1}{8}; 4,-\frac{13}{8}} +  \chi^{m6 \times \overline{m6}}_{4,\frac{13}{8}; 2,-\frac{1}{8}} +  \chi^{m6 \times \overline{m6}}_{7,\frac{1}{40}; 9,-\frac{21}{40}} +  \chi^{m6 \times \overline{m6}}_{9,\frac{21}{40}; 7,-\frac{1}{40}} 
 \end{split}
 \end{equation}
The various terms in each component of the partition function are conformal characters of the (6,5) minimal model. The expression $\chi^{m6\times \overline{m6}}_{a,h_a;\, b, -h_b}$ is a short-hand notation for the product of the left moving chiral conformal character associated with the primary operator labeled $a$ (set by an arbitrary indexing convention) with conformal weight $(h_a,0)$, and the right moving chiral conformal character associated with the primary operator labeled $b$ with conformal weight $(0,h_b)$. The superscript $m6 \times \overline{m6}$ indicates that both the left and right moving chiral conformal characters are picked from the same (6,5) minimal model.

Note that in the above multi-component ``SymTO-resolved" partition function, the $\one'$ sector contains the primary operators with odd charge under the $\Z^{\,}_2$ subgroups of $S^{\,}_3$. All of these operators have non-zero conformal spin since $h\neq \bar h$ for all of them. However, we can construct descendents with zero conformal spin. All such descendents have scaling dimension greater than 2 and hence must be irrelevant perturbations of the CFT.

\subsection{3-state Potts \texorpdfstring{$\boxtimes$}{x} Ising CFT boundary}
Similar to the previous subsection, we can perform a modular bootstrap numerical calculation to identify gapless boundaries of the 2+1d $S_3$ topological order, considered as a SymTO. 
The following multi-component partition function shows the various symmetry charge and symmetry twist sectors for the $S_3$ symmetry:
\small
\label{app:potts x ising cft partition function}
\begin{align*}
Z^{\eD(S_3)}_{\one} &= 
\chi^{m4 \times m6 \times \overline{m4} \times \overline{m6}}_{1,0; 1,0; 1,0; 1,0} 
+  \chi^{m4 \times m6 \times \overline{m4} \times \overline{m6}}_{1,0; 5,3; 1,0; 5,-3} 
+  \chi^{m4 \times m6 \times \overline{m4} \times \overline{m6}}_{1,0; 6,\frac{2}{5}; 1,0; 6,-\frac{2}{5}} 
+  \chi^{m4 \times m6 \times \overline{m4} \times \overline{m6}}_{1,0; 10,\frac{7}{5}; 1,0; 10,-\frac{7}{5}} 
\nonumber\\ 
& \quad +  \chi^{m4 \times m6 \times \overline{m4} \times \overline{m6}}_{2,\frac{1}{16}; 1,0; 2,-\frac{1}{16}; 5,-3} 
+  \chi^{m4 \times m6 \times \overline{m4} \times \overline{m6}}_{2,\frac{1}{16}; 5,3; 2,-\frac{1}{16}; 1,0} 
+  \chi^{m4 \times m6 \times \overline{m4} \times \overline{m6}}_{2,\frac{1}{16}; 6,\frac{2}{5}; 2,-\frac{1}{16}; 10,-\frac{7}{5}} 
+  \chi^{m4 \times m6 \times \overline{m4} \times \overline{m6}}_{2,\frac{1}{16}; 10,\frac{7}{5}; 2,-\frac{1}{16}; 6,-\frac{2}{5}} 
\nonumber\\
& \quad +  \chi^{m4 \times m6 \times \overline{m4} \times \overline{m6}}_{3,\frac{1}{2}; 1,0; 3,-\frac{1}{2}; 1,0} 
+  \chi^{m4 \times m6 \times \overline{m4} \times \overline{m6}}_{3,\frac{1}{2}; 5,3; 3,-\frac{1}{2}; 5,-3} 
+  \chi^{m4 \times m6 \times \overline{m4} \times \overline{m6}}_{3,\frac{1}{2}; 6,\frac{2}{5}; 3,-\frac{1}{2}; 6,-\frac{2}{5}} 
+  \chi^{m4 \times m6 \times \overline{m4} \times \overline{m6}}_{3,\frac{1}{2}; 10,\frac{7}{5}; 3,-\frac{1}{2}; 10,-\frac{7}{5}} 
 \nonumber \\
 \\
Z^{\eD(S_3)}_{\one'} &= 
\chi^{m4 \times m6 \times \overline{m4} \times \overline{m6}}_{1,0; 1,0; 1,0; 5,-3} 
+  \chi^{m4 \times m6 \times \overline{m4} \times \overline{m6}}_{1,0; 5,3; 1,0; 1,0} 
+  \chi^{m4 \times m6 \times \overline{m4} \times \overline{m6}}_{1,0; 6,\frac{2}{5}; 1,0; 10,-\frac{7}{5}} 
+  \chi^{m4 \times m6 \times \overline{m4} \times \overline{m6}}_{1,0; 10,\frac{7}{5}; 1,0; 6,-\frac{2}{5}} 
\nonumber\\
& \quad +  \chi^{m4 \times m6 \times \overline{m4} \times \overline{m6}}_{2,\frac{1}{16}; 1,0; 2,-\frac{1}{16}; 1,0} 
+  \chi^{m4 \times m6 \times \overline{m4} \times \overline{m6}}_{2,\frac{1}{16}; 5,3; 2,-\frac{1}{16}; 5,-3} 
+  \chi^{m4 \times m6 \times \overline{m4} \times \overline{m6}}_{2,\frac{1}{16}; 6,\frac{2}{5}; 2,-\frac{1}{16}; 6,-\frac{2}{5}} 
+  \chi^{m4 \times m6 \times \overline{m4} \times \overline{m6}}_{2,\frac{1}{16}; 10,\frac{7}{5}; 2,-\frac{1}{16}; 10,-\frac{7}{5}} 
\nonumber\\
& \quad +  \chi^{m4 \times m6 \times \overline{m4} \times \overline{m6}}_{3,\frac{1}{2}; 1,0; 3,-\frac{1}{2}; 5,-3} 
+  \chi^{m4 \times m6 \times \overline{m4} \times \overline{m6}}_{3,\frac{1}{2}; 5,3; 3,-\frac{1}{2}; 1,0} 
+  \chi^{m4 \times m6 \times \overline{m4} \times \overline{m6}}_{3,\frac{1}{2}; 6,\frac{2}{5}; 3,-\frac{1}{2}; 10,-\frac{7}{5}} 
+  \chi^{m4 \times m6 \times \overline{m4} \times \overline{m6}}_{3,\frac{1}{2}; 10,\frac{7}{5}; 3,-\frac{1}{2}; 6,-\frac{2}{5}} 
 \nonumber \\ \\
Z^{\eD(S_3)}_{\two} &= 
\chi^{m4 \times m6 \times \overline{m4} \times \overline{m6}}_{1,0; 3,\frac{2}{3}; 1,0; 3,-\frac{2}{3}} 
+  \chi^{m4 \times m6 \times \overline{m4} \times \overline{m6}}_{1,0; 8,\frac{1}{15}; 1,0; 8,-\frac{1}{15}} 
+  \chi^{m4 \times m6 \times \overline{m4} \times \overline{m6}}_{2,\frac{1}{16}; 3,\frac{2}{3}; 2,-\frac{1}{16}; 3,-\frac{2}{3}} 
+  \chi^{m4 \times m6 \times \overline{m4} \times \overline{m6}}_{2,\frac{1}{16}; 8,\frac{1}{15}; 2,-\frac{1}{16}; 8,-\frac{1}{15}} 
\nonumber\\
& \quad +  \chi^{m4 \times m6 \times \overline{m4} \times \overline{m6}}_{3,\frac{1}{2}; 3,\frac{2}{3}; 3,-\frac{1}{2}; 3,-\frac{2}{3}} 
+  \chi^{m4 \times m6 \times \overline{m4} \times \overline{m6}}_{3,\frac{1}{2}; 8,\frac{1}{15}; 3,-\frac{1}{2}; 8,-\frac{1}{15}} 
 \nonumber \\ \\
Z^{\eD(S_3)}_{r} &= 
\chi^{m4 \times m6 \times \overline{m4} \times \overline{m6}}_{1,0; 3,\frac{2}{3}; 1,0; 3,-\frac{2}{3}} 
+  \chi^{m4 \times m6 \times \overline{m4} \times \overline{m6}}_{1,0; 8,\frac{1}{15}; 1,0; 8,-\frac{1}{15}} 
+  \chi^{m4 \times m6 \times \overline{m4} \times \overline{m6}}_{2,\frac{1}{16}; 3,\frac{2}{3}; 2,-\frac{1}{16}; 3,-\frac{2}{3}} 
+  \chi^{m4 \times m6 \times \overline{m4} \times \overline{m6}}_{2,\frac{1}{16}; 8,\frac{1}{15}; 2,-\frac{1}{16}; 8,-\frac{1}{15}} 
\nonumber\\
& \quad +  \chi^{m4 \times m6 \times \overline{m4} \times \overline{m6}}_{3,\frac{1}{2}; 3,\frac{2}{3}; 3,-\frac{1}{2}; 3,-\frac{2}{3}} 
+  \chi^{m4 \times m6 \times \overline{m4} \times \overline{m6}}_{3,\frac{1}{2}; 8,\frac{1}{15}; 3,-\frac{1}{2}; 8,-\frac{1}{15}} 
\nonumber \\ \\
Z^{\eD(S_3)}_{r_1} &= 
\chi^{m4 \times m6 \times \overline{m4} \times \overline{m6}}_{1,0; 1,0; 1,0; 3,-\frac{2}{3}} 
+  \chi^{m4 \times m6 \times \overline{m4} \times \overline{m6}}_{1,0; 5,3; 1,0; 3,-\frac{2}{3}} 
+  \chi^{m4 \times m6 \times \overline{m4} \times \overline{m6}}_{1,0; 6,\frac{2}{5}; 1,0; 8,-\frac{1}{15}} 
+  \chi^{m4 \times m6 \times \overline{m4} \times \overline{m6}}_{1,0; 10,\frac{7}{5}; 1,0; 8,-\frac{1}{15}} 
\nonumber\\
& \quad +  \chi^{m4 \times m6 \times \overline{m4} \times \overline{m6}}_{2,\frac{1}{16}; 1,0; 2,-\frac{1}{16}; 3,-\frac{2}{3}} 
+  \chi^{m4 \times m6 \times \overline{m4} \times \overline{m6}}_{2,\frac{1}{16}; 5,3; 2,-\frac{1}{16}; 3,-\frac{2}{3}} 
+  \chi^{m4 \times m6 \times \overline{m4} \times \overline{m6}}_{2,\frac{1}{16}; 6,\frac{2}{5}; 2,-\frac{1}{16}; 8,-\frac{1}{15}} 
+  \chi^{m4 \times m6 \times \overline{m4} \times \overline{m6}}_{2,\frac{1}{16}; 10,\frac{7}{5}; 2,-\frac{1}{16}; 8,-\frac{1}{15}} 
\nonumber\\
& \quad +  \chi^{m4 \times m6 \times \overline{m4} \times \overline{m6}}_{3,\frac{1}{2}; 1,0; 3,-\frac{1}{2}; 3,-\frac{2}{3}} 
+  \chi^{m4 \times m6 \times \overline{m4} \times \overline{m6}}_{3,\frac{1}{2}; 5,3; 3,-\frac{1}{2}; 3,-\frac{2}{3}} 
+  \chi^{m4 \times m6 \times \overline{m4} \times \overline{m6}}_{3,\frac{1}{2}; 6,\frac{2}{5}; 3,-\frac{1}{2}; 8,-\frac{1}{15}} 
+  \chi^{m4 \times m6 \times \overline{m4} \times \overline{m6}}_{3,\frac{1}{2}; 10,\frac{7}{5}; 3,-\frac{1}{2}; 8,-\frac{1}{15}} 
 \nonumber \\ \\
Z^{\eD(S_3)}_{r_2} &= 
\chi^{m4 \times m6 \times \overline{m4} \times \overline{m6}}_{1,0; 3,\frac{2}{3}; 1,0; 1,0} 
+  \chi^{m4 \times m6 \times \overline{m4} \times \overline{m6}}_{1,0; 3,\frac{2}{3}; 1,0; 5,-3} 
+  \chi^{m4 \times m6 \times \overline{m4} \times \overline{m6}}_{1,0; 8,\frac{1}{15}; 1,0; 6,-\frac{2}{5}} 
+  \chi^{m4 \times m6 \times \overline{m4} \times \overline{m6}}_{1,0; 8,\frac{1}{15}; 1,0; 10,-\frac{7}{5}} 
\nonumber\\
& \quad +  \chi^{m4 \times m6 \times \overline{m4} \times \overline{m6}}_{2,\frac{1}{16}; 3,\frac{2}{3}; 2,-\frac{1}{16}; 1,0} 
+  \chi^{m4 \times m6 \times \overline{m4} \times \overline{m6}}_{2,\frac{1}{16}; 3,\frac{2}{3}; 2,-\frac{1}{16}; 5,-3} 
+  \chi^{m4 \times m6 \times \overline{m4} \times \overline{m6}}_{2,\frac{1}{16}; 8,\frac{1}{15}; 2,-\frac{1}{16}; 6,-\frac{2}{5}} 
+  \chi^{m4 \times m6 \times \overline{m4} \times \overline{m6}}_{2,\frac{1}{16}; 8,\frac{1}{15}; 2,-\frac{1}{16}; 10,-\frac{7}{5}} 
\nonumber\\
& \quad +  \chi^{m4 \times m6 \times \overline{m4} \times \overline{m6}}_{3,\frac{1}{2}; 3,\frac{2}{3}; 3,-\frac{1}{2}; 1,0} 
+  \chi^{m4 \times m6 \times \overline{m4} \times \overline{m6}}_{3,\frac{1}{2}; 3,\frac{2}{3}; 3,-\frac{1}{2}; 5,-3} 
+  \chi^{m4 \times m6 \times \overline{m4} \times \overline{m6}}_{3,\frac{1}{2}; 8,\frac{1}{15}; 3,-\frac{1}{2}; 6,-\frac{2}{5}} 
+  \chi^{m4 \times m6 \times \overline{m4} \times \overline{m6}}_{3,\frac{1}{2}; 8,\frac{1}{15}; 3,-\frac{1}{2}; 10,-\frac{7}{5}} 
 \nonumber \\ \\
Z^{\eD(S_3)}_{s} &= 
\chi^{m4 \times m6 \times \overline{m4} \times \overline{m6}}_{1,0; 2,\frac{1}{8}; 3,-\frac{1}{2}; 4,-\frac{13}{8}} 
+  \chi^{m4 \times m6 \times \overline{m4} \times \overline{m6}}_{1,0; 4,\frac{13}{8}; 3,-\frac{1}{2}; 2,-\frac{1}{8}} 
+  \chi^{m4 \times m6 \times \overline{m4} \times \overline{m6}}_{1,0; 7,\frac{1}{40}; 3,-\frac{1}{2}; 9,-\frac{21}{40}} 
+  \chi^{m4 \times m6 \times \overline{m4} \times \overline{m6}}_{1,0; 9,\frac{21}{40}; 3,-\frac{1}{2}; 7,-\frac{1}{40}} 
\nonumber\\
& \quad +  \chi^{m4 \times m6 \times \overline{m4} \times \overline{m6}}_{2,\frac{1}{16}; 2,\frac{1}{8}; 2,-\frac{1}{16}; 2,-\frac{1}{8}} 
+  \chi^{m4 \times m6 \times \overline{m4} \times \overline{m6}}_{2,\frac{1}{16}; 4,\frac{13}{8}; 2,-\frac{1}{16}; 4,-\frac{13}{8}} 
+  \chi^{m4 \times m6 \times \overline{m4} \times \overline{m6}}_{2,\frac{1}{16}; 7,\frac{1}{40}; 2,-\frac{1}{16}; 7,-\frac{1}{40}} 
+  \chi^{m4 \times m6 \times \overline{m4} \times \overline{m6}}_{2,\frac{1}{16}; 9,\frac{21}{40}; 2,-\frac{1}{16}; 9,-\frac{21}{40}} 
\nonumber\\
& \quad +  \chi^{m4 \times m6 \times \overline{m4} \times \overline{m6}}_{3,\frac{1}{2}; 2,\frac{1}{8}; 1,0; 4,-\frac{13}{8}} 
+  \chi^{m4 \times m6 \times \overline{m4} \times \overline{m6}}_{3,\frac{1}{2}; 4,\frac{13}{8}; 1,0; 2,-\frac{1}{8}} 
+  \chi^{m4 \times m6 \times \overline{m4} \times \overline{m6}}_{3,\frac{1}{2}; 7,\frac{1}{40}; 1,0; 9,-\frac{21}{40}} 
+  \chi^{m4 \times m6 \times \overline{m4} \times \overline{m6}}_{3,\frac{1}{2}; 9,\frac{21}{40}; 1,0; 7,-\frac{1}{40}} 
 \nonumber \\  \\
Z^{\eD(S_3)}_{s_1} &= 
\chi^{m4 \times m6 \times \overline{m4} \times \overline{m6}}_{1,0; 2,\frac{1}{8}; 3,-\frac{1}{2}; 2,-\frac{1}{8}} 
+  \chi^{m4 \times m6 \times \overline{m4} \times \overline{m6}}_{1,0; 4,\frac{13}{8}; 3,-\frac{1}{2}; 4,-\frac{13}{8}} 
+  \chi^{m4 \times m6 \times \overline{m4} \times \overline{m6}}_{1,0; 7,\frac{1}{40}; 3,-\frac{1}{2}; 7,-\frac{1}{40}} 
+  \chi^{m4 \times m6 \times \overline{m4} \times \overline{m6}}_{1,0; 9,\frac{21}{40}; 3,-\frac{1}{2}; 9,-\frac{21}{40}} 
\nonumber\\
& \quad +  \chi^{m4 \times m6 \times \overline{m4} \times \overline{m6}}_{2,\frac{1}{16}; 2,\frac{1}{8}; 2,-\frac{1}{16}; 4,-\frac{13}{8}} 
+  \chi^{m4 \times m6 \times \overline{m4} \times \overline{m6}}_{2,\frac{1}{16}; 4,\frac{13}{8}; 2,-\frac{1}{16}; 2,-\frac{1}{8}} 
+  \chi^{m4 \times m6 \times \overline{m4} \times \overline{m6}}_{2,\frac{1}{16}; 7,\frac{1}{40}; 2,-\frac{1}{16}; 9,-\frac{21}{40}} 
+  \chi^{m4 \times m6 \times \overline{m4} \times \overline{m6}}_{2,\frac{1}{16}; 9,\frac{21}{40}; 2,-\frac{1}{16}; 7,-\frac{1}{40}} 
\nonumber\\
& \quad +  \chi^{m4 \times m6 \times \overline{m4} \times \overline{m6}}_{3,\frac{1}{2}; 2,\frac{1}{8}; 1,0; 2,-\frac{1}{8}} 
+  \chi^{m4 \times m6 \times \overline{m4} \times \overline{m6}}_{3,\frac{1}{2}; 4,\frac{13}{8}; 1,0; 4,-\frac{13}{8}} 
+  \chi^{m4 \times m6 \times \overline{m4} \times \overline{m6}}_{3,\frac{1}{2}; 7,\frac{1}{40}; 1,0; 7,-\frac{1}{40}} 
+  \chi^{m4 \times m6 \times \overline{m4} \times \overline{m6}}_{3,\frac{1}{2}; 9,\frac{21}{40}; 1,0; 9,-\frac{21}{40}} 
\nonumber
\end{align*}
\normalsize

\smallskip

This multi-component, SymTO-resolved, partition function is constructed out of the conformal characters of the primary operators of the Ising (labeled as $m4$ indicating it is the (4,3) minimal model) and the Potts (labeled as $m6$ indicating it is the (6,5) minimal model) CFTs. Similar to the convention in Eq.\ \eqref{eq:potts mod bootstrap}, the expression $\chi^{m4 \times m6\times \overline{m4} \times  \overline{m6}}_{a,h_a,\al,h_\al;\, b, -h_b,\bt,h_\bt}$ is a short-hand notation for the product of the left moving chiral conformal characters associated with the primary operators of the Ising and Potts CFTs with conformal weights $(h_a,0)$ and $(h_\al,0)$ respectively, with the right moving chiral conformal characters associated with the primary operators of the Ising and Potts CFTs with conformal weights $(0,h_b)$ and $(0,h_\bt)$ respectively. 

The only sector relevant for the purposes of this paper is the ``vacuum" sector $Z_\one$, which contains the $S_3$ symmetric operators constructed out of the primary operators of Ising and Potts CFTs. Note that there are 3 relevant operators that have zero conformal spin. These have scaling dimensions $\frac45$, $1$, and $\frac95$. They correspond to the three relevant perturbations explored by the couplings $J_1-J_2, J_3-J_4$, and $J_{\perp}$ in \scn{sec:phase transitions S3}. Only $J_3-J_4$ is unchanged by the action of the $\Z_3$ KW duality.

%-------------------------

\section{Spin chain with \texorpdfstring{$G$}{G} symmetry and its gauged partner with \texorpdfstring{$\cRep(G)$}{Rep(G)} symmetry}
\label{app:G and Rep G}
In this appendix, we present a simple manifestation of the fact that upon gauging $ G $ symmetry of a spin chain, with on-site Hilbert space identical to the regular representation of $ G $, one obtains a dual spin chain, also with on-site Hilbert space identical to the regular representation of $ G $, that has $ \cRep(G) $ symmetry.

We consider a tensor product Hilbert space in one spatial dimension, where the 
on-site Hilbert spaces are $|G|$ dimensional, and spanned by orthonormal 
basis vectors labeled by group elements, \ie
\begin{equation}
\cH = \otimes_i \cH_i\, , \quad \cH_i = \text{span}\{\Ket{g}|g\in G\} 
\label{eq:HSonsite}
\end{equation}
Then, we can construct a spin chain symmetric under $G$ (0-form) symmetry, 
using two families of local symmetric operators.
\begin{equation}\label{eq:HG}
\widehat{H}_{G} = \sum_{i\in \text{sites}} \left ( \widehat{L}_i + J \hat{\delta}_{i,i+1}\right )
\end{equation}
where, $\widehat{L}_i = \sum_{h\in G} \widehat{L^h}_i$ and $\widehat{L^h}_i$ acts on a basis vector at site 
$i$ by left multiplication, \ie
\begin{equation} \label{eq:Lh}
\widehat{L^h}_i \Ket{g_i} \= \Ket{hg_i}.
\end{equation}
and 
\begin{equation}\label{eq:delta}
\hat{\delta}_{i,i+1}\= \sum_{\{g\}}\del_{g_i,g_{i+1}} \Ketbra{\{g\}}{\{g\}}
\end{equation}
%We can express the Hamiltonian $H_G$ more explicitly as
%\begin{equation}
%H_{G} = \sum_{i\in \text{sites}}   \left ( \sum_{h\in 
%G}\Ketbra{\dots,hg_i,\dots}{\dots,g_i,\dots} 
%+J \del_{g_i,g_{i+1}}  \right )
%\end{equation}
It is straightforward to check that this Hamiltonian has a $G$ symmetry that 
acts by left multiplication on the basis vectors,\footnote{It also has another independent $ G $ symmetry that acts by right-multiplication. We will ignore this symmetry in the present discussion. To be concrete, one can include additional terms in the Hamiltonian that explicitly break this symmetry, while preserving the $G$ symmetry acting by left-multiplications.}
\begin{equation}
\widehat{U_h}  \=  \prod_{i\in \text{sites}} \widehat{L^h}_i,
\end{equation}
where $h$ is any element of $G$. The $G$ symmetry is reflected in the fact 
that $H_G$ commutes with all $U_h$.

A dual model can be defined in terms of degrees of freedom on links instead 
of sites. The local Hilbert space on each link is isomorphic to the one described 
in Eq.\ \eqref{eq:HSonsite},
\begin{equation}
\cH = \otimes_{l\in \text{links}} \cH_l\, , \quad \cH_l = \text{span}\{\Ket{\t g}|\ \t 
g\in G\} \label{eq:HSlinks}.
\end{equation}
The Hamiltonian for this dual model is defined as
\begin{equation} \label{eq:HdualRepG}
\widehat{H}_{\cRep(G)} =  \sum_{i\in \text{sites}} \left (  \widehat{Q}_i + J \widehat{\Del}_{(i,i+1)}\right )
\end{equation}
where $\widehat{Q}_i = \sum_{h\in G} \widehat{Q^h}_i$, with $\widehat{Q^h}_i$ 
acting as
\begin{equation}\label{eq:Qh}
\widehat{Q^h}_i \Ket{\dots,\t g_{(i-1,i)},\t g_{(i,i+1)},\dots} \= \Ket{\dots,\t g_{(i-1,i)}h^{-1}, 
h \t g_{(i,i+1)},\dots},
\end{equation}
and $ \widehat{\Del}_{(i,i+1)} $ defined as
\begin{equation}\label{eq:Del}
\widehat{\Del}_{(i,i+1)}	\=  \sum_{\{\t g\}}  \del_{\t g_{(i,i+1)},e}  \Ketbra{\{\t g\}}{\{\t g\}},
\end{equation}
where $e$ is the identity element in $G$. In the equations above, we have parametrized the links with pairs of successive site indices 
$(i,i+1)$.
%We can rewrite $\widehat{H}_{\cRep(G)}$ more explicitly as
%\begin{equation}\label{eq:tHGexpl}
%\widehat{H}_{\cRep(G)} = \sum_{i\in \text{sites}}   \Bigg ( \sum_{h\in 
%G}\Ketbra{\dots,g_{(i-1,i)}h^{-1}, h \t g_{(i,i+1)},\dots}
%{\dots,g_{(i-1,i)}, \t g_{(i,i+1)},\dots} 
%+J \del_{\t g_{(i,i+1)},e}  \Bigg )
%\end{equation} 

Let us show that the model \eqref{eq:HdualRepG} can be obtained from the model \eqref{eq:HG} 
by gauging the symmetry $G$. To that end, we introduce link 
degrees of freedom and enlarge the Hilbert space to
\begin{equation}
\cH^{\text{large}} = \cH^{\text{sites}}\otimes \cH^{\text{links}}
\end{equation}
where $\cH^{\text{links}} = \otimes_{l\in \text{links}} \cH_l$, with $ \cH_l = 
\text{span}\{\Ket{g}|g\in G\}$.
Next, we 
minimally couple our gauge field degrees of freedom (on the links) to the site 
degrees of freedom by modifying the second term of \eqref{eq:HG} to
\begin{equation} \label{eq:mincp}
J \hat{\delta}'_{\, i,i+1} \= 
\sum_{\{g, \t g\}} 
J \del_{g_i^{-1} \t g_{(i,i+1)} g_{i+1},e}
\Ketbra{\{g,\t g\}}{\{g,\t g\}},
\end{equation}
where by $ \{g, \t g\} $, we refer to the basis vectors of $ \cH^{\text{large}} $, labeled by $ G $-variables on both sites and links.
Next, we impose the Gauss law constraints to project down to the smaller physical 
Hilbert space $\cH^{\text{phys}}$. We define the Gauss law operators 
via its action on the enlarged Hilbert space basis vectors,
\begin{equation}\label{eq:Gauss}
\widehat{\G^h}_j \Ket{\dots, \t g_{(j-1,j)},g_j,\t g_{(j,j+1)},\dots} \\
= \Ket{\dots, \t g_{(j-1,j)} h^{-1},h g_j ,h \t g_{(j,j+1)} ,\dots}.
\end{equation}
We can compactly express $\G^h_j$ in terms of the operators $\widehat{L^h}_j$ and $\widehat{Q^h}_j$ 
introduced in Eq.\ \eqref{eq:Lh} and Eq.\ \eqref{eq:Qh},
$ \widehat{\G^h}_j = \widehat{L^h}_j \, \widehat{Q^h}_j  $.
The physical Hilbert space is given by
\begin{equation}
\cH^{\text{phys}} \cong 
%{\left (\otimes_{j\in \text{sites}} \cH_j 	\otimes_{l\in \text{links}} \cH_l\right )}
\cH^{\text{large}}
/_{\{\G^h_{\,j} =1 \ \forall j \ \forall h\} }.
\end{equation}
On this reduced Hilbert space, $ \widehat{\G^h}_j $ acts as the 
identity operator, by definition. So as far as states in $ \cH^{\text{phys}} $ are 
concerned, $ \widehat{L^h}_j $ acts as $ \left (\widehat{Q^h}_j\right )^{-1} = \widehat{Q^{h^{-1}}}_j $. 
Let us now gauge-fix using the unitary operators,
\begin{equation}
\widehat{U}_{\{\ga_j\}} = \prod_{j \in \text{sites}} \widehat{\G^{\ga_j}}_j.
\end{equation}
In other words, we start with an arbitrary state
\begin{equation}
\Ket{\dots, \t g_{(j-1,j)}, g_j, \t g_{(j,j+1)}, \dots} \in \cH^{\text{large}}
\end{equation}
and gauge-fix by applying $ \widehat{U}_{\{\ga_j\}}$ with $\ga_j=g_j$, to end up 
with
\begin{equation}
\Ket{\dots, g_{j-1} \t g_{(j-1,j)} g_j^{-1}, e,  g_j \t g_{(j,j+1)} g_{j+1}^{-1}, \dots}.
\end{equation}
Thus our gauge-fixed states $\Ket{\dots, \t g_{(j-1,j)}^{\,'}, g_j^{\,'}, \t g_{(j,j+1)}^{\,'}, \dots}$ 
are given in terms of the original site and index labels by
\begin{equation}
\t g_{(j,j+1)}^{\,'} = g_j \t g_{(j,j+1)} g_{j+1}^{-1}\, , \quad  g_j^{\,'} =e,
\end{equation}
\ie the gauge-fixed states all have the site degrees of freedom labeled by the 
identity element of the group $G$. 
On these states, our minimal coupling term \eqref{eq:mincp} becomes $ J \widehat{\Delta}_{\, (i,i+1)} $ so that
the full gauge-fixed gauged Hamiltonian takes 
the form
\begin{equation*}\label{eq:Hgf}
\widehat{H}_{\cRep(G)}=\sum_{i\in \text{sites}} \ \widehat{Q}_i + J \widehat{\Del}_{ (i,i+1)},
\end{equation*}
thereby deriving Eq.\ \eqref{eq:HdualRepG}. We note that the gauge-fixed Hilbert space does have a tensor product structure, unlike $ \cH^{\text{phys}} $. Therefore, it makes sense to refer to this model as a ``spin chain".

Turns out, the Hamiltonian \eqref{eq:HdualRepG} has the non-invertible 0-form symmetry described by
the fusion category $\cRep(G)$.\footnote{This explains the subscript on the dual Hamiltonian \eqref{eq:HdualRepG}.} The associated symmetry transformations are implemented by the
operators
\begin{equation}
\widehat{W}^{\,}_R =\sum_{\{\t g_l\}} \Tr \ R \left ( \prod_{l\in \text{links}} \t g_l \right ) \Ketbra{\{\t g_l\}}{\{\t 
g_l\}}
\end{equation}
where $R$ takes values in the set of irreducible representations of $G$, 
namely the simple objects of $\cRep(G)$. 
The first term in Hamiltonian
\eqref{eq:HdualRepG} commutes with $ \widehat{W}_{R} $ due to the fact that $R(h^{-1})R(h) = R(h^{-1}h) = 
R(e)=\one$, on account of $ R $ being a representation of $ G $. The second term commutes as well since both $ \widehat{W}^{\,}_R $ and this term are diagonal in the product basis of the link degrees of freedom.

\section{Fermionic $\mathsf{SRep}(S^{\,}_{3})$ symmetry}
\label{app:fermion}
In Sec.\ \ref{sec:RepS3model}, we constructed the $\cRep(S^{\,}_{3})$-symmetric Hamiltonian
\eqref{eq:def Ham gen RepS3} form $S^{\,}_{3}$-symmetric 
Hamiltonian \eqref{eq:def Ham gen S3} by gauging the $\Z^{\,}_{2}$ subgroup of $S^{\,}_{3}$.
An alternative way to gauging this $\Z^{\,}_{2}$ symmetry is to introduce $\Z^{\,}_{2}$
link degrees of freedom that obeys fermionic statistics, which implements the so-called
Jordan-Wigner (JW) transformation~\cite{KT170108264,R180907757,KTC190205550}. 
Such a JW transformation is viable also for gauging the $\Z^{\,}_{2}$ subgroup of $S^{\,}_{3}$-symmetry
and delivers a super fusion category symmetry $\mathsf{SRep}(S^{\,}_{3})$ 
(see Ref.\ \cite{I220613159} for a discussion 
from topological quantum field theory perspective,
and Refs.\ \cite{BIT240509754,H240509611} for SymTO perspective).

\subsection{Jordan-Wigner duality and constructing $\mathsf{SRep}(S^{\,}_{3})$ symmetry}
\label{sec:constructing SRepS3}

We follow the strategy employed in Ref.\ \cite{AMFT230800743} and introduce two Majorana 
degrees of freedom $\left\{\hat{\eta}^{\,}_{i+1/2}, \hat{\xi}^{\,}_{i+1/2}\right\}$ 
on each link, which satisfy the fermionic anticommutation relations
\begin{align}
\left\{\hat{\eta}^{\,}_{i+1/2},\,\hat{\xi}^{\,}_{j+1/2}\right\}
=
0,
\qquad
\left\{\hat{\eta}^{\,}_{i+1/2},\,\hat{\eta}^{\,}_{j+1/2}\right\}
=
\left\{\hat{\xi}^{\,}_{i+1/2},\,\hat{\xi}^{\,}_{j+1/2}\right\}
=
2\delta^{\,}_{ij}.
\end{align}
Without loss of generality, we impose anti-periodic boundary conditions
on the fermionic degrees of freedom,\footnote{This choice corresponds to
the fermion parity \emph{untwisted} (in space direction) 
sector in the continuum limit. In contrast, the twisted sector
is given by choosing periodic boundary conditions.} 
\ie  $\hat{\eta}^{\,}_{i+L+1/2}=-\hat{\eta}^{\,}_{i+1/2}$
and $\hat{\xi}^{\,}_{i+L+1/2}=-\hat{\xi}^{\,}_{i+1/2}$ and for simplicity we assume $L$ is even.
To gauge the $\Z^{\,}_{2}$ subgroup of $S^{\,}_{3}$, we define the 
pairwise commuting Gauss operators
\begin{align}
\widehat{G}^{\mathrm{F}}_{i}
\=
\mathrm{i}
\hat{\xi}^{\,}_{i-1/2}\,
\hat{\sigma}^{x}_{i}\,
\hat{\tau}^{x}_{i}\,
\widehat{C}^{\,}_{i}\,
\hat{\eta}^{\,}_{i+1/2},
\qquad
\left[
\widehat{G}^{\mathrm{F}}_{i}
\right]^{2}
=
\hat{\mathbbm{1}},
\label{eq:Gaus operators fermions}
\end{align}
where, as opposed to the Gauss operator in Eq.\ \eqref{eq:Gauss operator Z2 gauging}
the local representative of $\widehat{U}^{\,}_{s}$ symmetry is sandwiched between fermionic operators.
Just as it was the case before, we define the gauge invariant subspace to be the one for which the 
Gauss operators are set to identity.

In a similar fashion to Sec.\ \ref{sec:RepS3model}, 
by minimally coupling the bond algebra \eqref{eq:def S3 algebra} 
we can construct a gauge invariant bond algebra.
To this end, we define the pairwise commuting local operators
\begin{align}
\hat{p}^{\,}_{i+1/2}
\=
\mathrm{i}
\hat{\xi}^{\,}_{i+1/2}\,
\hat{\eta}^{\,}_{i+1/2},
\qquad
\left[
\hat{p}^{\,}_{i+1/2},\,
\hat{p}^{\,}_{j+1/2}
\right]
=
0,
\quad
\hat{p}^{2}_{i+1/2}
=
\hat{\mathbbm{1}},
\end{align}
which are local at the links  $i+1/2$. 
The minimally coupled bond algebra is then
\begin{align}
\mathfrak{B}^{\mathrm{mc}}_{\mathrm{F}}
\=
\Big\langle
&
\hat{\sigma}^{z}_{i}\,
\hat{\tau}^{z}_{i},\,
\hat{\tau}^{z}_{i}\,
\hat{p}^{\,}_{i+1/2}\,
\hat{\sigma}^{z}_{i+1},\,
\hat{\sigma}^{x}_{i},\,
\hat{\tau}^{x}_{i},\,
\left(\widehat{X}^{\,}_{i}+\widehat{X}^{\dagger}_{i}\right),\,
\left(\widehat{Z}^{\hat{p}^{\,}_{i+1/2}}_{i}\,
\widehat{Z}^{\dagger}_{i+1}
+\widehat{Z}^{-\hat{p}^{\,}_{i+1/2}}_{i}\,
\widehat{Z}^{\,}_{i+1}\right),
\nonumber\\
&
\hat{\sigma}^{z}_{i}\,\left(\widehat{X}^{\,}_{i}-\widehat{X}^{\dagger}_{i}\right),\,
\hat{\tau}^{z}_{i}\,
\hat{p}^{\,}_{i+1/2}\,
\left(
\widehat{Z}^{\hat{p}^{\,}_{i+1/2}}_{i}\,
\widehat{Z}^{\dagger}_{i+1}
-\widehat{Z}^{-\hat{p}^{\,}_{i+1/2}}_{i}\,
\widehat{Z}^{\,}_{i+1}\right)
\Big\vert 
\,\,
\widehat{G}^{\mathrm{F}}_{i}=1,
\quad
i\in \Lambda
\Big\rangle,
\end{align}
where the local operator $\hat{p}^{\,}_{i+1/2}$ acts as a $\Z^{\,}_{2}$-valued 
bosonic gauge field. Physically, this operator measures the local fermion parity at link 
$i+1/2$. We implement an analogue of the unitary transformation 
\eqref{eq:Z2 gauging unitary} such that
\begin{align}
\begin{aligned}
&
\widehat{U}\,
\hat{\sigma}^{x}_{i}\,
\widehat{U}^{\dagger}
=
\mathrm{i}
\hat{\xi}^{\,}_{i-1/2}\,
\hat{\sigma}^{x}_{i}\,
\hat{\tau}^{x}_{i}\,
\widehat{C}^{\,}_{i}\,
\hat{\eta}^{\,}_{i+1/2},
\qquad
&&
\widehat{U}\,
\hat{\sigma}^{z}_{i}\,
\widehat{U}^{\dagger}
=
\hat{\sigma}^{z}_{i},
\\
&
\widehat{U}\,
\hat{\tau}^{x}_{i}\,
\widehat{U}^{\dagger}
=
\hat{\tau}^{x}_{i},
\qquad
&&
\widehat{U}\,
\hat{\tau}^{z}_{i}\,
\widehat{U}^{\dagger}
=
\hat{\tau}^{z}_{i}\,
\hat{\sigma}^{z}_{i},
\\
&
\widehat{U}\,
\widehat{X}^{\,}_{i}\,
\widehat{U}^{\dagger}
=
\widehat{X}^{\hat{\sigma}^{z}_{i}}_{i},
\qquad
&&
\widehat{U}\,
\widehat{Z}^{\,}_{i}\,
\widehat{U}^{\dagger}
=
\widehat{Z}^{\hat{\sigma}^{z}_{i}}_{i},
\\
&
\widehat{U}\,
\hat{\xi}^{\,}_{i+1/2}\,
\widehat{U}^{\dagger}
=
\hat{\xi}^{\,}_{i+1/2}\,
\hat{\sigma}^{z}_{i+1},
\qquad
&&
\widehat{U}\,
\hat{\eta}^{\,}_{i+1/2},
\widehat{U}^{\dagger}
=
\hat{\sigma}^{z}_{i}
\hat{\eta}^{\,}_{i+1/2},
\end{aligned}
\end{align}
which simplifies the Gauss operator to
$\widehat{U}\,\widehat{G}^{\mathrm{F}}_{i}\,\widehat{U}^{\dagger}=\hat{\sigma}^{x}_{i}$. 
Setting $\hat{\sigma}^{x}_{i}=1$ and shifting the fermionic link degrees of freedom to sites by
$i+1/2\mapsto i+1$ delivers the dual bond algebra
\begin{align}
\label{eq:def SRepS3 algebra}
\mathfrak{B}^{\,}_{\mathrm{F}}
\=
\Big\langle
&
\hat{\tau}^{z}_{i},\,
\hat{\tau}^{z}_{i}\,
\mathrm{i}\hat{\xi}^{\,}_{i+1}\,
\hat{\eta}^{\,}_{i+1},\,
\mathrm{i}
\hat{\xi}^{\,}_{i}\,
\hat{\tau}^{x}_{i}\,
\widehat{C}^{\,}_{i}\,
\hat{\eta}^{\,}_{i+1},\,
\hat{\tau}^{x}_{i},\,
\left(\widehat{X}^{\,}_{i}+\widehat{X}^{\dagger}_{i}\right),\,
\left(\widehat{Z}^{
\mathrm{i}\hat{\xi}^{\,}_{i+1}\,\hat{\eta}^{\,}_{i+1}}_{i}\,
\widehat{Z}^{\dagger}_{i+1}
+\widehat{Z}^{-\mathrm{i}\hat{\xi}^{\,}_{i+1}\,
	\hat{\eta}^{\,}_{i+1}}_{i}\,
\widehat{Z}^{\,}_{i+1}\right),
\nonumber\\
&
\left(\widehat{X}^{\,}_{i}-\widehat{X}^{\dagger}_{i}\right),\,
\hat{\tau}^{z}_{i}\,
\mathrm{i}\hat{\xi}^{\,}_{i+1}\,
\hat{\eta}^{\,}_{i+1}
\left(
\widehat{Z}^{\mathrm{i}\hat{\xi}^{\,}_{i+1}\,\hat{\eta}^{\,}_{i+1}}_{i}\,
\widehat{Z}^{\dagger}_{i+1}
-\widehat{Z}^{-
\mathrm{i}\hat{\xi}^{\,}_{i+1}\,
\hat{\eta}^{\,}_{i+1}}_{i}\,
\widehat{Z}^{\,}_{i+1}\right)
\Big\vert 
\,\,
i\in \Lambda
\Big\rangle.
\end{align}
By comparing with the bond algebra \eqref{eq:def S3/Z2 algebra} of $\cRep(S^{\,}_{3})$-symmetric operators, 
we conclude that the generators of the bond algebra \eqref{eq:def SRepS3 algebra} commute with the operators
\begin{subequations}
\label{eq:SRepS3 generators}
\begin{equation}
\begin{split}
\widehat{W}^{\,}_{\bm{1}}
&\=
\hat{\mathbbm{1}},
\\
\widehat{W}^{\,}_{p}
&\=
\prod_{i}^{L}
\mathrm{i}\hat{\xi}^{\,}_{i}\,
\hat{\eta}^{\,}_{i}
\equiv
\prod_{i}^{L}
\hat{p}^{\,}_{i},
\\
\widehat{W}^{\,}_{\bm{2}}
&\=
\frac{1}{2}
\left(
1
+
\prod_{i=1}^{L}
\mathrm{i}\hat{\xi}^{\,}_{i}\,
\hat{\eta}^{\,}_{i}
\right)
\left[
\prod_{i=1}^{L}
\widehat{X}^{\prod_{k=2}^{i}\mathrm{i}\hat{\xi}^{\,}_{k}\,\hat{\eta}^{\,}_{k}}_{i}
+
\widehat{X}^{-\prod_{k=2}^{i}\mathrm{i}\hat{\xi}^{\,}_{k}\,\hat{\eta}^{\,}_{k}}_{i}
\right].
\end{split}
\end{equation} 
These operator satisfy the fusion rules
\begin{align}
\widehat{W}^{\,}_{p}\,
\widehat{W}^{\,}_{p}
=
\widehat{W}^{\,}_{\bm{1}},
\qquad
\widehat{W}^{\,}_{p}\,
\widehat{W}^{\,}_{\bm{2}}
=
\widehat{W}^{\,}_{\bm{2}},
\qquad
\widehat{W}^{\,}_{\bm{2}}\,
\widehat{W}^{\,}_{\bm{2}}
=
\widehat{W}^{\,}_{\bm{1}}
+
\widehat{W}^{\,}_{p}
+
\widehat{W}^{\,}_{\bm{2}}.
\end{align}
\end{subequations}
Note that the operator $\widehat{W}^{\,}_{p}$ implements the
$\mathbb{Z}^{\mathrm{F}}_{2}$ fermion parity symmetry, which is special 
in the sense that it cannot be broken explicitly or spontaneously and is a symmetry of any fermionic
model. 
We call the symmetry generated by operators \eqref{eq:SRepS3 generators}
the super fusion category $\mathsf{SRep}(S^{\,}_{3})$, where the adjective super
signifies the non-trivial inclusion of fermion parity symmetry into the fusion category.

On the one hand, 
one verifies that the image of the gauged $\Z^{\,}_{2}$ symmetry $\widehat{U}^{\,}_{s}$
is
\begin{align}
\prod_{i=1}^{L}
\mathrm{i}
\hat{\xi}^{\,}_{i}\,
\hat{\eta}^{\,}_{i+1}
=
\prod_{i=1}^{L}
\mathrm{i}
\hat{\xi}^{\,}_{i}\,
\hat{\eta}^{\,}_{i},
\end{align}
where we used the facts that $L$ is even and 
we imposed antiperiodic boundary 
conditions on the fermionic degrees of freedom.
On the other hand, 
imposing antiperiodic boundary conditions on fermions and periodic 
ones on bosons 
implies that image of $\hat{\mathbbm{1}}=\prod_{i}\hat{\sigma}^{z}_{i}\,\hat{\sigma}^{z}_{i+1}$
is
\begin{align}
\prod_{i=1}^{L}
\mathrm{i}
\hat{\xi}^{\,}_{i}\,
\hat{\eta}^{\,}_{i}
\equiv
\widehat{W}^{\,}_{p}.
\end{align}
Therefore, we conclude that the duality between the bond algebras \eqref{eq:def S3 algebra}
and \eqref{eq:def SRepS3 algebra} holds in the subalgebras
\begin{align}
\mathfrak{B}^{\,}_{S^{\,}_{3}}\Big\vert^{\,}_{\widehat{U}^{\,}_{s}=+1}
\cong 
\mathfrak{B}^{\,}_{\mathrm{F}}\Big\vert^{\,}_{\widehat{W}^{\,}_{p}=+1}.
\end{align}

\subsection{Hamiltonian and its phase diagram}

Using this duality, we can construct the image of the Hamiltonian \eqref{eq:def Ham gen S3}
as
\begin{equation}
\label{eq:def Ham gen SRepS3}
\begin{split}
\widehat{H}^{\,}_{\mathsf{SRep}(S^{\,}_{3})}
\=
&-
J^{\,}_{1}\,
\sum_{i=1}^{L}
\left(
\widehat{Z}^{\mathrm{i}\hat{\xi}^{\,}_{i+1}\,\hat{\eta}^{\,}_{i+1}}_{i}\,
\widehat{Z}^{\dagger}_{i+1}
+
\widehat{Z}^{-\mathrm{i}\hat{\xi}^{\,}_{i+1}\,\hat{\eta}^{\,}_{i+1}}_{i}\,
\widehat{Z}^{\,}_{i+1}
\right)
-
J^{\,}_{2}\,
\sum_{i=1}^{L}
\left(
\widehat{X}^{\,}_{i}
+
\widehat{X}^{\dagger}_{i}
\right)
\\
&
-
J^{\,}_{3}\,
\sum_{i=1}^{L}
\left(
\hat{\tau}^{z}_{i}
+
\hat{\tau}^{z}_{i}\,
\mathrm{i}\hat{\xi}^{\,}_{i+1}\,\hat{\eta}^{\,}_{i+1}
\right)
-
J^{\,}_{4}\,
\sum_{i=1}^{L}
\left(
\mathrm{i}
\hat{\xi}^{\,}_{i}\,
\hat{\tau}^{x}_{i}\,
\widehat{C}^{\,}_{i}\,
\hat{\eta}^{\,}_{i+1}
+
\hat{\tau}^{x}_{i}
\right)
\\
&
-
J^{\,}_{5}\,
\sum_{i=1}^{L}
\mathrm{i}\,
\hat{\tau}^{z}_{i}\,
\mathrm{i}\hat{\xi}^{\,}_{i+1}\,\hat{\eta}^{\,}_{i+1}\,
\left(
\widehat{Z}^{\mathrm{i}\hat{\xi}^{\,}_{i+1}\,\hat{\eta}^{\,}_{i+1}}_{i}\,
\widehat{Z}^{\dagger}_{i+1}
-\widehat{Z}^{-\mathrm{i}\hat{\xi}^{\,}_{i+1}\,\hat{\eta}^{\,}_{i+1}}_{i}\,
\widehat{Z}^{\,}_{i+1}
\right)
\\
&
-
J^{\,}_{6}\,
\sum_{i=1}^{L}
\mathrm{i}\,
\left(
\widehat{X}^{\,}_{i}
-
\widehat{X}^{\dagger}_{i}
\right),
\end{split}
\end{equation}
which is symmetric under the $\mathsf{SRep}(S^{\,}_{3})$ symmetry generated by
operators \eqref{eq:SRepS3 generators}.
By duality, the phase diagram of this Hamiltonian has the same shape as 
that of Hamiltonian \eqref{eq:Ham gen S3 reparameterized}.
Without loss of generality, we set $J^{\,}_{5}=J^{\,}_{6}=0$
and identify the following ground states corresponding to four fixed-point gapped phases.
\begin{enumerate}[(i)]
\item
When $J^{\,}_{1}=J^{\,}_{4}=0$, 
Hamiltonian \eqref{eq:def Ham gen SRepS3} becomes
\begin{equation}
\label{eq:HSRepS3 J2, J3}
\widehat{H}^{\,}_{\mathsf{SRep}(S^{\,}_{3});2,3}
\= 	
-
J^{\,}_{2}\,
\sum_{i=1}^{L}
\left(
\widehat{X}^{\,}_{i}
+
\widehat{X}^{\dagger}_{i}
\right)
-
J^{\,}_{3}\,
\sum_{i=1}^{L}
\left(
\hat{\tau}^{z}_{i}
+
\hat{\tau}^{z}_{i}\,
\mathrm{i}\hat{\xi}^{\,}_{i+1}\,\hat{\eta}^{\,}_{i+1}
\right)
\end{equation}
There is a single nondegenerate gapped ground state
\begin{align}
\label{eq:GS srepS3 triv}
\ket{\mathrm{GS}^{\,}_{\mathrm{Triv}}}
\=
\bigotimes_{i=1}^{L}
\ket{\tau^{z}_{i}=1,\,\mathrm{i}\xi^{\,}_{i}\,\eta^{\,}_{i}=1,\, X^{\,}_{i}=1},
\end{align}
which is symmetric under the entire $\mathsf{SRep}(S^{\,}_{3})$ symmetry. 
This ground state carries even fermion parity and is a trivial 
invertible fermionic topological state. For that reason we call the phase 
trivial $\mathsf{SRep}(S^{\,}_{3})$-symmetric phase.

\item 
When $J^{\,}_{2}=J^{\,}_{4}=0$, 
Hamiltonian \eqref{eq:def Ham gen SRepS3} becomes
\begin{align}
\label{eq:HSRepS3 J1, J3}
\widehat{H}^{\,}_{\cRep(S^{\,}_{3});1,3}
\= 	
&
-
J^{\,}_{1}\,
\sum_{i=1}^{L}
\left(
\widehat{Z}^{\mathrm{i}\hat{\xi}^{\,}_{i+1}\,\hat{\eta}^{\,}_{i+1}}_{i}\,
\widehat{Z}^{\dagger}_{i+1}
+
\widehat{Z}^{-\mathrm{i}\hat{\xi}^{\,}_{i+1}\,\hat{\eta}^{\,}_{i+1}}_{i}\,
\widehat{Z}^{\,}_{i+1}
\right)
\nonumber\\
&
-
J^{\,}_{3}\,
\sum_{i=1}^{L}
\left(
\hat{\tau}^{z}_{i}
+
\hat{\tau}^{z}_{i}\,
\mathrm{i}\hat{\xi}^{\,}_{i+1}\,\hat{\eta}^{\,}_{i+1}
\right).
\end{align}
There are three degenerate ground states 
\begin{align}
\label{eq:GS Wp}
\ket{\mathrm{GS}^{\alpha}_{\mathrm{Triv}}}
\=
\bigotimes_{i=1}^{L}
\ket{\tau^{z}_{i}=1,\,\mathrm{i}\xi^{\,}_{i}\,\eta^{\,}_{i}=1,\, Z^{\,}_{i}=\omega^{\alpha}}.
\end{align}
These ground states preserve the $\mathbb{Z}^{\mathrm{F}}_{2}$ fermion parity
symmetry generated by 
$\widehat{W}^{\,}_{p}$ while they break the non-invertible
$\widehat{W}^{\,}_{\two}$ symmetry. 
Under the latter each ground state is mapped to equal superposition
of the the other two, i.e.,
\begin{equation}
\begin{split}
&
\widehat{W}^{\,}_{\two}\,
\ket{\mathrm{GS}^{1}_{p}}
=
\ket{\mathrm{GS}^{2}_{p}}
+
\ket{\mathrm{GS}^{3}_{p}},
\\
&
\widehat{W}^{\,}_{\two}\,
\ket{\mathrm{GS}^{2}_{p}}
=
\ket{\mathrm{GS}^{3}_{p}}
+
\ket{\mathrm{GS}^{1}_{p}},
\\
&
\widehat{W}^{\,}_{\two}\,
\ket{\mathrm{GS}^{3}_{p}}
=
\ket{\mathrm{GS}^{1}_{p}}
+
\ket{\mathrm{GS}^{2}_{p}}.
\end{split}
\end{equation}
Each ground state realize a trivial invertible fermionic state.
We call this the trivial $\mathsf{SRep}(S^{\,}_{3})/\mathbb{Z}^{\mathrm{F}}_{2}$
SSB phase.

\item 
When $J^{\,}_{1}=J^{\,}_{3}=0$, 
the Hamiltonian \eqref{eq:def Ham gen SRepS3} becomes
\begin{equation}
\label{eq:HSRepS3 J2, J4}
\widehat{H}^{\,}_{\mathsf{SRep}(S^{\,}_{3});2,4}
\= 	
-
J^{\,}_{2}\,
\sum_{i=1}^{L}
\left(
\widehat{X}^{\,}_{i}
+
\widehat{X}^{\dagger}_{i}
\right)
-
J^{\,}_{4}\,
\sum_{i=1}^{L}
\left(
\mathrm{i}
\hat{\xi}^{\,}_{i}\,
\hat{\tau}^{x}_{i}\,
\widehat{C}^{\,}_{i}\,
\hat{\eta}^{\,}_{i+1}
+
\hat{\tau}^{x}_{i}
\right).
\end{equation}
There is a single nondegenerate ground state 
\begin{align}
\label{eq:GS Kitaev}
\ket{\mathrm{GS}^{\,}_{\mathrm{Kitaev}}}
\=
\bigotimes_{i=1}^L
\ket{\tau^{x}_{i}=1,\,
\mathrm{i}\xi^{\,}_{i}\,\eta^{\,}_{i+1}= 1,\, X^{\,}_{i}=1}.
\end{align}
As opposed to the ground state \eqref{eq:GS srepS3 triv},
the fermions in this ground state realize a non-trivial invertible phase of 
matter, \ie the ground state of the Kitaev chain~\cite{K0131}.
When open boundary conditions are imposed, the ground states become twofold degenerate
with unpaired Majorana degrees of freedom at each end of the chain. 

Ground state is invariant under ${\mathsf{SRep}(S^{\,}_{3})}$
symmetry as it transforms by
\begin{align}
\widehat{W}^{\,}_{p}\,
\ket{\mathrm{GS}^{\,}_{\mathrm{Kitaev}}}
=
+
\ket{\mathrm{GS}^{\,}_{\mathrm{Kitaev}}},
\qquad
\widehat{W}^{\,}_{\two}\,
\ket{\mathrm{GS}^{\,}_{\mathrm{Kitaev}}}
=
2\ket{\mathrm{GS}^{\,}_{\mathrm{Kitaev}}}.
\end{align}
Therefore, we call this the 
Kitaev $\mathsf{SRep}(S^{\,}_{3})$-symmetric phase.

\item
When $J^{\,}_{2}=J^{\,}_{3}=0$, 
the Hamiltonian \eqref{eq:def Ham gen SRepS3} becomes
\begin{align}
\label{eq:HSRepS3 J1, J4}
\widehat{H}^{\,}_{\mathsf{SRep}(S^{\,}_{3});1,4} 
:= 
&
-
J^{\,}_{1}\,
\sum_{i=1}^{L}
\left(
\widehat{Z}^{\mathrm{i}\hat{\xi}^{\,}_{i+1}\,\hat{\eta}^{\,}_{i+1}}_{i}\,
\widehat{Z}^{\dagger}_{i+1}
+
\widehat{Z}^{-\mathrm{i}\hat{\xi}^{\,}_{i+1}\,\hat{\eta}^{\,}_{i+1}}_{i}\,
\widehat{Z}^{\,}_{i+1}
\right)
\nonumber\\
&
-
J^{\,}_{4}\,
\sum_{i=1}^{L}
\left(
\mathrm{i}
\hat{\xi}^{\,}_{i}\,
\hat{\tau}^{x}_{i}\,
\widehat{C}^{\,}_{i}\,
\hat{\eta}^{\,}_{i+1}
+
\hat{\tau}^{x}_{i}
\right).
\end{align}
There are two degenerate ground states. 
First, there is a ground state obtained by setting $\widehat{Z}^{\,}_{i}=1$
for all sites that is given by
\begin{subequations}
\label{eq:GS 1 2-fold}
\begin{equation}
\label{eq:GS 1 Kitaev}
\Ket{\text{GS}^{\mathrm{K}}_{\one}} 
\= 
\bigotimes_{i=1}^{L} \Ket{\tau_i^x=1,
\mathrm{i}\xi^{\,}_{i}\,\eta^{\,}_{i+1}= 1, Z^{\,}_i=1}.
\end{equation}
The second ground state is~\footnote{
Much like the ground state \eqref{eq:gs3},
it is not obvious that $
\Ket{\text{GS}^{\mathrm{Triv}}_{\one}}$
is shot-range entangled. 
However, there exists a  finite depth local unitary circuit
that prepares this state from the product state 
$
\bigotimes_{i=1}^L 
\left [
\Ket{\tau^x_i = 1, \mathrm{i}\hat{\xi}^{\,}_{i}\,\hat{\eta}^{\,}_{i} = 1}
\otimes \frac{1}{\sqrt 2} \left (\Ket{ Z^{\,}_i = \om} + \Ket{ Z^{\,}_i = \om^*} \right )
\right ]$.
Namely, 
$\Ket{\text{GS}^{\mathrm{Triv}}_{\one}} 
= 
\prod_{j=1}^{L} 
\widehat{C}^{\mathrm{F}}_{j}\,
\bigotimes_{i=1}^L 
\left[
\Ket{\tau^x_i = 1, \mathrm{i}\hat{\xi}^{\,}_{i}\,\hat{\eta}^{\,}_{i+1} = 1}
\otimes \frac{1}{\sqrt 2} \left (\Ket{ Z^{\,}_i = \om} + \Ket{ Z^{\,}_i = \om^*} \right )
\right]$
where $\widehat{C}^{\mathrm{F}}_{j}$ is a kind of CZ operator that acts as the identity operator if 
$ Z^{\,}_j = \om $ and as $\mathrm{i}\hat{\xi}^{\,}_{j-1}\hat{\eta}^{\,}_{j}$ if $ Z^{\,}_j=\om^* $.
}
\begin{equation}
\label{eq:GS 1 Triv}
\Ket{\text{GS}^{\mathrm{Triv}}_{\one}} 
\= 
\frac{1}{2^{L/2}}
\sum_{\{s_i = \pm 1 \}} 
s^{\,}_{1}\,
\left(
\bigotimes_{i=1}^{L} 
\Ket{\tau_i^x=1, 
\mathrm{i}\xi^{\,}_{i}\,\eta^{\,}_{i}=  s^{\,}_{i}\,s^{\,}_{i-1},
Z^{\,}_i=\om^{s_i} }
\right).
\end{equation}
\end{subequations}
Fermionic degrees of freedom in the former ground state \eqref{eq:GS 1 Kitaev}
realize the Kitaev phase, while they are in the trivial phase for
the latter ground state \eqref{eq:GS 1 Triv}.
These two ground states transform under $\mathsf{SRep}(S^{\,}_{3})$
symmetry as
\begin{equation}
\begin{split}
&
\widehat{W}^{\,}_{p}\,
\Ket{\text{GS}^{\mathrm{K}}_{\one}} 
=
+
\Ket{\text{GS}^{\mathrm{K}}_{\one}},
\qquad
\widehat{W}^{\,}_{p}\,
\Ket{\text{GS}^{\mathrm{Triv}}_{\one}} 
=
+
\Ket{\text{GS}^{\mathrm{Triv}}_{\one}},
\\
&
\widehat{W}^{\,}_{\two}\,
\Ket{\text{GS}^{\mathrm{K}}_{\one}} 
=
2\,
\Ket{\text{GS}^{\mathrm{K}}_{\one}} ,
\qquad
\widehat{W}^{\,}_{\two}\,
\Ket{\text{GS}^{\mathrm{Triv}}_{\one}} 
=
-
\Ket{\text{GS}^{\mathrm{Triv}}_{\one}}.
\end{split}
\end{equation}
Since fermionic degrees of freedom realize trivial and non-trivial invertible 
fermionic states, we call this phase mixed  
$\mathsf{SRep}(S^{\,}_{3})/\mathbb{Z}^{\mathrm{F}}_{2}$ 
SSB phase.
\end{enumerate}

\begin{figure}[t!]
\centering
\begin{subfigure}{.45\textwidth}
\centering
\includegraphics[width=\linewidth]{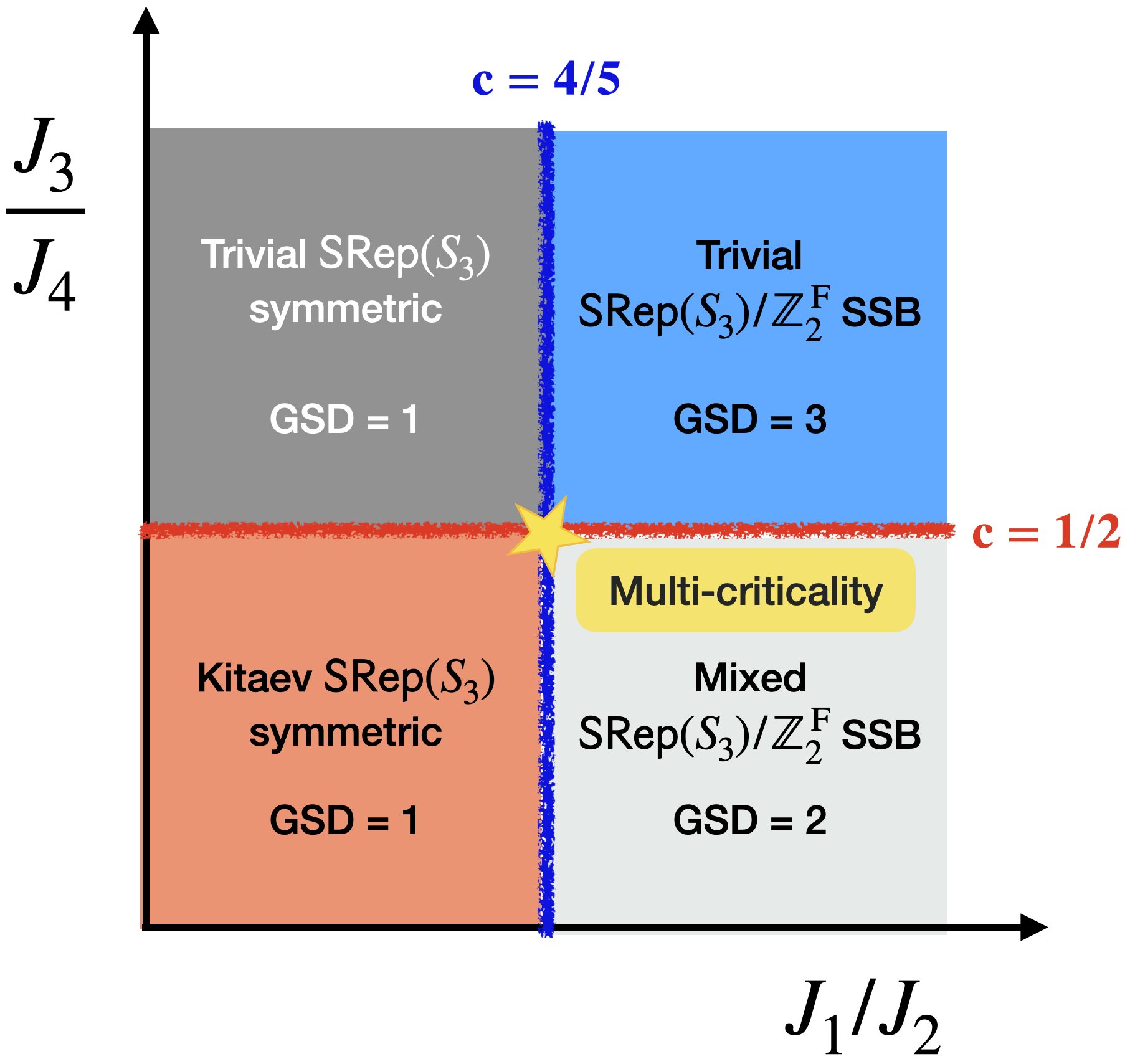}  
\vspace*{0.7mm} %added to balance the vertical alignment of (a) and (b)
\caption{}
\label{fig:sreps3_pd analytical}
\end{subfigure}
\begin{subfigure}{.45\textwidth}
\centering
\includegraphics[width=\linewidth]{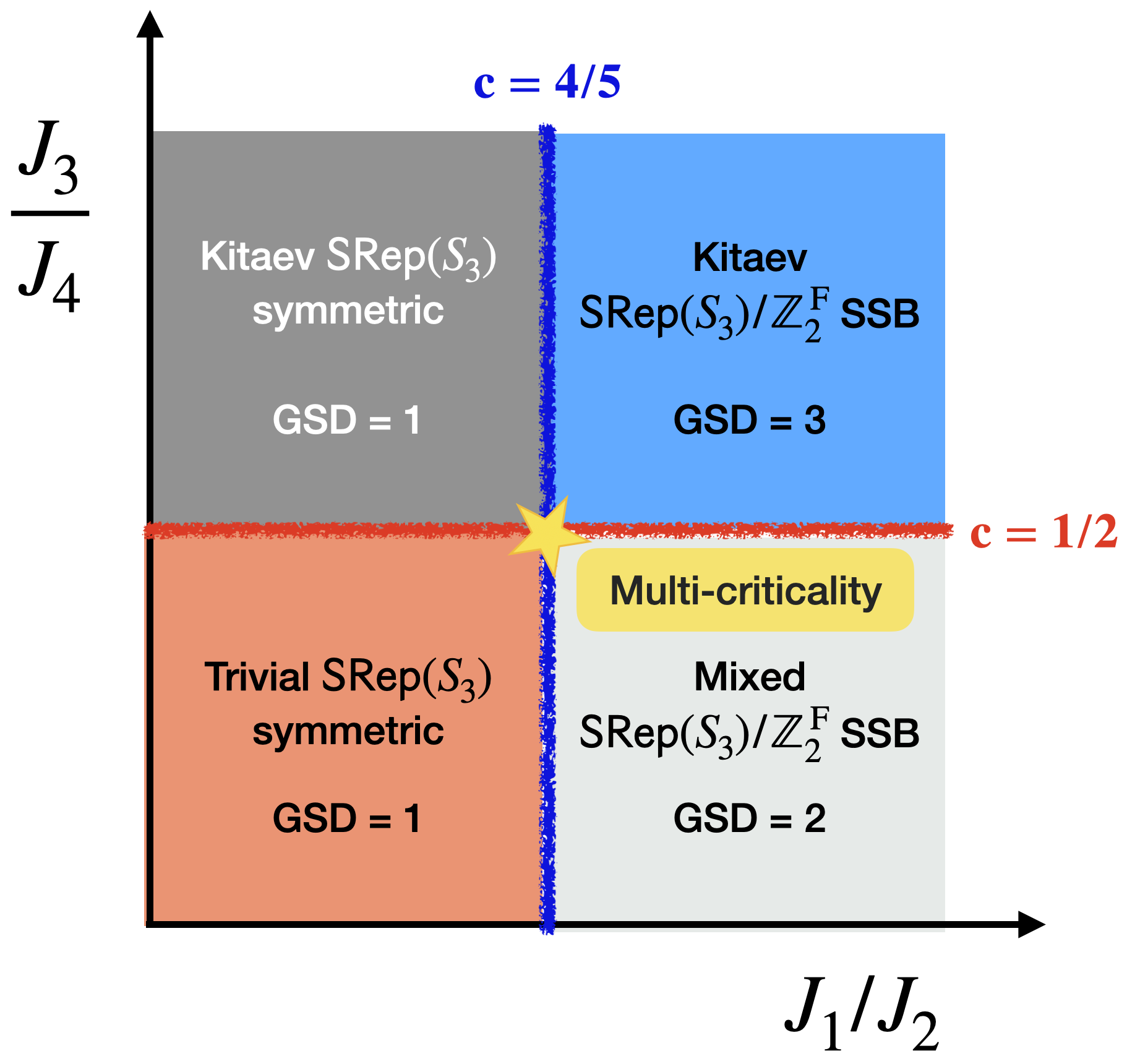}  
\vspace*{0.7mm} %added to balance the vertical alignment of (a) and (b)
\caption{}
\label{fig:sreps3_pd analytical half translation}
\end{subfigure}
\caption{(a) 
The phase diagram of Hamiltonian \eqref{eq:def Ham gen SRepS3}.
Since the fermion parity symmetry $\mathsf{SRep}(S^{\,}_{3})$ 
cannot be broken (explicitly or spontaneously), in each gapped phase 
$\mathsf{SRep}(S^{\,}_{3})/\mathbb{Z}^{\mathrm{F}}_{2}$ is broken if SSB takes place. 
There are three distinct SSB patterns that are distinguished by ground state degeneracies
and whether if the fermionic degrees of freedom are in trivial or Kitaev phase.
The corresponding fixed-point ground states are given in Eqs.\ 
\eqref{eq:GS srepS3 triv}, \eqref{eq:GS Wp}, \eqref{eq:GS Kitaev}, and
\eqref{eq:GS 1 2-fold}.
(b)
The phase diagram of Hamiltonian \eqref{eq:def Ham gen SRepS3 2}
that is equivalent to Hamiltonian \eqref{eq:def Ham gen SRepS3}
by half translation \eqref{eq:half-translation of Majoranas}.
Because of the unitary equivalence the phase diagram of the two Hamiltonians 
and corresponding SSB patterns are the same. Half-translation 
\eqref{eq:half-translation of Majoranas} corresponds to stacking each ground state with 
the Kitaev state, which results in an exchange of labels trivial and Kitaev.
\label{fig:SRepS3 pd}
}
\end{figure}

We identified four gapped fixed-points and constructed 
the corresponding ground states. We can deduce the shape of the phase diagram using 
the duality between Hamiltonians \eqref{eq:def Ham gen S3} and 
\eqref{eq:def Ham gen SRepS3}. 
In Fig.\ \ref{fig:sreps3_pd analytical}, 
we show the phase diagram of \eqref{eq:def Ham gen SRepS3} when $J^{\,}_{5}=J^{\,}_{6}=0$.
We deduce the continuous phase transitions also using the fact that the duality 
transformation does not change the central charge. We replace the Ising CFT with
Majorana CFT which are known to be dual to each other under JW transformation 
we implemented.

We note that since fermion parity symmetry $\mathbb{Z}^{\mathrm{F}}_{2}$ 
cannot be spontaneously broken, in all gapped phase the only possible symmetry 
that can be broken is $\mathsf{SRep}(S^{\,}_{3})/\mathbb{Z}^{\mathrm{F}}_{2}$.
We observe this in two of the four gapped phases. In each of the 
these two gapped phases, there is a distinct 
$\mathsf{SRep}(S^{\,}_{3})/\mathbb{Z}^{\mathrm{F}}_{2}$ 
SSB pattern showcasing a rich possibility of phase diagrams 
when non-invertible symmetries are spontaneously broken. 
We distinguish these symmetry breaking patterns by ground state degeneracy 
and whether the degenerate states realize trivial state or Kitaev state 
(see Fig.\ \ref{fig:sreps3_pd analytical}).

\subsection{Alternative Jordan-Wigner duality}

There is a second way to gauge the $\mathbb{Z}^{\,}_{2}$ subgroup of 
$S^{\,}_{3}$ symmetry using fermionic gauge fields, which also delivers 
an $\mathsf{SRep}(S^{\,}_{3})$. This alternative way differs from the discussion in Sec.\
\ref{sec:constructing SRepS3} by the choice of Gauss operator. 
Namely, we can define 
\begin{align}
\widehat{G}^{\mathrm{F}'}_{i}
\=
\hat{\sigma}^{x}_{i}\,
\hat{\tau}^{x}_{i}\,
\widehat{C}^{\,}_{i}\,
\mathrm{i}
\hat{\eta}^{\,}_{i+1/2}\,
\hat{\xi}^{\,}_{i+1/2},
\qquad
\left[
\widehat{G}^{\mathrm{F}'}_{i}
\right]^{2}
=
\hat{\mathbbm{1}},
\end{align}
which, as opposed to the Gauss operator \eqref{eq:Gaus operators fermions}
acts on only a single link. The two ways of gauging are related 
by a half-translation on Majorana operators
\begin{align}
\left(
\hat{\xi}^{\,}_{i+1/2},\,\,
\hat{\eta}^{\,}_{i+1/2}
\right)
\mapsto
\left(
\hat{\eta}^{\,}_{i+3/2},\,\,
\hat{\xi}^{\,}_{i+1/2}
\right)
\label{eq:half-translation of Majoranas}
\end{align}
which can also be interpreted as ``stacking'' 
the resulting fermionic theory with Kitaev
chain~\cite{I220613159,AMFT230800743,SS230702534}.
Alternatively, this half-translation corresponds to first gauging $\mathbb{Z}^{\,}_{2}$
subgroup using bosonic gauge fields, as we have done in Sec.\ \ref{sec:RepS3model},
and then gauging the $\mathbb{Z}^{\,}_{2}$ subgroup of resulting $\cRep(S^{\,}_{3})$
symmetry. 

Under this second way of implementing JW duality, 
the $\cRep(S^{\,}_{3})$ symmetric bond algebra is 
\small
\begin{align}
\label{eq:def SRepS3 algebra 2}
\mathfrak{B}^{'}_{\mathrm{F}}
\=
&\Big\langle
\hat{\tau}^{z}_{i},\,
\hat{\tau}^{z}_{i}\,
\mathrm{i}
\hat{\xi}^{\,}_{i+1}
\hat{\eta}^{\,}_{i+2},\,
\hat{\tau}^{x}_{i}\,
\widehat{C}^{\,}_{i}\,
\mathrm{i}
\hat{\xi}^{\,}_{i+1}\,
\hat{\eta}^{\,}_{i+1},\,
\hat{\tau}^{x}_{i},\,
\left(\widehat{X}^{\,}_{i}+\widehat{X}^{\dagger}_{i}\right),\,
\left(\widehat{Z}^{
-\mathrm{i}\hat{\xi}^{\,}_{i+1}\,\hat{\eta}^{\,}_{i+2}}_{i}\,
\widehat{Z}^{\dagger}_{i+1}
+\widehat{Z}^{+\mathrm{i}\hat{\xi}^{\,}_{i+1}\,
\hat{\eta}^{\,}_{i+2}}_{i}\,
\widehat{Z}^{\,}_{i+1}\right),
\nonumber\\
&
\left(\widehat{X}^{\,}_{i}-\widehat{X}^{\dagger}_{i}\right),\,
\hat{\tau}^{z}_{i}\,
\mathrm{i}\hat{\xi}^{\,}_{i+1}\,
\hat{\eta}^{\,}_{i+2}
\left(
\widehat{Z}^{
-\mathrm{i}\hat{\xi}^{\,}_{i+1}\,\hat{\eta}^{\,}_{i+2}}_{i}\,
\widehat{Z}^{\dagger}_{i+1}
-\widehat{Z}^{
	+\mathrm{i}\hat{\xi}^{\,}_{i+1}\,\hat{\eta}^{\,}_{i+2}}_{i}\,
\widehat{Z}^{\,}_{i+1}\right)
\Big\vert 
\,\,
i\in \Lambda
\Big\rangle.
\end{align}
\normalsize
Since half-translation operator anticommutes with the fermion parity,
we find the symmetries of this bond algebra to be generated by
\begin{subequations}
\label{eq:SRepS3 generators 2}
\begin{equation}
\begin{split}
\widehat{W}^{'}_{\bm{1}}
&\=
\hat{\mathbbm{1}},
\\
\widehat{W}^{'}_{p}
&\=
\prod_{i}^{L}
\mathrm{i}\hat{\xi}^{\,}_{i}\,
\hat{\eta}^{\,}_{i}
\equiv
\prod_{i}^{L}
\hat{p}^{\,}_{i},
\\
\widehat{W}^{'}_{\bm{2}}
&\=
\frac{1}{2}
\left(
1
+
\prod_{i=1}^{L}
\mathrm{i}\hat{\xi}^{\,}_{i}\,
\hat{\eta}^{\,}_{i}
\right)
\left[
\prod_{i=1}^{L}
\widehat{X}^{\prod_{k=2}^{i}\mathrm{i}\hat{\eta}^{\,}_{k+1}\,\hat{\xi}^{\,}_{k}}_{i}
+
\widehat{X}^{-\prod_{k=2}^{i}\mathrm{i}\hat{\eta}^{\,}_{k+1}\,\hat{\xi}^{\,}_{k}}_{i}
\right].
\end{split}
\end{equation} 
which satisfy the fusion rules
\begin{align}
\widehat{W}^{'}_{p}\,
\widehat{W}^{'}_{p}
=
\widehat{W}^{'}_{\bm{1}},
\qquad
\widehat{W}^{'}_{p}\,
\widehat{W}^{'}_{\bm{2}}
=
\widehat{W}^{'}_{\bm{2}},
\qquad
\widehat{W}^{'}_{\bm{2}}\,
\widehat{W}^{'}_{\bm{2}}
=
\widehat{W}^{'}_{\bm{1}}
+
\widehat{W}^{'}_{p}
+
\widehat{W}^{'}_{\bm{2}}.
\end{align}
\end{subequations}
The duality between bond algebras \eqref{eq:def S3 algebra}
and \eqref{eq:def SRepS3 algebra 2} holds in the subalgebras
\begin{align}
\mathfrak{B}^{\,}_{S^{\,}_{3}}\Big\vert^{\,}_{\widehat{U}^{\,}_{s}=+1}
\cong 
\mathfrak{B}^{'}_{\mathrm{F}}\Big\vert^{\,}_{\widehat{W}^{\,}_{p}=+1}.
\end{align}
Under the half-translation \eqref{eq:half-translation of Majoranas},
the Hamiltonian \eqref{eq:def Ham gen SRepS3} becomes
\begin{equation}
\label{eq:def Ham gen SRepS3 2}
\begin{split}
\widehat{H}^{'}_{\mathsf{SRep}(S^{\,}_{3})}
\=
&-
J^{\,}_{1}\,
\sum_{i=1}^{L}
\left(
\widehat{Z}^{-\mathrm{i}\hat{\xi}^{\,}_{i+1}\,\hat{\eta}^{\,}_{i+2}}_{i}\,
\widehat{Z}^{\dagger}_{i+1}
+
\widehat{Z}^{+\mathrm{i}\hat{\xi}^{\,}_{i+1}\,\hat{\eta}^{\,}_{i+2}}_{i}\,
\widehat{Z}^{\,}_{i+1}
\right)
-
J^{\,}_{2}\,
\sum_{i=1}^{L}
\left(
\widehat{X}^{\,}_{i}
+
\widehat{X}^{\dagger}_{i}
\right)
\\
&
-
J^{\,}_{3}\,
\sum_{i=1}^{L}
\left(
\hat{\tau}^{z}_{i}
+
\hat{\tau}^{z}_{i}\,
\mathrm{i}\hat{\eta}^{\,}_{i+1}\,\hat{\xi}^{\,}_{i+2}
\right)
-
J^{\,}_{4}\,
\sum_{i=1}^{L}
\left(
\hat{\tau}^{x}_{i}\,
\widehat{C}^{\,}_{i}\,
\mathrm{i}
\hat{\eta}^{\,}_{i+1}\,
\hat{\xi}^{\,}_{i+1}
+
\hat{\tau}^{x}_{i}
\right)
\\
&
+
J^{\,}_{5}\,
\sum_{i=1}^{L}
\mathrm{i}\,
\hat{\tau}^{z}_{i}\,
\mathrm{i}\hat{\xi}^{\,}_{i+1}\,\hat{\eta}^{\,}_{i+2}\,
\left(
\widehat{Z}^{-\mathrm{i}\hat{\xi}^{\,}_{i+1}\,\hat{\eta}^{\,}_{i+2}}_{i}\,
\widehat{Z}^{\dagger}_{i+1}
-\widehat{Z}^{+\mathrm{i}\hat{\xi}^{\,}_{i+1}\,\hat{\eta}^{\,}_{i+2}}_{i}\,
\widehat{Z}^{\,}_{i+1}
\right)
\\
&
-
J^{\,}_{6}\,
\sum_{i=1}^{L}
\mathrm{i}\,
\left(
\widehat{X}^{\,}_{i}
-
\widehat{X}^{\dagger}_{i}
\right).
\end{split}
\end{equation}
Being unitarily equivalent to Hamiltonian \eqref{eq:def Ham gen SRepS3}, this
Hamiltonian shares the same phase diagram, which is shown in Fig.\
\ref{fig:sreps3_pd analytical half translation} when $J^{\,}_{5}=J^{\,}_{6}=0$.
The only difference between the two Hamiltonians is that the labels trivial and
Kitaev that denote the fermionic sector of the ground states are exchanged.
This is expected since half-translation \eqref{eq:half-translation of
Majoranas} effectively stacks a Kitaev chain on top of each ground state. Since
this is a unitary transformation the SSB patterns do not change.  The two
Hamiltonians are no longer unitarily equivalent when open boundary conditions
are imposed, since the unitary equivalence under  half-translation
\eqref{eq:half-translation of Majoranas} relies on the translation invariance
which is broken by open boundary conditions. This inequivalence is reflected by
the differing ground state degeneracies of the two Hamiltonians with open
boundary conditions.

\section{The gapless states for the self-dual $S_3$-symmetric model}

To obtain the potential gapless states for the self-dual $S_3$-symmetric model,
we ask, for example, do we have gapless states described by $m4\times
\overline{m4}$ minimal model CFT?  To answer this question, we use the holographic modular bootstrap \cite{CZ190312334,JW191213492,CW220506244}, \ie try to use the
conformal characters of $m4\times \overline{m4}$ model CFT to construct the
25-component partition functions that transform covariantly under the modular
transformation generated by the $S,T$-matrices of the $\mathrm{JK}_4\boxtimes
\overline{\mathrm{JK}}_4$ SymTO.
We obtain the following solution

\begin{align}
\label{Zm4}
 Z_{\mathbf{1}; \mathbf{1}}^{JK_4 \boxtimes \overline{JK}_4} &={\color{blue} \chi^{m4 \times \overline{m4}}_{1,0; 1,0}} +  {\color{red} \chi^{m4 \times \overline{m4}}_{3,\frac{1}{2}; 3,-\frac{1}{2}} }
 \nonumber \\ 
Z_{\mathbf{1}; \bar e}^{JK_4 \boxtimes \overline{JK}_4} &= \chi^{m4 \times \overline{m4}}_{2,\frac{1}{16}; 2,-\frac{1}{16}} 
 \nonumber \\ 
Z_{e; \mathbf{1}}^{JK_4 \boxtimes \overline{JK}_4} &= \chi^{m4 \times \overline{m4}}_{2,\frac{1}{16}; 2,-\frac{1}{16}} 
 \nonumber \\ 
Z_{e; \bar e}^{JK_4 \boxtimes \overline{JK}_4} &={\color{blue} \chi^{m4 \times \overline{m4}}_{1,0; 1,0}} +  \chi^{m4 \times \overline{m4}}_{3,\frac{1}{2}; 3,-\frac{1}{2}} 
 \nonumber \\ 
Z_{m; \bar m}^{JK_4 \boxtimes \overline{JK}_4} &= \chi^{m4 \times \overline{m4}}_{2,\frac{1}{16}; 2,-\frac{1}{16}} 
 \nonumber \\ 
Z_{m; \bar m_1}^{JK_4 \boxtimes \overline{JK}_4} &= \chi^{m4 \times \overline{m4}}_{1,0; 3,-\frac{1}{2}} +  \chi^{m4 \times \overline{m4}}_{3,\frac{1}{2}; 1,0} 
 \nonumber \\ 
Z_{m_1; \bar m}^{JK_4 \boxtimes \overline{JK}_4} &= \chi^{m4 \times \overline{m4}}_{1,0; 3,-\frac{1}{2}} +  \chi^{m4 \times \overline{m4}}_{3,\frac{1}{2}; 1,0} 
 \nonumber \\ 
Z_{m_1; \bar m_1}^{JK_4 \boxtimes \overline{JK}_4} &= \chi^{m4 \times \overline{m4}}_{2,\frac{1}{16}; 2,-\frac{1}{16}} 
 \nonumber \\ 
Z_{q; \bar q}^{JK_4 \boxtimes \overline{JK}_4} &={\color{blue} \chi^{m4 \times \overline{m4}}_{1,0; 1,0}} +  \chi^{m4 \times \overline{m4}}_{2,\frac{1}{16}; 2,-\frac{1}{16}} +  \chi^{m4 \times \overline{m4}}_{3,\frac{1}{2}; 3,-\frac{1}{2}} 
\nonumber\\
\text{others} &= 0.
 \end{align}
From $ Z_{\mathbf{1}; \mathbf{1}}^{JK_4 \boxtimes \overline{JK}_4} $, we see that
there is one symmetric relevant perturbation of scaling dimension $(h,\bar
h)=(\frac12,\frac12)$.   From the
position of $\chi^{m4 \times \overline{m4}}_{1,0; 1,0}$, we see that the above
gapless state is induced by condensation $\cA_1=\one \otimes e\bar e\otimes
q\bar q$.

In the following, we list the gapless phases (\ie the gapless states with no
symmetric relevant operators) of self-dual $S_3$ symmetric model that we find
using the above method. The self-dual $\cRep(S_3)$ symmetric model have the same
gapless phases.  The SymTO $JK_4 \boxtimes \overline{JK}_4$ may be spontaneously broken
in those gapless phases. We also list the correspond condensation $\cA$
that reduces the SymTO $JK_4 \boxtimes \overline{JK}_4$ to a smaller one
$\eM_{/\cA}$.

\begingroup
\allowdisplaybreaks

\begin{align}
\label{m6-52}
 Z_{\mathbf{1}; \mathbf{1}}^{JK_4 \boxtimes \overline{JK}_4} &={\color{blue} \chi^{m6 \times \overline{m6}}_{1,0; 1,0}} +  \chi^{m6 \times \overline{m6}}_{10,\frac{7}{5}; 10,-\frac{7}{5}} 
 \nonumber \\ 
Z_{\mathbf{1}; \bar e}^{JK_4 \boxtimes \overline{JK}_4} &= \chi^{m6 \times \overline{m6}}_{1,0; 5,-3} +  \chi^{m6 \times \overline{m6}}_{10,\frac{7}{5}; 6,-\frac{2}{5}} 
 \nonumber \\ 
Z_{\mathbf{1}; \bar m}^{JK_4 \boxtimes \overline{JK}_4} &= \chi^{m6 \times \overline{m6}}_{5,3; 2,-\frac{1}{8}} +  \chi^{m6 \times \overline{m6}}_{6,\frac{2}{5}; 9,-\frac{21}{40}} 
 \nonumber \\ 
Z_{\mathbf{1}; \bar m_1}^{JK_4 \boxtimes \overline{JK}_4} &= \chi^{m6 \times \overline{m6}}_{5,3; 4,-\frac{13}{8}} +  \chi^{m6 \times \overline{m6}}_{6,\frac{2}{5}; 7,-\frac{1}{40}} 
 \nonumber \\ 
Z_{\mathbf{1}; \bar q}^{JK_4 \boxtimes \overline{JK}_4} &= \chi^{m6 \times \overline{m6}}_{1,0; 3,-\frac{2}{3}} +  \chi^{m6 \times \overline{m6}}_{10,\frac{7}{5}; 8,-\frac{1}{15}} 
 \nonumber \\ 
Z_{e; \mathbf{1}}^{JK_4 \boxtimes \overline{JK}_4} &= \chi^{m6 \times \overline{m6}}_{5,3; 1,0} +  \chi^{m6 \times \overline{m6}}_{6,\frac{2}{5}; 10,-\frac{7}{5}} 
 \nonumber \\ 
Z_{e; \bar e}^{JK_4 \boxtimes \overline{JK}_4} &= \chi^{m6 \times \overline{m6}}_{5,3; 5,-3} +  \chi^{m6 \times \overline{m6}}_{6,\frac{2}{5}; 6,-\frac{2}{5}} 
 \nonumber \\ 
Z_{e; \bar m}^{JK_4 \boxtimes \overline{JK}_4} &= \chi^{m6 \times \overline{m6}}_{1,0; 2,-\frac{1}{8}} +  \chi^{m6 \times \overline{m6}}_{10,\frac{7}{5}; 9,-\frac{21}{40}} 
 \nonumber \\ 
Z_{e; \bar m_1}^{JK_4 \boxtimes \overline{JK}_4} &= \chi^{m6 \times \overline{m6}}_{1,0; 4,-\frac{13}{8}} +  \chi^{m6 \times \overline{m6}}_{10,\frac{7}{5}; 7,-\frac{1}{40}} 
 \nonumber \\ 
Z_{e; \bar q}^{JK_4 \boxtimes \overline{JK}_4} &= \chi^{m6 \times \overline{m6}}_{5,3; 3,-\frac{2}{3}} +  \chi^{m6 \times \overline{m6}}_{6,\frac{2}{5}; 8,-\frac{1}{15}} 
 \nonumber \\ 
Z_{m; \mathbf{1}}^{JK_4 \boxtimes \overline{JK}_4} &= \chi^{m6 \times \overline{m6}}_{2,\frac{1}{8}; 5,-3} +  \chi^{m6 \times \overline{m6}}_{9,\frac{21}{40}; 6,-\frac{2}{5}} 
 \nonumber \\ 
Z_{m; \bar e}^{JK_4 \boxtimes \overline{JK}_4} &= \chi^{m6 \times \overline{m6}}_{2,\frac{1}{8}; 1,0} +  \chi^{m6 \times \overline{m6}}_{9,\frac{21}{40}; 10,-\frac{7}{5}} 
 \nonumber \\ 
Z_{m; \bar m}^{JK_4 \boxtimes \overline{JK}_4} &= \chi^{m6 \times \overline{m6}}_{4,\frac{13}{8}; 4,-\frac{13}{8}} +  \chi^{m6 \times \overline{m6}}_{7,\frac{1}{40}; 7,-\frac{1}{40}} 
 \nonumber \\ 
Z_{m; \bar m_1}^{JK_4 \boxtimes \overline{JK}_4} &= \chi^{m6 \times \overline{m6}}_{4,\frac{13}{8}; 2,-\frac{1}{8}} +  \chi^{m6 \times \overline{m6}}_{7,\frac{1}{40}; 9,-\frac{21}{40}} 
 \nonumber \\ 
Z_{m; \bar q}^{JK_4 \boxtimes \overline{JK}_4} &= \chi^{m6 \times \overline{m6}}_{2,\frac{1}{8}; 3,-\frac{2}{3}} +  \chi^{m6 \times \overline{m6}}_{9,\frac{21}{40}; 8,-\frac{1}{15}} 
 \nonumber \\ 
Z_{m_1; \mathbf{1}}^{JK_4 \boxtimes \overline{JK}_4} &= \chi^{m6 \times \overline{m6}}_{4,\frac{13}{8}; 5,-3} +  \chi^{m6 \times \overline{m6}}_{7,\frac{1}{40}; 6,-\frac{2}{5}} 
 \nonumber \\ 
Z_{m_1; \bar e}^{JK_4 \boxtimes \overline{JK}_4} &= \chi^{m6 \times \overline{m6}}_{4,\frac{13}{8}; 1,0} +  \chi^{m6 \times \overline{m6}}_{7,\frac{1}{40}; 10,-\frac{7}{5}} 
 \nonumber \\ 
Z_{m_1; \bar m}^{JK_4 \boxtimes \overline{JK}_4} &= \chi^{m6 \times \overline{m6}}_{2,\frac{1}{8}; 4,-\frac{13}{8}} +  \chi^{m6 \times \overline{m6}}_{9,\frac{21}{40}; 7,-\frac{1}{40}} 
 \nonumber \\ 
Z_{m_1; \bar m_1}^{JK_4 \boxtimes \overline{JK}_4} &= \chi^{m6 \times \overline{m6}}_{2,\frac{1}{8}; 2,-\frac{1}{8}} +  \chi^{m6 \times \overline{m6}}_{9,\frac{21}{40}; 9,-\frac{21}{40}} 
 \nonumber \\ 
Z_{m_1; \bar q}^{JK_4 \boxtimes \overline{JK}_4} &= \chi^{m6 \times \overline{m6}}_{4,\frac{13}{8}; 3,-\frac{2}{3}} +  \chi^{m6 \times \overline{m6}}_{7,\frac{1}{40}; 8,-\frac{1}{15}} 
 \nonumber \\ 
Z_{q; \mathbf{1}}^{JK_4 \boxtimes \overline{JK}_4} &= \chi^{m6 \times \overline{m6}}_{3,\frac{2}{3}; 1,0} +  \chi^{m6 \times \overline{m6}}_{8,\frac{1}{15}; 10,-\frac{7}{5}} 
 \nonumber \\ 
Z_{q; \bar e}^{JK_4 \boxtimes \overline{JK}_4} &= \chi^{m6 \times \overline{m6}}_{3,\frac{2}{3}; 5,-3} +  \chi^{m6 \times \overline{m6}}_{8,\frac{1}{15}; 6,-\frac{2}{5}} 
 \nonumber \\ 
Z_{q; \bar m}^{JK_4 \boxtimes \overline{JK}_4} &= \chi^{m6 \times \overline{m6}}_{3,\frac{2}{3}; 2,-\frac{1}{8}} +  \chi^{m6 \times \overline{m6}}_{8,\frac{1}{15}; 9,-\frac{21}{40}} 
 \nonumber \\ 
Z_{q; \bar m_1}^{JK_4 \boxtimes \overline{JK}_4} &= \chi^{m6 \times \overline{m6}}_{3,\frac{2}{3}; 4,-\frac{13}{8}} +  \chi^{m6 \times \overline{m6}}_{8,\frac{1}{15}; 7,-\frac{1}{40}} 
 \nonumber \\ 
Z_{q; \bar q}^{JK_4 \boxtimes \overline{JK}_4} &= \chi^{m6 \times \overline{m6}}_{3,\frac{2}{3}; 3,-\frac{2}{3}} +  \chi^{m6 \times \overline{m6}}_{8,\frac{1}{15}; 8,-\frac{1}{15}} 
 \nonumber \\ 
\cA & = \one .
\end{align}

\begin{align}
\label{m6-12}
 Z_{\mathbf{1}; \mathbf{1}}^{JK_4 \boxtimes \overline{JK}_4} &={\color{blue} \chi^{m6 \times \overline{m6}}_{1,0; 1,0}} +  \chi^{m6 \times \overline{m6}}_{10,\frac{7}{5}; 10,-\frac{7}{5}} 
 \nonumber \\ 
Z_{\mathbf{1}; \bar e}^{JK_4 \boxtimes \overline{JK}_4} &= \chi^{m6 \times \overline{m6}}_{1,0; 5,-3} +  \chi^{m6 \times \overline{m6}}_{10,\frac{7}{5}; 6,-\frac{2}{5}} 
 \nonumber \\ 
Z_{\mathbf{1}; \bar m}^{JK_4 \boxtimes \overline{JK}_4} &= \chi^{m6 \times \overline{m6}}_{1,0; 2,-\frac{1}{8}} +  \chi^{m6 \times \overline{m6}}_{10,\frac{7}{5}; 9,-\frac{21}{40}} 
 \nonumber \\ 
Z_{\mathbf{1}; \bar m_1}^{JK_4 \boxtimes \overline{JK}_4} &= \chi^{m6 \times \overline{m6}}_{1,0; 4,-\frac{13}{8}} +  \chi^{m6 \times \overline{m6}}_{10,\frac{7}{5}; 7,-\frac{1}{40}} 
 \nonumber \\ 
Z_{\mathbf{1}; \bar q}^{JK_4 \boxtimes \overline{JK}_4} &= \chi^{m6 \times \overline{m6}}_{1,0; 3,-\frac{2}{3}} +  \chi^{m6 \times \overline{m6}}_{10,\frac{7}{5}; 8,-\frac{1}{15}} 
 \nonumber \\ 
Z_{e; \mathbf{1}}^{JK_4 \boxtimes \overline{JK}_4} &= \chi^{m6 \times \overline{m6}}_{5,3; 1,0} +  \chi^{m6 \times \overline{m6}}_{6,\frac{2}{5}; 10,-\frac{7}{5}} 
 \nonumber \\ 
Z_{e; \bar e}^{JK_4 \boxtimes \overline{JK}_4} &= \chi^{m6 \times \overline{m6}}_{5,3; 5,-3} +  \chi^{m6 \times \overline{m6}}_{6,\frac{2}{5}; 6,-\frac{2}{5}} 
 \nonumber \\ 
Z_{e; \bar m}^{JK_4 \boxtimes \overline{JK}_4} &= \chi^{m6 \times \overline{m6}}_{5,3; 2,-\frac{1}{8}} +  \chi^{m6 \times \overline{m6}}_{6,\frac{2}{5}; 9,-\frac{21}{40}} 
 \nonumber \\ 
Z_{e; \bar m_1}^{JK_4 \boxtimes \overline{JK}_4} &= \chi^{m6 \times \overline{m6}}_{5,3; 4,-\frac{13}{8}} +  \chi^{m6 \times \overline{m6}}_{6,\frac{2}{5}; 7,-\frac{1}{40}} 
 \nonumber \\ 
Z_{e; \bar q}^{JK_4 \boxtimes \overline{JK}_4} &= \chi^{m6 \times \overline{m6}}_{5,3; 3,-\frac{2}{3}} +  \chi^{m6 \times \overline{m6}}_{6,\frac{2}{5}; 8,-\frac{1}{15}} 
 \nonumber \\ 
Z_{m; \mathbf{1}}^{JK_4 \boxtimes \overline{JK}_4} &= \chi^{m6 \times \overline{m6}}_{2,\frac{1}{8}; 1,0} +  \chi^{m6 \times \overline{m6}}_{9,\frac{21}{40}; 10,-\frac{7}{5}} 
 \nonumber \\ 
Z_{m; \bar e}^{JK_4 \boxtimes \overline{JK}_4} &= \chi^{m6 \times \overline{m6}}_{2,\frac{1}{8}; 5,-3} +  \chi^{m6 \times \overline{m6}}_{9,\frac{21}{40}; 6,-\frac{2}{5}} 
 \nonumber \\ 
Z_{m; \bar m}^{JK_4 \boxtimes \overline{JK}_4} &= \chi^{m6 \times \overline{m6}}_{2,\frac{1}{8}; 2,-\frac{1}{8}} +  \chi^{m6 \times \overline{m6}}_{9,\frac{21}{40}; 9,-\frac{21}{40}} 
 \nonumber \\ 
Z_{m; \bar m_1}^{JK_4 \boxtimes \overline{JK}_4} &= \chi^{m6 \times \overline{m6}}_{2,\frac{1}{8}; 4,-\frac{13}{8}} +  \chi^{m6 \times \overline{m6}}_{9,\frac{21}{40}; 7,-\frac{1}{40}} 
 \nonumber \\ 
Z_{m; \bar q}^{JK_4 \boxtimes \overline{JK}_4} &= \chi^{m6 \times \overline{m6}}_{2,\frac{1}{8}; 3,-\frac{2}{3}} +  \chi^{m6 \times \overline{m6}}_{9,\frac{21}{40}; 8,-\frac{1}{15}} 
 \nonumber \\ 
Z_{m_1; \mathbf{1}}^{JK_4 \boxtimes \overline{JK}_4} &= \chi^{m6 \times \overline{m6}}_{4,\frac{13}{8}; 1,0} +  \chi^{m6 \times \overline{m6}}_{7,\frac{1}{40}; 10,-\frac{7}{5}} 
 \nonumber \\ 
Z_{m_1; \bar e}^{JK_4 \boxtimes \overline{JK}_4} &= \chi^{m6 \times \overline{m6}}_{4,\frac{13}{8}; 5,-3} +  \chi^{m6 \times \overline{m6}}_{7,\frac{1}{40}; 6,-\frac{2}{5}} 
 \nonumber \\ 
Z_{m_1; \bar m}^{JK_4 \boxtimes \overline{JK}_4} &= \chi^{m6 \times \overline{m6}}_{4,\frac{13}{8}; 2,-\frac{1}{8}} +  \chi^{m6 \times \overline{m6}}_{7,\frac{1}{40}; 9,-\frac{21}{40}} 
 \nonumber \\ 
Z_{m_1; \bar m_1}^{JK_4 \boxtimes \overline{JK}_4} &= \chi^{m6 \times \overline{m6}}_{4,\frac{13}{8}; 4,-\frac{13}{8}} +  \chi^{m6 \times \overline{m6}}_{7,\frac{1}{40}; 7,-\frac{1}{40}} 
 \nonumber \\ 
Z_{m_1; \bar q}^{JK_4 \boxtimes \overline{JK}_4} &= \chi^{m6 \times \overline{m6}}_{4,\frac{13}{8}; 3,-\frac{2}{3}} +  \chi^{m6 \times \overline{m6}}_{7,\frac{1}{40}; 8,-\frac{1}{15}} 
 \nonumber \\ 
Z_{q; \mathbf{1}}^{JK_4 \boxtimes \overline{JK}_4} &= \chi^{m6 \times \overline{m6}}_{3,\frac{2}{3}; 1,0} +  \chi^{m6 \times \overline{m6}}_{8,\frac{1}{15}; 10,-\frac{7}{5}} 
 \nonumber \\ 
Z_{q; \bar e}^{JK_4 \boxtimes \overline{JK}_4} &= \chi^{m6 \times \overline{m6}}_{3,\frac{2}{3}; 5,-3} +  \chi^{m6 \times \overline{m6}}_{8,\frac{1}{15}; 6,-\frac{2}{5}} 
 \nonumber \\ 
Z_{q; \bar m}^{JK_4 \boxtimes \overline{JK}_4} &= \chi^{m6 \times \overline{m6}}_{3,\frac{2}{3}; 2,-\frac{1}{8}} +  \chi^{m6 \times \overline{m6}}_{8,\frac{1}{15}; 9,-\frac{21}{40}} 
 \nonumber \\ 
Z_{q; \bar m_1}^{JK_4 \boxtimes \overline{JK}_4} &= \chi^{m6 \times \overline{m6}}_{3,\frac{2}{3}; 4,-\frac{13}{8}} +  \chi^{m6 \times \overline{m6}}_{8,\frac{1}{15}; 7,-\frac{1}{40}} 
 \nonumber \\ 
Z_{q; \bar q}^{JK_4 \boxtimes \overline{JK}_4} &= \chi^{m6 \times \overline{m6}}_{3,\frac{2}{3}; 3,-\frac{2}{3}} +  \chi^{m6 \times \overline{m6}}_{8,\frac{1}{15}; 8,-\frac{1}{15}} 
 \nonumber \\ 
\cA &= \one .
\end{align}

\begin{align}
\label{m6-1e}
 Z_{\mathbf{1}; \mathbf{1}}^{JK_4 \boxtimes \overline{JK}_4} &={\color{blue} \chi^{m6 \times \overline{m6}}_{1,0; 1,0}} +  \chi^{m6 \times \overline{m6}}_{5,3; 1,0} +  \chi^{m6 \times \overline{m6}}_{6,\frac{2}{5}; 10,-\frac{7}{5}} +  \chi^{m6 \times \overline{m6}}_{10,\frac{7}{5}; 10,-\frac{7}{5}} 
 \nonumber \\ 
Z_{\mathbf{1}; \bar e}^{JK_4 \boxtimes \overline{JK}_4} &= \chi^{m6 \times \overline{m6}}_{1,0; 5,-3} +  \chi^{m6 \times \overline{m6}}_{5,3; 5,-3} +  \chi^{m6 \times \overline{m6}}_{6,\frac{2}{5}; 6,-\frac{2}{5}} +  \chi^{m6 \times \overline{m6}}_{10,\frac{7}{5}; 6,-\frac{2}{5}} 
 \nonumber \\ 
Z_{\mathbf{1}; \bar m}^{JK_4 \boxtimes \overline{JK}_4} &= \chi^{m6 \times \overline{m6}}_{1,0; 2,-\frac{1}{8}} +  \chi^{m6 \times \overline{m6}}_{5,3; 2,-\frac{1}{8}} +  \chi^{m6 \times \overline{m6}}_{6,\frac{2}{5}; 9,-\frac{21}{40}} +  \chi^{m6 \times \overline{m6}}_{10,\frac{7}{5}; 9,-\frac{21}{40}} 
 \nonumber \\ 
Z_{\mathbf{1}; \bar m_1}^{JK_4 \boxtimes \overline{JK}_4} &= \chi^{m6 \times \overline{m6}}_{1,0; 4,-\frac{13}{8}} +  \chi^{m6 \times \overline{m6}}_{5,3; 4,-\frac{13}{8}} +  \chi^{m6 \times \overline{m6}}_{6,\frac{2}{5}; 7,-\frac{1}{40}} +  \chi^{m6 \times \overline{m6}}_{10,\frac{7}{5}; 7,-\frac{1}{40}} 
 \nonumber \\ 
Z_{\mathbf{1}; \bar q}^{JK_4 \boxtimes \overline{JK}_4} &= \chi^{m6 \times \overline{m6}}_{1,0; 3,-\frac{2}{3}} +  \chi^{m6 \times \overline{m6}}_{5,3; 3,-\frac{2}{3}} +  \chi^{m6 \times \overline{m6}}_{6,\frac{2}{5}; 8,-\frac{1}{15}} +  \chi^{m6 \times \overline{m6}}_{10,\frac{7}{5}; 8,-\frac{1}{15}} 
 \nonumber \\ 
Z_{e; \mathbf{1}}^{JK_4 \boxtimes \overline{JK}_4} &={\color{blue} \chi^{m6 \times \overline{m6}}_{1,0; 1,0}} +  \chi^{m6 \times \overline{m6}}_{5,3; 1,0} +  \chi^{m6 \times \overline{m6}}_{6,\frac{2}{5}; 10,-\frac{7}{5}} +  \chi^{m6 \times \overline{m6}}_{10,\frac{7}{5}; 10,-\frac{7}{5}} 
 \nonumber \\ 
Z_{e; \bar e}^{JK_4 \boxtimes \overline{JK}_4} &= \chi^{m6 \times \overline{m6}}_{1,0; 5,-3} +  \chi^{m6 \times \overline{m6}}_{5,3; 5,-3} +  \chi^{m6 \times \overline{m6}}_{6,\frac{2}{5}; 6,-\frac{2}{5}} +  \chi^{m6 \times \overline{m6}}_{10,\frac{7}{5}; 6,-\frac{2}{5}} 
 \nonumber \\ 
Z_{e; \bar m}^{JK_4 \boxtimes \overline{JK}_4} &= \chi^{m6 \times \overline{m6}}_{1,0; 2,-\frac{1}{8}} +  \chi^{m6 \times \overline{m6}}_{5,3; 2,-\frac{1}{8}} +  \chi^{m6 \times \overline{m6}}_{6,\frac{2}{5}; 9,-\frac{21}{40}} +  \chi^{m6 \times \overline{m6}}_{10,\frac{7}{5}; 9,-\frac{21}{40}} 
 \nonumber \\ 
Z_{e; \bar m_1}^{JK_4 \boxtimes \overline{JK}_4} &= \chi^{m6 \times \overline{m6}}_{1,0; 4,-\frac{13}{8}} +  \chi^{m6 \times \overline{m6}}_{5,3; 4,-\frac{13}{8}} +  \chi^{m6 \times \overline{m6}}_{6,\frac{2}{5}; 7,-\frac{1}{40}} +  \chi^{m6 \times \overline{m6}}_{10,\frac{7}{5}; 7,-\frac{1}{40}} 
 \nonumber \\ 
Z_{e; \bar q}^{JK_4 \boxtimes \overline{JK}_4} &= \chi^{m6 \times \overline{m6}}_{1,0; 3,-\frac{2}{3}} +  \chi^{m6 \times \overline{m6}}_{5,3; 3,-\frac{2}{3}} +  \chi^{m6 \times \overline{m6}}_{6,\frac{2}{5}; 8,-\frac{1}{15}} +  \chi^{m6 \times \overline{m6}}_{10,\frac{7}{5}; 8,-\frac{1}{15}} 
 \nonumber \\ 
Z_{m; \mathbf{1}}^{JK_4 \boxtimes \overline{JK}_4} &=0
 \nonumber \\ 
Z_{m; \bar e}^{JK_4 \boxtimes \overline{JK}_4} &=0
 \nonumber \\ 
Z_{m; \bar m}^{JK_4 \boxtimes \overline{JK}_4} &=0
 \nonumber \\ 
Z_{m; \bar m_1}^{JK_4 \boxtimes \overline{JK}_4} &=0
 \nonumber \\ 
Z_{m; \bar q}^{JK_4 \boxtimes \overline{JK}_4} &=0
 \nonumber \\ 
Z_{m_1; \mathbf{1}}^{JK_4 \boxtimes \overline{JK}_4} &=0
 \nonumber \\ 
Z_{m_1; \bar e}^{JK_4 \boxtimes \overline{JK}_4} &=0
 \nonumber \\ 
Z_{m_1; \bar m}^{JK_4 \boxtimes \overline{JK}_4} &=0
 \nonumber \\ 
Z_{m_1; \bar m_1}^{JK_4 \boxtimes \overline{JK}_4} &=0
 \nonumber \\ 
Z_{m_1; \bar q}^{JK_4 \boxtimes \overline{JK}_4} &=0
 \nonumber \\ 
Z_{q; \mathbf{1}}^{JK_4 \boxtimes \overline{JK}_4} &= 2\chi^{m6 \times \overline{m6}}_{3,\frac{2}{3}; 1,0} +  2\chi^{m6 \times \overline{m6}}_{8,\frac{1}{15}; 10,-\frac{7}{5}} 
 \nonumber \\ 
Z_{q; \bar e}^{JK_4 \boxtimes \overline{JK}_4} &= 2\chi^{m6 \times \overline{m6}}_{3,\frac{2}{3}; 5,-3} +  2\chi^{m6 \times \overline{m6}}_{8,\frac{1}{15}; 6,-\frac{2}{5}} 
 \nonumber \\ 
Z_{q; \bar m}^{JK_4 \boxtimes \overline{JK}_4} &= 2\chi^{m6 \times \overline{m6}}_{3,\frac{2}{3}; 2,-\frac{1}{8}} +  2\chi^{m6 \times \overline{m6}}_{8,\frac{1}{15}; 9,-\frac{21}{40}} 
 \nonumber \\ 
Z_{q; \bar m_1}^{JK_4 \boxtimes \overline{JK}_4} &= 2\chi^{m6 \times \overline{m6}}_{3,\frac{2}{3}; 4,-\frac{13}{8}} +  2\chi^{m6 \times \overline{m6}}_{8,\frac{1}{15}; 7,-\frac{1}{40}} 
 \nonumber \\ 
Z_{q; \bar q}^{JK_4 \boxtimes \overline{JK}_4} &= 2\chi^{m6 \times \overline{m6}}_{3,\frac{2}{3}; 3,-\frac{2}{3}} +  2\chi^{m6 \times \overline{m6}}_{8,\frac{1}{15}; 8,-\frac{1}{15}} 
 \nonumber \\ 
\cA &= \one \oplus e .
\end{align}

\begin{align}
\label{m6-1be}
 Z_{\mathbf{1}; \mathbf{1}}^{JK_4 \boxtimes \overline{JK}_4} &={\color{blue} \chi^{m6 \times \overline{m6}}_{1,0; 1,0}} +  \chi^{m6 \times \overline{m6}}_{1,0; 5,-3} +  \chi^{m6 \times \overline{m6}}_{10,\frac{7}{5}; 6,-\frac{2}{5}} +  \chi^{m6 \times \overline{m6}}_{10,\frac{7}{5}; 10,-\frac{7}{5}} 
 \nonumber \\ 
Z_{\mathbf{1}; \bar e}^{JK_4 \boxtimes \overline{JK}_4} &={\color{blue} \chi^{m6 \times \overline{m6}}_{1,0; 1,0}} +  \chi^{m6 \times \overline{m6}}_{1,0; 5,-3} +  \chi^{m6 \times \overline{m6}}_{10,\frac{7}{5}; 6,-\frac{2}{5}} +  \chi^{m6 \times \overline{m6}}_{10,\frac{7}{5}; 10,-\frac{7}{5}} 
 \nonumber \\ 
Z_{\mathbf{1}; \bar m}^{JK_4 \boxtimes \overline{JK}_4} &=0
 \nonumber \\ 
Z_{\mathbf{1}; \bar m_1}^{JK_4 \boxtimes \overline{JK}_4} &=0
 \nonumber \\ 
Z_{\mathbf{1}; \bar q}^{JK_4 \boxtimes \overline{JK}_4} &= 2\chi^{m6 \times \overline{m6}}_{1,0; 3,-\frac{2}{3}} +  2\chi^{m6 \times \overline{m6}}_{10,\frac{7}{5}; 8,-\frac{1}{15}} 
 \nonumber \\ 
Z_{e; \mathbf{1}}^{JK_4 \boxtimes \overline{JK}_4} &= \chi^{m6 \times \overline{m6}}_{5,3; 1,0} +  \chi^{m6 \times \overline{m6}}_{5,3; 5,-3} +  \chi^{m6 \times \overline{m6}}_{6,\frac{2}{5}; 6,-\frac{2}{5}} +  \chi^{m6 \times \overline{m6}}_{6,\frac{2}{5}; 10,-\frac{7}{5}} 
 \nonumber \\ 
Z_{e; \bar e}^{JK_4 \boxtimes \overline{JK}_4} &= \chi^{m6 \times \overline{m6}}_{5,3; 1,0} +  \chi^{m6 \times \overline{m6}}_{5,3; 5,-3} +  \chi^{m6 \times \overline{m6}}_{6,\frac{2}{5}; 6,-\frac{2}{5}} +  \chi^{m6 \times \overline{m6}}_{6,\frac{2}{5}; 10,-\frac{7}{5}} 
 \nonumber \\ 
Z_{e; \bar m}^{JK_4 \boxtimes \overline{JK}_4} &=0
 \nonumber \\ 
Z_{e; \bar m_1}^{JK_4 \boxtimes \overline{JK}_4} &=0
 \nonumber \\ 
Z_{e; \bar q}^{JK_4 \boxtimes \overline{JK}_4} &= 2\chi^{m6 \times \overline{m6}}_{5,3; 3,-\frac{2}{3}} +  2\chi^{m6 \times \overline{m6}}_{6,\frac{2}{5}; 8,-\frac{1}{15}} 
 \nonumber \\ 
Z_{m; \mathbf{1}}^{JK_4 \boxtimes \overline{JK}_4} &= \chi^{m6 \times \overline{m6}}_{2,\frac{1}{8}; 1,0} +  \chi^{m6 \times \overline{m6}}_{2,\frac{1}{8}; 5,-3} +  \chi^{m6 \times \overline{m6}}_{9,\frac{21}{40}; 6,-\frac{2}{5}} +  \chi^{m6 \times \overline{m6}}_{9,\frac{21}{40}; 10,-\frac{7}{5}} 
 \nonumber \\ 
Z_{m; \bar e}^{JK_4 \boxtimes \overline{JK}_4} &= \chi^{m6 \times \overline{m6}}_{2,\frac{1}{8}; 1,0} +  \chi^{m6 \times \overline{m6}}_{2,\frac{1}{8}; 5,-3} +  \chi^{m6 \times \overline{m6}}_{9,\frac{21}{40}; 6,-\frac{2}{5}} +  \chi^{m6 \times \overline{m6}}_{9,\frac{21}{40}; 10,-\frac{7}{5}} 
 \nonumber \\ 
Z_{m; \bar m}^{JK_4 \boxtimes \overline{JK}_4} &=0
 \nonumber \\ 
Z_{m; \bar m_1}^{JK_4 \boxtimes \overline{JK}_4} &=0
 \nonumber \\ 
Z_{m; \bar q}^{JK_4 \boxtimes \overline{JK}_4} &= 2\chi^{m6 \times \overline{m6}}_{2,\frac{1}{8}; 3,-\frac{2}{3}} +  2\chi^{m6 \times \overline{m6}}_{9,\frac{21}{40}; 8,-\frac{1}{15}} 
 \nonumber \\ 
Z_{m_1; \mathbf{1}}^{JK_4 \boxtimes \overline{JK}_4} &= \chi^{m6 \times \overline{m6}}_{4,\frac{13}{8}; 1,0} +  \chi^{m6 \times \overline{m6}}_{4,\frac{13}{8}; 5,-3} +  \chi^{m6 \times \overline{m6}}_{7,\frac{1}{40}; 6,-\frac{2}{5}} +  \chi^{m6 \times \overline{m6}}_{7,\frac{1}{40}; 10,-\frac{7}{5}} 
 \nonumber \\ 
Z_{m_1; \bar e}^{JK_4 \boxtimes \overline{JK}_4} &= \chi^{m6 \times \overline{m6}}_{4,\frac{13}{8}; 1,0} +  \chi^{m6 \times \overline{m6}}_{4,\frac{13}{8}; 5,-3} +  \chi^{m6 \times \overline{m6}}_{7,\frac{1}{40}; 6,-\frac{2}{5}} +  \chi^{m6 \times \overline{m6}}_{7,\frac{1}{40}; 10,-\frac{7}{5}} 
 \nonumber \\ 
Z_{m_1; \bar m}^{JK_4 \boxtimes \overline{JK}_4} &=0
 \nonumber \\ 
Z_{m_1; \bar m_1}^{JK_4 \boxtimes \overline{JK}_4} &=0
 \nonumber \\ 
Z_{m_1; \bar q}^{JK_4 \boxtimes \overline{JK}_4} &= 2\chi^{m6 \times \overline{m6}}_{4,\frac{13}{8}; 3,-\frac{2}{3}} +  2\chi^{m6 \times \overline{m6}}_{7,\frac{1}{40}; 8,-\frac{1}{15}} 
 \nonumber \\ 
Z_{q; \mathbf{1}}^{JK_4 \boxtimes \overline{JK}_4} &= \chi^{m6 \times \overline{m6}}_{3,\frac{2}{3}; 1,0} +  \chi^{m6 \times \overline{m6}}_{3,\frac{2}{3}; 5,-3} +  \chi^{m6 \times \overline{m6}}_{8,\frac{1}{15}; 6,-\frac{2}{5}} +  \chi^{m6 \times \overline{m6}}_{8,\frac{1}{15}; 10,-\frac{7}{5}} 
 \nonumber \\ 
Z_{q; \bar e}^{JK_4 \boxtimes \overline{JK}_4} &= \chi^{m6 \times \overline{m6}}_{3,\frac{2}{3}; 1,0} +  \chi^{m6 \times \overline{m6}}_{3,\frac{2}{3}; 5,-3} +  \chi^{m6 \times \overline{m6}}_{8,\frac{1}{15}; 6,-\frac{2}{5}} +  \chi^{m6 \times \overline{m6}}_{8,\frac{1}{15}; 10,-\frac{7}{5}} 
 \nonumber \\ 
Z_{q; \bar m}^{JK_4 \boxtimes \overline{JK}_4} &=0
 \nonumber \\ 
Z_{q; \bar m_1}^{JK_4 \boxtimes \overline{JK}_4} &=0
 \nonumber \\ 
Z_{q; \bar q}^{JK_4 \boxtimes \overline{JK}_4} &= 2\chi^{m6 \times \overline{m6}}_{3,\frac{2}{3}; 3,-\frac{2}{3}} +  2\chi^{m6 \times \overline{m6}}_{8,\frac{1}{15}; 8,-\frac{1}{15}} 
 \nonumber \\ 
\cA &= \one \oplus \bar e .
\end{align}

\begin{align}
\label{m4m4U1}
 Z_{\mathbf{1}; \mathbf{1}}^{JK_4 \boxtimes \overline{JK}_4} &={\color{blue} \chi^{m4 \times m4 \times \overline{U1}_4}_{1,0; 1,0; 1,0}} + {\color{red} \chi^{m4 \times m4 \times \overline{U1}_4}_{1,0; 3,\frac{1}{2}; 3,-\frac{1}{2}}} + {\color{red} \chi^{m4 \times m4 \times \overline{U1}_4}_{3,\frac{1}{2}; 1,0; 3,-\frac{1}{2}}} +  \chi^{m4 \times m4 \times \overline{U1}_4}_{3,\frac{1}{2}; 3,\frac{1}{2}; 1,0} 
 \nonumber \\ 
Z_{\mathbf{1}; \bar e}^{JK_4 \boxtimes \overline{JK}_4} &= \chi^{m4 \times m4 \times \overline{U1}_4}_{2,\frac{1}{16}; 2,\frac{1}{16}; 2,-\frac{1}{8}} +  \chi^{m4 \times m4 \times \overline{U1}_4}_{2,\frac{1}{16}; 2,\frac{1}{16}; 4,-\frac{9}{8}} 
 \nonumber \\ 
Z_{\mathbf{1}; \bar m}^{JK_4 \boxtimes \overline{JK}_4} &=0
 \nonumber \\ 
Z_{\mathbf{1}; \bar m_1}^{JK_4 \boxtimes \overline{JK}_4} &=0
 \nonumber \\ 
Z_{\mathbf{1}; \bar q}^{JK_4 \boxtimes \overline{JK}_4} &=0
 \nonumber \\ 
Z_{e; \mathbf{1}}^{JK_4 \boxtimes \overline{JK}_4} &= \chi^{m4 \times m4 \times \overline{U1}_4}_{2,\frac{1}{16}; 2,\frac{1}{16}; 2,-\frac{1}{8}} +  \chi^{m4 \times m4 \times \overline{U1}_4}_{2,\frac{1}{16}; 2,\frac{1}{16}; 4,-\frac{9}{8}} 
 \nonumber \\ 
Z_{e; \bar e}^{JK_4 \boxtimes \overline{JK}_4} &={\color{blue} \chi^{m4 \times m4 \times \overline{U1}_4}_{1,0; 1,0; 1,0}} +  \chi^{m4 \times m4 \times \overline{U1}_4}_{1,0; 3,\frac{1}{2}; 3,-\frac{1}{2}} +  \chi^{m4 \times m4 \times \overline{U1}_4}_{3,\frac{1}{2}; 1,0; 3,-\frac{1}{2}} +  \chi^{m4 \times m4 \times \overline{U1}_4}_{3,\frac{1}{2}; 3,\frac{1}{2}; 1,0} 
 \nonumber \\ 
Z_{e; \bar m}^{JK_4 \boxtimes \overline{JK}_4} &=0
 \nonumber \\ 
Z_{e; \bar m_1}^{JK_4 \boxtimes \overline{JK}_4} &=0
 \nonumber \\ 
Z_{e; \bar q}^{JK_4 \boxtimes \overline{JK}_4} &=0
 \nonumber \\ 
Z_{m; \mathbf{1}}^{JK_4 \boxtimes \overline{JK}_4} &=0
 \nonumber \\ 
Z_{m; \bar e}^{JK_4 \boxtimes \overline{JK}_4} &=0
 \nonumber \\ 
Z_{m; \bar m}^{JK_4 \boxtimes \overline{JK}_4} &= \chi^{m4 \times m4 \times \overline{U1}_4}_{2,\frac{1}{16}; 2,\frac{1}{16}; 2,-\frac{1}{8}} +  \chi^{m4 \times m4 \times \overline{U1}_4}_{2,\frac{1}{16}; 2,\frac{1}{16}; 4,-\frac{9}{8}} 
 \nonumber \\ 
Z_{m; \bar m_1}^{JK_4 \boxtimes \overline{JK}_4} &= \chi^{m4 \times m4 \times \overline{U1}_4}_{1,0; 1,0; 3,-\frac{1}{2}} +  \chi^{m4 \times m4 \times \overline{U1}_4}_{1,0; 3,\frac{1}{2}; 1,0} +  \chi^{m4 \times m4 \times \overline{U1}_4}_{3,\frac{1}{2}; 1,0; 1,0} +  \chi^{m4 \times m4 \times \overline{U1}_4}_{3,\frac{1}{2}; 3,\frac{1}{2}; 3,-\frac{1}{2}} 
 \nonumber \\ 
Z_{m; \bar q}^{JK_4 \boxtimes \overline{JK}_4} &=0
 \nonumber \\ 
Z_{m_1; \mathbf{1}}^{JK_4 \boxtimes \overline{JK}_4} &=0
 \nonumber \\ 
Z_{m_1; \bar e}^{JK_4 \boxtimes \overline{JK}_4} &=0
 \nonumber \\ 
Z_{m_1; \bar m}^{JK_4 \boxtimes \overline{JK}_4} &= \chi^{m4 \times m4 \times \overline{U1}_4}_{1,0; 1,0; 3,-\frac{1}{2}} +  \chi^{m4 \times m4 \times \overline{U1}_4}_{1,0; 3,\frac{1}{2}; 1,0} +  \chi^{m4 \times m4 \times \overline{U1}_4}_{3,\frac{1}{2}; 1,0; 1,0} +  \chi^{m4 \times m4 \times \overline{U1}_4}_{3,\frac{1}{2}; 3,\frac{1}{2}; 3,-\frac{1}{2}} 
 \nonumber \\ 
Z_{m_1; \bar m_1}^{JK_4 \boxtimes \overline{JK}_4} &= \chi^{m4 \times m4 \times \overline{U1}_4}_{2,\frac{1}{16}; 2,\frac{1}{16}; 2,-\frac{1}{8}} +  \chi^{m4 \times m4 \times \overline{U1}_4}_{2,\frac{1}{16}; 2,\frac{1}{16}; 4,-\frac{9}{8}} 
 \nonumber \\ 
Z_{m_1; \bar q}^{JK_4 \boxtimes \overline{JK}_4} &=0
 \nonumber \\ 
Z_{q; \mathbf{1}}^{JK_4 \boxtimes \overline{JK}_4} &=0
 \nonumber \\ 
Z_{q; \bar e}^{JK_4 \boxtimes \overline{JK}_4} &=0
 \nonumber \\ 
Z_{q; \bar m}^{JK_4 \boxtimes \overline{JK}_4} &=0
 \nonumber \\ 
Z_{q; \bar m_1}^{JK_4 \boxtimes \overline{JK}_4} &=0
 \nonumber \\ 
Z_{q; \bar q}^{JK_4 \boxtimes \overline{JK}_4} &={\color{blue} \chi^{m4 \times m4 \times \overline{U1}_4}_{1,0; 1,0; 1,0}} +  \chi^{m4 \times m4 \times \overline{U1}_4}_{1,0; 3,\frac{1}{2}; 3,-\frac{1}{2}} +  \chi^{m4 \times m4 \times \overline{U1}_4}_{2,\frac{1}{16}; 2,\frac{1}{16}; 2,-\frac{1}{8}} +  \chi^{m4 \times m4 \times \overline{U1}_4}_{2,\frac{1}{16}; 2,\frac{1}{16}; 4,-\frac{9}{8}}
\nonumber\\ &\ \ \ \ \
 +  \chi^{m4 \times m4 \times \overline{U1}_4}_{3,\frac{1}{2}; 1,0; 3,-\frac{1}{2}} +  \chi^{m4 \times m4 \times \overline{U1}_4}_{3,\frac{1}{2}; 3,\frac{1}{2}; 1,0} 
 \nonumber \\ 
\cA &= \one \oplus e \bar e  \oplus q\bar q.
\end{align}

In the following, we list the gapless critical states with only one symmetric
relevant operators of self-dual $S_3$ symmetric model or the self-dual
$\cRep(S_3)$ symmetric model.

\begin{align}
\label{m4-eeqq}
 Z_{\mathbf{1}; \mathbf{1}}^{JK_4 \boxtimes \overline{JK}_4} &={\color{blue} \chi^{m4 \times \overline{m4}}_{1,0; 1,0}} + {\color{red} \chi^{m4 \times \overline{m4}}_{3,\frac{1}{2}; 3,-\frac{1}{2}}} 
 \nonumber \\ 
Z_{\mathbf{1}; \bar e}^{JK_4 \boxtimes \overline{JK}_4} &= \chi^{m4 \times \overline{m4}}_{2,\frac{1}{16}; 2,-\frac{1}{16}} 
 \nonumber \\ 
Z_{\mathbf{1}; \bar m}^{JK_4 \boxtimes \overline{JK}_4} &=0
 \nonumber \\ 
Z_{\mathbf{1}; \bar m_1}^{JK_4 \boxtimes \overline{JK}_4} &=0
 \nonumber \\ 
Z_{\mathbf{1}; \bar q}^{JK_4 \boxtimes \overline{JK}_4} &=0
 \nonumber \\ 
Z_{e; \mathbf{1}}^{JK_4 \boxtimes \overline{JK}_4} &= \chi^{m4 \times \overline{m4}}_{2,\frac{1}{16}; 2,-\frac{1}{16}} 
 \nonumber \\ 
Z_{e; \bar e}^{JK_4 \boxtimes \overline{JK}_4} &={\color{blue} \chi^{m4 \times \overline{m4}}_{1,0; 1,0}} +  \chi^{m4 \times \overline{m4}}_{3,\frac{1}{2}; 3,-\frac{1}{2}} 
 \nonumber \\ 
Z_{e; \bar m}^{JK_4 \boxtimes \overline{JK}_4} &=0
 \nonumber \\ 
Z_{e; \bar m_1}^{JK_4 \boxtimes \overline{JK}_4} &=0
 \nonumber \\ 
Z_{e; \bar q}^{JK_4 \boxtimes \overline{JK}_4} &=0
 \nonumber \\ 
Z_{m; \mathbf{1}}^{JK_4 \boxtimes \overline{JK}_4} &=0
 \nonumber \\ 
Z_{m; \bar e}^{JK_4 \boxtimes \overline{JK}_4} &=0
 \nonumber \\ 
Z_{m; \bar m}^{JK_4 \boxtimes \overline{JK}_4} &= \chi^{m4 \times \overline{m4}}_{2,\frac{1}{16}; 2,-\frac{1}{16}} 
 \nonumber \\ 
Z_{m; \bar m_1}^{JK_4 \boxtimes \overline{JK}_4} &= \chi^{m4 \times \overline{m4}}_{1,0; 3,-\frac{1}{2}} +  \chi^{m4 \times \overline{m4}}_{3,\frac{1}{2}; 1,0} 
 \nonumber \\ 
Z_{m; \bar q}^{JK_4 \boxtimes \overline{JK}_4} &=0
 \nonumber \\ 
Z_{m_1; \mathbf{1}}^{JK_4 \boxtimes \overline{JK}_4} &=0
 \nonumber \\ 
Z_{m_1; \bar e}^{JK_4 \boxtimes \overline{JK}_4} &=0
 \nonumber \\ 
Z_{m_1; \bar m}^{JK_4 \boxtimes \overline{JK}_4} &= \chi^{m4 \times \overline{m4}}_{1,0; 3,-\frac{1}{2}} +  \chi^{m4 \times \overline{m4}}_{3,\frac{1}{2}; 1,0} 
 \nonumber \\ 
Z_{m_1; \bar m_1}^{JK_4 \boxtimes \overline{JK}_4} &= \chi^{m4 \times \overline{m4}}_{2,\frac{1}{16}; 2,-\frac{1}{16}} 
 \nonumber \\ 
Z_{m_1; \bar q}^{JK_4 \boxtimes \overline{JK}_4} &=0
 \nonumber \\ 
Z_{q; \mathbf{1}}^{JK_4 \boxtimes \overline{JK}_4} &=0
 \nonumber \\ 
Z_{q; \bar e}^{JK_4 \boxtimes \overline{JK}_4} &=0
 \nonumber \\ 
Z_{q; \bar m}^{JK_4 \boxtimes \overline{JK}_4} &=0
 \nonumber \\ 
Z_{q; \bar m_1}^{JK_4 \boxtimes \overline{JK}_4} &=0
 \nonumber \\ 
Z_{q; \bar q}^{JK_4 \boxtimes \overline{JK}_4} &={\color{blue} \chi^{m4 \times \overline{m4}}_{1,0; 1,0}} +  \chi^{m4 \times \overline{m4}}_{2,\frac{1}{16}; 2,-\frac{1}{16}} +  \chi^{m4 \times \overline{m4}}_{3,\frac{1}{2}; 3,-\frac{1}{2}} 
 \nonumber \\ 
\cA &= \one \oplus e \bar e  \oplus q\bar q.
\end{align}

\begin{align}
\label{m6-ee}
 Z_{\mathbf{1}; \mathbf{1}}^{JK_4 \boxtimes \overline{JK}_4} &={\color{blue} \chi^{m6 \times \overline{m6}}_{1,0; 1,0}} +  \chi^{m6 \times \overline{m6}}_{5,3; 5,-3} + {\color{red} \chi^{m6 \times \overline{m6}}_{6,\frac{2}{5}; 6,-\frac{2}{5}}} +  \chi^{m6 \times \overline{m6}}_{10,\frac{7}{5}; 10,-\frac{7}{5}} 
 \nonumber \\ 
Z_{\mathbf{1}; \bar e}^{JK_4 \boxtimes \overline{JK}_4} &= \chi^{m6 \times \overline{m6}}_{1,0; 5,-3} +  \chi^{m6 \times \overline{m6}}_{5,3; 1,0} +  \chi^{m6 \times \overline{m6}}_{6,\frac{2}{5}; 10,-\frac{7}{5}} +  \chi^{m6 \times \overline{m6}}_{10,\frac{7}{5}; 6,-\frac{2}{5}} 
 \nonumber \\ 
Z_{\mathbf{1}; \bar m}^{JK_4 \boxtimes \overline{JK}_4} &=0
 \nonumber \\ 
Z_{\mathbf{1}; \bar m_1}^{JK_4 \boxtimes \overline{JK}_4} &=0
 \nonumber \\ 
Z_{\mathbf{1}; \bar q}^{JK_4 \boxtimes \overline{JK}_4} &= \chi^{m6 \times \overline{m6}}_{1,0; 3,-\frac{2}{3}} +  \chi^{m6 \times \overline{m6}}_{5,3; 3,-\frac{2}{3}} +  \chi^{m6 \times \overline{m6}}_{6,\frac{2}{5}; 8,-\frac{1}{15}} +  \chi^{m6 \times \overline{m6}}_{10,\frac{7}{5}; 8,-\frac{1}{15}} 
 \nonumber \\ 
Z_{e; \mathbf{1}}^{JK_4 \boxtimes \overline{JK}_4} &= \chi^{m6 \times \overline{m6}}_{1,0; 5,-3} +  \chi^{m6 \times \overline{m6}}_{5,3; 1,0} +  \chi^{m6 \times \overline{m6}}_{6,\frac{2}{5}; 10,-\frac{7}{5}} +  \chi^{m6 \times \overline{m6}}_{10,\frac{7}{5}; 6,-\frac{2}{5}} 
 \nonumber \\ 
Z_{e; \bar e}^{JK_4 \boxtimes \overline{JK}_4} &={\color{blue} \chi^{m6 \times \overline{m6}}_{1,0; 1,0}} +  \chi^{m6 \times \overline{m6}}_{5,3; 5,-3} +  \chi^{m6 \times \overline{m6}}_{6,\frac{2}{5}; 6,-\frac{2}{5}} +  \chi^{m6 \times \overline{m6}}_{10,\frac{7}{5}; 10,-\frac{7}{5}} 
 \nonumber \\ 
Z_{e; \bar m}^{JK_4 \boxtimes \overline{JK}_4} &=0
 \nonumber \\ 
Z_{e; \bar m_1}^{JK_4 \boxtimes \overline{JK}_4} &=0
 \nonumber \\ 
Z_{e; \bar q}^{JK_4 \boxtimes \overline{JK}_4} &= \chi^{m6 \times \overline{m6}}_{1,0; 3,-\frac{2}{3}} +  \chi^{m6 \times \overline{m6}}_{5,3; 3,-\frac{2}{3}} +  \chi^{m6 \times \overline{m6}}_{6,\frac{2}{5}; 8,-\frac{1}{15}} +  \chi^{m6 \times \overline{m6}}_{10,\frac{7}{5}; 8,-\frac{1}{15}} 
 \nonumber \\ 
Z_{m; \mathbf{1}}^{JK_4 \boxtimes \overline{JK}_4} &=0
 \nonumber \\ 
Z_{m; \bar e}^{JK_4 \boxtimes \overline{JK}_4} &=0
 \nonumber \\ 
Z_{m; \bar m}^{JK_4 \boxtimes \overline{JK}_4} &= \chi^{m6 \times \overline{m6}}_{2,\frac{1}{8}; 2,-\frac{1}{8}} +  \chi^{m6 \times \overline{m6}}_{4,\frac{13}{8}; 4,-\frac{13}{8}} +  \chi^{m6 \times \overline{m6}}_{7,\frac{1}{40}; 7,-\frac{1}{40}} +  \chi^{m6 \times \overline{m6}}_{9,\frac{21}{40}; 9,-\frac{21}{40}} 
 \nonumber \\ 
Z_{m; \bar m_1}^{JK_4 \boxtimes \overline{JK}_4} &= \chi^{m6 \times \overline{m6}}_{2,\frac{1}{8}; 4,-\frac{13}{8}} +  \chi^{m6 \times \overline{m6}}_{4,\frac{13}{8}; 2,-\frac{1}{8}} +  \chi^{m6 \times \overline{m6}}_{7,\frac{1}{40}; 9,-\frac{21}{40}} +  \chi^{m6 \times \overline{m6}}_{9,\frac{21}{40}; 7,-\frac{1}{40}} 
 \nonumber \\ 
Z_{m; \bar q}^{JK_4 \boxtimes \overline{JK}_4} &=0
 \nonumber \\ 
Z_{m_1; \mathbf{1}}^{JK_4 \boxtimes \overline{JK}_4} &=0
 \nonumber \\ 
Z_{m_1; \bar e}^{JK_4 \boxtimes \overline{JK}_4} &=0
 \nonumber \\ 
Z_{m_1; \bar m}^{JK_4 \boxtimes \overline{JK}_4} &= \chi^{m6 \times \overline{m6}}_{2,\frac{1}{8}; 4,-\frac{13}{8}} +  \chi^{m6 \times \overline{m6}}_{4,\frac{13}{8}; 2,-\frac{1}{8}} +  \chi^{m6 \times \overline{m6}}_{7,\frac{1}{40}; 9,-\frac{21}{40}} +  \chi^{m6 \times \overline{m6}}_{9,\frac{21}{40}; 7,-\frac{1}{40}} 
 \nonumber \\ 
Z_{m_1; \bar m_1}^{JK_4 \boxtimes \overline{JK}_4} &= \chi^{m6 \times \overline{m6}}_{2,\frac{1}{8}; 2,-\frac{1}{8}} +  \chi^{m6 \times \overline{m6}}_{4,\frac{13}{8}; 4,-\frac{13}{8}} +  \chi^{m6 \times \overline{m6}}_{7,\frac{1}{40}; 7,-\frac{1}{40}} +  \chi^{m6 \times \overline{m6}}_{9,\frac{21}{40}; 9,-\frac{21}{40}} 
 \nonumber \\ 
Z_{m_1; \bar q}^{JK_4 \boxtimes \overline{JK}_4} &=0
 \nonumber \\ 
Z_{q; \mathbf{1}}^{JK_4 \boxtimes \overline{JK}_4} &= \chi^{m6 \times \overline{m6}}_{3,\frac{2}{3}; 1,0} +  \chi^{m6 \times \overline{m6}}_{3,\frac{2}{3}; 5,-3} +  \chi^{m6 \times \overline{m6}}_{8,\frac{1}{15}; 6,-\frac{2}{5}} +  \chi^{m6 \times \overline{m6}}_{8,\frac{1}{15}; 10,-\frac{7}{5}} 
 \nonumber \\ 
Z_{q; \bar e}^{JK_4 \boxtimes \overline{JK}_4} &= \chi^{m6 \times \overline{m6}}_{3,\frac{2}{3}; 1,0} +  \chi^{m6 \times \overline{m6}}_{3,\frac{2}{3}; 5,-3} +  \chi^{m6 \times \overline{m6}}_{8,\frac{1}{15}; 6,-\frac{2}{5}} +  \chi^{m6 \times \overline{m6}}_{8,\frac{1}{15}; 10,-\frac{7}{5}} 
 \nonumber \\ 
Z_{q; \bar m}^{JK_4 \boxtimes \overline{JK}_4} &=0
 \nonumber \\ 
Z_{q; \bar m_1}^{JK_4 \boxtimes \overline{JK}_4} &=0
 \nonumber \\ 
Z_{q; \bar q}^{JK_4 \boxtimes \overline{JK}_4} &= 2\chi^{m6 \times \overline{m6}}_{3,\frac{2}{3}; 3,-\frac{2}{3}} +  2\chi^{m6 \times \overline{m6}}_{8,\frac{1}{15}; 8,-\frac{1}{15}} 
 \nonumber \\ 
\cA &= \one \oplus e \bar e  .
\end{align}

\begin{align}
\label{m6-ebe}
 Z_{\mathbf{1}; \mathbf{1}}^{JK_4 \boxtimes \overline{JK}_4} &={\color{blue} \chi^{m6 \times \overline{m6}}_{1,0; 1,0}} +  \chi^{m6 \times \overline{m6}}_{1,0; 5,-3} +  \chi^{m6 \times \overline{m6}}_{5,3; 1,0} +  \chi^{m6 \times \overline{m6}}_{5,3; 5,-3} + {\color{red} \chi^{m6 \times \overline{m6}}_{6,\frac{2}{5}; 6,-\frac{2}{5}}}
\nonumber\\ &\ \ \ \ 
 +  \chi^{m6 \times \overline{m6}}_{6,\frac{2}{5}; 10,-\frac{7}{5}} +  \chi^{m6 \times \overline{m6}}_{10,\frac{7}{5}; 6,-\frac{2}{5}} +  \chi^{m6 \times \overline{m6}}_{10,\frac{7}{5}; 10,-\frac{7}{5}} 
 \nonumber \\ 
Z_{\mathbf{1}; \bar e}^{JK_4 \boxtimes \overline{JK}_4} &={\color{blue} \chi^{m6 \times \overline{m6}}_{1,0; 1,0}} +  \chi^{m6 \times \overline{m6}}_{1,0; 5,-3} +  \chi^{m6 \times \overline{m6}}_{5,3; 1,0} +  \chi^{m6 \times \overline{m6}}_{5,3; 5,-3} +  \chi^{m6 \times \overline{m6}}_{6,\frac{2}{5}; 6,-\frac{2}{5}}
\nonumber\\ &\ \ \ \ 
 +  \chi^{m6 \times \overline{m6}}_{6,\frac{2}{5}; 10,-\frac{7}{5}} +  \chi^{m6 \times \overline{m6}}_{10,\frac{7}{5}; 6,-\frac{2}{5}} +  \chi^{m6 \times \overline{m6}}_{10,\frac{7}{5}; 10,-\frac{7}{5}} 
 \nonumber \\ 
Z_{\mathbf{1}; \bar m}^{JK_4 \boxtimes \overline{JK}_4} &=0
 \nonumber \\ 
Z_{\mathbf{1}; \bar m_1}^{JK_4 \boxtimes \overline{JK}_4} &=0
 \nonumber \\ 
Z_{\mathbf{1}; \bar q}^{JK_4 \boxtimes \overline{JK}_4} &= 2\chi^{m6 \times \overline{m6}}_{1,0; 3,-\frac{2}{3}} +  2\chi^{m6 \times \overline{m6}}_{5,3; 3,-\frac{2}{3}} +  2\chi^{m6 \times \overline{m6}}_{6,\frac{2}{5}; 8,-\frac{1}{15}} +  2\chi^{m6 \times \overline{m6}}_{10,\frac{7}{5}; 8,-\frac{1}{15}} 
 \nonumber \\ 
Z_{e; \mathbf{1}}^{JK_4 \boxtimes \overline{JK}_4} &={\color{blue} \chi^{m6 \times \overline{m6}}_{1,0; 1,0}} +  \chi^{m6 \times \overline{m6}}_{1,0; 5,-3} +  \chi^{m6 \times \overline{m6}}_{5,3; 1,0} +  \chi^{m6 \times \overline{m6}}_{5,3; 5,-3} +  \chi^{m6 \times \overline{m6}}_{6,\frac{2}{5}; 6,-\frac{2}{5}}
\nonumber\\ &\ \ \ \ 
 +  \chi^{m6 \times \overline{m6}}_{6,\frac{2}{5}; 10,-\frac{7}{5}} +  \chi^{m6 \times \overline{m6}}_{10,\frac{7}{5}; 6,-\frac{2}{5}} +  \chi^{m6 \times \overline{m6}}_{10,\frac{7}{5}; 10,-\frac{7}{5}} 
 \nonumber \\ 
Z_{e; \bar e}^{JK_4 \boxtimes \overline{JK}_4} &={\color{blue} \chi^{m6 \times \overline{m6}}_{1,0; 1,0}} +  \chi^{m6 \times \overline{m6}}_{1,0; 5,-3} +  \chi^{m6 \times \overline{m6}}_{5,3; 1,0} +  \chi^{m6 \times \overline{m6}}_{5,3; 5,-3} +  \chi^{m6 \times \overline{m6}}_{6,\frac{2}{5}; 6,-\frac{2}{5}}
\nonumber\\ &\ \ \ \ 
 +  \chi^{m6 \times \overline{m6}}_{6,\frac{2}{5}; 10,-\frac{7}{5}} +  \chi^{m6 \times \overline{m6}}_{10,\frac{7}{5}; 6,-\frac{2}{5}} +  \chi^{m6 \times \overline{m6}}_{10,\frac{7}{5}; 10,-\frac{7}{5}} 
 \nonumber \\ 
Z_{e; \bar m}^{JK_4 \boxtimes \overline{JK}_4} &=0
 \nonumber \\ 
Z_{e; \bar m_1}^{JK_4 \boxtimes \overline{JK}_4} &=0
 \nonumber \\ 
Z_{e; \bar q}^{JK_4 \boxtimes \overline{JK}_4} &= 2\chi^{m6 \times \overline{m6}}_{1,0; 3,-\frac{2}{3}} +  2\chi^{m6 \times \overline{m6}}_{5,3; 3,-\frac{2}{3}} +  2\chi^{m6 \times \overline{m6}}_{6,\frac{2}{5}; 8,-\frac{1}{15}} +  2\chi^{m6 \times \overline{m6}}_{10,\frac{7}{5}; 8,-\frac{1}{15}} 
 \nonumber \\ 
Z_{m; \mathbf{1}}^{JK_4 \boxtimes \overline{JK}_4} &=0
 \nonumber \\ 
Z_{m; \bar e}^{JK_4 \boxtimes \overline{JK}_4} &=0
 \nonumber \\ 
Z_{m; \bar m}^{JK_4 \boxtimes \overline{JK}_4} &=0
 \nonumber \\ 
Z_{m; \bar m_1}^{JK_4 \boxtimes \overline{JK}_4} &=0
 \nonumber \\ 
Z_{m; \bar q}^{JK_4 \boxtimes \overline{JK}_4} &=0
 \nonumber \\ 
Z_{m_1; \mathbf{1}}^{JK_4 \boxtimes \overline{JK}_4} &=0
 \nonumber \\ 
Z_{m_1; \bar e}^{JK_4 \boxtimes \overline{JK}_4} &=0
 \nonumber \\ 
Z_{m_1; \bar m}^{JK_4 \boxtimes \overline{JK}_4} &=0
 \nonumber \\ 
Z_{m_1; \bar m_1}^{JK_4 \boxtimes \overline{JK}_4} &=0
 \nonumber \\ 
Z_{m_1; \bar q}^{JK_4 \boxtimes \overline{JK}_4} &=0
 \nonumber \\ 
Z_{q; \mathbf{1}}^{JK_4 \boxtimes \overline{JK}_4} &= 2\chi^{m6 \times \overline{m6}}_{3,\frac{2}{3}; 1,0} +  2\chi^{m6 \times \overline{m6}}_{3,\frac{2}{3}; 5,-3} +  2\chi^{m6 \times \overline{m6}}_{8,\frac{1}{15}; 6,-\frac{2}{5}} +  2\chi^{m6 \times \overline{m6}}_{8,\frac{1}{15}; 10,-\frac{7}{5}} 
 \nonumber \\ 
Z_{q; \bar e}^{JK_4 \boxtimes \overline{JK}_4} &= 2\chi^{m6 \times \overline{m6}}_{3,\frac{2}{3}; 1,0} +  2\chi^{m6 \times \overline{m6}}_{3,\frac{2}{3}; 5,-3} +  2\chi^{m6 \times \overline{m6}}_{8,\frac{1}{15}; 6,-\frac{2}{5}} +  2\chi^{m6 \times \overline{m6}}_{8,\frac{1}{15}; 10,-\frac{7}{5}} 
 \nonumber \\ 
Z_{q; \bar m}^{JK_4 \boxtimes \overline{JK}_4} &=0
 \nonumber \\ 
Z_{q; \bar m_1}^{JK_4 \boxtimes \overline{JK}_4} &=0
 \nonumber \\ 
Z_{q; \bar q}^{JK_4 \boxtimes \overline{JK}_4} &= 4\chi^{m6 \times \overline{m6}}_{3,\frac{2}{3}; 3,-\frac{2}{3}} +  4\chi^{m6 \times \overline{m6}}_{8,\frac{1}{15}; 8,-\frac{1}{15}} 
 \nonumber \\ 
\cA &= \one \oplus e \oplus \bar e  \oplus e \bar e  .
\end{align}

\begin{align}
\label{m7-82}
 Z_{\mathbf{1}; \mathbf{1}}^{JK_4 \boxtimes \overline{JK}_4} &={\color{blue} \chi^{m7 \times \overline{m7}}_{1,0; 1,0}} + {\color{red} \chi^{m7 \times \overline{m7}}_{3,\frac{5}{7}; 3,-\frac{5}{7}}} +  \chi^{m7 \times \overline{m7}}_{5,\frac{22}{7}; 5,-\frac{22}{7}} 
 \nonumber \\ 
Z_{\mathbf{1}; \bar e}^{JK_4 \boxtimes \overline{JK}_4} &= \chi^{m7 \times \overline{m7}}_{2,\frac{1}{7}; 5,-\frac{22}{7}} +  \chi^{m7 \times \overline{m7}}_{4,\frac{12}{7}; 3,-\frac{5}{7}} +  \chi^{m7 \times \overline{m7}}_{6,5; 1,0} 
 \nonumber \\ 
Z_{\mathbf{1}; \bar m}^{JK_4 \boxtimes \overline{JK}_4} &= \chi^{m7 \times \overline{m7}}_{8,\frac{1}{56}; 2,-\frac{1}{7}} +  \chi^{m7 \times \overline{m7}}_{10,\frac{33}{56}; 4,-\frac{12}{7}} +  \chi^{m7 \times \overline{m7}}_{12,\frac{23}{8}; 6,-5} 
 \nonumber \\ 
Z_{\mathbf{1}; \bar m_1}^{JK_4 \boxtimes \overline{JK}_4} &= \chi^{m7 \times \overline{m7}}_{7,\frac{3}{8}; 6,-5} +  \chi^{m7 \times \overline{m7}}_{9,\frac{5}{56}; 4,-\frac{12}{7}} +  \chi^{m7 \times \overline{m7}}_{11,\frac{85}{56}; 2,-\frac{1}{7}} 
 \nonumber \\ 
Z_{\mathbf{1}; \bar q}^{JK_4 \boxtimes \overline{JK}_4} &= \chi^{m7 \times \overline{m7}}_{13,\frac{4}{3}; 1,0} +  \chi^{m7 \times \overline{m7}}_{14,\frac{10}{21}; 5,-\frac{22}{7}} +  \chi^{m7 \times \overline{m7}}_{15,\frac{1}{21}; 3,-\frac{5}{7}} 
 \nonumber \\ 
Z_{e; \mathbf{1}}^{JK_4 \boxtimes \overline{JK}_4} &= \chi^{m7 \times \overline{m7}}_{1,0; 6,-5} +  \chi^{m7 \times \overline{m7}}_{3,\frac{5}{7}; 4,-\frac{12}{7}} +  \chi^{m7 \times \overline{m7}}_{5,\frac{22}{7}; 2,-\frac{1}{7}} 
 \nonumber \\ 
Z_{e; \bar e}^{JK_4 \boxtimes \overline{JK}_4} &= \chi^{m7 \times \overline{m7}}_{2,\frac{1}{7}; 2,-\frac{1}{7}} +  \chi^{m7 \times \overline{m7}}_{4,\frac{12}{7}; 4,-\frac{12}{7}} +  \chi^{m7 \times \overline{m7}}_{6,5; 6,-5} 
 \nonumber \\ 
Z_{e; \bar m}^{JK_4 \boxtimes \overline{JK}_4} &= \chi^{m7 \times \overline{m7}}_{8,\frac{1}{56}; 5,-\frac{22}{7}} +  \chi^{m7 \times \overline{m7}}_{10,\frac{33}{56}; 3,-\frac{5}{7}} +  \chi^{m7 \times \overline{m7}}_{12,\frac{23}{8}; 1,0} 
 \nonumber \\ 
Z_{e; \bar m_1}^{JK_4 \boxtimes \overline{JK}_4} &= \chi^{m7 \times \overline{m7}}_{7,\frac{3}{8}; 1,0} +  \chi^{m7 \times \overline{m7}}_{9,\frac{5}{56}; 3,-\frac{5}{7}} +  \chi^{m7 \times \overline{m7}}_{11,\frac{85}{56}; 5,-\frac{22}{7}} 
 \nonumber \\ 
Z_{e; \bar q}^{JK_4 \boxtimes \overline{JK}_4} &= \chi^{m7 \times \overline{m7}}_{13,\frac{4}{3}; 6,-5} +  \chi^{m7 \times \overline{m7}}_{14,\frac{10}{21}; 2,-\frac{1}{7}} +  \chi^{m7 \times \overline{m7}}_{15,\frac{1}{21}; 4,-\frac{12}{7}} 
 \nonumber \\ 
Z_{m; \mathbf{1}}^{JK_4 \boxtimes \overline{JK}_4} &= \chi^{m7 \times \overline{m7}}_{2,\frac{1}{7}; 8,-\frac{1}{56}} +  \chi^{m7 \times \overline{m7}}_{4,\frac{12}{7}; 10,-\frac{33}{56}} +  \chi^{m7 \times \overline{m7}}_{6,5; 12,-\frac{23}{8}} 
 \nonumber \\ 
Z_{m; \bar e}^{JK_4 \boxtimes \overline{JK}_4} &= \chi^{m7 \times \overline{m7}}_{1,0; 12,-\frac{23}{8}} +  \chi^{m7 \times \overline{m7}}_{3,\frac{5}{7}; 10,-\frac{33}{56}} +  \chi^{m7 \times \overline{m7}}_{5,\frac{22}{7}; 8,-\frac{1}{56}} 
 \nonumber \\ 
Z_{m; \bar m}^{JK_4 \boxtimes \overline{JK}_4} &= \chi^{m7 \times \overline{m7}}_{7,\frac{3}{8}; 7,-\frac{3}{8}} +  \chi^{m7 \times \overline{m7}}_{9,\frac{5}{56}; 9,-\frac{5}{56}} +  \chi^{m7 \times \overline{m7}}_{11,\frac{85}{56}; 11,-\frac{85}{56}} 
 \nonumber \\ 
Z_{m; \bar m_1}^{JK_4 \boxtimes \overline{JK}_4} &= \chi^{m7 \times \overline{m7}}_{8,\frac{1}{56}; 11,-\frac{85}{56}} +  \chi^{m7 \times \overline{m7}}_{10,\frac{33}{56}; 9,-\frac{5}{56}} +  \chi^{m7 \times \overline{m7}}_{12,\frac{23}{8}; 7,-\frac{3}{8}} 
 \nonumber \\ 
Z_{m; \bar q}^{JK_4 \boxtimes \overline{JK}_4} &= \chi^{m7 \times \overline{m7}}_{13,\frac{4}{3}; 12,-\frac{23}{8}} +  \chi^{m7 \times \overline{m7}}_{14,\frac{10}{21}; 8,-\frac{1}{56}} +  \chi^{m7 \times \overline{m7}}_{15,\frac{1}{21}; 10,-\frac{33}{56}} 
 \nonumber \\ 
Z_{m_1; \mathbf{1}}^{JK_4 \boxtimes \overline{JK}_4} &= \chi^{m7 \times \overline{m7}}_{2,\frac{1}{7}; 11,-\frac{85}{56}} +  \chi^{m7 \times \overline{m7}}_{4,\frac{12}{7}; 9,-\frac{5}{56}} +  \chi^{m7 \times \overline{m7}}_{6,5; 7,-\frac{3}{8}} 
 \nonumber \\ 
Z_{m_1; \bar e}^{JK_4 \boxtimes \overline{JK}_4} &= \chi^{m7 \times \overline{m7}}_{1,0; 7,-\frac{3}{8}} +  \chi^{m7 \times \overline{m7}}_{3,\frac{5}{7}; 9,-\frac{5}{56}} +  \chi^{m7 \times \overline{m7}}_{5,\frac{22}{7}; 11,-\frac{85}{56}} 
 \nonumber \\ 
Z_{m_1; \bar m}^{JK_4 \boxtimes \overline{JK}_4} &= \chi^{m7 \times \overline{m7}}_{7,\frac{3}{8}; 12,-\frac{23}{8}} +  \chi^{m7 \times \overline{m7}}_{9,\frac{5}{56}; 10,-\frac{33}{56}} +  \chi^{m7 \times \overline{m7}}_{11,\frac{85}{56}; 8,-\frac{1}{56}} 
 \nonumber \\ 
Z_{m_1; \bar m_1}^{JK_4 \boxtimes \overline{JK}_4} &= \chi^{m7 \times \overline{m7}}_{8,\frac{1}{56}; 8,-\frac{1}{56}} +  \chi^{m7 \times \overline{m7}}_{10,\frac{33}{56}; 10,-\frac{33}{56}} +  \chi^{m7 \times \overline{m7}}_{12,\frac{23}{8}; 12,-\frac{23}{8}} 
 \nonumber \\ 
Z_{m_1; \bar q}^{JK_4 \boxtimes \overline{JK}_4} &= \chi^{m7 \times \overline{m7}}_{13,\frac{4}{3}; 7,-\frac{3}{8}} +  \chi^{m7 \times \overline{m7}}_{14,\frac{10}{21}; 11,-\frac{85}{56}} +  \chi^{m7 \times \overline{m7}}_{15,\frac{1}{21}; 9,-\frac{5}{56}} 
 \nonumber \\ 
Z_{q; \mathbf{1}}^{JK_4 \boxtimes \overline{JK}_4} &= \chi^{m7 \times \overline{m7}}_{1,0; 13,-\frac{4}{3}} +  \chi^{m7 \times \overline{m7}}_{3,\frac{5}{7}; 15,-\frac{1}{21}} +  \chi^{m7 \times \overline{m7}}_{5,\frac{22}{7}; 14,-\frac{10}{21}} 
 \nonumber \\ 
Z_{q; \bar e}^{JK_4 \boxtimes \overline{JK}_4} &= \chi^{m7 \times \overline{m7}}_{2,\frac{1}{7}; 14,-\frac{10}{21}} +  \chi^{m7 \times \overline{m7}}_{4,\frac{12}{7}; 15,-\frac{1}{21}} +  \chi^{m7 \times \overline{m7}}_{6,5; 13,-\frac{4}{3}} 
 \nonumber \\ 
Z_{q; \bar m}^{JK_4 \boxtimes \overline{JK}_4} &= \chi^{m7 \times \overline{m7}}_{8,\frac{1}{56}; 14,-\frac{10}{21}} +  \chi^{m7 \times \overline{m7}}_{10,\frac{33}{56}; 15,-\frac{1}{21}} +  \chi^{m7 \times \overline{m7}}_{12,\frac{23}{8}; 13,-\frac{4}{3}} 
 \nonumber \\ 
Z_{q; \bar m_1}^{JK_4 \boxtimes \overline{JK}_4} &= \chi^{m7 \times \overline{m7}}_{7,\frac{3}{8}; 13,-\frac{4}{3}} +  \chi^{m7 \times \overline{m7}}_{9,\frac{5}{56}; 15,-\frac{1}{21}} +  \chi^{m7 \times \overline{m7}}_{11,\frac{85}{56}; 14,-\frac{10}{21}} 
 \nonumber \\ 
Z_{q; \bar q}^{JK_4 \boxtimes \overline{JK}_4} &= \chi^{m7 \times \overline{m7}}_{13,\frac{4}{3}; 13,-\frac{4}{3}} +  \chi^{m7 \times \overline{m7}}_{14,\frac{10}{21}; 14,-\frac{10}{21}} +  \chi^{m7 \times \overline{m7}}_{15,\frac{1}{21}; 15,-\frac{1}{21}} 
 \nonumber \\ 
\cA &= \one   .
\end{align}

\begin{align}
\label{m7-85}
 Z_{\mathbf{1}; \mathbf{1}}^{JK_4 \boxtimes \overline{JK}_4} &={\color{blue} \chi^{m7 \times \overline{m7}}_{1,0; 1,0}} + {\color{red} \chi^{m7 \times \overline{m7}}_{3,\frac{5}{7}; 3,-\frac{5}{7}}} +  \chi^{m7 \times \overline{m7}}_{5,\frac{22}{7}; 5,-\frac{22}{7}} 
 \nonumber \\ 
Z_{\mathbf{1}; \bar e}^{JK_4 \boxtimes \overline{JK}_4} &= \chi^{m7 \times \overline{m7}}_{2,\frac{1}{7}; 5,-\frac{22}{7}} +  \chi^{m7 \times \overline{m7}}_{4,\frac{12}{7}; 3,-\frac{5}{7}} +  \chi^{m7 \times \overline{m7}}_{6,5; 1,0} 
 \nonumber \\ 
Z_{\mathbf{1}; \bar m}^{JK_4 \boxtimes \overline{JK}_4} &= \chi^{m7 \times \overline{m7}}_{8,\frac{1}{56}; 5,-\frac{22}{7}} +  \chi^{m7 \times \overline{m7}}_{10,\frac{33}{56}; 3,-\frac{5}{7}} +  \chi^{m7 \times \overline{m7}}_{12,\frac{23}{8}; 1,0} 
 \nonumber \\ 
Z_{\mathbf{1}; \bar m_1}^{JK_4 \boxtimes \overline{JK}_4} &= \chi^{m7 \times \overline{m7}}_{7,\frac{3}{8}; 1,0} +  \chi^{m7 \times \overline{m7}}_{9,\frac{5}{56}; 3,-\frac{5}{7}} +  \chi^{m7 \times \overline{m7}}_{11,\frac{85}{56}; 5,-\frac{22}{7}} 
 \nonumber \\ 
Z_{\mathbf{1}; \bar q}^{JK_4 \boxtimes \overline{JK}_4} &= \chi^{m7 \times \overline{m7}}_{13,\frac{4}{3}; 1,0} +  \chi^{m7 \times \overline{m7}}_{14,\frac{10}{21}; 5,-\frac{22}{7}} +  \chi^{m7 \times \overline{m7}}_{15,\frac{1}{21}; 3,-\frac{5}{7}} 
 \nonumber \\ 
Z_{e; \mathbf{1}}^{JK_4 \boxtimes \overline{JK}_4} &= \chi^{m7 \times \overline{m7}}_{1,0; 6,-5} +  \chi^{m7 \times \overline{m7}}_{3,\frac{5}{7}; 4,-\frac{12}{7}} +  \chi^{m7 \times \overline{m7}}_{5,\frac{22}{7}; 2,-\frac{1}{7}} 
 \nonumber \\ 
Z_{e; \bar e}^{JK_4 \boxtimes \overline{JK}_4} &= \chi^{m7 \times \overline{m7}}_{2,\frac{1}{7}; 2,-\frac{1}{7}} +  \chi^{m7 \times \overline{m7}}_{4,\frac{12}{7}; 4,-\frac{12}{7}} +  \chi^{m7 \times \overline{m7}}_{6,5; 6,-5} 
 \nonumber \\ 
Z_{e; \bar m}^{JK_4 \boxtimes \overline{JK}_4} &= \chi^{m7 \times \overline{m7}}_{8,\frac{1}{56}; 2,-\frac{1}{7}} +  \chi^{m7 \times \overline{m7}}_{10,\frac{33}{56}; 4,-\frac{12}{7}} +  \chi^{m7 \times \overline{m7}}_{12,\frac{23}{8}; 6,-5} 
 \nonumber \\ 
Z_{e; \bar m_1}^{JK_4 \boxtimes \overline{JK}_4} &= \chi^{m7 \times \overline{m7}}_{7,\frac{3}{8}; 6,-5} +  \chi^{m7 \times \overline{m7}}_{9,\frac{5}{56}; 4,-\frac{12}{7}} +  \chi^{m7 \times \overline{m7}}_{11,\frac{85}{56}; 2,-\frac{1}{7}} 
 \nonumber \\ 
Z_{e; \bar q}^{JK_4 \boxtimes \overline{JK}_4} &= \chi^{m7 \times \overline{m7}}_{13,\frac{4}{3}; 6,-5} +  \chi^{m7 \times \overline{m7}}_{14,\frac{10}{21}; 2,-\frac{1}{7}} +  \chi^{m7 \times \overline{m7}}_{15,\frac{1}{21}; 4,-\frac{12}{7}} 
 \nonumber \\ 
Z_{m; \mathbf{1}}^{JK_4 \boxtimes \overline{JK}_4} &= \chi^{m7 \times \overline{m7}}_{1,0; 12,-\frac{23}{8}} +  \chi^{m7 \times \overline{m7}}_{3,\frac{5}{7}; 10,-\frac{33}{56}} +  \chi^{m7 \times \overline{m7}}_{5,\frac{22}{7}; 8,-\frac{1}{56}} 
 \nonumber \\ 
Z_{m; \bar e}^{JK_4 \boxtimes \overline{JK}_4} &= \chi^{m7 \times \overline{m7}}_{2,\frac{1}{7}; 8,-\frac{1}{56}} +  \chi^{m7 \times \overline{m7}}_{4,\frac{12}{7}; 10,-\frac{33}{56}} +  \chi^{m7 \times \overline{m7}}_{6,5; 12,-\frac{23}{8}} 
 \nonumber \\ 
Z_{m; \bar m}^{JK_4 \boxtimes \overline{JK}_4} &= \chi^{m7 \times \overline{m7}}_{8,\frac{1}{56}; 8,-\frac{1}{56}} +  \chi^{m7 \times \overline{m7}}_{10,\frac{33}{56}; 10,-\frac{33}{56}} +  \chi^{m7 \times \overline{m7}}_{12,\frac{23}{8}; 12,-\frac{23}{8}} 
 \nonumber \\ 
Z_{m; \bar m_1}^{JK_4 \boxtimes \overline{JK}_4} &= \chi^{m7 \times \overline{m7}}_{7,\frac{3}{8}; 12,-\frac{23}{8}} +  \chi^{m7 \times \overline{m7}}_{9,\frac{5}{56}; 10,-\frac{33}{56}} +  \chi^{m7 \times \overline{m7}}_{11,\frac{85}{56}; 8,-\frac{1}{56}} 
 \nonumber \\ 
Z_{m; \bar q}^{JK_4 \boxtimes \overline{JK}_4} &= \chi^{m7 \times \overline{m7}}_{13,\frac{4}{3}; 12,-\frac{23}{8}} +  \chi^{m7 \times \overline{m7}}_{14,\frac{10}{21}; 8,-\frac{1}{56}} +  \chi^{m7 \times \overline{m7}}_{15,\frac{1}{21}; 10,-\frac{33}{56}} 
 \nonumber \\ 
Z_{m_1; \mathbf{1}}^{JK_4 \boxtimes \overline{JK}_4} &= \chi^{m7 \times \overline{m7}}_{1,0; 7,-\frac{3}{8}} +  \chi^{m7 \times \overline{m7}}_{3,\frac{5}{7}; 9,-\frac{5}{56}} +  \chi^{m7 \times \overline{m7}}_{5,\frac{22}{7}; 11,-\frac{85}{56}} 
 \nonumber \\ 
Z_{m_1; \bar e}^{JK_4 \boxtimes \overline{JK}_4} &= \chi^{m7 \times \overline{m7}}_{2,\frac{1}{7}; 11,-\frac{85}{56}} +  \chi^{m7 \times \overline{m7}}_{4,\frac{12}{7}; 9,-\frac{5}{56}} +  \chi^{m7 \times \overline{m7}}_{6,5; 7,-\frac{3}{8}} 
 \nonumber \\ 
Z_{m_1; \bar m}^{JK_4 \boxtimes \overline{JK}_4} &= \chi^{m7 \times \overline{m7}}_{8,\frac{1}{56}; 11,-\frac{85}{56}} +  \chi^{m7 \times \overline{m7}}_{10,\frac{33}{56}; 9,-\frac{5}{56}} +  \chi^{m7 \times \overline{m7}}_{12,\frac{23}{8}; 7,-\frac{3}{8}} 
 \nonumber \\ 
Z_{m_1; \bar m_1}^{JK_4 \boxtimes \overline{JK}_4} &= \chi^{m7 \times \overline{m7}}_{7,\frac{3}{8}; 7,-\frac{3}{8}} +  \chi^{m7 \times \overline{m7}}_{9,\frac{5}{56}; 9,-\frac{5}{56}} +  \chi^{m7 \times \overline{m7}}_{11,\frac{85}{56}; 11,-\frac{85}{56}} 
 \nonumber \\ 
Z_{m_1; \bar q}^{JK_4 \boxtimes \overline{JK}_4} &= \chi^{m7 \times \overline{m7}}_{13,\frac{4}{3}; 7,-\frac{3}{8}} +  \chi^{m7 \times \overline{m7}}_{14,\frac{10}{21}; 11,-\frac{85}{56}} +  \chi^{m7 \times \overline{m7}}_{15,\frac{1}{21}; 9,-\frac{5}{56}} 
 \nonumber \\ 
Z_{q; \mathbf{1}}^{JK_4 \boxtimes \overline{JK}_4} &= \chi^{m7 \times \overline{m7}}_{1,0; 13,-\frac{4}{3}} +  \chi^{m7 \times \overline{m7}}_{3,\frac{5}{7}; 15,-\frac{1}{21}} +  \chi^{m7 \times \overline{m7}}_{5,\frac{22}{7}; 14,-\frac{10}{21}} 
 \nonumber \\ 
Z_{q; \bar e}^{JK_4 \boxtimes \overline{JK}_4} &= \chi^{m7 \times \overline{m7}}_{2,\frac{1}{7}; 14,-\frac{10}{21}} +  \chi^{m7 \times \overline{m7}}_{4,\frac{12}{7}; 15,-\frac{1}{21}} +  \chi^{m7 \times \overline{m7}}_{6,5; 13,-\frac{4}{3}} 
 \nonumber \\ 
Z_{q; \bar m}^{JK_4 \boxtimes \overline{JK}_4} &= \chi^{m7 \times \overline{m7}}_{8,\frac{1}{56}; 14,-\frac{10}{21}} +  \chi^{m7 \times \overline{m7}}_{10,\frac{33}{56}; 15,-\frac{1}{21}} +  \chi^{m7 \times \overline{m7}}_{12,\frac{23}{8}; 13,-\frac{4}{3}} 
 \nonumber \\ 
Z_{q; \bar m_1}^{JK_4 \boxtimes \overline{JK}_4} &= \chi^{m7 \times \overline{m7}}_{7,\frac{3}{8}; 13,-\frac{4}{3}} +  \chi^{m7 \times \overline{m7}}_{9,\frac{5}{56}; 15,-\frac{1}{21}} +  \chi^{m7 \times \overline{m7}}_{11,\frac{85}{56}; 14,-\frac{10}{21}} 
 \nonumber \\ 
Z_{q; \bar q}^{JK_4 \boxtimes \overline{JK}_4} &= \chi^{m7 \times \overline{m7}}_{13,\frac{4}{3}; 13,-\frac{4}{3}} +  \chi^{m7 \times \overline{m7}}_{14,\frac{10}{21}; 14,-\frac{10}{21}} +  \chi^{m7 \times \overline{m7}}_{15,\frac{1}{21}; 15,-\frac{1}{21}} 
 \nonumber \\ 
\cA &= \one    .
\end{align}

\begin{align}
\label{m7-1be}
 Z_{\mathbf{1}; \mathbf{1}}^{JK_4 \boxtimes \overline{JK}_4} &={\color{blue} \chi^{m7 \times \overline{m7}}_{1,0; 1,0}} +  \chi^{m7 \times \overline{m7}}_{2,\frac{1}{7}; 5,-\frac{22}{7}} + {\color{red} \chi^{m7 \times \overline{m7}}_{3,\frac{5}{7}; 3,-\frac{5}{7}}} +  \chi^{m7 \times \overline{m7}}_{4,\frac{12}{7}; 3,-\frac{5}{7}} +  \chi^{m7 \times \overline{m7}}_{5,\frac{22}{7}; 5,-\frac{22}{7}} +  \chi^{m7 \times \overline{m7}}_{6,5; 1,0} 
 \nonumber \\ 
Z_{\mathbf{1}; \bar e}^{JK_4 \boxtimes \overline{JK}_4} &={\color{blue} \chi^{m7 \times \overline{m7}}_{1,0; 1,0}} +  \chi^{m7 \times \overline{m7}}_{2,\frac{1}{7}; 5,-\frac{22}{7}} +  \chi^{m7 \times \overline{m7}}_{3,\frac{5}{7}; 3,-\frac{5}{7}} +  \chi^{m7 \times \overline{m7}}_{4,\frac{12}{7}; 3,-\frac{5}{7}} +  \chi^{m7 \times \overline{m7}}_{5,\frac{22}{7}; 5,-\frac{22}{7}} +  \chi^{m7 \times \overline{m7}}_{6,5; 1,0} 
 \nonumber \\ 
Z_{\mathbf{1}; \bar m}^{JK_4 \boxtimes \overline{JK}_4} &=0
 \nonumber \\ 
Z_{\mathbf{1}; \bar m_1}^{JK_4 \boxtimes \overline{JK}_4} &=0
 \nonumber \\ 
Z_{\mathbf{1}; \bar q}^{JK_4 \boxtimes \overline{JK}_4} &= 2\chi^{m7 \times \overline{m7}}_{13,\frac{4}{3}; 1,0} +  2\chi^{m7 \times \overline{m7}}_{14,\frac{10}{21}; 5,-\frac{22}{7}} +  2\chi^{m7 \times \overline{m7}}_{15,\frac{1}{21}; 3,-\frac{5}{7}} 
 \nonumber \\ 
Z_{e; \mathbf{1}}^{JK_4 \boxtimes \overline{JK}_4} &= \chi^{m7 \times \overline{m7}}_{1,0; 6,-5} +  \chi^{m7 \times \overline{m7}}_{2,\frac{1}{7}; 2,-\frac{1}{7}} +  \chi^{m7 \times \overline{m7}}_{3,\frac{5}{7}; 4,-\frac{12}{7}} +  \chi^{m7 \times \overline{m7}}_{4,\frac{12}{7}; 4,-\frac{12}{7}} +  \chi^{m7 \times \overline{m7}}_{5,\frac{22}{7}; 2,-\frac{1}{7}} +  \chi^{m7 \times \overline{m7}}_{6,5; 6,-5} 
 \nonumber \\ 
Z_{e; \bar e}^{JK_4 \boxtimes \overline{JK}_4} &= \chi^{m7 \times \overline{m7}}_{1,0; 6,-5} +  \chi^{m7 \times \overline{m7}}_{2,\frac{1}{7}; 2,-\frac{1}{7}} +  \chi^{m7 \times \overline{m7}}_{3,\frac{5}{7}; 4,-\frac{12}{7}} +  \chi^{m7 \times \overline{m7}}_{4,\frac{12}{7}; 4,-\frac{12}{7}} +  \chi^{m7 \times \overline{m7}}_{5,\frac{22}{7}; 2,-\frac{1}{7}} +  \chi^{m7 \times \overline{m7}}_{6,5; 6,-5} 
 \nonumber \\ 
Z_{e; \bar m}^{JK_4 \boxtimes \overline{JK}_4} &=0
 \nonumber \\ 
Z_{e; \bar m_1}^{JK_4 \boxtimes \overline{JK}_4} &=0
 \nonumber \\ 
Z_{e; \bar q}^{JK_4 \boxtimes \overline{JK}_4} &= 2\chi^{m7 \times \overline{m7}}_{13,\frac{4}{3}; 6,-5} +  2\chi^{m7 \times \overline{m7}}_{14,\frac{10}{21}; 2,-\frac{1}{7}} +  2\chi^{m7 \times \overline{m7}}_{15,\frac{1}{21}; 4,-\frac{12}{7}} 
 \nonumber \\ 
Z_{m; \mathbf{1}}^{JK_4 \boxtimes \overline{JK}_4} &= \chi^{m7 \times \overline{m7}}_{1,0; 12,-\frac{23}{8}} +  \chi^{m7 \times \overline{m7}}_{2,\frac{1}{7}; 8,-\frac{1}{56}} +  \chi^{m7 \times \overline{m7}}_{3,\frac{5}{7}; 10,-\frac{33}{56}} +  \chi^{m7 \times \overline{m7}}_{4,\frac{12}{7}; 10,-\frac{33}{56}} +  \chi^{m7 \times \overline{m7}}_{5,\frac{22}{7}; 8,-\frac{1}{56}} +  \chi^{m7 \times \overline{m7}}_{6,5; 12,-\frac{23}{8}} 
 \nonumber \\ 
Z_{m; \bar e}^{JK_4 \boxtimes \overline{JK}_4} &= \chi^{m7 \times \overline{m7}}_{1,0; 12,-\frac{23}{8}} +  \chi^{m7 \times \overline{m7}}_{2,\frac{1}{7}; 8,-\frac{1}{56}} +  \chi^{m7 \times \overline{m7}}_{3,\frac{5}{7}; 10,-\frac{33}{56}} +  \chi^{m7 \times \overline{m7}}_{4,\frac{12}{7}; 10,-\frac{33}{56}} +  \chi^{m7 \times \overline{m7}}_{5,\frac{22}{7}; 8,-\frac{1}{56}} +  \chi^{m7 \times \overline{m7}}_{6,5; 12,-\frac{23}{8}} 
 \nonumber \\ 
Z_{m; \bar m}^{JK_4 \boxtimes \overline{JK}_4} &=0
 \nonumber \\ 
Z_{m; \bar m_1}^{JK_4 \boxtimes \overline{JK}_4} &=0
 \nonumber \\ 
Z_{m; \bar q}^{JK_4 \boxtimes \overline{JK}_4} &= 2\chi^{m7 \times \overline{m7}}_{13,\frac{4}{3}; 12,-\frac{23}{8}} +  2\chi^{m7 \times \overline{m7}}_{14,\frac{10}{21}; 8,-\frac{1}{56}} +  2\chi^{m7 \times \overline{m7}}_{15,\frac{1}{21}; 10,-\frac{33}{56}} 
 \nonumber \\ 
Z_{m_1; \mathbf{1}}^{JK_4 \boxtimes \overline{JK}_4} &= \chi^{m7 \times \overline{m7}}_{1,0; 7,-\frac{3}{8}} +  \chi^{m7 \times \overline{m7}}_{2,\frac{1}{7}; 11,-\frac{85}{56}} +  \chi^{m7 \times \overline{m7}}_{3,\frac{5}{7}; 9,-\frac{5}{56}} +  \chi^{m7 \times \overline{m7}}_{4,\frac{12}{7}; 9,-\frac{5}{56}} +  \chi^{m7 \times \overline{m7}}_{5,\frac{22}{7}; 11,-\frac{85}{56}} +  \chi^{m7 \times \overline{m7}}_{6,5; 7,-\frac{3}{8}} 
 \nonumber \\ 
Z_{m_1; \bar e}^{JK_4 \boxtimes \overline{JK}_4} &= \chi^{m7 \times \overline{m7}}_{1,0; 7,-\frac{3}{8}} +  \chi^{m7 \times \overline{m7}}_{2,\frac{1}{7}; 11,-\frac{85}{56}} +  \chi^{m7 \times \overline{m7}}_{3,\frac{5}{7}; 9,-\frac{5}{56}} +  \chi^{m7 \times \overline{m7}}_{4,\frac{12}{7}; 9,-\frac{5}{56}} +  \chi^{m7 \times \overline{m7}}_{5,\frac{22}{7}; 11,-\frac{85}{56}} +  \chi^{m7 \times \overline{m7}}_{6,5; 7,-\frac{3}{8}} 
 \nonumber \\ 
Z_{m_1; \bar m}^{JK_4 \boxtimes \overline{JK}_4} &=0
 \nonumber \\ 
Z_{m_1; \bar m_1}^{JK_4 \boxtimes \overline{JK}_4} &=0
 \nonumber \\ 
Z_{m_1; \bar q}^{JK_4 \boxtimes \overline{JK}_4} &= 2\chi^{m7 \times \overline{m7}}_{13,\frac{4}{3}; 7,-\frac{3}{8}} +  2\chi^{m7 \times \overline{m7}}_{14,\frac{10}{21}; 11,-\frac{85}{56}} +  2\chi^{m7 \times \overline{m7}}_{15,\frac{1}{21}; 9,-\frac{5}{56}} 
 \nonumber \\ 
Z_{q; \mathbf{1}}^{JK_4 \boxtimes \overline{JK}_4} &= \chi^{m7 \times \overline{m7}}_{1,0; 13,-\frac{4}{3}} +  \chi^{m7 \times \overline{m7}}_{2,\frac{1}{7}; 14,-\frac{10}{21}} +  \chi^{m7 \times \overline{m7}}_{3,\frac{5}{7}; 15,-\frac{1}{21}} +  \chi^{m7 \times \overline{m7}}_{4,\frac{12}{7}; 15,-\frac{1}{21}} +  \chi^{m7 \times \overline{m7}}_{5,\frac{22}{7}; 14,-\frac{10}{21}} +  \chi^{m7 \times \overline{m7}}_{6,5; 13,-\frac{4}{3}} 
 \nonumber \\ 
Z_{q; \bar e}^{JK_4 \boxtimes \overline{JK}_4} &= \chi^{m7 \times \overline{m7}}_{1,0; 13,-\frac{4}{3}} +  \chi^{m7 \times \overline{m7}}_{2,\frac{1}{7}; 14,-\frac{10}{21}} +  \chi^{m7 \times \overline{m7}}_{3,\frac{5}{7}; 15,-\frac{1}{21}} +  \chi^{m7 \times \overline{m7}}_{4,\frac{12}{7}; 15,-\frac{1}{21}} +  \chi^{m7 \times \overline{m7}}_{5,\frac{22}{7}; 14,-\frac{10}{21}} +  \chi^{m7 \times \overline{m7}}_{6,5; 13,-\frac{4}{3}} 
 \nonumber \\ 
Z_{q; \bar m}^{JK_4 \boxtimes \overline{JK}_4} &=0
 \nonumber \\ 
Z_{q; \bar m_1}^{JK_4 \boxtimes \overline{JK}_4} &=0
 \nonumber \\ 
Z_{q; \bar q}^{JK_4 \boxtimes \overline{JK}_4} &= 2\chi^{m7 \times \overline{m7}}_{13,\frac{4}{3}; 13,-\frac{4}{3}} +  2\chi^{m7 \times \overline{m7}}_{14,\frac{10}{21}; 14,-\frac{10}{21}} +  2\chi^{m7 \times \overline{m7}}_{15,\frac{1}{21}; 15,-\frac{1}{21}} 
 \nonumber \\ 
\cA &= \one \oplus \bar e    .
\end{align}

\begin{align}
\label{m7-1e}
 Z_{\mathbf{1}; \mathbf{1}}^{JK_4 \boxtimes \overline{JK}_4} &={\color{blue} \chi^{m7 \times \overline{m7}}_{1,0; 1,0}} +  \chi^{m7 \times \overline{m7}}_{1,0; 6,-5} + {\color{red} \chi^{m7 \times \overline{m7}}_{3,\frac{5}{7}; 3,-\frac{5}{7}}} +  \chi^{m7 \times \overline{m7}}_{3,\frac{5}{7}; 4,-\frac{12}{7}} +  \chi^{m7 \times \overline{m7}}_{5,\frac{22}{7}; 2,-\frac{1}{7}} +  \chi^{m7 \times \overline{m7}}_{5,\frac{22}{7}; 5,-\frac{22}{7}} 
 \nonumber \\ 
Z_{\mathbf{1}; \bar e}^{JK_4 \boxtimes \overline{JK}_4} &= \chi^{m7 \times \overline{m7}}_{2,\frac{1}{7}; 2,-\frac{1}{7}} +  \chi^{m7 \times \overline{m7}}_{2,\frac{1}{7}; 5,-\frac{22}{7}} +  \chi^{m7 \times \overline{m7}}_{4,\frac{12}{7}; 3,-\frac{5}{7}} +  \chi^{m7 \times \overline{m7}}_{4,\frac{12}{7}; 4,-\frac{12}{7}} +  \chi^{m7 \times \overline{m7}}_{6,5; 1,0} +  \chi^{m7 \times \overline{m7}}_{6,5; 6,-5} 
 \nonumber \\ 
Z_{\mathbf{1}; \bar m}^{JK_4 \boxtimes \overline{JK}_4} &= \chi^{m7 \times \overline{m7}}_{8,\frac{1}{56}; 2,-\frac{1}{7}} +  \chi^{m7 \times \overline{m7}}_{8,\frac{1}{56}; 5,-\frac{22}{7}} +  \chi^{m7 \times \overline{m7}}_{10,\frac{33}{56}; 3,-\frac{5}{7}} +  \chi^{m7 \times \overline{m7}}_{10,\frac{33}{56}; 4,-\frac{12}{7}} +  \chi^{m7 \times \overline{m7}}_{12,\frac{23}{8}; 1,0} +  \chi^{m7 \times \overline{m7}}_{12,\frac{23}{8}; 6,-5} 
 \nonumber \\ 
Z_{\mathbf{1}; \bar m_1}^{JK_4 \boxtimes \overline{JK}_4} &= \chi^{m7 \times \overline{m7}}_{7,\frac{3}{8}; 1,0} +  \chi^{m7 \times \overline{m7}}_{7,\frac{3}{8}; 6,-5} +  \chi^{m7 \times \overline{m7}}_{9,\frac{5}{56}; 3,-\frac{5}{7}} +  \chi^{m7 \times \overline{m7}}_{9,\frac{5}{56}; 4,-\frac{12}{7}} +  \chi^{m7 \times \overline{m7}}_{11,\frac{85}{56}; 2,-\frac{1}{7}} +  \chi^{m7 \times \overline{m7}}_{11,\frac{85}{56}; 5,-\frac{22}{7}} 
 \nonumber \\ 
Z_{\mathbf{1}; \bar q}^{JK_4 \boxtimes \overline{JK}_4} &= \chi^{m7 \times \overline{m7}}_{13,\frac{4}{3}; 1,0} +  \chi^{m7 \times \overline{m7}}_{13,\frac{4}{3}; 6,-5} +  \chi^{m7 \times \overline{m7}}_{14,\frac{10}{21}; 2,-\frac{1}{7}} +  \chi^{m7 \times \overline{m7}}_{14,\frac{10}{21}; 5,-\frac{22}{7}} +  \chi^{m7 \times \overline{m7}}_{15,\frac{1}{21}; 3,-\frac{5}{7}} +  \chi^{m7 \times \overline{m7}}_{15,\frac{1}{21}; 4,-\frac{12}{7}} 
 \nonumber \\ 
Z_{e; \mathbf{1}}^{JK_4 \boxtimes \overline{JK}_4} &={\color{blue} \chi^{m7 \times \overline{m7}}_{1,0; 1,0}} +  \chi^{m7 \times \overline{m7}}_{1,0; 6,-5} +  \chi^{m7 \times \overline{m7}}_{3,\frac{5}{7}; 3,-\frac{5}{7}} +  \chi^{m7 \times \overline{m7}}_{3,\frac{5}{7}; 4,-\frac{12}{7}} +  \chi^{m7 \times \overline{m7}}_{5,\frac{22}{7}; 2,-\frac{1}{7}} +  \chi^{m7 \times \overline{m7}}_{5,\frac{22}{7}; 5,-\frac{22}{7}} 
 \nonumber \\ 
Z_{e; \bar e}^{JK_4 \boxtimes \overline{JK}_4} &= \chi^{m7 \times \overline{m7}}_{2,\frac{1}{7}; 2,-\frac{1}{7}} +  \chi^{m7 \times \overline{m7}}_{2,\frac{1}{7}; 5,-\frac{22}{7}} +  \chi^{m7 \times \overline{m7}}_{4,\frac{12}{7}; 3,-\frac{5}{7}} +  \chi^{m7 \times \overline{m7}}_{4,\frac{12}{7}; 4,-\frac{12}{7}} +  \chi^{m7 \times \overline{m7}}_{6,5; 1,0} +  \chi^{m7 \times \overline{m7}}_{6,5; 6,-5} 
 \nonumber \\ 
Z_{e; \bar m}^{JK_4 \boxtimes \overline{JK}_4} &= \chi^{m7 \times \overline{m7}}_{8,\frac{1}{56}; 2,-\frac{1}{7}} +  \chi^{m7 \times \overline{m7}}_{8,\frac{1}{56}; 5,-\frac{22}{7}} +  \chi^{m7 \times \overline{m7}}_{10,\frac{33}{56}; 3,-\frac{5}{7}} +  \chi^{m7 \times \overline{m7}}_{10,\frac{33}{56}; 4,-\frac{12}{7}} +  \chi^{m7 \times \overline{m7}}_{12,\frac{23}{8}; 1,0} +  \chi^{m7 \times \overline{m7}}_{12,\frac{23}{8}; 6,-5} 
 \nonumber \\ 
Z_{e; \bar m_1}^{JK_4 \boxtimes \overline{JK}_4} &= \chi^{m7 \times \overline{m7}}_{7,\frac{3}{8}; 1,0} +  \chi^{m7 \times \overline{m7}}_{7,\frac{3}{8}; 6,-5} +  \chi^{m7 \times \overline{m7}}_{9,\frac{5}{56}; 3,-\frac{5}{7}} +  \chi^{m7 \times \overline{m7}}_{9,\frac{5}{56}; 4,-\frac{12}{7}} +  \chi^{m7 \times \overline{m7}}_{11,\frac{85}{56}; 2,-\frac{1}{7}} +  \chi^{m7 \times \overline{m7}}_{11,\frac{85}{56}; 5,-\frac{22}{7}} 
 \nonumber \\ 
Z_{e; \bar q}^{JK_4 \boxtimes \overline{JK}_4} &= \chi^{m7 \times \overline{m7}}_{13,\frac{4}{3}; 1,0} +  \chi^{m7 \times \overline{m7}}_{13,\frac{4}{3}; 6,-5} +  \chi^{m7 \times \overline{m7}}_{14,\frac{10}{21}; 2,-\frac{1}{7}} +  \chi^{m7 \times \overline{m7}}_{14,\frac{10}{21}; 5,-\frac{22}{7}} +  \chi^{m7 \times \overline{m7}}_{15,\frac{1}{21}; 3,-\frac{5}{7}} +  \chi^{m7 \times \overline{m7}}_{15,\frac{1}{21}; 4,-\frac{12}{7}} 
 \nonumber \\ 
Z_{m; \mathbf{1}}^{JK_4 \boxtimes \overline{JK}_4} &=0
 \nonumber \\ 
Z_{m; \bar e}^{JK_4 \boxtimes \overline{JK}_4} &=0
 \nonumber \\ 
Z_{m; \bar m}^{JK_4 \boxtimes \overline{JK}_4} &=0
 \nonumber \\ 
Z_{m; \bar m_1}^{JK_4 \boxtimes \overline{JK}_4} &=0
 \nonumber \\ 
Z_{m; \bar q}^{JK_4 \boxtimes \overline{JK}_4} &=0
 \nonumber \\ 
Z_{m_1; \mathbf{1}}^{JK_4 \boxtimes \overline{JK}_4} &=0
 \nonumber \\ 
Z_{m_1; \bar e}^{JK_4 \boxtimes \overline{JK}_4} &=0
 \nonumber \\ 
Z_{m_1; \bar m}^{JK_4 \boxtimes \overline{JK}_4} &=0
 \nonumber \\ 
Z_{m_1; \bar m_1}^{JK_4 \boxtimes \overline{JK}_4} &=0
 \nonumber \\ 
Z_{m_1; \bar q}^{JK_4 \boxtimes \overline{JK}_4} &=0
 \nonumber \\ 
Z_{q; \mathbf{1}}^{JK_4 \boxtimes \overline{JK}_4} &= 2\chi^{m7 \times \overline{m7}}_{1,0; 13,-\frac{4}{3}} +  2\chi^{m7 \times \overline{m7}}_{3,\frac{5}{7}; 15,-\frac{1}{21}} +  2\chi^{m7 \times \overline{m7}}_{5,\frac{22}{7}; 14,-\frac{10}{21}} 
 \nonumber \\ 
Z_{q; \bar e}^{JK_4 \boxtimes \overline{JK}_4} &= 2\chi^{m7 \times \overline{m7}}_{2,\frac{1}{7}; 14,-\frac{10}{21}} +  2\chi^{m7 \times \overline{m7}}_{4,\frac{12}{7}; 15,-\frac{1}{21}} +  2\chi^{m7 \times \overline{m7}}_{6,5; 13,-\frac{4}{3}} 
 \nonumber \\ 
Z_{q; \bar m}^{JK_4 \boxtimes \overline{JK}_4} &= 2\chi^{m7 \times \overline{m7}}_{8,\frac{1}{56}; 14,-\frac{10}{21}} +  2\chi^{m7 \times \overline{m7}}_{10,\frac{33}{56}; 15,-\frac{1}{21}} +  2\chi^{m7 \times \overline{m7}}_{12,\frac{23}{8}; 13,-\frac{4}{3}} 
 \nonumber \\ 
Z_{q; \bar m_1}^{JK_4 \boxtimes \overline{JK}_4} &= 2\chi^{m7 \times \overline{m7}}_{7,\frac{3}{8}; 13,-\frac{4}{3}} +  2\chi^{m7 \times \overline{m7}}_{9,\frac{5}{56}; 15,-\frac{1}{21}} +  2\chi^{m7 \times \overline{m7}}_{11,\frac{85}{56}; 14,-\frac{10}{21}} 
 \nonumber \\ 
Z_{q; \bar q}^{JK_4 \boxtimes \overline{JK}_4} &= 2\chi^{m7 \times \overline{m7}}_{13,\frac{4}{3}; 13,-\frac{4}{3}} +  2\chi^{m7 \times \overline{m7}}_{14,\frac{10}{21}; 14,-\frac{10}{21}} +  2\chi^{m7 \times \overline{m7}}_{15,\frac{1}{21}; 15,-\frac{1}{21}} 
 \nonumber \\ 
\cA &= \one \oplus  e    .
\end{align}



Self-dual $S_3$ symmetric model also has the gapless states described by
$m5\times\overline{m5}$ CFT.  But those gapless states have two or more
symmetric relevant operators.

\endgroup

\end{appendix}

%\bibliography{./local.bib,../../../bib/all,../../../bib/allnew,../../../bib/publst,../../../bib/wencross}

{\small
\bibliography{references}
}

\nolinenumbers

\end{document}